\documentclass{article}
\usepackage[utf8]{inputenc}
\usepackage{pgfplots}
\pgfplotsset{compat=newest}
\usepgfplotslibrary{fillbetween}
\usetikzlibrary{arrows.meta}
\usepackage[english]{babel}
\usepackage{amsmath,amsthm}
\usepackage{graphicx}
\usepackage{amssymb}
\usepackage{enumerate}
\usepackage{mathtools}
\usepackage{natbib}
\usepackage[T1]{fontenc}
\usepackage[multiple]{footmisc}
\usepackage{lmodern}
\usepackage{textgreek}
\usepackage[autostyle]{csquotes}
\usepackage{setspace}
\usepackage{appendix}
\usepackage{tikz-cd}
\usepackage[normalem]{ulem}
\usepackage[breaklinks]{hyperref}
\theoremstyle{definition}
\newtheorem{definition}{Definition}
\theoremstyle{plain}
\newtheorem{theorem}{Theorem}

\newtheorem{corollary}{Corollary}
\newtheorem{assumption}{Assumption}
\newtheorem{remark}{Remark}

\newtheorem{proposition}{Proposition}

\newtheorem{error}{Error}
\newtheorem{decomposition}{Decomposition}
\newtheorem{principle}{Principle}

\DeclareMathAlphabet{\mathpzc}{OT1}{pzc}{m}{it}
\DeclareMathOperator*{\plim}{plim}
\title{Stochastic Equilibrium the Lucas Critique and Keynesian Economics}
\author{David Staines}
\date{\today}
\title{Stochastic Equilibrium the Lucas Critique and Keynesian Economics}
\author{David Staines}
\date{\today}
\begin{document}
\maketitle
\begin{abstract}
In this paper, a mathematically rigorous solution overturns existing wisdom regarding New Keynesian Dynamic Stochastic General Equilibrium. I develop a formal concept of stochastic equilibrium. I prove uniqueness and necessity, when agents are patient, with general application. Existence depends on appropriately specified eigenvalue conditions. Otherwise, no solution of any kind exists. I construct the equilibrium with Calvo pricing. I provide novel comparative statics with the non-stochastic model of analytical significance. I uncover a bifurcation between neighbouring stochastic systems and approximations taken from the
Zero Inflation Non-Stochastic Steady State (ZINSS). The correct Phillips curve agrees with the zero limit from the trend inflation framework. It contains a large lagged inflation coefficient and a small response to expected inflation. Price dispersion can be first or second order depending how shocks are scaled. The response to the output gap is always muted and is zero at standard parameters. A neutrality result is presented to explain why and align Calvo with Taylor pricing. Present and lagged demand shocks enter the Phillips curve so there is no Divine Coincidence and the system is identified from structural shocks alone.
The lagged inflation slope is increasing in the inflation response, embodying substantive policy trade-offs. The Taylor principle is reversed, inactive settings are necessary, pointing towards inertial policy. 
The observational equivalence
idea of the Lucas critique is disproven. The bifurcation results from the breakdown of the constraints implied by lagged nominal rigidity, associated with
cross-equation cancellation possible only at ZINSS. There is a dual relationship between restrictions on the econometrician and constraints on 
repricing firms. Thus, if the model is correct, goodness of fit will jump.
\end{abstract}
\newpage
\tableofcontents
\section{Introduction}
\par Empirical understanding in macroeconomics has made great strides over recent years. A growing body of model-free evidence shows that price and wage rigidity are ubiquitous. This is borne out by many microeconometric studies of price adjustment such as \cite{alvarez2006sticky}, \cite{dhyne2006price}, \cite{gagnon2009price}, \cite{klenow2010microeconomic}, \cite{vermeulen2012price}, \cite{berardi2015more} and \cite{kehoe2015prices},\footnote{This is also true of online prices, (see \cite{cavallo2016billion}, \cite{cavallo2017online}, \cite{gorodnichenko2017price}, \cite{gorodnichenko2018price} and \cite{cavallo2018scraped}).} whilst individual wages adjust slowly with particular resistance to nominal reductions, as found for example by \cite{fehr2005robustness}, \cite{dickens2007wages}, \cite{barattieri2014some}, \cite{kaur2019nominal} and \cite{grigsby2021aggregate}.
A variety of quasi-natural experiments reveal that aggregate demand shocks are quantitatively important business cycle drivers (see \newline \cite{auerbach2012measuring},
\cite{auerbach2012fiscal}, \cite{auerbach2013output}, \cite{bils2013testing}, \cite{acconcia2014mafia}, \cite{mian2014explains}, \cite{chodorow2016cyclicality} and \cite{chodorow2019macro}). Monetary policy shocks can have large and possibly long-lasting impacts on output- (see \cite{christiano1999monetary}, \cite{romer2004new}, \cite{velde2009chronicle}, \cite{gertler2015monetary}, \cite{jorda2020effects}, \cite{jorda2020long},
and \cite{palma2022real}). 
\par This work has bolstered the \emph{New Neo-Classical Synthesis} discussed in \cite{goodfriend1997new} and \cite{snowdon2005modern}. The idea is to analyze monetary policy by adding sticky prices or frictional adjustment into otherwise standard optimizing models from the Real Business Cycle (RBC) tradition. The goal has been to generate credible micro-founded complements to the aggregate demand and aggregate supply equations of \emph{Old Keynesian} economics.
\par In this goal the literature has so far failed. Inflation, surely the key variable in New Keynesian economics, is excessively forward-looking, in fact under the current benchmark model it has no intrinsic persistence and is not identified without entirely ad hoc shocks. 
Moreover, there appears no short run trade-off between inflation and employment or output stabilization christened the \emph{Divine Coincidence} by \cite{blanchard2007real} so central banks can "fine-tune" away any inefficient fluctuations.\footnote{The term is frequently attributed to Walter Heller, Chief Economic Adviser to President Kennedy (see for example http://connection.ebscohost.com/c/reference-entries/40422478/fine-tuning-1960s-economics). It referred originally to fiscal policy in an "Old Keynesian" setup. Scepticism about the concept was focal to monetarist opposition to traditional Keynesian macroeconomics, see for example \cite{friedman1968role} and \cite{snowdon2005modern}.} I show that all these findings are a figment of erroneous solutions to 
the benchmark Calvo model. The task of this paper is to construct the simplest possible correct formulation of New Keynesian economics that can serve as a platform for future theoretical, empirical and mathematical work. 
\par The paper has three complementary tasks. 
The first is to uncover the dynamic properties of the New Keynesian Phillips curve. The second is to sketch its major econometric implications and theoretical rationale. Lastly, I will analyze whether and under what conditions a solution to a DSGE exists and provide precise parametric answers for the benchmark model. 
\par 
The benchmark New Keynesian Phillips obtained by linearizing the Calvo model \emph{at} ZINSS does not represent the dynamical properties of the underlying stochastic system however small shocks are. Except for price dispersion dynamics at ZINSS are forward-looking to all orders of approximation. By contrast, the underlying non-linear system is persistent with probability one, featuring lags of inflation and shock terms. It is already known that a hybrid system pertains when there is non-zero trend inflation- (see \cite{damjanovic2010relative}, \cite{coibion2011monetary}, \cite{ascari2014macroeconomics} and \cite{kurozumi2017trend} and \cite{qureshi2021cost}). This can be formalized as the \emph{Non-Stochastic Bifurcation} by 
considering limits of the linear approximation as inflation goes to zero. I demonstrate \emph{Stochastic Bifurcation} by showing that when inflation is precisely zero neighboring stochastic systems have equivalent dynamic properties, so trend inflation describes behavior \emph{around} ZINSS. 
\par The correct solution requires backward substitution steps, which is where the lag terms appear. The difference between the two systems is called a wall-crossing singularity. Formally, the boundary between the two solutions of the benchmark model 
is a two dimensional surface  
with a one dimensional wall leading into a three dimensional hole. Informally, it is formed by equating the differing steps between the two solutions and integrating. The wall is the inflation equality that is necessary to place the economy inside the singular surface. The 
second component is an equality 
connecting inflation, lagged and present marginal costs, reflecting a backward substitution step missing at ZINSS. It reflects the inter-temporal substitution motive present under staggered optimization and thus a change in the monetary transmission mechanism. It introduces a cost-push channel into the Phillips curve, so the Central Bank or financial conditions directly influence the pricing decisions of firms. The third dimension of the (inner) hole represents the cancellation of the demand shock with its lag.
\par These changes seem to fit a wide variety of existing evidence. \cite{barth2001cost}, \cite{gaiotti2006there} and \cite{chowdhury2006inflation} provide impressive support for a cost channel of monetary policy. These forces seem to be particularly strong for firms under financial distress according to \cite{antoun2015firms},  \cite{gilchrist2017inflation}, \cite{meinen2018sign}, \cite{palmen2020inflation}, \cite{abbate2023financial} and \cite{montero2021markup}.\footnote{On the inter-temporal substitution front there is substantial support for an inter-temporal aggregate demand equation in both subjective beliefs and revealed preferences (see \cite{coibion2023forward}, \cite{drager2021consumers}, \cite{duca2021inflation}), as well as discussion of previous microeconometric evidence in Appendix H.1.} 
The impact on the coefficients of the Phillips curve are dramatic. The responsiveness of inflation to the output gap drops, typically 
from close to one to around zero.
The lagged inflation coefficient is always larger than that of expected inflation. These results accord with recent empirical estimates including 
\cite{fuhrer2006intrinsic}, \cite{mavroeidis2014empirical}, \cite{ball2020nonpuzzling}, \cite{hindrayanto2019phillips}, \cite{bobeica2019missing}, 
\cite{hooper2020prospects}, \cite{zobl2021condemned} and \cite{ball2021phillips}.\footnote{None of the specifications are directly comparable to the theoretical model, which underlines the importance of structural estimation, although, \cite{fuhrer2006intrinsic} is the closest. Many use unemployment rather than the output gap as a forcing variable. This has an august pedigree going back to \cite{phillips1958relation} and \cite{phelps1968money}. It is popular since there is better data. The two can be connected via Okun's law- the business cycle relationship between unemployment and output deviations, which appears empirically strong (\cite{ball2013okun}). Nevertheless, it remains a priority to derive and test micro-founded wage Phillips curves.} \footnote{Many of these studies are able to demonstrate stable relationships over recent times, despite structural change and the effects of the recent financial crisis (see also \cite{stock2020slack} and \cite{candia2021inflation}). Substantial inflation persistence is robust to different levels of aggregation, policy regime and plausible assumptions about trends in other macroeconomic variables, for evidence consult for example \cite{clark2006disaggregate}, \cite{altissimo2009can}, \cite{vaona2012regional}; \cite{o2005has}, \cite{beechey2012rise}, \cite{gerlach2012inflation}; \cite{cogley2002evolving},  \cite{stock2016core} and \cite{kejriwal2020robust}.}
\par All the analysis is underpinned by a rigorous theory of stochastic equilibrium derived from ergodic theory, the branch of mathematics concerned with the long-term behavior of dynamical systems. A stochastic equilibrium is the state of the economy today that implies the probability of any future event is equal to its long-run (time) average. Crucially, I am able to construct this equilibrium explicitly. 
\par  A large literature exists in economics and related disciplines that uses the theory of ergodic processes to prove existence and often uniqueness also, of general equilibrium (see for example \cite{stokey1989recursive}, \cite{hopenhayn1992entry}, 
\cite{hopenhayn1992stochastic}, \cite{stachurski2002stochastic},
 \cite{li2014solving}, 
\cite{kamihigashi2016seeking}, \cite{brumm2017recursive}, 
\cite{accikgoz2018existence}, \cite{borovivcka2020necessary},  \cite{marinacci2019unique}, \cite{kirkby2019bewley}, \cite{hu2019unique}, \cite{light2022mean} and \cite{pohl2023existence}). All of these results are confined to small 
models lacking features, such as endogenous capital and labor accumulation or nominal rigidity, 
of primary interest to applied macroeconomists.\footnote{Results of most applied interest have 
concerned the \cite{krusell1998income} benchmark model of capital accumulation under incomplete markets and aggregate risk, such as \cite{cao2020recursive} and \cite{prohl2023existence}. However, these contain a degree of contingency (and are therefore not full existence proofs).}
This paper is the first to precisely define the long-run equilibrium conditions for a nonlinear stochastic model with no closed form solution, in variables of immediate economic interest. None have addressed criticality- the possibility that if a model does not meet certain conditions it does not have any solution. 
\par I supply wide-ranging comparative statics under weak assumptions. Unlike previous results, that rely on global restrictions, mine leverage the stochastic equilibrium which allows me to deduce global characteristics from local properties 
of the steady state.\footnote{A literature on monotone comparative statics has developed around \cite{milgrom1994comparing}, \cite{milgrom1994monotone}, \cite{milgrom2002envelope} and \cite{athey2002monotone}. More recently, convexity conditions have been popular following \cite{acemoglu2013aggregate} and \cite{jensen2018distributional}. In fact, once the stochastic equilibrium has been derived the subsequent steps are often considerably easier than these results or recent extensions, with the promises of applications to a wider set of models.} Moreover, I am able to undertake novel experiments, comparing any stochastic steady state with its non-stochastic counterpart. Its mathematical significance is addressed in the next section. More progress in this area may come as a byproduct of stronger ties with mathematics.
\par Macroeconomists have long been aware that approximations taken at non-stochastic steady states might not be accurate or dynamically representative, in particular when considering financial markets and risk premia. Since \cite{coeurdacier2011risky} and \cite{juillard2011local} it has been common to include the effect of higher order deviation terms when calculating equilibrium from which perturbations are analyzed.
 The main focus has naturally been on financial markets and the movements of risk premia in particular. \cite{ascari2018welfare} is the most notable New Keynesian thus far. The concept of stochastic equilibrium here formalizes and clarifies these ideas. 
\par Stochastic equilibrium constitutes a full probabilistic description of an economy. This gives them an intrinsic uniqueness property for systems driven by continuous shocks. This extends to the the probabilistic future path, which corresponds with the definition of recursive equilibrium in previous New Keynesian economics and the infinite time solution of mean-field game with common noise in mathematics. I also prove an equivalent finite dimensional state space form exists for benchmark New Keynesian models, comparable to  previous definitions from classical economics (see \cite{prescott1980recursive} and \cite{mehra2006recursive}). This technique is crucial to the analysis here and is bound to have wide application. 
\par This striking result comes however with a powerful converse. Models that macroeconomists previously thought had multiple equilibria in fact have none. In particular for a standard class of DSGE models, when the limiting value of a linear approximation around its stochastic equilibrium is indeterminate, what is actually happening is that the expectations of one or more of the forward-looking variables in the underlying non-linear model is exploding. This is because the underlying  optimization will blow up and welfare will collapse whenever the model is expected to reach one of the boundaries. 
This point is more obvious when there are too many eigenvalues outside the unit circle and a state variable drives the blow-up. In fact, this is the only barrier to a solution of the Calvo model around ZINSS, overturning current wisdom. In these situations we should think of the DSGE model as being misspecified, in the same way that econometricians see a correlation between non-stationary variables as
statistically spurious.\footnote{Consult \cite{kennedy2003guide} for insightful discussion, and \cite{hamilton1995time} for technical exposition. \cite{granger1974spurious} and \cite{phillips1986understanding} are prominent original papers.} 
\par
The result reinterprets the well-known \cite{blanchard1980solution} eigenvalue conditions. The requirement for existence of a solution to the non-linear model is the uniqueness condition evaluated around the stochastic steady state. My approach generalizes existing linearization techniques. The non-stochastic linear approximation is the limit of the stochastic approximation as shocks become arbitrarily small. Therefore existing linearization techniques are generically (away from bifurcation points) correct, confirming the existing intuition from a different asymptotic experiment. Away from the small noise limit, the stochastic linear approximation features stochastic coefficients. 
This is because the derivative at the stochastic steady state features expectations of non-linear functions. This creates 
novel technical challenges, which I discuss. Simplicity and comparability with past work motivate focus on small noise limits, in the quantitative part.
\par Contrary to current opinion, 
Rotemberg and Calvo are never equivalent in stochastic equilibrium. In fact, there are no settings for the standard policy rule where both exist. This is because Rotemberg is unaffected by the singularity impacting Calvo. This equivalence was the primary motivation for using Rotemberg pricing, although, it does mean that results derived under the existing benchmark could be viewed as pertaining to the Rotemberg model. This, rather than the effect of high order terms, is the likely explanation for differences in performance under global solution methods (see \cite{leith2016inflation}).  
\par A double limit system emerges depending on whether price dispersion is regarded as first or second order around ZINSS. This situation is called \emph{polydromy}. First order price dispersion $(\Delta)$ reflects a volatile policy regime. The \emph{non-volatile regime} 
emerges when average shocks are very small. More broadly, $\Delta$ can be viewed as real rigidity- in the sense of \cite{ball1990real}- the effect of price rigidity on the flexible economy. This suggests that nominal rigidity alone might be sufficient for modelling inflation dynamics, overturning a current of thought starting with \cite{ball1991sticky}.
\par If we are prepared to focus on the neighborhood of ZINSS then an advantage of my approach over trend inflation is that I can downplay or remove entirely the role of price dispersion. At empirically relevant values, price dispersion is increasing and convex in inflation. Stochastic equilibrium worsens the problem,
consistent with the empirical estimates from DSGE models (see \cite{ascari2018welfare}). However, microeconomic evidence typically suggests a much more muted relationship between price dispersion and inflation at low positive levels (see \cite{gagnon2009price}, \cite{coibion2015cyclicality}, \cite{wulfsberg2016inflation},  
\cite{nakamura2018elusive}, \cite{alvarez2018hyperinflation}, \cite{sheremirov2020price},
\cite{anayi2022firming} and \cite{adam2023inflation}). Broadly, consistent with these results, if price dispersion arises in the linear approximation around ZINSS it is independent of the first order dynamics of inflation. Furthermore, the approximation will be less forward-looking than its positive trend inflation counterpart (\cite{ascari2014macroeconomics}). 
A small literature has developed that looks at extensions of the benchmark model that give less extreme predictions concerning nominal dispersion (see  \cite{bakhshi2007new}, \cite{kurozumi2016endogenous}, \cite{kurozumi2016kinked}  and \cite{hahn2022price}).
Together they may justify considering an inflation rate closer to zero than for example the common Central Bank inflation target of two percent, without resorting to counterfactual considerations, such as full nominal price indexation.\footnote{It is in fact possible to remove the impact of price dispersion on the dynamics of marginal costs by assuming that labor is firm-specific (see \cite{coibion2008monetary} and \cite{eggertsson2019log}). I find this result useful for theoretical expositions, although not fully convincing. It does not however, remedy the economic distortion on consumer welfare. Moreover, nominal dispersion would naturally return if wage rigidity were included in a Calvo fashion. Finally, utilities, unionized or minimum wage labor and general purpose technologies like IT, transport, logistics and office infrastructure surely imply a sizeable common component of firms' cost base.}  
\par The analysis here yields deep econometric implications. Firstly, the structural model and the demand shocks are identified with a standard policy rule under the null hypothesis that the model is correct.\footnote{This is because under the null the rational expectations model is white noise so there are no valid instruments. Subjective expectations data solve this problem.} 
This remedies a fundamental inconsistency in the existing framework. Additional persistence and a degree of symmetry in the coefficients should improve small sample properties.
\par Existing work is biased. Under the null hypothesis that the model is correct, there is \emph{Econometric Duality}; an equivalence between constraints on the re-optimization of the representative firm and statistical restrictions on the econometrician seeking to fit the Phillips curve model to the data.  
Stochastic equilibrium offers new sources of identification. This new theory promises new challenges with a tighter link between macroeconomic modelling and econometric theory. 
\par The paper carries powerful results for all aspects of the \cite{lucas1976econometric} critique. On the one hand, the original equivalence result is wrong. Keynesian models contain lagged variables reflecting staggered optimization that the neoclassical framework does not. 
On the other hand, the notion of a trade-off where monetary activism causes an adverse movement in the Phillips curve schedule is borne out by the analysis of the coefficients. In fact, I show that the benchmark price Phillips curve is spurious in the sense that its slope is zero at a standard parametization and can be negative, which fits the message of his paper. The problem lies with the transmission mechanism which is entirely inter-temporal. 
\par This is the essence of 
\emph{Output Neutrality}. Where inter-temporal forces fall away, current inflation- as determined by optimal pricing- depends only on past and present pricing incentives, reflected in the lag and the expectation of the lead of inflation. Indeed, down this limit, inflation equals a half its lag and a half its future. These features agree with Taylor pricing, where inflation is determined by a weighted average of its lagged and future values, also with equal weight on past and future inflation. Consistent with output neutrality, the coefficients on marginal costs (or the output gap) sum to zero, as under Calvo. This creates a sturdy bridge between microeconomics and macroeconomics, as well as clear commonality between alternative models. \par Furthermore, the focus on mapping between micro and macroeconomic behavior returns to the fore in the guise of the \emph{a priori} bifurcation analysis. This implies that the previous framework was not truly reflecting its underlying microfoundations, reflected in the re-optimization constraints. Finally, we can see the instances of non-existence through this lens, as they imply non-trivial barriers between microeconomic and macroeconomic inference. 
\par The demise of Divine Coincidence brings wide-ranging benefits. It breaks the previous reliance on unintuitive mark-up shocks to shift the Phillips curve (see \cite{le2011much} and \cite{fratto2020accounting}). These shocks are widely dismissed as credible explanations for inflationary bursts by surveys of leading economists (see \cite{vaitilingum2022inflation}), along with the natural policy prescription price controls.\footnote{The article and the actual answers are available at the following addresses: \\ https://voxeu.org/article/inflation-market-power-and-price-controls-igm-forum-survey \\ 
https://www.igmchicago.org/surveys/inflation-market-power-and-price-controls/ \\ \cite{aparicio2021targeted} analyze a set of price controls on Argentinian supermarkets and find that they had a limited effect on inflation that was reversed once they were removed.} It will allow macroeconomists to avoid invoking an Effective Lower Bound (ELB) on nominal interest rates to generate 
policy trade-offs. This is particularly relevant following recent increases in interest rates in major economies. It chimes with empirical evidence that ELB was not important in the last decade because Quantitative Easing (QE) seemed to mimic interest rate cuts, structural macroeconomic relationships appeared stable and deflation was missing particularly in the UK,\footnote{The Fisherian channel where deflation drives up the real interest rate has been missing in the recent low interest rate spell. The extreme example of this lack of deflation phenomenon is the United Kingdom. From February 2009, just before the Bank cut its headline rate to a then record low of $0.5\%$ to July 2018, immediately before the next time the base rate exceeded this level, the Consumer Price Index grew at an average annual of $2.55\%$ in excess of the $2\%$ mandated target.  
Data for price levels and policy changes are available from the sites below, results are robust to plausible changes of dating \\ https://www.ons.gov.uk/economy/inflationandpriceindices/timeseries/d7bt/mm23
\\
https://www.bankofengland.co.uk/boeapps/database/Bank-Rate.asp} (see for example \cite{wu2016measuring}, \cite{dahlhaus2018international}, \cite{kuttner2018outside}, \cite{dell2018unconventional}, \cite{wu2018credit}, \cite{matousek2019effectiveness}, \cite{di2020quantitative} and \cite{weale2022financial} (QE); \cite{auerbach2017fiscal}, \cite{garin2019supply}, \cite{debortoli2019empirical} and \cite{mertens2021expect} (structural).\footnote{Any friction could in principle break Divine Coincidence, \cite{blanchard2007real} use real wage rigidity. It is not clear whether Central Banks are concerned or able to correct real market failures over and above their stabilization objectives. It is debatable whether these alternative frictions are really first order at business cycle frequency. The best candidate is financial frictions. This has been a particular focus since the Great Recession. However, this recent interest has been accompanied by alternative instruments (see \cite{clement2010term}, \cite{hanson2011macroprudential}, \cite{duncan2015objectives}, \cite{aikman2019would} and \cite{kashyap2020my}). Financial concerns were not a significant in monetary policymaking at major Central Banks previously, according to \cite{baxa2013time}, \cite{rotemberg2013shifts}, \cite{rotemberg2015federal} and \cite{oet2017does}.
In fact, financial shocks do not seem to be very important outside of crisis times, where they seem to operate like standard demand shocks (see \cite{mian2014explains}, \cite{muir2017financial}, \cite{mian2018finance}, \cite{huber2018disentangling}, \cite{gertler2018happened}, \cite{benguria2020after} and \cite{haque2021uncertainty}).} Central Bank independence has surely been one of the most successful policy experiments in economic history (see for instance \cite{alesina1993central}, \cite{cukierman1993central},
\cite{bernanke1999inflation}
\cite{acemoglu2008does}, \cite{balls2017twenty},  and \cite{garriga2020more}). It is high time an intuitive benchmark model were provided to guide day-to-day deliberations. The solution here is a first step in this regard, although more work is needed to understand the effect of supply shocks. 
\par Finally, there are significant implications of the possibility for inactive policy. From a microeconomic point of view it can be seen as indicating stability of general equilibrium without aggregate shocks, against the rigidity of individual prices. This might help explain why policymakers have traditionally been unconcerned by microeconomic shocks. 
\par On the macroeconomic ledger it has implications for both current and historical policy regimes. Macroeconomists often see stabilization through the prism of the Taylor principle- which states that inflation is controlled by raising the real interest rate in response to deviations of inflation from equilibrium. 
The policy rule derived here implies that it is not possible to immediately adjust interest rates to control inflation. This necessitates gradual changes of policy stance. 
This is in keeping with Central Bank best practice so-called \emph{"coarse-tuning"} (\cite{lindbeck1992macroeconomic}). It is usually implemented via \emph{Inflation Forecast-Targeting} (\cite{kohn2009policy}, \cite{svensson2010inflation} and \cite{svensson2012evaluating}). This is where policy and projections for future policy are adjusted to yield a desirable expected path for inflation and real activity, consistent with \emph{medium term} stability. This is usually defined as forecast inflation and output gap sufficiently close to target after a time frame of 18 months to 3 years.\footnote{Time series methods and policymakers' wisdom suggest that it takes between 18 months and two years for a change in monetary policy to have its maximum impact on inflation. This result seems to be robust across changes in policy regimes according to \cite{bernanke1999inflation} (see p 315-320), \cite{batini2001lag}, \cite{gerlach2003money} and recent analysis by \cite{goodhart2023snapshot}.
On its website the Bank of England advises the general public that: "Monetary policy operates with a time lag of about two years." (http://www.bankofengland.co.uk/monetarypolicy/Pages/overview.aspx) However, the Bank publishes forecasts three years ahead and frequently talks about "inflation returning to target by the three year horizon" consistent with a longer view of the stabilization and empirical work by \cite{havranek2013}.  (http://www.bankofengland.co.uk/publications/Pages/inflationreport/infrep.aspx) Practices are similar at other leading inflation targeting Central Banks.} 
\par Lastly, the policy result is of significance to our understanding of economic history. It helps to rationalize the survival of non-interventionist regimes like the Gold Standard where the "Rules of the Game" forbade active stabilization policy (see \cite{barsky1988gibson}, \cite{bayoumi1997modern} and \cite{bordo2009retrospective}). Secondly, it should make it easier to analyze hypotheses like secular stagnation (\cite{hansen1939economic} and \cite{summers2015demand}) which require long-lasting liquidity traps not possible in the existing representative agent setup. Lastly, it should support credible quantitative assessment of the benefits of modern macroeconomic management. 
\par The subsequent section gives a more detailed description of the proof techniques suitable for a technical audience. It also sets the paper within the mathematical literature. Although, designed to be accessible, it could be skipped by less mathematically inclined readers. 
\section{Mathematical Arguments}
This paper has two mathematical objectives, one analytic the other algebraic. The first is to analyze the most basic qualitative and quantitative properties governing existence of equilibrium. The second to describe the algebraic structures around ZINSS that give rise to the erroneous conclusions holding back modern macroeconomics. The subsequent discussion is heuristic rather than technically precise.\footnote{It is helpful to define a few terms from mathematical analysis for the benefit of economists. A quantitative estimate is where an unknown function or constant is bounded using only the model's primitives. Mathematical rigidity is where a collection of objects can be described from a smaller set of their properties. Universality is where models with different formulations display common behavior or structures. They will appear repeatedly in this section.} 
\par The central analytic contribution is a powerful fixed-point result. In fact, Theorem 3 provides necessary and sufficient conditions for any solution to exist, in a wide class of DSGE models, for a popular limiting case. Away from this limit, the condition is a requirement for standard statistical inference. 
\par The proof has two parts. The first step uses an argument from the Markov chains literature to show that the model has to mix. This follows easily from two standard features of these models: firstly, the existence of a sufficiently smooth cocycle in the mathematical expectation (coming from rational expectations) and secondly, the blow up on the boundary of the relevant objective functions of the optimizing agents, along a popular limit. 
\par The second step is motivated by control theory of linear rational expectations systems. 
It employs the local regularity properties from a stochastic Grobman-Hartman theorem of \cite{coayla2007hartman} to prove that only a unique stable solution 
is consistent with 
equilibrium existence. This arises where the number of eigenvalues inside the unit circle matches the number of jump variables and the number outside the total predetermined variables. Standard calculations yield functional conditions containing non-linear functions evaluated at the ergodic invariant measure. 
\par Applications follow swiftly. Theorem 4 uses basic ergodic theory to demonstrate that the probabilistic trajectory of the mean field system is unique, overturning previous non-rigorous work. The main focus of the paper is on limits where the noise is small. This allows me to derive quantitative estimates of the structural parameters supporting equilibrium existence using elementary methods. Polydromy arises because one variable, which is second order around the non-stochastic steady state, is first order around the stochastic steady state, although existence conditions are unaffected.
\par For future functional analysis, I provide a mathematical rigidity result stemming from ergodicity. The comparative statics in Theorem 2 can also be interpreted in this light as (sharp) \emph{a priori quantitative moment estimates}, reflecting model-specific equilibrium restrictions. They could surely be profitably combined in future to yield quantitative descriptions of the function space supporting this and other DSGE. Moreover, in Theorem 8, I prove a basic regularity result pertaining to the optimal policy problem that should be more amenable to existing functional analysis goals and techniques.  
\par The De Rham cohomology underpins the bifurcation analysis. In Decomposition 1, I calculate the singular surface around ZINSS. Theorem 6 and extensions connect the structure of the singular surface to the efficiency properties of ZINSS and the absence of distortions from large shocks. The disproof of Divine Coincidence, in Theorem 8, confirms the constraints interpretation. 
\par More abstract techniques including schemes and categories help to generalize and interpret these bifurcations. In Theorem 7, scheme theory is used to show that cross-equation cancellations are the ultimate source of all bifurcations in DSGE systems (with well-defined solutions) and admits speedy generalization. Furthermore, the idea of "near solutions", constructed via scheme theory, allow me to formalize \emph{Stochastic Bifurcation} and to explain how the effect attributed to trend inflation 
conceals an error in the Phillips curve approximated at ZINSS. 
Econometric duality results from the confluence of constrained optimization, scheme and cohomology theory. Finally, categories allow me to axiomize an aspect of the Lucas critique.
\par The paper contributes to 
two strands of mathematical and statistical literature. The first concerns mean field games. These are a joint endeavour between mathematics, economics, engineering and other disciplines. Some prominent papers include \cite{jovanovic1988anonymous}, \cite{caines2006large}, \cite{lasry2006jeux}, \cite{lasry2007mean}, \cite{gueant2011mean} and \cite{cardaliaguet2019master}. 
\par The closest to macroeconomics include \cite{lions2007large}, \cite{nourian2013}, \cite{bensoussan2016mean}, \cite{lasry2018mean} and \cite{cardaliaguet2020remarks} that consider games with small and large players. Macroeconomic applications, so far, include \cite{achdou2022income} who construct a mean field solution to an incomplete markets model and \cite{porretta2022traveling} who study economic growth in the context of knowledge diffusion. Lastly, \cite{alvarez2023price} focus on a restricted variant of the Calvo model studied here.\footnote{They use the price gap formulation of the period profit function (\cite{woodford2009information}), which avoids the singularities in the existing approximation, at the expense of simplifying dynamics.} \footnote{\cite{bilal2023solving} develops perturbation techniques using the master equation but does not prove existence results. His focus on reflecting boundaries suits his main application to job ladders and unemployment but contrasts with the blow-up condition here.} 
\par Results in adjacent areas include \cite{carmona2016mean}, \cite{lacker2016general} and \cite{carmona2018probabilistic}, who study common noise (aggregate shocks) but in finite time or with restrictive monotonicity conditions, as in \cite{bertucci2021master}.\footnote{Monotonicity conditions are popular in the mean field game literature but are restrictive because macroeconomic time series typically display non-monotone responses particularly when there are shocks to monetary policy. They are often accompanied by assumptions that keep the objective function uniformly bounded.} 
\cite{gomes2010discrete},  \cite{chau2017discrete}, \cite{saldi2018markov}, \cite{saldi2020approximate} and \cite{saldi2023discrete} are 
formulated in discrete time. \cite{delarue2024intrinsic} use noise to achieve uniqueness. Finally,
\cite{cardaliaguet2022monotone} attack a (representative agent) model with no idiosyncratic noise. The use of stationary noise is a distinction between my work and the Brownian motion standard in this literature, as are the Inada conditions that are used to rule out passages to the boundary of the state space.
\par The closest contributions to my own concern ergodicity and non-existence. In ergodic mean field games, players maximize a long-term average objective function. 
\cite{cao2022stationary} show that discounted mean field games approach their ergodic counterpart as time preference recedes. This parallels my focus on stochastic equilibrium here. Secondly, \cite{cirant2022existence} 
prove a non-existence result. 
The culprit is a strong complementarity (externality), in their case, it is effectively in the economy's growth process. Decomposition 2 has this flavour here, albeit with a different source. With direct macroeconomic application, \cite{alvarez2023price} uncovers non-existence, when strategic complementarities in pricing cross a critical threshold. Both results focus on the limit of finite time, with a boundary condition designed to represent long-run equilibrium, rather than a true infinite horizon with an endogenous destination.
\par A unique feature of my approach is an emphasis on game-theoretic constructive methods. The limiting construct is inspired by the folk theorem and the eigenvalue condition functions like a dynamic incentive compatibility constraint for the entire economy. This paper constitutes the first progress on constructing rigorous solutions to a general class of mean field games with common noise over infinite time - which is surely the major open problem of the whole literature (\cite{achdou2014partial}). 
\par This paper is formulated in discrete time, partly to reduce the technical burden on readers. 
Nevertheless, I am confident that techniques and concepts will carry over into continuous time, under some circumstances. An example, discussed in Section 6, shows that there may be differences in qualitative behavior between discrete and continuous time. It should be possible to analyze this \emph{time-scaling} behavior, both to help economists decide which solutions to use and for its mathematical beauty.
\par Both the \emph{a priori} duality result and the \emph{ex post} blow ups are connected to 
a small literature on phase transition of test statistics, which studies conditions where statistical estimators cease to converge. Examples include \cite{cover1965geometrical}, \cite{silvapulle1981existence}, \cite{albert1984existence}, \cite{santner1986note}, \cite{sur2019likelihood}, \cite{candes2020phase} and \cite{mccracken2020diverging}. I am confident statistical theory will expand along many frontiers to accommodate the deep and multi-faceted predictions of stochastic equilibrium theory. 
\par In my epsilon management I will deploy the asymptotic notation $x<<y$ to mean "$x$ is negligible relative to $y$".\footnote{This corresponds to $x=o(y)$ in the notation of \cite{tao2012cheap}, where further background can be found.} Unless otherwise stated, all background results are available in \cite{aliprantis2007infinite} or \cite{royden2010real}. \cite{jacobson2009basic} and \cite{nathan1980basic} might be helpful references for the abstract algebra with \cite{eisenbud2006geometry} useful for schemes. The main text will emphasize intuition with more formality and some preliminaries in the Appendices.
\section{Outline and Preview} 
This final introductory section has two parts: the first summarizes the Phillips curve argument, whilst the second sets out the structure of the rest of the paper.
\subsection{Phillips Curve}
A typical textbook New Keynesian Phillips curve currently looks like this 
\begin{equation} \pi_{t}= 
0.833 \hat{y}^{e}_{t} + \mathbb{E}_{t}\pi_{t+1} \end{equation}
where $y^{e}_{t}$ is an appropriate definition of the output gap and the final coefficient is a limiting case. Now the correct version of the 
$\sqrt{\varepsilon}$ Calvo Phillips curve around ZINSS is in fact 
\begin{equation} \pi_{t}=0.575 \pi_{t-1}  + 0.3\mathbb{E}_{t}\pi_{t+1} +0.25(\hat{\psi}_{t}-\hat{\psi}_{t-1})\end{equation}
The reason why is that as we approach ZINSS the Phillips curve is punctured by the following system of three singularities.  
\begin{equation} \pi_{t}=\pi_{t-1}\end{equation}
\begin{equation} \hat{\psi}_{t}=\hat{\psi}_{t-1}\end{equation}
\begin{equation} \pi_{t}= -0.607 \pi_{t-1} + 0.264\hat{y}^{e}_{t}\end{equation}
The first is the wall of the crossing, which is necessary to the bifurcation. The second is a statistical restriction that destroys the error term. The third
reflects the inter-temporal substitution incentive of the resetting firms. The crucial point is that substituting the singularities into the correct Phillips curve (2) simplifies to yield (1) the incorrect counterpart. The nature of the wall-crossing is that it is not possible to move in the other direction. This explains where macroeconomists went wrong for so many years. 
\par The precise parametric form of all coefficients are detailed later in the paper. The output gap term vanishing is specific to the parametization choices, which is discussed in the Appendix H alongside robustness exercises. The only non-standard selection was a lower value for the inflation response, in the policy rule, in line with the anti-Taylor principle. 
\subsection{Roadmap}
The remainder of the paper is organized as follows. Section 4 sets out the New Keynesian framework. Section 5 focuses on existing solutions and details the Lucas critique. Section 6 lays out the stochastic equilibrium and \emph{a priori} results. Section 7 derives the general Phillips curve.
Section 8 proves the theoretical conditions governing existence and non-existence of equilibrium. 
Section 9 rigorously develops the bifurcation theory. Section 10 covers theoretical and econometric implications. Section 11 is devoted to the policy rule. Section 12 is for discussion. 
\section{Calvo Framework}
This section lays out the principal model used throughout the paper featuring Calvo pricing and a Taylor rule. The assumptions concerning error terms present throughout the paper are discussed. The section winds up with a crucial finite dimensional characterization of the system. 
\subsection{Household's Problem}
This part lays out the households optimization problem that underpins the demand side of the economy. There is a single representative household\footnote{You may be surprised with a model where the world is organized into one happy family. It simplifies the dynamic exposition. \cite{werning2016incomplete} provides useful aggregation results, however none survive a full stochastic setting. The bifurcation analysis, the backbone of the Phillips curve, goes through without reference to a specific Euler.} that chooses consumption $C$ and labor supply $L$, so as to maximize the following objective function
\begin{equation}\max_{C_{t}, \, L_{t}} \: U_{t}= {\mathbb{E}_{t}\sum_{T=t}^{\infty}\beta^{T-t}\,\bigg[u(C_{T})-
\nu(L_{T})\bigg ]\,\psi_{T}}\end{equation}
subject to the budget constraint 
\begin{equation} P_{t}C_{t}+B_{t+1}=(1+i_{t-1})B_{t}+P_{t}(1-\tau_{t})W_{t}L_{t}+ \int_{0}^{1}{\Pi_{t}(i)}\,{d}i +T_{t} \end{equation}
$\beta \in (0,\, 1)$ is the discount factor, $u$ the utility function and $\nu$ the cost of work. $\psi_{T}$ is the demand shock. It is a preference shock, such that a higher value induces the household to demand more consumption today and less tomorrow. It can be taken to encompass financial shocks.\,$B$ refers to the holding of one period risk free nominal bonds.\,$i_{T}$ is the risk-free nominal interest rate paid at the end of period $T$ on the bond.\,$P$ is the price level - bonds are the numeraire here.\,$W$ is the real wage. \par $T_{t}$ is a lump sum tax that can be used to fund a lump sum subsidy on wages $\tau_{t}$. This unrealistic feature is designed to counteract the effect of price mark ups that will arise under imperfect competition so as to allow me to talk about the Divine Coincidence. They do not drive dynamic results.
\par There is a unit continuum of firms. $\Pi(i)$ is profit from an individual firm $i$ given by
\begin{equation}
\Pi_{t}(i)=p_{t}(i)\,y_{t}(i)-W_{t}l_{t}(i)
\end{equation}
\par Note in a stochastic environment firms need not make the same profits even with a symmetric equilibrium. Since when price rigidity is introduced, firms with the same demand curve will charge different prices, depending on when they last re-optimized and can therefore make different levels of profit.\footnote{In fact 
non-resetters need not have positive profit expectations or even a non-negative stock price, although, I will show this does not apply in our quantitative setting. Alternatively, one could swerve around this problem by imagining there are a continuum of households who each own one firm and an efficient long-term insurance market with contracts set arbitrarily far back in the past (so that there is no conditioning on pricing history). 
In this case the market will absorb all idiosyncratic risk associated with non-optimal price setting behavior.}
The budget constraint states that the uses for nominal income (consumption and saving) must be equal to the sources of income (wealth, labor and dividend income).
\par Finally, there are two constraints
\begin{equation}B_{T}=0
\end{equation}
\begin{equation}\lim_{T \rightarrow \infty}  \beta^{T-t} \mathbb{E}_{t} \psi_{T}u'(C_{T})\geq 0\end{equation}
The first is a bond market clearing condition. It is simplification designed to allow me to ignore the effect of government spending or debt management. It is not important at this stage because the analysis focuses on limiting cases where, 
without nominal frictions, Ricardian equivalence ensures that debt levels do not matter (\cite{barro1974government}).
The second is a transversality condition. It is necessary for a well-defined optimization problem.
It forces the household to honor their debts. Otherwise they would seek to borrow an unbounded amount and never repay. \cite{giglio2016no} provides model-independent support. 
\subsection{Preferences}The focus is on well-behaved models that readily generate unique interior solutions, to this end, I make several intuitive restrictions on preferences. To do so, I make several intuitive restrictions. To this end, $u'>0$ so agents always wish to consume more and the transversality condition will bind with equality. $u''<0$ to incentivize consumption smoothing.  It is costly for agents to work $\nu'>0$,  $\nu''>0$ encourages the agent to balance work and leisure.\footnote{Alternatively, we could imagine leisure is a good in demand that represents an opportunity cost of working. It will be convenient to do so in one 
proof discussed in the Appendix (C.1.2).} Together these conditions ensure uniqueness of interior solutions to the agent's optimization at any point in time with a non-stochastic background. 
\par Additional conditions are required to rule out boundary solutions. The standard Inada condition for consumption is
\begin{equation}\lim_{C \rightarrow 0}u'(C)= \infty \end{equation} along with zero net wealth. This ensures the representative household will always work. 
To force them to take leisure 
\begin{equation}\lim_{L \rightarrow \bar{L}} \nu( L_{t}) \rightarrow \infty \end{equation}
where $\bar{L}$ is the maximum possible labour supply. 
The key parameters are 
\begin{equation}\sigma=\frac{-Cu''}{u'}>0 \end{equation}
\begin{equation}\eta=\frac{L\nu''}{\nu'}>0\end{equation}
$\sigma$ represents the coefficient of relative risk aversion. It is also the inverse of the elasticity of inter-temporal substitution. $\eta$ is the inverse Frisch elasticity of labor supply. For the empirical part I will work with the popular functional forms. This ensures these parameters are structural and do not vary with income allowing comparison with standard econometric estimates.\footnote{Eagle-eyed readers will notice that there is no finite $\bar{L}$ such that the second condition is met. This will not matter because we are working in the small noise limit where any influence of the boundary will become vanishingly small. Alternatively, one could imagine that the household could hire in outside labor at a cost in extra wages that would have to be passed onto the whole labor market where there is perfect competition. Similar functional forms are available for non-limiting cases.  }
\begin{equation} u(C)=\begin{cases}
C^{1-\sigma}/(1-\sigma), & \sigma \neq 1 \\ 
\log(C), & \sigma =1 
\end{cases}\end{equation}
\begin{equation}
\nu(L)=\frac{L^{1+\eta}}{1+\eta}
\end{equation}
Each firm produces an individual variety for which demand is given by 
\begin{equation}y_{t}(i)=\bigg(\frac{p_{t}(i)}{P_{t}}\bigg)^{-\theta} Y_{t}\end{equation}
$\theta$ is the elasticity of demand. Appendix B.1.1 details the underlying optimization problem.
\subsection{Household Equilibrium Conditions}
There are three partial equilibrium relations stemming from the household's optimization. \begin{equation}u'(C_{t})=\beta(1+i_{t}) \mathbb{E}_{t}u'(C_{t+1})\frac{\psi_{t+1}}{\psi_{t}}\frac{P_{t}}{P_{t+1}}\end{equation} 
\begin{equation}
u'(C_{t})W_{t}= \nu'(L_{t})\end{equation}
The consumption Euler is the household's inter-temporal optimization condition balancing the marginal utility return to consumption today with that of the next period. Second is the intra-temporal optimal labor supply constraint equalizing the value of extra consumption with the marginal cost of working. Interest rates and wage rates are equilibriating mechanisms. The goods market clearing condition is simply 
\begin{equation}C_{t}=Y_{t}\end{equation}
\subsection{Price-Setting Problem}
\cite{calvo1983staggered} pricing is the most popular approach to inject nominal rigidity into a DSGE model. Re-optimization is governed by a stochastic process common across firms. With probability $1-\alpha$ each firm is free to reset its price (at no cost), whilst with probability $\alpha$ it keeps its price fixed and meets demand at its existing price. Firms reset their prices to maximize the expected present value of profits through the lifetime of the price as follows: 
\begin{equation}\max_{p_{t}^{*}(i)}\mathbb{E}_{t}\sum_{T=t}^{\infty}\alpha^{T-t}Q_{t,\, T}\bigg[\frac{p_{t}(i)}{P_{T}}y_{T}(i)-C (y_{T})(i)\bigg]  \end{equation}
subject to the individual demand (17). Here  \begin{equation}Q_{t,\, t+k}=\beta ^k \frac{\psi_{t+k}\, u'(C_{t+k})}{\psi_{t}\, u'(C_{t})}\Pi_{t,\, t+k}^{\theta}
\end{equation}
represents the real stochastic discount factor (SDF). It is the risk-adjusted present value of future consumption $k$ periods ahead which depends on the gross rate of inflation
\begin{equation}\Pi_{t,\, t+k}=\frac{P_{t+k}}{P_{t}}=(1+\pi_{t+1}) \cdots (1+\pi_{t+k})\end{equation} between today time $t$ and a future time $T > t$. The reset price problem is non-recursive since the firms' choice variable $p_{t}^{*}$ depends not just on the state of the economy next period $t+1$ but its whole future.\footnote{Recall that a recursive problem is one where the optimization can be rendered as a dynamic programming problem  
$$\max_{z_{t}} V_{t}(z_{t}, \, \psi_{t})=u(x_{t}, \, z_{t},\, \psi_{t})+ \beta \mathbb{E}_{t}V_{t+1}(z_{t+1},\, \psi_{t})$$  where $V_{t}$ is today's objective function which depends upon today's value of $z_{t}$ the state variable (one whose expectation is set in the previous period so $z_{t+1}=E_{t}z_{t+1}$), $x_{t}$ is the control or jump variable determined in each period and $\psi_{t}$ a random disturbance. $u$ and $\mathbb{E}_{t}V_{t+1}$ are known respectively as the instantaneous and continuation pay-off. In the canonical consumption-savings problem, which underpins the RBC framework, $z$ is capital, $x$ consumption and $u$ the household's pay-off from contemporaneous consumption.} This is the source of \emph{endogenous persistence} in the benchmark New Keynesian framework. Non-recursive optimization problems are common to all settings in which nominal rigidity arises through some firms not changing their price every period. The first order condition is
\begin{equation}\mathbb{E}_{t}\sum_{T=t}^{\infty}\alpha^{T-t}Q_{t,\, T}\frac{\partial y_{T}(i)}{\partial p_{t}(i)}\bigg[MR_{T}(y_{T}(i))-MC_{T}(y_{T}(i))\bigg]=0
\end{equation}
It states that optimal pricing sets a weighted stream of marginal revenues equal to a weighted stream of marginal costs, which in turn implies a similar relationship between (real) price and marginal costs 
\begin{equation}
\mathbb{E}_{t}\sum_{T=t}^{\infty} (1-\alpha)^{T-t}Q_{t,\, T}\bigg(\frac{p_{t}^{*}}{P_{T}}\bigg)^{-\theta}Y_{T}\bigg[\frac{p_{t}^{*}}{P_{T}}-\frac{\theta}{\theta-1}MC_{T}(y_{T}(i))\bigg]=0
\end{equation} 
By strict concavity of the optimization problem, the equilibrium price $p^{*}_{t}(i)$ is unique for each firm so there is a unique optimal reset price each period $p^{*}_{t}$. Therefore, by a law of large numbers argument (see \cite{anderson1991non}), the price level evolves as follows: 
\begin{equation}P_{t}^{1-\theta}=\alpha P_{t-1}^{1-\theta}+(1-\alpha)(p_{t}^{*})^{1-\theta}
\end{equation}
The persistence of the price level depends on $\alpha$ the degree of price rigidity. The reset price can be expressed as
\begin{equation}\frac{p^{*}_{t}}{P_{t}}=\frac{\theta}{\theta -1} \frac{\aleph_{t}}{\beth_{t}}\end{equation}
where 
\begin{equation}\aleph_{t}=\mathbb{E}_{t}\sum_{T=t}^{\infty}(\alpha \beta)^{T-t}\, \Pi^{\theta}_{t,\, T}\, \psi_{T}u'(C_{T})Y_{T}MC_{T} \end{equation}
\begin{equation}
\beth_{t}=\mathbb{E}_{t}\sum_{T=t}^{\infty}(\alpha \beta)^{T-t}\, \Pi^{\theta-1}_{t,\, T}\, \psi_{T}u'(C_{T})Y_{T}
\end{equation}
both numerator and denominator have recursive forms 
\begin{equation}
\aleph_{t}=\psi_{t}u'(C_{t})Y_{t}MC_{t}+\alpha \beta \mathbb{E}_{t}(1+\pi_{t+1})^{\theta}\, \aleph_{t+1}
\end{equation}
\begin{equation}
\beth_{t}=\psi_{t}u'(C_{t})Y_{t}+\alpha \beta \mathbb{E}_{t}(1+\pi_{t+1})^{\theta-1}\, \beth_{t+1}
\end{equation}
Intuitively, $\aleph_{t}$ is a scale-weighted measure of marginal costs with a discounting scheme that reflects the expected age of the price, whilst $\beth_{t}$ is a similarly weighted measure of the scale of future demand. It reflects how the real value of a rigid nominal price varies at different horizons according to the rate of inflation. In fact using (24) we can view the relative reset price as a weighted average of the (desired) \emph{flexible price} the firm would charge in the absence of nominal rigidity.\footnote{This flexible price is often called the desired price to distinguish it from the price selected by the current resetters $p^{*}_{t}$ which although costlessly adjusted will be set in expectation of future nominal rigidity. In inflationary environments it will always move more than the flexible price would. This is demonstrated in Appendix B.2.} Nominal rigidity is where the actual price differs from this flexible price. 
ZINSS is the unique position where nominal rigidity disappears. Appendix B.2 contains a formalization of these basic averaging concepts here. 
\par The first elementary result settles the connection between the relative reset price and the rate of inflation, central to the interpretation of the bifurcation. 
\begin{proposition}
There is an isomorphism (homeomorphism) between the relative reset price $p^{*}_{t}/P_{t}$ and the inflation rate $\pi_{t}$. 
\end{proposition}
\begin{proof} Using (26) establishes that 
\begin{equation} \frac{p^{*}_{t}}{P_{t}}= 
\frac{(1-\alpha)^{1/(\theta-1)}}{(1-\alpha (1+\pi_{t})^{\theta-1})^{1/(\theta-1)}}\end{equation}
The result follows from taking and signing the derivative 
$$\frac{\partial(p^{*}_{t}/P_{t})}{\partial \pi_{t}}= \frac{(1-\alpha)^{1/(\theta-1)}(1+\pi_{t})^{\theta-2}}{(1-\alpha (1+\pi_{t})^{\theta-1})^{\theta/(\theta-1)}}>0 $$
\end{proof}
\subsection{Nominal Equilibrium Conditions}
Nominal rigidity generates real distortions through the dispersion term. The demand aggregator 
\begin{equation}\Delta= \int_{i}{\bigg(\frac{p_{i}}{P}\bigg)}^{-\theta}\; \mathrm{d}\mu_{i}\end{equation}
represents price dispersion since
\begin{proposition}$\Delta \geq 1$ with $\Delta=1$ if and only if $p_{t}(i)=P_{t},\;\forall \;i$.\end{proposition}
\begin{proposition}$\Delta$ is second order when approximated from ZINSS. \end{proposition}
The first result is a global property of price dispersion, which will be called upon when analyzing the boundary conditions in the fixed-point theorem. The second is a local condition that justifies the $\sqrt{\varepsilon}$ limiting construction. Neither result is novel so proofs are relegated to Appendix B.3. 
The intuition is that consumers prefer variety and it is therefore costly to substitute between high and low price goods. Therefore they cannot achieve the same utility when prices are dispersed which will always arise when prices are rigid and inflation variable.
Here with Calvo pricing, $\Delta$ evolves according to the following relationship: 
\begin{equation}\Delta_{t}=(1-\alpha){\bigg(\frac{p_{t}^{*}}{P_{t}}\bigg)}^{-\theta}+ \; \alpha {(1+ \pi_{t})}^{\theta} \Delta_{t-1}\end{equation}
Using Proposition 1, I can solve for the reset price to give a recursion in inflation \begin{equation} \Delta_{t}=\frac{({1-\alpha(1+\pi_{t})^{\theta-1})}^{\theta/(\theta-1)}}{(1-\alpha)^{1/(\theta-1)}} + 
\alpha(1+\pi_{t})^{\theta}\Delta_{t-1} \end{equation}
Finally, there is the monetary policy rule. 
\begin{equation}i_{t}=i^{*}_{t} + a_{\pi}\hat{\pi}_{t}+ a_{y}\hat{y}^{e}_{t} 
\end{equation}
This so-called Taylor rule is an ad hoc stabilization condition motivated by \cite{taylor1993discretion}. His idea was that the Central Bank should respond to 
contemporaneous deviations in inflation and the output gap to drive the economy back to equilibrium. This motivates the restriction on the reaction coefficients $(a_{\pi}, \, a_{y})\geq \bf{0}$.
\par When selecting such a simple rule, my key concerns were tractability and comparability with previous work. In the next part, I confirm that including lags or leads would not be justified by welfare concerns in a purely forward-looking model. This fits with my ultimate goal of justifying inertial policy from an optimal policy standpoint. My 
policy analysis should prove immune to timing issues and information available to the Central Bank. This is because the boundary case of inactivity is common to all possible policy rules, whilst in a persistent stochastic setting even the expectations of future variables would induce inertia.\footnote{Taylor's actual proposal was somewhat closer to my analysis. His rule used an inflation measure averaged over the previous four quarters. Although, his motivation was different, he envisaged that the lagged inflation rate would serve as a proxy for expected inflation. In the same conference (\cite{henderson1993comparison}) came up with a very similar formulation.}
\par  The output gap is defined so as to be consistent with the Phillips curve and the objects of welfare analysis. $i^{*}_{t}$ is the natural rate of interest, the nominal interest rate consistent with equilibrium in the benchmark classical model. The idea is that the rule automatically adjusts to accommodate real shocks. 
In Appendix C.2.3 an alternative situation where this is not possible is discussed.  
\subsection{Real Equilibrium Conditions}
There are constant returns to scale in production. Therefore 
\begin{equation}Y_{t}=A_{t}L_{t}
\end{equation}
The firms' cost minimization condition covered in 
Appendix A.4, alongside suitable generalizations, yields
\begin{equation}MC_{t}=\frac{W_{t}}{A_{t}}\end{equation}
Hence the general equilibrium condition obtained from (20) and the integration over the individual demand curve (17) is
\begin{equation}\Delta_{t}C_{t}=A_{t}L_{t}
\end{equation}
\subsection{Shocks}
There are two sources of disturbance in the structural model, the demand shock $\psi$ and the supply shock $A$. Both processes will be continuously distributed. The productivity process will be autoregressive in deviations 
\begin{equation}\hat{a}_{t}= \rho \hat{a}_{t} + \hat{e}^{A}_{t}\end{equation}
where $\rho \in (0,1)$ and the innovation $e^{A}_{t}$ is white noise. However, persistent Euler errors are more difficult to motivate. Serial correlation is traditionally seen as a sign of misspecification. This was the case in initial studies of the Euler equation like \cite{hall1978stochastic} and \cite{hansen1983stochastic}, as well as more recent efforts discussed in \cite{jappelli2017economics}. Whereas, it is in principle possible to measure the persistence of productivity due to oil, technology or pandemic shocks, it is more difficult to measure persistent taste shocks.
It is possible to motivate persistent shocks with more sophisticated descriptions of financial markets and complicated contracts. However, in a benchmark model we are interested in what happens when price rigidity is the sole or dominant friction without any ad hoc features.
\par Finally, it is worth noting that several quantitative estimates and comparative statics require that error distributions be independent of the endogenous variables which rules out higher moments of shocks depending on the choices of consumers and firms,   
for example inflation directly effecting the volatility of the demand shock.\footnote{There is strong evidence dating back to \cite{engle1982autoregressive} that inflation, in particular, is hetroscedastic with second moment dependence on other business cycle variables. This can be seen as a sign of misspecification of a structural model \cite{kennedy2003guide}. It will be absent in the small noise limits in this paper. Stochastic equilibrium theory presents an ideal framework for first principled modelling of high order dynamics.}
\subsection{Recursive Equilibrium} 
I finish by characterizing the state space form of the non-linear model. It is not possible to find a closed form but I am able to find a recursive form by inverting expressions for the recursive term $\aleph$. It reveals that economic dynamics are generically persistent. Present inflation depends on its lag and that of both the structural shocks. This is because the non-linear New Keynesian has a hybrid form. This is true $\mu$ almost everywhere. This motivates the bifurcation analysis. \begin{proposition}The non-linear model can be written in canonical form as
$\mathbb{E}_{t}Z_{t+1}=f(Z_{t}, \, U_{t}, \, \gamma)$
$\mu$ a.e. with smoothness properties inherited from the primitive functions. The endogenous variables are $Z_{t}=(\pi_{t}, \,  y_{t}, \, \pi_{t-1}, \, \Delta_{t})$, $U_{t}=(\psi_{t}, \, A_{t}, \, \psi_{t-1}, \, A_{t-1})$ are the errors, whilst 
$\gamma \in \Gamma$ contains all the parameters of the model, assumed to include details of the stochastic process omitted here. \end{proposition}
Hence we can decompose the model into jump variables that have to be solved forward and state variables that have to be solved backwards as follows:\footnote{Technically an individual jump $X (j)$ can be written as $X_{t}(j)=f(X_{t}(-j), \, \mathbb{E}_{t}X_{t+1}, \, \cdot)$ where the latter dependence is nontrivial. Likewise, for a state variable $X(s)$ we can write $X_{t}(s)=f(X_{t}(-s), \,  X_{t-1}, \, \cdot)$ with same functional condition. Note that $\Delta_{t}$ is a state variable even though $\mathbb{E}_{t}\Delta_{t+1}$ is stochastic, since it can be written as a function of inflation history. This is consistent with the definitions in \cite{buiter1982predetermined}.} $Z^{J}_{t}=(\pi_{t}, \, y_{t})$ and $Z^{S}_{t}=(\pi_{t-1}, \, \Delta_{t})$. 
The lengthy but relatively straightforward proof is housed in
Appendix A.5.
\begin{remark}This is an application of the original recursive equilibrium formulation of \cite{prescott1980recursive} and \cite{mehra2006recursive}, with the canonical form playing the role of the policy function, associated with a planners problem Crucially, this formal approach has hitherto been missing from 
New Keynesian economics.\end{remark}
\begin{remark}
The transversality condition is used to resist additional terms in bubbly assets entering.
\end{remark}
\begin{remark}This result 
means infinite dimensional methods will be unnecessary to solve or approximate this model and should extend to other DSGE. 
\end{remark}
Subsequent business will focus on which of these variables are first order around ZINSS and why approximations at ZINSS differ. 
\section{Existing Solutions}
This section sets out the existing perturbation solutions, taken from ZINSS. There is a natural focus on the first order approximation, although I am able to generalize salient results to non-linear solutions. I set out familiar problems with their predictions concerning persistence and optimal policy. 
\par Next, I set out two alternative models of nominal rigidity the Lucas and Rotemberg Phillips curves. Up to parametric conditions, I connect these three 
models in a pseudo-observational equivalence, which I later disprove. This allows me to link this wrong claim back to a particular argument in \cite{lucas1976econometric}. Finally, I review all aspects of this famous paper. 
\subsection{Singular Phillips Curve}
This subsection flags up the subtle error in the standard approximation that brings about a purely forward-looking solution.
Linearizing the two components of the Phillips curve at 
ZINSS reveals a special structure, 
in particular, there is a common root $\mathbb{L}=1/\alpha \beta$ in both lag polynomials.
\begin{equation}\hat{\aleph}_{t}=(1-\alpha \beta)\hat{mc}_{t}+(1-\alpha \beta) \hat{\psi}_{t} +(1-\alpha \beta)(1-\sigma)\hat{y}_{t}+\alpha \beta \mathbb{E}_{t}\hat{\aleph}_{t+1}\end{equation}
\begin{equation}\hat{\beth}_{t}=(1-\alpha \beta) \hat{\psi}_{t}+(1-\alpha \beta)(1-\sigma)\hat{y}_{t} +\alpha \beta \mathbb{E}_{t}\hat{\beth}_{t+1}\end{equation}
This means the system can be solved by elimination. Therefore, we do not have to lag one of the recursions to perform a substitution introducing lagged terms. The next result confirms that this is unrepresentative of the dynamics of any stochastic system because it only applies when inflation is at steady state where the approximation is irrelevant. Let $\mathbb{L}_{\aleph}$ equal the zero of the lag polynomial in $\aleph$.
\begin{proposition} In the 
$\epsilon$-neighborhood of ZINSS it is the case that $\mathbb{L}_{\aleph} < \mathbb{L}_{\beth}$.\end{proposition}
\begin{proof} By elementary manipulation it is clear that this amounts to proving that 
$$\mathbb{E}_{t}\pi_{t+1}(1+\pi_{t+1})^{\theta-1}>0$$
This follows immediately from Chebyshev's correlation inequality (Lemma 1) which states that 
$$\mathbb{E}AB \geq \mathbb{E}A \, \mathbb{E} B$$ for two strictly increasing functions with equality if and only the measure is degenerate. 
\end{proof}
ZINSS is an example of a (measure zero) singularity not covered by Proposition 3. In fact it is the wall of the crossing which gives rise to (1). Bifurcation and common roots will be formally connected in Section 9. Thus we arrive at the incorrect step. 
\begin{error} Cross Equation Cancellation in the Phillips curve 
\end{error}
This yields the forward-looking relationship. 
\begin{equation}\pi_{t}= \kappa \hat{mc}_{t} + \beta \mathbb{E}_{t}\pi_{t+1} \end{equation}
where the slope parameter is
\begin{equation}\kappa= (1-\alpha)(1- \alpha \beta)/\alpha\end{equation}
to derive an output gap expression first linearize the marginal cost function (38) \begin{equation}\hat{mc}_{t}=\hat{w}_{t}-\hat{a}_{t}\end{equation}
Next linearize the production function (36)
\begin{equation}\hat{a}_{t}=\hat{y}_{t}-\hat{l}_{t}+ \hat{\Delta}_{t}\end{equation}
now using labor supply constraint and market clearing (19) and (20) \begin{equation}\hat{mc}_{t}= ( \sigma + \eta )\bigg[\hat{y}_{t}-\frac{1+\eta}{ \sigma  + \eta}\hat{a}_{t}+\frac{\eta}{\sigma + \eta} \hat{\Delta}_{t}\bigg]
\end{equation}
The behaviour of price dispersion is governed by the linearization of (35)
\begin{multline}\hat{\Delta}_{t}=\frac{\alpha \theta (1+\pi)^{\theta-2}}{\Delta 
(1-\alpha)^{1/(\theta-1)}}\bigg[(1+\pi)\Delta (1-\alpha)^{1/(\theta-1)}\,-\,(1-\alpha(1+\pi)^{\theta-1})^{1/(\theta-1)}\bigg]\hat{\pi}_{t} \,+\, \\  \alpha (1+\pi)^{\theta}\hat{\Delta}_{t-1}\end{multline}
substituting in $(\pi, \Delta)=(0,\, 1)$ negates the first term leaving \begin{equation}\hat{\Delta}_{t}=\alpha \hat{\Delta}_{t-1}\end{equation}
For the remainder of this Section, I will use the $\sqrt{\varepsilon}$ limit, where $\Delta$ vanishes thanks to Proposition 3. Continuing the derivation, without price dispersion there is no distortion to production efficiency coming from price rigidity. This means that actual output $\hat{y}_{t}$ moves one for one with output in a comparable flexible price economy $\hat{y}^{f}_{t}$. The difference is called the efficient output gap \begin{equation}\hat{y}^{e}_{t}=\hat{y}_{t}-\hat{y}^{f}_{t}
\end{equation} 
I can use the expression for ZINSS, given in Appendix D, to express the flexible price output in terms of the technology shock as
\begin{equation}\hat{y}_{t}^{f}=\frac{1+\eta}{\sigma + \eta }\hat{a}_{t} 
\end{equation}
Substituting into (47) reveals a proportional relationship between changes in marginal costs and the efficient output gap. 
\begin{equation}\hat{mc}_{t}=(\sigma +\eta)\hat{y}_{t}^{e} \end{equation}
Substituting into (43) yields the final form of the Phillips curve 
\begin{equation}\pi_{t}= \omega \hat{y}_{t}^{e} + \beta \mathbb{E}_{t}\pi_{t+1}
\end{equation}
where the form of the composite parameter is
\begin{equation} \omega=(\sigma +\eta)\frac{(1-\alpha)(1-\alpha \beta )}{\alpha}\end{equation}
Finally, the approximate Euler equation takes the form
\begin{equation}\hat{y}^{e}_{t}=-\frac{1}{\sigma}\hat{i}_{t}+ \mathbb{E}_{t+1}\hat{y}^{e}_{t}-\frac{1}{\sigma}\mathbb{E}_{t}\pi_{t+1}-\frac{1}{\sigma}\hat{\psi}_{t}
\end{equation}
\subsection{Persistence and Policy Puzzles}
This subsection begins with observations concerning the stochastic properties or lack thereof of the Phillips curve. I lay out Divine Coincidence. When I add suitable error terms, I am able to extend these results to rule out persistence. 
Moreover, under a different plausible timing condition there can be no meaningful policy change so optimal policy implements \emph{laissez-faire}. The second subsection shows that these properties extend to non-linear approximations either directly or under a suitable special case. 
\subsubsection{Linearization}
\begin{remark}The New Keynesian Phillips curve (53) and (54) is stochastically singular.
It cannot be statistically estimated without an additional non-structural disturbance i.e. unrelated to model primitives.This justifies the criticism of "dubious structural shocks" by \cite{chari2009new} in particular the role ascribed to mark-up shocks in the Phillips curves of leading New Keynesian models based on the current linearization. \end{remark}
This gives rise to the following policy puzzle. 
\begin{proposition}(\textbf{Divine Coincidence})
Suppose a benevolent planner faced the linearized system in deviations around ZINSS represented by (53)-(55).
Allow that policy can be conditioned on the present shock. Furthermore, they were able to levy lump sum taxes and provide proportional subsidies to firms, then a Pareto efficient allocation could be implemented.\end{proposition}
Its proof, contained in Appendix C.1.2, is a modification of Theorem 5.5 in \cite{acemoglu2009introduction}. It features the infinite horizon budget constraint, from which some additional economic intuition is garnered. The result intrinsically concerns linear approximations, trade-offs will arise through price dispersion in the non-linear setting.
\par Subsequent results set out related problems with policy and persistence that will extend to alternative models with similar properties to the bifurcated Calvo model. Turning to the Phillips curve, to derive econometric interpretations it is necessary to add non-structural errors. The preferred interpretation is that this represents an expectation error. Future work should explore the  distinct role and implications of structured and unstructured features of error terms.
\begin{proposition}Any free error process $\{\hat{e}_{t}\}$ introduced additively into the \\ Phillips curve must be white noise. Otherwise, it is observationally equivalent to a model with adaptive expectations.
\end{proposition}
\begin{proof} Any Phillips curve with additive error can be written as, $\mathbb{E}^{\mathcal{S}}_{t}\pi_{t+1}=\mathbb{E}_{t}\pi_{t+1} +e_{t}$.
where $\mathbb{E}^{\mathcal{S}}$ is the subjective expectation operator assumed common across resetters. The rational expectations hypothesis stipulates that individual expectation errors must be serially uncorrelated.\footnote{This formulation is a \emph{weak} form of the rational expectations hypothesis, as opposed to the \emph{strict} form where the individuals fully understand the structural model which here would take us back to stochastic singularity.} Therefore, 
\begin{equation}\newcommand{\Cov}{\mathrm{Cov}}
\Cov(\mathbb{E}^{\mathcal{S}}_{t}\pi_{t+1}-\mathbb{E}_{t}\pi_{t+1},\, \mathbb{E}^{\mathcal{S}}_{t-1}\pi_{t}-\mathbb{E}_{t-1}\pi_{t}) =
\Cov(e_{t},\, e_{t-1})=0 \end{equation}
However, if we allowed a non-white noise error then $\newcommand{\Cov}{\mathrm{Cov}}\Cov(e_{t},\, e_{t-1})\neq 0$, contradicting the rational expectations hypothesis. \end{proof}
Unexplained serial correlation in econometrics is an indication of misspecification, as discussed in \cite{kennedy2003guide}. This principle has underpinned empirical investigations of the Phillips curve, such as \cite{gali1999inflation} and others discussed in \cite{mavroeidis2014empirical}.\footnote{Similar sentiments and goals appear in \cite{rotemberg1996real} and \cite{nakamura2010monetary} among others.} The lack of viable persistence sources can have stark implications. 
\begin{proposition}\textbf{(No Persistence)}
The economy given by (36) and  
(53)-(55) endowed with unstructured rational expectation consistent error terms is white noise and is unidentified under the null hypothesis. It is correctly specified, provided that the following determinacy condition is met $a_{\pi}+ (1-\beta)a_{y}/ \kappa > 1$ \end{proposition}
The system is not statistically identified for lack of valid instruments. Intuitively, when there is no persistence there is no way of instrumenting for future expectations, in particular for inflation. It should not be surprising that attempts to do so have been an empirical failure. This result implies that these estimates are in a strict sense ill-defined with respect to the underlying theory. The current model of the Phillips curve is not fit for policy purposes, where theory consistent econometric investigation is 
paramount. 
\begin{remark} Determinacy would become existence conditions if one were to interpret the bifurcated model as an approximation of a non-linear model based on its singular surface. \end{remark}
\par The final two results here concern policymaking. The first originated with \cite{woodford2001taylor}. 
\begin{corollary}\textbf{(Taylor Principle)} Suppose that $a_{y}=0$ or $\beta \rightarrow 1$ then $a_{\pi}>1$ is required for a unique stable solution.\end{corollary}
This familiar result states that monetary policymakers must "lean against the wind" sufficiently or they would lose control of inflation expectations. It is a figment of an entirely forward-looking model. The final result extends the Divine Coincidence idea to a stochastic environment. Its proof, similar to Proposition 8 is omitted. 
\begin{proposition} Suppose the monetary policy maker has a pay-off function that can be expressed in the form 
$\mathcal{W}_{t} =\sum_{T=t}^{\infty}\beta^{T-t} \, w(\pi_{T}, \, Y_{T})$, where the period pay-off $w$ is maximized at $\pi=0, \,Y=y^{f}$ and the policymaker moves in advance of the realization of shocks, then the optimal policy tracks the natural rate $i_{t}=\bar{i}_{t}$ \end{proposition}
If there is no (ad hoc) backward-looking component to the objective function and the Central Bank is interested in stabilizing the economy, it never changes the interest rate, in response to business cycle conditions. This justifies my decision to neglect backward-looking terms in the policy function. It should only respond to financial market conditions and long-run equilibrium movements. These were arguably features present in the Classical Gold Standard,
as argued in \cite{bayoumi1997modern}. Indeed, they could work automatically through financial market equilibrium forces. 
\par This "do nothing" policy has a long tradition in classical and monetarist economics, (consult \cite{friedman1960prog}, \cite{de2000triumph} and \cite{chari2009new}). However, it is not Keynesian and does not reflect modern policy setting anywhere with an inflation targeting framework. The persistence I derive will give a rationale for active monetary policy. The nature of optimal policy will be left for future work.
\subsubsection{Non-Linear Approximations}
It is become more feasible and increasingly common for economists to use non-linear techniques, such as higher order approximations in their work. 
It is important to know that my conclusions are not a figment of the linearization. The first result shows that all but the most extreme of the previous results applies to a refined version of the non-linear model. 
Firm-specific labor supply is a popular simplifying assumption that allows us to divorce price dispersion from the optimizing decisions of the firm. It is isomorphic to setting $\Delta=1$. It would show up elsewhere in the optimizing problem but we will ignore this as Divine Coincidence is a linear phenomenon.\footnote{Alternatively we could imagine an ad hoc demand system which will be shown in the next subsection.} 
\begin{proposition}
Suppose the system has a unique solution as part of a wider DSGE model without additional persistence. Furthermore, let the following assumptions hold (i) Firm Specific Labour (ii) $\sigma =1$, then the Phillips curve has no persistence or statistical identification under the null with respect to $\{ y^{e}_{t}, \, \pi_{t}\}$ as defined previously.
\end{proposition}
\begin{proof} The proof consists of simply expanding out the non-linear Phillips curve without loss of generality. I can work with analytic functions (since they are dense in a standard function space)\footnote{Consider the subset of functions on $\mathbb{R}^{n}$ defined on our state space and apply the Weierstrass-Stone approximation theorem.} Proceed once more by subtracting (41) from (42). Focus first on the bank of terms in $\psi_{t}$ and $MC_{t}$ and use the labor supply curve formed from (19) and (38). However, it is no longer stochastically singular. 
    $$\kappa \bigg( \hat{\psi}_{t} + (1+ \psi \hat{\psi}_{t} )\sum^{\infty}_{k=1}\prod^{\infty}_{k=1} (1+\eta) \cdots (1+\eta -\{k-1\})(\hat{y}^{e}_{t})^{k} -\hat{\psi}_{t}\bigg) $$
The central points are that $\sigma=1$ rules out output terms not entering through marginal costs, whilst cross-product terms remove stochastic singularity. The second bank of terms is 
\begin{multline} (1-\alpha)\beta\bigg(\aleph (1+\mathbb{E}_{t}\hat{\aleph}_{t+1})\prod_{k=1}^{\infty}\theta \cdots (\theta - \{ k-1\})\mathbb{E}_{t}\pi^{k}_{t+1} - \\  \beth (1+\mathbb{E}_{t}\hat{\beth}_{t+1})\prod_{k=1}^{\infty}(\theta-1) \cdots (\theta -  k)\mathbb{E}_{t}\pi^{k}_{t+1} +  \aleph \mathbb{E}_{t}\hat{\aleph}_{t+1}- \beth \mathbb{E}_{t}\hat{\beth}_{t+1} \bigg)\end{multline}
Using the reset price derived from (27) moved forward one period and the fact that with $\sigma=1$ $\mathbb{E}_{t}\beth_{t+1}=0$ yields the final form of the Phillips curve
\begin{multline} \pi_{t}= \kappa  
(1+ \psi \hat{\psi}_{t} )\sum^{\infty}_{k=1}\prod^{\infty}_{k=1} (1+\eta) \cdots (1+\eta -\{k-1\})(\hat{y}^{e}_{t})^{k} + \beta \mathbb{E}_{t}\pi_{t+1} + \\
 \bigg((1-\alpha)\beta \prod^{\infty}_{k=2}k(\theta-1) \cdots (\theta -\{k -1\}) +  \alpha \beta \prod^{\infty}_{k=2}\theta \cdots (\theta -\{k -2\}) \bigg)\mathbb{E}_{t}\pi^{k}_{t+1}
 \end{multline}
 Under the wider DSGE hypothesis, it is now possible to repeat the results of Proposition 8 setting $\mathbb{E}_{t}\pi^{k}_{t+1}=0$ for this forward-looking system. This means that ${y}^{e}_{t}$ will depend purely on the contemporaneous demand shocks. This will give rise to no persistence and thus no identification, under the maintained assumption these are white noise. 
\end{proof}
Appendix C.2.1 confirms that $\sigma=1$ is in fact necessary, otherwise technology shocks enter. 
Hence the singular surface of the original model can be characterized by the function $f^{sing}(\pi_{t}, \, y^{e}_{t}
, \, \Delta_{t}, \, \psi_{t}, \, A_{t})$ and $f^{sing}_{\sigma=1}(\pi_{t}, \, y^{e}_{t}, \, \Delta_{t}, \, \psi_{t})$.
This confirms that the absence of lagged terms is a topological feature of the singular surface around ZINSS.\footnote{These high order approximations are not the focus here but in application terms in $\psi$ and $A$ would be normalized. Technology shocks can enter first order dynamics if there is uncertainty or disagreement about the output gap measure. However, this persistence does not carry over to inflation response to a demand shock (see C.2.2).} 
\subsection{Two Comparison Models}
This subsection introduces two alternative models featuring monetary non-neutrality. They will be used in the next section and throughout the paper as a point of comparison with solutions of the Calvo model. The first model is classical, in the sense that it features perfectly flexible prices. The second is Keynesian because prices are not able to adjust fully to inflation. Along with the menu cost model discussed in Section 8, it is an example of a state-dependent pricing model. Crucially, it does not feature staggered optimization. Prices will typically change in every period. I layout the equivalence between the existing solution of Calvo and the Rotemberg model, I extend this equivalence to the original Lucas model, allowing me to view it as part of the observational equivalence claim of \cite{lucas1976econometric}. Finally, I briefly review the rest of the critique.
\subsubsection{Lucas Model}
\cite{lucas1972expectations} is the original rational expectations monetary model. Markets are perfectly competitive and anticipated monetary shocks are impotent. Monetary non-neutrality arises because of imperfect information rather than sticky prices. Nevertheless, by adding a money demand function, I am able to derive surprising similarity to New Keynesian models with price rigidity. \par There are a collection of price-taking households.\footnote{It is not common to have differentiated products and price-taking behavior. For motivation imagine there were actually multiple households producing the same good and therefore competition between them or the possibility to set-up firms with a consequent free entry condition. Recall the Bertrand paradox that two competing firms can implement a competitive equilibrium.} Each produces a single good, using just its own labor.\footnote{This formulation comes from \cite{romer2012advanced}. The idea is to avoid the households using the economy-wide labor market to deduce the aggregate price level. In the original paper, Lucas used an overlapping generations structure where households produced in one period then sold output in the next. Recent literature has developed more sophisticated stories about dispersed private information and strategic interaction. I do not pursue this approach as I am looking for a benchmark model.} The choice problem simplifies to a static labor-leisure trade-off
\begin{equation}
u_{t}(i)=c_{t}(i)-{l}^{1 + \eta}_{t}(i)/(1 + \eta)
\end{equation}
as before, there are constant returns to scale and common technology.
Hence, the firms objective function can be expressed in the firms output $y_{t}(i)$ as 
\begin{equation}
\frac{p_{t}(i)}{P_{t}}y_{t}(i) - y^{1 +\eta}_{t}(i)/(1+ \eta)
\end{equation}
The first order condition balances marginal revenue and marginal cost so
\begin{equation}
\frac{p_{t}(i)}{P_{t}} -(y_{t}(i))^{\eta} = 0
\end{equation}
Solving and log-linearizing gives an expression for an individual firm's supply curve
\begin{equation}
\hat{y}_{t}(i)=\frac{1}{\eta}(\hat{p}_{t}(i)-\hat{p}_{t})
\end{equation}
For parsimony, a reduced form aggregate demand curve replaces the Euler equation (18). \begin{equation}
\hat{y}_{t}= \hat{m}_{t}- \hat{p}_{t}
\end{equation}
Idiosyncratic taste shocks, reflected by the disturbance
$z_{t}(i)$ in the individual firms demand schedule, are essential to the Lucas model. I assume they are independent and identically distributed across producers and time, with tail restrictions such that they will cancel out in aggregate almost surely. Here is the log-linear form \begin{equation}
\hat{y}_{t}(i)= \hat{y}_{t} + \hat{z}_{t}(i) - \theta(\hat{p}_{t}(i)- \hat{p}_{t}) 
\end{equation}
There is linear aggregation so 
\begin{equation}
\hat{p}_{t}=\bar{\hat{p}}_{t}(i)
\end{equation}
\begin{equation}
\hat{y}_{t}=\bar{\hat{y}}_{t}(i)
\end{equation}
whilst \begin{equation} \pi_{t}=\hat{p}_{t}-\hat{p}_{t-1}\end{equation} 
Monetary non-neutrality is entirely driven by unexpected shocks. There is no role for price dispersion here, since the firms 
are price-takers and the distribution of relative price is unaffected by inflation.
Unlike its Keynesian counterparts, the Lucas model would be unaffected by trend inflation.
\par The driving force for this model is an inability for producers to distinguish changes in the aggregate price level $P_{t}$ from their idiosyncratic shock $z_{t}(i)$. Instead it only observes the price of its own good $p_{t}(i)$. A decomposition of the individual  price change into real and nominal factors is instructive \begin{equation}
\hat{p}_{t}(i)= \hat{p}_{t} + (\hat{p}_{t}(i)- \hat{p}_{t})\equiv \pi_{t} + r_{t}(i)
\end{equation}
where $r_{t}(i)=\hat{p}_{t}(i)-\hat{p}_{t}$ is relative price inflation - the change in the relative price of good $i$ from period $t-1$ to the current period $t$. Thus the producer observes not current inflation but the sum of relative price change and general inflation. \par There is a \emph{Signal Extraction} problem here, with full information or absent uncertainty about either monetary policy or consumer preferences, the firm would base its production decision on relative prices alone. However, the producer does not observe $\hat{r}_{t}(i)$ but must estimate it given the observable own price $p_{t}(i)$.\footnote{Recall that the firm is owned by a single household. If the household knew others' prices through making purchases, it could deduce $p_{t}(i)$ and hence $r_{t}(i)$. There are several ways to rule this out. The most common is to assume each household can be divided into a "producer" and a "shopper" who do not communicate. Alternatively, with Lucas' original overlapping generation approach the problem is avoided because the individual produces in the first period and shops in the second. \cite{jonung1981perceived}, \cite{huber2011frequency}, \cite{ashton2012real}, \cite{del2016perceived} and \cite{detmeister2016inflation} provide direct empirical support for the signal extraction hypothesis. In particular, behavioral bias coming from the accessibility heuristic (that individuals over-weight common items when estimating their personal inflation rate) can mimic signal extraction, even when information about the aggregate price level is in fact easily available.}
Hence (64) becomes 
\begin{equation}\hat{y}_{t}(i)= \frac{1}{\eta}\mathbb{E}_{t}[r_{t}(i) \; \vert \; \hat{p}_{t}(i)]\end{equation}
Finally, Lucas imposed functional form restrictions on the monetary shock $\hat{m}$ and $\hat{z}_{t}(i)$. They are normally distributed with mean $0$ and variances $V_{m}$ and $V_{z}$ and they are independent. The solution is obtained by guessing and verifying that $\pi$ and $r_{i}$ are normally distributed and independent. The hypothesis yields
\begin{align}  \nonumber 
\mathbb{E}_{t}[r_{t}(i) \; \vert \; \hat{p}_{t}(i)]&=\mathbb{E}_{t}[r_{t}(i)] + \frac{V_{r}}{V_{r} + V_{\pi}}(\hat{p}_{t}(i)-\mathbb{E}_{t}(\hat{p}_{t}(i))) \\ &=\frac{V_{r}}{V_{r} + V_{\pi}}(\hat{p}_{t}(i)-\mathbb{E}_{t}(\hat{p}_{t}(i)))
\end{align}
where $V_{r}$ and $V_{\pi}$ are the strictly positive variances of relative prices and inflation respectively. Note that the ratio of $V_{r}$ to $V_{\pi}$ is the \emph{signal-to-noise ratio}. Finally, substituting (68) into (69), using definition (66) and aggregating with (64) and (65)
yields  the famous Lucas supply curve \begin{equation}
\hat{y}_{t}=b(\pi- \mathbb{E}_{t} \pi_{t})
\end{equation}
where  
\begin{equation}b=\frac{1}{\eta} \frac{V_{r}}{V_{r} +V_{\pi}}\end{equation} To finish it is necessary to solve out for $V_{r}$ and $V_{\pi}$ in terms of the structural parameters of the model $V_{m}$ and $V_{z}$. The approach is as follows: solve for $\pi$ and $y^{e}$ using aggregate demand curve (63) with inflation 
(67) and Lucas supply curve (70) and (71) yields
\begin{equation}
\pi_{t}= \frac{1}{1+b}\hat{m}_{t} + \frac{b}{1+b}\mathbb{E}_{t}\pi_{t}
\end{equation}
\begin{equation}
\hat{y}_{t}= \frac{b}{1+b} \hat{m}_{t} - \frac{b}{1+b}\mathbb{E}_{t}\pi_{t}
\end{equation}
Passing expectations through (68) gives 
\begin{equation}
\mathbb{E}_{t}\pi_{t}= \mathbb{E}_{t}\hat{m}_{t}
\end{equation}
Using (70) and the fact that $\hat{m}_{t}=\mathbb{E}_{t}\hat{m}_{t} + (\hat{m}_{t}- \mathbb{E}_{t}\hat{m}_{t})$, 
(68) and (69) can be rewritten as 
\begin{equation}
\pi_{t}= \mathbb{E}_{t}\hat{m}_{t} + \frac{1}{1+b} (\hat{m}_{t}- \mathbb{E}_{t}\hat{m}_{t})
\end{equation}
\begin{equation}
\hat{y}_{t}=\frac{b}{1+b}(\hat{m}_{t}- \mathbb{E}_{t}\hat{m}_{t})
\end{equation}
Using the individual supply curve (62), demand curve (64), steady state deviation (68) and the Lucas supply curve, (71) and (72) implies the expression for relative price deviation
\begin{equation}
\hat{p}_{t}(i)- \hat{p}_{t}=\frac{\hat{z}_{t}(i)}{\theta + b}
\end{equation}
\begin{equation}V_{r}=\frac{V_{z}}{(\theta + b)^{2}}\end{equation} 
and from (69) and (70) 
\begin{equation}V_{p}= \frac{V_{m}}{(1+b)^{2}}\end{equation}
 This allows me to derive the following implicit formulation final form for the slope of the Lucas surprise Phillips curve 
 \begin{equation}
 b=\frac{1}{\eta}\bigg[\frac{(1+b)^{2}V_{z}}{(1+b)^{2}V_{z} + (\theta +b)^{2} V_{m}}\bigg] \end{equation}
 Lucas focused on the limiting case $\lim\, \theta \rightarrow 1$, which yields 
 $b=V_{z}/ \eta(V_{z} + V_{m})$.
 To arrive at the final form of the Phillips curve, relabel the output gap (there is no steady state market failure) and remove the inflation expectation by imposing that the monetary shock is white noise. This could be justified by comparability with the other models of this section, that do not feature explicit policy shocks.\footnote{More formally, it would arise from a monetary targeting analagous to a Taylor rule. Known as a McCallum rule, it stipulates changing output and inflation from trend, (see for example \cite{cooper1972simulations}, \cite{mccallum1988robustness} and \cite{mccallum1993specification}). In reality, the money supply proved more difficult to control than interest rates (\cite{mishkin2001monetary}).} It could be defended with arguments rehearsed in Proposition 9.
 \begin{equation}
 \pi_{t}=b^{-1}\, y^{e}_{t}
 \end{equation}
 \par The internal logic of the model is communicated through equations (69) and (70) that determine the relationship between the two key macroeconomic variables: the output gap $y^{e}$ and inflation  $\pi$ and the policy instrument, in this case the money supply (in deviation form) $\hat{m}$. The component of aggregate demand that is observed, $\mathbb{E}\, {\hat{m}}$ affects only prices, passing straight through into inflation. However, the part that is unobserved $\hat{m}- \mathbb{E}_{m}$ has real effects. Consider a positive shock to the money supply $\hat{m}$. This increases aggregate demand, producing an outward shift to each individual producers' demand curve. Since aggregate demand is not observed, each supplier's best guess is that some portion of the rise in product demand represents a shock (positive in this case) to individual demand through the relative price. Thus producers increase output. 
 \par The effect of an observed increase in $\hat{m}$ is completely different. Suppose there is an anticipated increase in aggregate demand so $\mathbb{E} \, \hat{m}$ is increased with $m - \mathbb{E} \, \hat{m}$ held constant. Now each producer attributes the rise in demand for their product solely to the expansion of the money supply and thus there is no change in aggregate supply. Therefore observed movements in aggregate demand affect only prices. 
 \par The policy implications of this model, as pointed out by \cite{sargent1975rational} and \cite{barro1976rational}, are stark. Monetary policy understood as systematic movements in aggregate demand, reflected here in $\mathbb{E}\,  \hat{m}$, will have no effect on the path of real variables, summarized here by $\hat{y}$, because agents with rational expectations will anticipate changes in demand and see through them. Only the unpredictable part of demand $m - \mathbb{E} \, m$ will matter for real output. However, this roll of a dice is not what is understood by monetary policy. In particular, counter-cyclical policies based on the common sense, that Central Banks should lean against the macroeconomic cycle, cutting rates during a downturn and raising them when there is an unsustainable boom, are ineffective. It would therefore prove surprising if it could be equivalent to any Keynesian model or proposed approximation to it. 
 Of course it is not: the approximation, not Keynesian economics, is faulty. 
 \subsubsection{Rotemberg Pricing}
This model \cite{rotemberg1982monopolistic} is perhaps the simplest 
formulation of state dependent pricing and the one most commonly compared to Calvo. The idea is that firms face a convex cost of changing prices and therefore they do not adjust immediately to their optimal flexible price. 
The problem is that convex costs of price changing imply firms change their prices every period through small adjustments. This runs contrary to an overwhelming body of evidence of 
infrequent large adjustment discussed in Appendix H.1.3.
\par The household problem, technology and policy rule will be as before. The difference lies in the source of price rigidity. The firm faces a convex cost of changing each price usually parametized as a quadratic and scaled by $c_{p}$
\begin{equation}
C^{a}_{t}(i)=\frac{c_{p}}{2}\bigg(\frac{p_{t}(i)}{p_{t-1}(i)}-1\bigg)^{2}
\end{equation}
Profit maximization requires firms: 
\begin{equation}
\max_{\{p_{T}(i), \, l_{T}(i)\}} \mathbb{E}_{t}\sum_{T=t}^{\infty}Q_{t,\, T}\bigg[\frac{p_{T}(i)}{P_{T}}y_{T}(i)-W_{T}l_{T}(i) -\frac{c_{p}}{2}\bigg(\frac{p_{t}(i)}{p_{t-1}(i)}-1\bigg)^{2}y_{T}(i)\, \bigg]
\end{equation}
Note, unlike with Calvo pricing, the firms' problem is recursive in present discounted profits with $$V_{t}(p_{t-1}(i))=\frac{p_{t}(i)}{P_{t}}y_{t}(i)-W_{t}l_{t}(i) -\frac{c_{p}}{2}\bigg(\frac{p_{t}(i)}{p_{t-1}(i)}-1\bigg)^{2}y_{t}(i) + \mathbb{E}_{t}[Q_{t,\, t+1}V_{t+1}]$$
The first order conditions for an individual firm is
\begin{multline}
y_{t}(i)- \theta p^{*}_{t}(i)  \bigg(\frac{p^{*}_{t}(i)}{P_{t}}\bigg)^{-(\theta +1)}\frac{Y_{t}}{P_{t}} + \theta \bigg(\frac{p^{*}_{t}(i)}{P_{t}}\bigg)^{-(\theta +1)} MC_{t}Y_{t} - \\ c_{p}\bigg(\frac{p^{*}_{t}(i)}{p_{t-1}(i)}-1\bigg)\frac{P_{t}}{p^{*}_{t-1}(i)}Y_{t} + c_{p} \mathbb{E}_{t}Q_{t, \, t+1}\bigg(\frac{p^{*}_{t+1}(i)}{p^{*}_{t}(i)}-1\bigg)Y_{t+1}\frac{p^{*}_{t+1}(i)}{(p^{*}_{t}(i))^{2}}P_{t+1}=0
\end{multline}
where the top line is the marginal revenue net of the cost of price changing and the second reflects the effect on present and future marginal costs of an adjustment today. The dynamics are considerably simpler with 
\begin{equation}\frac{p_{t}(i)}{p_{t-1}(i)}=1+ \pi_{t}\end{equation}
\begin{equation}\Delta =1 \end{equation}
Therefore the first order conditions simplify to 
\begin{equation}
(1- \theta)  + \theta  MC_{t} 
-  c_{p}\pi_{t} (1+\pi_{t}) +  c_{p} \mathbb{E}_{t}Q_{t, \, t+1}\pi_{t+1} (1+\pi_{t+1})\frac{Y_{t+1}}{Y_{t}}=0
\end{equation}
Finally, the resource constraint reflecting a wedge between production and consumption caused by the cost of price changing replaces (20)
\begin{equation}
Y_{t}=C_{t} + \frac{1}{2}c_{p}\pi^{2}_{t}Y_{t}
\end{equation}
Therefore around ZINSS $\hat{c}_{t}=\hat{y}^{e}_{t}$, now using the steady state marginal cost relationship and (52)
I arrive at   
\begin{equation}\pi_{t}=\tilde{\omega}\hat{y}^{e}_{t} + \beta \mathbb{E}_{t}\pi_{t+1} \end{equation}
where 
\begin{equation}
\tilde{\omega}=\frac{(\sigma + \eta)(\theta -1)}{c_{p}}
\end{equation}
\subsection{Lucas Critique}
This part begins by explaining the Lucas critique and its various interpretations. The second subsection sets out observational equivalences and relates them to the Lucas critique framework. This should be viewed as a failure of existing New Keynesian economics. I will go onto demonstrate that the approximation I construct "passes" this crucial aspect of the Lucas critique by breaking this equivalence. In Appendix
C.2.3, I confirm that there are (correctly solved) multiple models of recent vintage that are still ensnared. 
\subsubsection{Interpretations}
\begin{enumerate}
\item \bf{Mapping Micro to Macro} \\ \textnormal{\cite{lucas1976econometric} heralded a new approach to macroeconomics that emphasised that models should be based on aggregation of the optimizing behaviour of agents. This methodology has become ubiquitous in modern research, providing an underlying motivation for all the New Keynesian models in this text. Researchers however have neglected to check solutions to DSGE models preserve the mapping from micro to macro. This paper analyses and exemplifies two threats to this connection. There are \emph{a priori} reduced form approximations used in econometric work that do not represent neighboring stochastic systems 
because of failure to account for singularities. Principle 1 links this phenomenon back to the 
breakdown of constraints in individual optimization problems, akin to the common understanding of microfoundations. Secondly, there arise instances where the local approximation would be correct but \emph{ex post} the dynamics of the reduced form are not consistent with the existence of nearby non-stochastic systems. In fact, these two forces combine to obscure any connection between the behavior of approximation solutions and the developments and mechanisms of an underlying non-linear New Keynesian economy. }
 \item \bf{Observational Equivalence}
\\ \textnormal{The sole formal contribution of \cite{lucas1976econometric} was to demonstrate the equivalence between a new classical Phillips curve and a then standard Keynesian formulation of the relationship. It is a seminal result of macroeconometrics. I am able to show that it still applies to popular New Keynesian formulations but breaks down when Calvo is solved correctly. It demonstrates a fundamental advantage of Keynesian over Classical monetary models. } 
\item \bf{Breakdown of Statistical Relationships}
 \\ \textnormal{Lucas argued that statistical relationships like the Phillips curve\footnote{He had in mind the original Phillips curve, a negative relationship between unemployment and inflation estimated by \cite{phillips1958relation}.} would break down when exploited for policy purposes.\footnote{Similar ideas were advanced in the context of monetary targeting by \cite{goodhart1975problems} without formal mathematical underpinning. The principles are of wide application throughout economics.} This argument gained credence in the 1970s when there was stagflation (simultaneous increases in unemployment and inflation). The idea was central to the rise of New Classical economics, indeed it spurred the creation of DSGE. The result is only partially correct. He was right that the Phillips curve is not policy invariant, as the policy rule parameters affected the optimal supply decisions with drawbacks to monetary activism. He was also correct that there need not be an upward sloping relationship between inflation and output in equilibrium. On the other hand, this does not rule out systematic countercylical policy consistent with the Keynesian consensus (see \cite{snowdon2005modern}). 
 Lastly, the comparative statics, part of Theorem 2, can be viewed as an analysis of policy regime changes of the kind considered by Lucas.} 
 \end{enumerate} 
 \subsubsection{Observational Equivalence}
 This part formalizes the traditional interpretation of the Lucas critique and proves that, subject to parametric conditions, it extends to the current benchmark solution of Calvo and the Rotemberg Phillips curves. 
 The equivalence between 
 these approximations is not novel, although the link back to Lucas is.
 \begin{proposition}{\bf{(Lucas Critique)}}
  There can arise an observational equivalence between the Singular Calvo, Rotemberg and Lucas Phillips curves provided that $\tilde{\omega} > \eta$.
 \end{proposition}
 \begin{proof} Proposition 8 ensures the expectation terms vanish from the two forward-looking models. The rest of the proof is constructive. Equating (54) and (91) yields 
$$c_{p}=\frac{\alpha (\theta - 1)}{(1-\alpha)(1-\alpha \beta)}$$
which sets Rotemberg equal to Calvo. Deploying (81) and equating shows equivalence to the Lucas model when 
$$\frac{V_{m}}{V_{z}}=\frac{( \tilde{\omega}-\eta)}{\eta}\frac{(1+b)^{2}}{(\theta+b)^{2}}$$ the bound follows because these variances have to be strictly positive. 
 \end{proof}
 The interpretation is that provided the Phillips curve is sufficiently steep, then the methods of macroeconometrics cannot be used to distinguish between these three underlying designs for the Phillips curve. As an empirical matter, this is doubtful. Empirical specifications of the Phillips curve strongly favor a flat slope $(\tilde{\omega}<1)$, 
 whilst microeconometric evidence  strongly supports an inelastic labor supply $(\eta>1)$.\footnote{An alternative interpretation is that the cost of nominal rigidity $c_{p}$ is too great to support a sufficiently steep sloping Phillips curve.} Furthermore, there appears to be greater volatility of monetary shocks relative to technology shocks than the model would support. These issues are taken up in detail, in Appendices H and I.
\par The broader interpretation is that these solutions are just barely "passing" the observational equivalence aspect of the Lucas critique. There are many dynamic variants of the Lucas model, featuring more sophisticated information frictions (see \cite{angeletos2021imperfect} and \cite{mackowiak2023rational}). They typically generating flatter inflation-output trade-offs (\cite{afrouzi2021dynamic}). Thus with respect to these versions, forward-looking Phillips curves may still be ensnared by the critique. Conversely, by overturning the singular Calvo solution, I can rule out all these models too. We can do better than this. \par The Divine Coincidence and lack of dynamics have helped motivate a move away from DSGE back to ad hoc models. With new methods and better results, I anticipate a reversal.  
Moreover, armed with corrected approximations to Calvo around ZINSS, it should be possible to search for similar observational equivalences, in a wider class of DSGE. The link with Taylor pricing, established in Section 10, is a starting point. 
\section{Stochastic Equilibrium}
This section defines and develops the stochastic equilibrium methodology that forms the backbone of this paper. There are four subsections. The first provides a short background in the underlying mathematical disciplines of ergodic theory and random dynamical systems, with immediate application to our context. The second explicitly constructs the equilibrium. 
The third compares it with existing informal concepts in the literature. The fourth shows how stochastic equilibrium surfaces can be used to derive global properties of economic and mathematical importance. 
\subsection{Ergodic Theory and Random Dynamical Systems}
I will use ergodic theory, the branch of mathematics that studies the long run behavior of dynamical systems to build the novel equilibrium concept. It is necessary to present several definitions integral to ergodic theory and dynamical systems, as well as clarifying aspects of the mathematical environment. 
\subsubsection{Random Dynamical Systems}
This part is prerequisite for understanding ergodic theory. It sheds light on some of the errors discussed in the previous section, which revealed popular confusion about the properties of the underlying probabilistic system. 
\begin{definition}A measure-preserving dynamical system is defined as a probability space with a measure-preserving transform. $$(X, \,  \mathbb{T}, \, \Sigma , \, \mu, \, T)$$ such that \begin{enumerate} [(i)] \item $X=\prod \limits_{i=1}^{k} X_{i}$ is a reflexive space
and $\mathbb{T}$ the time index \item $\Sigma_{i}$ is a $\sigma$-algebra over $X_{i}$ and $\Sigma=\prod \limits_{i=1}^{k}\Sigma_{i}$ is the $\sigma$-algebra of the product measure on $X$ 
\item $X$ is measurable with respect to $\mu$ with $\mu_{i}: \Sigma_{i} \times X_{-i} \rightarrow [0,1]$ so $\mu_{i}(\emptyset_{i})=0$ and $\mu_{i}(X_{i})=1$ and $\mu=\prod \limits_{i=1}^{k}\mu_{i}$  
\item $T: X \rightarrow X$ is a measurable transformation.
\end{enumerate}
\end{definition}
To analyze the time evolution of the system I introduce 
\begin{definition}
A \textbf{trajectory} of a random dynamical system $T: X \rightarrow X$ from an arbitrary initial position $X_{0}$ is given by the set 
$$\mathcal{O}_{T}=\{X_{0}, \, T(X_{0}),\, T^{2}(X_{0}), \, \cdots \}$$. 
\end{definition}
\begin{remark} The potentially set-valued inverse is called a \emph{path} $T^{-1}$. Non-uniqueness will not cause problems because it arises with zero probability and we will be able to treat special cases with a sandwiching argument. \end{remark}
\begin{remark}Reflexivity is the minimal requirement to admit the existence of (mathematical) expectations (for self-maps). This is critical as the focus here will be on trajectories of the form $T(X)=\mathbb{E}_{t}X_{t+1}$. Consult \cite{conway2013course} for an exposition of reflexive spaces. We will usually be working with manifolds often with boundaries.\end{remark}
\subsubsection{Ergodicity}
The following attributes characterize ergodicity: 
 \begin{definition}
 Let $(X, \, \Sigma, \, \mu)$ and $T: X \rightarrow X$ be a measure-preserving transformation. We say that $T$ is ergodic with respect to $\mu$ (or alternatively that $\mu$ is \textbf{ergodic} with respect to $T$) if one of the following equivalent statements is true 
 \begin{enumerate}[(i)]
 \item for every $E \in \Sigma$ with $T^{-1}(E)=E$ either $\mu(E)=0$ or $\mu(E)=1$
 \item for every $E \in \Sigma$ with $\mu \big(T^{-1}(E)\, \Delta \, E\big)$ we have $\mu(E)=0$ or $\mu(E)=1$ where $\Delta$ denotes the symmetric difference $A \,\Delta \, B=(A \cup B) / (A \cap B) $
 \item for every $E \in \Sigma$ with positive measure we have $\mu \bigg(\mathop{\bigcup}\limits_{s=1}^{\infty}T ^{-s}(E)\bigg)=1$
 \item for every two $A$ and $B$ of positive measure, there exists an $s > 0$ such that $\mu\big({T}^{-s}(A)\cap B\big) > 0$
 \item Every measurable function (formally $j \in L^{1}(\mu)) $  with $j \circ T= j$ is \emph{almost surely} (a.s) constant.
 \end{enumerate}  \end{definition}
 Intuitively, ergodicity is the property that a system forgets its initial position and that it has well-defined long-run behavior. It is a property of the system as a whole or a subset that behaves as a component of the whole process. Conditions (ii)-(iv) state that if you move far enough back or forward in time (almost) any two positions of the system will occur in probability. \cite{aliprantis2007infinite} Theorem 20.7 provides a proof that these conditions are equivalent. 
 \par For Markov chains, ergodic measures have a powerful uniqueness property. Although, a state space may have multiple ergodic sets, a given ergodic set can only possess one ergodic measure, for proof consult \cite{hairer2018ergodic} Theorem 5.7 p. 40-42. Part (iv) can be interpreted as a weak restriction on hysteresis, that is intuitive in a business cycle context. The final part is the long run equilibrium property. It implies that the ergodic measure sets all moments to their long term values. The "almost sure" aspect allows us to associate it with a point in space as we would a non-stochastic steady state. 
 \begin{definition} The \textbf{Stochastic Equilibrium} of $\mathbb{E}_{t}X_{t+1}=f(X_{t}, \, \gamma )$ is an ergodic fixed point of the dual operator 
$\Upsilon: \mathcal{P}\rightarrow \mathcal{P}$ where $\mathcal{P}$ is the space of probability measures.\end{definition}
In fact, we can interpret $\mathcal{P}$ as the space of probability distributions here. Therefore, its evolution can be represented by the stochastic kernel $p(X_{t+1})=\int p(X_{t}, \, X_{t+1})\,\mathrm{d}X_{t}$, where $p(X_{t}, \, X_{t+1})$ is the joint distribution of the current and next period state of the economy.\footnote{To track the individual distributions, we can use Sklar's theorem to uniquely decompose the joint distribution into marginals and a copula describing the interdependence, as follows $p(X)=(p_{1}(X_{1}),\, \cdots \, , \, p_{n}(X_{n}),\, C(X))$ (see 
\cite{nelsen2006introduction}).} 
\footnote{Note that since $\textbf{p}(X_{t}) \in L^{1}(\mu)$, we can view the ergodic distribution $\textbf{p}(X)$ as a suitable non-linear average (GSM) of the sequence of probability distribution functions $\textbf{p}(X_{t})$. It conforms with the GSM part 1 of Definition A.1 in Appendix B.1.2. The first condition is met because $\mu$ is a valid probability measure and therefore integrates to one. The second condition is met by monotonicity, (uniform) continuity of the Lebesgue integral and the properties of a measure-preserving transform.} 
\par To analyze dynamics and stability properties of
equilibrium it is necessary to introduce stronger restrictions known as mixing conditions:
\begin{definition}For a measure-preserving transform $T: X \rightarrow X$ the following are defining conditions for every $A, B \in \Sigma$ \begin{enumerate}[i]
\item  \textbf{Strong Mixing} $$\lim_{s \rightarrow \infty}\mu(A \cap T^{-s}(B))=\mu(A)\mu(B)$$
 \item \textbf{Weak Mixing}
$$ \lim_{s \rightarrow \infty}\frac{1}{s}\sum_{t=0}^{s}\lvert \mu(A \cap T^{-s}(B))-\mu(A)\mu(B)\rvert \rightarrow 0$$
\item \textbf{Ergodic}
$$\lim_{s \rightarrow \infty} \frac{1}{s}\sum_{t=0}^{s}  \mu(A \cap T^{-s}(B)) \rightarrow
\mu(A)\mu(B)$$
\end{enumerate}
\end{definition}
\begin{proposition}{\textbf{(Ergodic Hierarchy)}}\\
Strong Mixing $\Rightarrow$ Weak Mixing $\Rightarrow$ Ergodicity
\end{proposition}
\begin{proof}
Provided by Theorem 20.11 and Corollary 20.12 (p 662-664) in \cite{aliprantis2007infinite}.
\end{proof}
The most demanding condition, strong mixing, corresponds to stochastic equilibrium for the recursive equilibrium. It means that we can consider a stochastic steady state as probabilistically equivalent to its non-stochastic counterpart. Weak mixing allows for fluctuations that die out on average. The realised value given by the mapping $X_{t+1}=f'(X_{t}, \, \gamma, \, U_{t})$ would be weak mixing, assuming there were persistence, as in the Calvo but there would be strong mixing under Rotemberg.\footnote{In fact, under these conditions it would be true for any purely forward-looking model provided the innovations were independently and identically distributed.} Ergodicity without mixing admits fluctuations that cancel out but do not die away. An economic example would be seasonal variation, not present here. 
\par The final result clarifies that an ergodic measure sends all (non-divergent) moments to their long run equilibrium values. It should be of general economic interest and application. 
\begin{theorem}(\textbf{Birkhoff}) For a measure-preserving transform $T: X \rightarrow X$ on $(X, \Sigma, \mu )$ and a function $j(X_{t}) \in L^{1}(\mu)$ with time average \ $$\mathcal{A}_{\mathbb{T}}(j(X_{0}))=\lim_{s \rightarrow \infty}\frac{1}{s}\sum_{t=0}^{s} {j(T^{s}X_{0})}$$ and ensemble average $\mathcal{A}_{n}=\int j \, \mathrm {d}\mu$ then the two coincide for $\mu$ a.e. $X$. \end{theorem}
Consult \cite{hairer2018ergodic} Theorem 5.2 p 38-40 for a proof. 
\subsection{Equilibrium Construction}
Here I characterize the stochastic equilibrium of the Calvo New Keynesian economy. The main tool is Birkhoff's ergodic theorem. The stochastic steady state resembles its non-stochastic counterpart,
except there are expectation terms attached to future realizations of non-linear functions of the variables 
\newline $(\pi, \, y, \,  \Delta, \, \psi, \, A)$ 
underpinning the recursive equilibrium, described back in Section 4.8 and Proposition 4. Current values of this set can be interpreted as their long run mathematical expectation. Existence results will be established in Section 8. 
\par Starting with the Phillips curve, in particular the  numerator of the reset price relationship 
from Proposition 1 given by combining infinite horizon expression
(28) and recursive form (30)
\begin{multline}
\aleph(\pi,\, Y,\, \Delta)=\psi u'(Y)Y\frac{\nu'(\Delta Y/A)}{A} \\ + \alpha \beta \mathbb{E}\sum_{T=t+1}^{\infty}(\alpha \beta(1+\pi)^{\theta})^{T-(t+1)}\psi u'(Y)Y\frac{\nu'(\Delta Y/A)}{A}
\end{multline}
The second term $\mathbb{E}_{t}\aleph_{t+1}$ evaluated at the ergodic measure can be expressed as the following iterated integral, thanks to the stochastic equilibrium property\footnote{This integral is technically a Bochner or strong integral, the natural extension of Lebesgue integral to infinite dimensional space, whereas that of (28) is a more general Pettis integral.}
\begin{equation}
 \sum_{T=t+1}^{\infty}(\alpha \beta)^{T-(t+1)}\int \cdots \int (1+\pi)^{\theta(T-(t+1))}\psi u'(Y)Y\frac{\nu'(\Delta Y/A)}{A}\mathrm{d}\mu^{T-(t+1)}
\end{equation}
Now applying Tonelli's theorem 
and the fixed point property of the ergodic measure,
we can write the term from date $T-t$ term as 
\begin{equation}\bigg((\alpha \beta)\int (1+\pi)^{\theta}\mathrm{d}\mu\bigg)^{T-(t+1)}\bigg(\alpha \beta \int (1+\pi)^{\theta}\psi u'(Y)Y\frac{\nu'(\Delta Y/A)}{A}\mathrm{d}\mu \bigg)\end{equation}
where the left hand term represents expected gross inflation cumulated from the present up to period $T-(t+1)$, whilst the right hand term is the contribution of marginal costs at time $T-t$, both evaluated at the ergodic measure. This forms a geometric progression, with initial value equal to the right hand term and ratio $\alpha \beta \mathbb{E}(1+\pi)^{\theta}<1$, otherwise the sum diverges and the steady state does not exist.
Hence, the expression for $\aleph$ in stochastic steady state is
\begin{equation}
\aleph(Y, \, \pi, \, \Delta)=\psi u'(Y)Y\frac{\nu'(\Delta Y/A)}{A}+\alpha \beta \frac{\mathbb{E} (1+\pi)^{\theta}\psi u'(Y)Y\nu'(\Delta Y/A)/A}{1-\alpha \beta \mathbb{E}(1+\pi)^{\theta}}
\end{equation}
by the same method for the weighting equation 
\begin{equation}
\beth(Y,\, \pi,\, \Delta)= \psi u'(Y)Y + \frac{\alpha \beta\mathbb{E}(1+\pi)^{\theta-1}\psi u'(Y)Y}{1-\alpha \beta \mathbb{E}(1+\pi)^{\theta-1}}
\end{equation}
combining gives the basis for the Stochastic Equilibrium Phillips curve. 
\begin{multline} \bigg(\frac{1-\alpha}{1-\alpha(1+\pi)^{\theta-1}}\bigg)^{1/(\theta-1)}= \\ \frac{\theta}{\theta-1}\bigg(\psi u'(Y)Y\frac{\nu'(\Delta Y/A)}{A} +\alpha \beta \frac{\mathbb{E} (1+\pi)^{\theta}\psi u'(Y)Y\nu'(\Delta Y/A)/A}{1-\alpha \beta \mathbb{E}(1+\pi)^{\theta}}\bigg) \\ \Bigg/\bigg(\psi u'(Y)Y + \frac{\alpha \beta\mathbb{E}(1+\pi)^{\theta-1}\psi u'(Y)Y}{1-\alpha \beta \mathbb{E}(1+\pi)^{\theta-1}}\bigg) \end{multline}
Price dispersion evolves as follows 
 \begin{equation}
\Delta=\mathbb{E}\Delta=\frac{1}{(1-\alpha)^{1/(\theta-1)}}\frac{\mathbb{E}(1-\alpha(1+\pi)^{\theta-1})^{\theta/(\theta-1)}}{(1-\alpha \mathbb{E} (1+\pi)^{\theta})}
\end{equation}
It is clear that the asymptote in all these equations corresponds to the case where $\Delta$ grows unbounded. The Euler equation takes the form 
\begin{equation} \psi u'(Y)= \beta (1+i) \mathbb{E}\frac{\psi u'(Y)}{(1+\pi)}\end{equation}
Hence the equilibrium real interest rate, known as the natural rate of interest rate, is
\begin{equation}r \equiv \mathbb{E}\frac{(1+i)}{1+\pi}-1 = \frac{\mathbb{E}\psi u'(Y)}{\mathbb{E}\psi u'(Y)/(1+\pi)}-1\end{equation}
Similar expressions could be written for the quantities $U$ and $\Pi$ but are suppressed in the interest of space. 
\subsection{Literature Comparison}
This third subsection sets the new equilibrium concept in the context of the existing literature. This allows me to draw out salient features and delineate its applicability.
\cite{meyer2014risky} succinctly summarizes existing equilibrium notions in the following framework 
\begin{equation}x_{t}=g(x_{t-1}, \, k \epsilon_{t}, \, k)\end{equation}
where $k$ is a dummy variable: where $k=1$ indicates the world is stochastic and $k=0$ that it is non-stochastic. The three equilibrium concepts called  deterministic steady state, stochastic steady state and ergodic mean are defined as follows 
\begin{equation}\bar{x}^{det}_{t}=g(\bar{x}^{det}, \, 0, \, 0)\end{equation}
\begin{equation}\bar{x}^{stoch}_{t}=g(\bar{x}^{stoch}, \, 0, \, 1)\end{equation}
\begin{equation}\bar{x}^{mean}_{t}=\mathbb{E} x= \mathbb{E}g(x, \, 0, \, 1)\end{equation}
Stochastic equilibrium has the property that 
\begin{equation}\mathbb{E}Z= f(\mathbb{E}Z, \, 0, \, 1)\end{equation}
Thus for the variables defining the recursive equilibrium (and affine combinations) stochastic equilibrium unites stochastic steady state and ergodic mean. Note that nonlinear functions of defining variables will not generally be at their mean value in the present period. For example,
away from $\sigma=1$, $MC(Z) \neq \mathbb{E}MC$ since in general $MC(\mathbb{E}Z) \neq \mathbb{E}MC$.\footnote{Z is unique up to Borel isomorphism (see \cite{dudley2018real}). Therefore, you could redefine the cocycle with $Z'$ that included MC but left out one more of the variables in $Z$ defined in Section 4. You would obtain an alternative map $f'$ in which you could interpret that $MC = \mathbb{E}MC$, although in return you would have to give up the possibility of giving this interpretation to another variable left out of $Z'$. With $MC$ there would not be a problem but if higher moments were included you would have to verify the relevant expectations existed.} 
\par However, results need not carry over to continuous time. For example, \cite{fernandez2023financial} solve a heterogeneous agent model with non-linear interactions between asset prices and precautionary saving. They find three stochastic steady states (corresponding to $\mathrm{d}Z_{t}=0$), of which two are locally stable. The reason is that in continuous time there is almost surely no incremental uncertainty\footnote{This could change if there were jump processes.} (as the future, present and past are arbitrarily close together). Therefore, there can be multiple stochastic steady states, in the way that there can be multiple non-stochastic steady states in discrete time. However, there is a unique ergodic invariant measure, with multiple steady states corresponding to a multi-modal distribution.
The issue is that the model does not possess strong mixing; just because a fixed point is locally stable does not mean the economy is expected to stay there in the long-run.\footnote{Formally, the incremental generating function from which the expectation is derived is weakly rather than strongly mixing, unlike $f$ the corresponding object in the discrete time case constructed in Proposition 4.}
 It would be interesting to see how findings here carry over to models with multiple steady states and whether these techniques offer any traction in challenging continuous time environments. Beyond mathematical interest, it might help applied researchers decide between continuous and discrete time in specific applications. 
 \par Finally, a concrete formulation of stochastic equilibrium may be of deeper theoretical interest. It has the desirable property that it is a fixed point concept associated with a single point in space, that is nevertheless determined entirely by out of equilibrium dynamics. This is clearly an improvement on previous disequilibrium modelling.
Typically contributions have been static. Dynamic analysis has tended to use temporary equilibrium, where expectations are exogenous, as in \cite{barro1971general}, \cite{dreze1975existence}, \cite{hahn1978non}, \cite{benassy1990non} and \cite{grandmont1982temporary}. In fact, it can be seen as a synthesis of Keynesian adjustment and traditional general equilibrium theory.\footnote{The construction here speaks to Smale's eighth problem about the inclusion of explicit price adjustment in general equilibrium. His inclusion of this economic problem in his list of major problems for twenty first century mathematics was esoteric (\cite{smale2000mathematical}.) It harked back to Hilbert's problems whose sixth referred to theoretical physics (\cite{hilbert1902mathematical}.) I suspect it will receive extensive mathematical study. }
\subsection{Equilibrium Analysis}
This final part sets out local and global properties learned from applying basic analysis techniques and then moves onto discuss their economic and mathematical significance. 
\begin{theorem}(\textbf{Surface Estimates / Comparative Statics})\par The following relationships exist between stochastic and non-stochastic values 
\begin{enumerate}[i]
    \item $\mathbb{E}i < i^{NSS}$
    \\ suppose further that $\pi > \underline{\underline {\pi}}<0$ and $\theta > 2 $ then
    \item $\Delta > \Delta^{NSS}$
    \\ moreover if $\sigma \geq 1$ then 
    \item  $Y < Y^{NSS}$ 
   \newline  where all parametric bounds are sharp. 
\end{enumerate}
\end{theorem}
Proofs typically involve applications of Jensen's and covariance inequalities, along with monotonicity arguments. They are contained in the Appendix A. 
The next result has a comparative statics flavour but a different style proof. It will be used to specify the distributional consequences of moving from a non-stochastic to a stochastic environment. Beforehand, it is necessary to set out a couple of assumptions defended later on. 
\begin{assumption} One of the following statements is true: (i) the kernel of the stochastic discount factor $M=\mathbb{E}u'(Y)/u'(Y)$ covaries positively with profits (ii) all shocks are small (iii) discount factor shocks are small and profits are procyclical (in the sense that $\Pi$ is stochastically monotone with respect to $Y$)\end{assumption}
The first condition has been extensively tested in the financial econometrics literature- examples include \cite{fama1992cross}, \cite{ait1998nonparametric}, \cite{alvarez2005using}, \cite{gourieroux2007econometric}, \cite{hansen2009pricing}, \cite{fama2015five} and  \cite{christensen2017nonparametric}. Despite anomalies it has received broad support across a variety of asset markets. 
The second case is almost trivial. The idea behind the third case is that we could isomorphically replace the discount factor shock with a disturbance to the policy rule.\footnote{This applies even if we expand our definition of a preference shock to include any disturbance to the financial system. Changes in policy-making have arguably been more significant in many periods- as suggested by studies such as \cite{stock2002has}, \cite{muir2017financial} and \cite{ascari2019walk}.}
Recall that (positive) stochastic monotonicity is where $Y(\Pi_{1})$ stochastically dominates $Y(\Pi_{2})$ if and only if $\Pi_{1} \leq \Pi_{2}$.\footnote{This means that $\mathbb{P}(Y(\Pi_{1})\leq y) \leq \mathbb{P}(Y(\Pi_{2})\leq y)$ if and only if $\Pi_{1} \leq \Pi_{2}$.} It is discussed extensively in \cite{shaked2007stochastic} where it is called the "usual order". It is designed to control all aspects of the distribution by a single parameter in this case profit. Thereby stopping for example 
profits at the top of the distribution moving in different directions to the main course of the business cycle (central moment) or those of the least profitable firms. It has some intuitive economic appeal.
\begin{assumption}
$\mathbb{E}A$ is sufficiently large relative to $\Delta$.
\end{assumption}
\begin{proposition} $0< \mathbb{E} \Pi < \Pi^{NSS}$, where the lower bound claim uses Assumption 1 and $\beta \rightarrow 1$, whilst the upper bound supposes that Assumption 2 holds, along with the parametric conditions from Theorem 2. \end{proposition}
The first part follows from the construction of stochastic equilibrium, a suitable comparison of the resetters problem and the distribution of profits at a point in time. The proof of the second part has a microeconomic flavour, involving a suitable Slutsky decomposition. The subsequent remarks clarify the role and significance of the assumptions.
\begin{remark}Suppose the economy could accommodate a positive growth rate then we could guarantee that the normalized value of profits with shocks $\mathbb{E}\Pi/A$ would be below its non-stochastic counterpart. \end{remark}
This idea is fleshed out in 
Appendix D.2.3.
Alternatively we could scale down the units of the labor endowment to make $A$ arbitrarily large. 
\begin{remark} 
Assumption 1(i) and (iii) the more appealing alternatives, are not  primitive restrictions, thus Proposition 13 is not a true quantitative estimate. Nevertheless, I suspect that a more appealing primitive restriction may arise as a bi-product of rigorous global analysis of the model. \end{remark}
Macroeconomists have traditionally grouped variables into pro and counter-cyclical, in order to summarize information about the economy. The assumption that gross profits are pro-cyclical is a very weak one. Sales are pro-cyclical (in this model) by definition. Markups would have to be strongly countercyclical to overcome this. In fact recent evidence suggests they are pro-cyclical.\footnote{\cite{nekarda2020cyclical} show that the inverse labor share measure directly pertaining to this model is pro-cyclical. Using detailed micro data, \cite{anderson2023markups} conclude retail mark ups are weakly pro-cyclical. \cite{domowitz1986business} uncovered similar patterns for manufacturing. \cite{farinas2003profit} and \cite{macallan2008cyclicality} provide corroborating evidence for a wider set of countries and industries.  \cite{nekarda2020cyclical} describe how procedures involving more sophisticated production functions can lead to different results that are not relevant in the present setting. Finally, they note how cyclicality would be expected to depend on the distribution of shocks. With Keynesian intuition and their econometric strategy, they suggest that markups were pro-cyclical in response to technology shocks but counter-cyclical when there were shocks to investment.}
Hence, these conditions suffice from an empirical point of view but not from a mathematical vantage, as we have not demonstrated this prediction is consistent with the model. Nevertheless, the approach may be of interest to economic theorists, as it allows one to derive rigorous predictions, without the complexities of in-depth mathematical analysis.\footnote{This strategy should combine profitably with simulations to verify the assumptions.}
\par Finally, I am able to confirm that in the patient limit the serious pathology of negative stock prices does not arise.
\begin{proposition} The value of every firm $\mathcal{V}^{\Pi}_{t}(i) >0$ as $\beta \rightarrow 1$ \end{proposition}
This problem can afflict alternative models with costly price changes. The intuition behind the proof is that as discounting dies away, the value of the firm is dominated by prices in the distant future, by which time the firm will in expectation have had an opportunity to reset its prices and from the point of view of the present, uncertainty will recede as stochastic equilibrium is approached. The result confirms the reconciliation of classical theories of entry and exit with sticky prices- thus supporting neoclassical synthesis and completing this round of comparative analysis.
\par Therefore, the likely distributional consequences of incorporating uncertainty can be summarized as follows. Relative price distortions will decrease the value of output, which will decrease wages and real marginal costs. In equilibrium, output will fall and labor input will have to increase to make up for some but not all the loss of effective productivity. Profitability will fall. Welfare is lower, indicating that volatility is costly.  Finally, uncertainty will cause the interest rate to drop. These findings feel economically intuitive and for the most part should stand up to alternative model details, although the labor supply prediction might change with additional frictions.
\par Moreover, the new construction offers further applications. 
\begin{corollary} When $\beta \rightarrow 1$ the real interest rate  $r=i-\pi<0$ \end{corollary}
\begin{proof} Follows immediately from Theorem 3 (i) and sending $\beta \rightarrow 1 $, whilst keeping the shock size fixed. \end{proof}
Very low interest rates were widely documented from the 
2008 Global Financial Crisis  until the early stages of the COVID pandemic. Long run rates indicate these were expected to continue for a long time.\footnote{On July 11th 2020 \emph{The Economist} reported that Austrian 100 year bonds were yielding just $0.7\%$. At the ECB's official target of $2\%$ and assuming a trend growth rate of $2\%$ then it would be reasonable to calculate $r-g = -3\%$.}
Prominent studies, such as \cite{del2019global}, estimated 
a negative natural rate and estimates fell further during the acute stage of the pandemic (see for example \cite{van2021benefits}).
However, non-stochastic growth models cannot support real rates below growth rates.
\par It is intuitive that in a stochastic environment, government debt without default risk can command a safety (negative risk) premium. 
\cite{brunnermeier2020fiscal} are able to demonstrate this by deriving a closed form solution to a model, with a financial sector and a bubble in the government debt, solved in continuous time. I am able to dispense with these esoteric assumptions. I therefore show, that this prediction is in fact a general consequence of risk aversion, a fundamental economic mechanism.  
\par In general, the analysis mediates in favor of discrete over continuous time for the majority of macroeconomic applications. 
Macroeconomists typically adopt continuous time in search of closed form solutions. These require strong functional form assumptions, typically Brownian motion (also called a Wiener process).
Furthermore, unless you have high frequency data, it makes convincing econometrics more challenging. On the other hand, discrete time stochastic equilibrium analysis allows one to obtain quasi-closed forms without specifying particular error distributions, with easy to interpret existence conditions. It also avoids challenges with ascertaining whether particular derivatives exist, which may not be intuitive in many economic situations. 
\par Finally, stochastic surface solutions provide a promising avenue for parametric comparative statics. Under the limiting assumption justifying the use of standard linearization techniques, we can simply transfer results from the non-stochastic steady state to neighboring stochastic environments. It is instructive to look beyond this case at the effect of larger shocks on comparative statics.
I focus on the simplest relationship between price dispersion and the inflation target $\pi$. 
\begin{proposition}(\textbf{Trend Inflation}) Price dispersion $\Delta$ is strictly increasing for
$\underline{\underline{\pi}}<\underline{\pi}<0$
\end{proposition}
Geometrically the inflation-price dispersion schedule is shifted up and translated to the left in the range of primary interest.
 Price dispersion will be minimized at a negative rate of trend inflation. This might provide a rationale for deflationary policy.\footnote{On this theme, \cite{yun2005optimal} uses price dispersion to justify temporary deflation in a non-stochastic environment. \cite{schmitt2010optimal} discusses alternative rationale for negative inflation. 
On the other hand, \cite{adam2019optimal} are able to rationalize positive inflation by adding heterogeniety in firm level productivity growth into our Calvo framework.}
This also has a dynamic interpretation. Stochastic equilibrium pulls down dynamic behavior of price dispersion from a positive trend inflation environment to zero and nearby negative rates of inflation.
It suggests there need not be a difference in dynamic behavior between low trend inflation and the low rates of trend deflation typically observed. \cite{hirose2020estimated} is the only paper I know of to date that looks at trend deflation, in the context of Japan, and he works with a prior trend centered at just $-0.214 \%$ (see their seventeenth footnote).
\par Lastly, the analysis here could be of general mathematical and statistical interest. Firstly, the Theorem is an example of mathematical rigidity, in the sense that I have deduced important statistical properties of the solution by analyzing just a single neighborhood. Secondly, the results constitute sharp \emph{a priori} quantitative estimates of important expectation functionals. They are sharp because the non-stochastic steady state is the limit of the stochastic steady state. They are \emph{a priori} because we have not yet proven the solution exists; in some instances we will see that a solution does not exist.
\par In econometric terms, they represent over-identifying conditions. They constitute restrictions on the solution associated with the need for infinite horizon optimization problems to exist. They are the analogy of the steady state that provide additional restrictions on the coefficients of any reduced form representation, like 
the one presented in the next section. These kind of conditions are rarely seen in statistics where exact identification is predominant or mathematical physics where under-identification is ubiquitous.\footnote{For example, in quantum theory Heisenberg's uncertainty principle (\cite{heisenberg1927anschaulichen}) tells us that we cannot know the position and the velocity of an electron, so given the velocity of an electron, its position is unidentified. Conversely, given the speed the position is unidentified - hence the system as a whole is under-identified. String theorists are interested by mirror symmetry where different shaped universes are observationally equivalent through certain topological lenses (see \cite{greene2000elegant} and \cite{hartnett2018mathematicians}). There are exceptions, for example, researchers studying small sets of rigid bodies that are over-determined. Nevertheless, the concept of over-identification, central to structural econometrics, would seem foreign to the majority of mathematical physicists and the particular kind here are surely novel. This likely explains the failure of mean field game theory to gain traction over DSGE.} 
\section{General Linearized Phillips Curve} 
\par This section sketches the derivation of the Phillips curve log-linearized around any stochastic equilibrium. It rests on an application of two stochastic Grobman-Hartman theorems. Its formal justification will be supplied in the next two sections. The first part focuses on the slope coefficients and is supported by Appendix E, whilst the second solves out the structural errors around ZINSS. 
A full analysis of the properties of the error coefficients elsewhere is beyond the scope of this paper.
\par The surprising feature of this new linear approximation
is that although it is certainty equivalent in deviations, this is not true of the whole system because the coefficients represent higher moments of the distributions. This novel aspect will allow this approximation to summarize critical aspects of the dynamical system. This approximation framework should prove a natural setting to analyze the dynamic and statistical properties models with idiosyncratic or large aggregate risks. This will require new computational and econometric routines, which I will touch on here.
\par The main quantitative work will focus on limiting cases comparable or in some cases equivalent to existing linearization designs. The solution will feature extensively manipulations with the lag operator. These will reappear prominently in the bifurcation analysis of Section 9, where I will also draw precise connections between terms in the Phillips curve and singularities like those discussed in Section 3.
\par During the course of the derivation it is essential to keep track of which expressions are to be evaluated at their present certainty equivalent value and those which are evaluated at their expectation under the equilibrium measure. As the derivation progresses, these will become mixed up inside the coefficient expressions. In future applications, if one is only interested in small noise limiting applications it is possible to simplify these steps by effectively evaluating all the stochastic coefficients at their non-stochastic values and working directly with the efficient output gap $\hat{y}^{e}$, as I do here in 6.1.3 and 6.2.2. 
\subsection{Slope Coefficients} The first subsection has the derivation and the second houses the main results. 
\subsubsection{Derivation}
It is most convenient to start with (30) and (31) then apply Theorem 4 from \cite{coayla2007hartman}, appropriate for (convergent) infinite horizon trajectories \begin{multline}
\hat{\aleph}_{t}=\frac{ \psi \nu'(\Delta Y/A)Y}{A \aleph}\hat{\psi}_{t} + \eta\frac{ \psi \nu'(\Delta Y/A)Y}{A \aleph} \hat{\Delta}_{t}+(1+\eta)\frac{ \psi \nu'(\Delta Y/A)Y}{A \aleph}\hat{y}_{t}- \\ (1+\eta)\frac{ \psi \nu'(\Delta Y/A)Y}{A \aleph}\hat{a}_{t} +  \frac{\alpha \beta \theta}{\aleph}\mathbb{E}(1+\pi)^{\theta-1}\aleph\, \mathbb{E}_{t}\hat{\pi}_{t+1} + \alpha \beta \mathbb{E}(1+\pi)^{\theta}\, \mathbb{E}_{t}\hat{\aleph}_{t+1} 
\end{multline}
\begin{multline}
\hat{\beth}_{t}= \frac{\psi u'(Y)Y}{\beth}\hat{\psi}_{t} + \frac{\psi u'(Y)(1-\sigma)}{\beth}\hat{y}_{t} + \frac{\alpha \beta (\theta-1)}{\beth}\mathbb{E}(1+\pi)^{\theta-2}\, \beth \, \mathbb{E}_{t}\hat{\pi}_{t+1} + \\  \alpha \beta \mathbb{E}(1+\pi)^{\theta-1}\mathbb{E}_{t}\hat{\beth}_{t+1} 
\end{multline}
where I have used the following substitution derived from (19) and (31) $$u'(Y)Y MC =  \frac{\nu'(\Delta Y /A)Y}{A}$$
\par The comparative statics are instructive, $\hat{\aleph}_{t}$ the business cycle deviation of weighted marginal costs is increasing in $\hat{\Delta}_{t}$ and $\hat{y}_{t}$, as these place upward pressure on real wages, whilst greater technical efficiency $\hat{a}_{t}$ has the opposite effect. The state of expected inflation $\mathbb{E}_{t}{\hat{\pi}}_{t+1}$ influences weighted marginal costs through the level of future sales and the next periods expectation $\mathbb{E}_{t}\hat{\aleph}_{t+1}$ reflects the recursive evolution of the supply side of the economy under Calvo pricing. Conversely, $\hat{\beth}_{t}$ gives the deviation of the weights for the resetters' marginal revenue. There is no role for price dispersion or technology here. There are forward looking terms similar to the supply side.
\par The effect of changes in output on marginal revenues depends on the propensity for inter-temporal substitution reflected by $\sigma$ (the inverse of the elasticity of substitution.) If $\sigma > 1$ the propensity to smooth consumption dominates the incentive to substitution over time, such that higher output today implies higher expected future marginal revenues- conversely, if $\sigma < 1$ the substitution exceeds the smoothing incentive so higher current output is associated with lower future marginal revenues. When $\sigma =1$ the two forces balance out. Appendix D presents an intuitive argument for $\sigma =1$ valid for this benchmark formulation.
\par To solve out the model, I use the lag and expectation operators to condense the expressions for $\hat{\aleph}$ and $\hat{\beth}$ to 
\begin{multline}(1-\alpha \beta \mathbb{E} (1+\pi)^{\theta}\,\mathbb{L}^{-1})
\hat{\aleph}_{t}=
\eta\frac{ \psi \nu'(\Delta Y/A)Y}{A \aleph} \hat{\Delta}_{t}+(1+\eta)\frac{ \psi \nu'(\Delta Y/A)Y}{A \aleph}\hat{y}_{t}\\ -(1+\eta)\frac{ \psi \nu'(\Delta Y/A)Y}{A \aleph}\hat{a}_{t} +\alpha \beta \theta \frac{\mathbb{E}(1+\pi)^{\theta-1}\, \aleph}{\aleph} \mathbb{L}^{-1}\hat{\pi}_{t}+ v_{0}(\hat{u}_{t},\, \hat{u}_{t+1})
\end{multline}
\begin{multline}
(1-\alpha \beta \mathbb{E} (1+\pi)^{\theta-1}\, \mathbb{L}^{-1})\hat{\beth}_{t}= \frac{\psi u'(Y)Y(1-\sigma)}{\beth}\hat{y}_{t} 
+ \\  \alpha \beta (\theta-1)\frac{\mathbb{E}(1+\pi)^{\theta-2}\, \beth}{\beth} \mathbb{L}^{-1}\hat{\pi}_{t} + v_{1}(\hat{u}_{t}, \, \hat{u}_{t+1})
\end{multline}
The error terms above reflect the difference between the actual and expected values of $\pi$, $\aleph$ and $\beth$ in period $t+1$ that comes about because the future value of the structural jump errors are unknown and the model is linear in percentage deviation form. Now using the reset price equation to remove $\aleph$ and $\beth$, the price level construction equation to express the reset price in terms of inflation and then manipulating terms in the lag operator yields
\begin{multline}
(\mathbb{L}-\alpha\beta\mathbb{E}(1+\pi)^{\theta})(1-\alpha \beta \mathbb{E}(1+\pi)^{\theta-1}\mathbb{L}^{-1})\hat{\pi}_{t}=\frac{(1-\alpha(1+\pi)^{\theta-1})}{\alpha (1+\pi)^{\theta-2}}\\ \times \bigg[(\mathbb{L}-\alpha \beta \mathbb{E} (1+\pi)^{\theta-1})\bigg(\eta \frac{\nu'(\Delta Y/A)Y}{A \aleph} \hat{\Delta}_{t}+(1+\eta)\frac{ \nu'(\Delta Y/A)Y}{A \aleph}\hat{y}_{t} \\ -(1+\eta)\frac{\nu'(\Delta Y/A)Y}{A \aleph}\hat{a}_{t} +\alpha \beta \theta \frac{\mathbb{E}(1+\pi)^{\theta-1}\, \aleph}{\aleph}\mathbb{L}^{-1}\hat{\pi}_{t} \bigg)\\ -(\mathbb{L}-\alpha \beta \mathbb{E}(1+\pi)^{\theta})\bigg(\frac{\psi u'(Y)Y(1-\sigma)}{\beth}\hat{y}_{t}+\alpha \beta (\theta-1)\frac{\mathbb{E} (1+\pi)^{\theta-2} \, \beth}{\beth}\mathbb{L}^{-1}\hat{\pi}_{t}\bigg)\bigg] \\ + 
v_{2}(\hat{u}_{t-1}, \, \hat{u}_{t},\, \hat{u}_{t+1})
\end{multline}
Expanding the lag operator, collecting terms and passing expectations from time $t$ yields 
\begin{multline}
\hat{\pi}_{t-1}-\alpha \beta \, \mathbb{E}(1+\pi)^{\theta-1}\, 
(2+\pi)\hat{\pi}_{t} +(\alpha \beta)^{2}\, \mathbb{E}(1+\pi)^{\theta-1}\, \mathbb{E}(1+\pi)^{\theta}\, \mathbb{E}_{t}\hat{\pi}_{t+1}=\\ \frac{1-\alpha(1+\pi)^{\theta-1}}{\alpha (1+\pi)^{\theta-2}}\frac{\eta \nu'(\Delta Y/A)Y}{A \aleph}\hat{\Delta}_{t-1} \\ + \frac{1-\alpha(1+\pi)^{\theta-1}}{\alpha (1+\pi)^{\theta-2}}\frac{(1+\eta )\nu'(\Delta Y/A)Y}{A \aleph}\hat{y}_{t-1} \\ -\frac{1-\alpha(1+\pi)^{\theta-1}}{\alpha (1+\pi)^{\theta-2}}\frac{(1+\eta ) \nu'(\Delta Y/A)Y}{A \aleph}\hat{a}_{t-1}\\ + \beta \theta(1-\alpha (1+\pi)^{\theta-1})\frac{\mathbb{E}(1+\pi)^{\theta-1}\, \aleph}{(1+\pi)^{\theta-2}\, \aleph} \hat{\pi}_{t}\\-\beta  (1-\alpha(1+\pi)^{\theta-1})\frac{\mathbb{E}(1+\pi)^{\theta-1}}{(1+\pi)^{\theta-2}}\frac{\eta \nu'(\Delta Y/A)Y}{A\aleph}\hat{\Delta}_{t}\\- \beta (1-\alpha(1+\pi)^{\theta-1})\frac{\mathbb{E}(1+\pi)^{\theta-1}}{(1+\pi)^{\theta-2}}\frac{(1+\eta)\nu'(\Delta Y/A)Y}{A\aleph}\hat{y}_{t}\\+ \beta (1-\alpha(1+\pi)^{\theta-1})\frac{\mathbb{E}(1+\pi)^{\theta-1}}{(1+\pi)^{\theta-2}}\frac{(1+\eta) \nu'(\Delta Y/A)Y}{A\aleph}\hat{a}_{t}\\ -\alpha \beta^{2} \theta (1-\alpha(1+\pi)^{\theta-1}) \mathbb{E}(1+\pi)^{\theta-1}\frac{\mathbb{E}(1+\pi)^{\theta-1} \, \aleph}{(1+\pi)^{\theta-2} \, \aleph}\mathbb{E}_{t}\hat{\pi}_{t+1}\\-\frac{1-\alpha(1+\pi)^{\theta-1}}{\alpha (1+\pi)^{\theta-2}}\frac{(1-\sigma)\psi u'(Y)Y}{\beth}\hat{y}_{t-1}\\-\beta (\theta-1)(1-\alpha (1+\pi)^{\theta-1})\frac{\mathbb{E}(1+\pi)^{\theta-2}\, \beth}{(1+\pi)^{\theta-2}\, \beth}\hat{\pi}_{t}\\+ \beta (1-\sigma)(1-\alpha (1+\pi)^{\theta-1})\frac{\psi u'(Y)Y}{\beth}\frac{\mathbb{E}(1+\pi)^{\theta}}{(1+\pi)^{\theta-2}}\, \hat{y}_{t}\\ + \alpha\beta^{2}(\theta-1)(1-\alpha (1+\pi)^{\theta-1})\mathbb{E}(1+\pi)^{\theta}\frac{\mathbb{E}(1+\pi)^{\theta-2}\, \beth}{(1+\pi)^{\theta-2}\, \beth}\mathbb{E}_{t}\hat{\pi}_{t+1} + v_{3}(\hat{u}_{t-1}, \, \hat{u}_{t})
\end{multline}
\par Note that passing expectations has replaced future error terms- so the dependence notation can be dropped henceforth.
Now removing terms in $\aleph$ and $\beth$ via substitution and several applications of Tonelli's theorem, then simplifying coefficients
\begin{multline}
  \beta \bigg(\alpha \mathbb{E}(1+\pi)^{\theta-1}(2+\pi)+ \frac{(1-\alpha (1+\pi)^{\theta-1})}{(1+\pi)^{\theta-2}} \times \\ \bigg[ \frac{\theta \mathbb{E}(1+\pi)^{\theta-1} \nu'(\Delta Y/A)Y/A(1-\alpha \beta \mathbb{E}(1+\pi)^{\theta})} {\nu'(\Delta Y/A)Y/A + \alpha \beta \mathbb{E}(1+\pi)^{\theta}\nu'(\Delta Y/A)Y/A(1-\alpha \beta \mathbb{E}(1+\pi)^{\theta})}\\ -(\theta-1)
\frac{\mathbb{E}(1+\pi)^{\theta-2}\psi u'(Y)Y/(1-\alpha \beta \mathbb{E}(1+\pi)^{\theta-1})}{\psi u'(Y)Y + \alpha \beta \mathbb{E}(1+\pi)^{\theta-1}\psi u'(Y)Y/(1-\alpha  \beta \mathbb{E}(1+\pi)^{\theta-1})} \bigg] \bigg)\hat{\pi}_{t}=  \\ \hat{\pi}_{t-1}  -\frac{\eta  (1-\alpha(1+\pi)^{\theta-1})\nu'(\Delta Y/A)Y/A \alpha (1+\pi)^{\theta-2}}{ \nu'(\Delta Y/A)Y/A + \alpha \beta \mathbb{E}(1+\pi)^{\theta}\nu'(\Delta Y/A)Y/A(1-\alpha \beta \mathbb{E}(1+\pi)^{\theta})}\hat{\Delta}_{t-1} \\
-\frac{(1-\alpha (1+\pi)^{\theta-1})\nu'(\Delta Y/A)Y/ A \alpha(1+\pi)^{\theta-2}}{\nu'(\Delta Y/A)Y/A + \alpha \beta \mathbb{E}(1+\pi)^{\theta}\nu'(\Delta Y/A)Y/A(1-\alpha \beta \mathbb{E}(1+\pi)^{\theta})}\times \\ \bigg[(1+\eta)\nu'(\Delta Y/A)\frac{Y}{A} -(1-\sigma)\psi u'(Y)Y\frac{(\theta-1)}{\theta} \frac{(1-\alpha)^{1/(\theta-1)}}{(1-\alpha (1+\pi)^{\theta-1})^{1/(\theta-1)}}\bigg]\hat{y}_{t-1}
\\  + \frac{\beta  (1-\alpha(1+\pi)^{\theta-1})\{\mathbb{E}(1+\pi)^{\theta-1}\eta \nu'(\Delta Y/A)Y/A\}/(1+\pi)^{\theta-2} }{\nu'(\Delta Y/A)Y/A + \alpha \beta \mathbb{E}(1+\pi)^{\theta}\nu'(\Delta Y/A)Y/A(1-\alpha \beta \mathbb{E}(1+\pi)^{\theta})}\hat{\Delta}_{t} \\ + \frac{\beta(1-\alpha (1+\pi)^{\theta-1})/(1+\pi)^{\theta-2}}{\nu'(\Delta Y/A)Y/A + \alpha \beta \mathbb{E}(1+\pi)^{\theta}\nu'(\Delta Y/A)Y/A(1-\alpha \beta \mathbb{E}(1+\pi)^{\theta})}\times \\ \bigg[(1+\eta)\nu'(\Delta Y/A)\frac{Y}{A}\mathbb{E}(1+\pi)^{\theta-1}\\ -(1-\sigma)\psi u'(Y)Y \frac{\theta-1}{\theta} \frac{(1-\alpha)^{1/(\theta-1)}}{(1-\alpha (1+\pi)^{\theta-1})^{1/(\theta-1)}}\mathbb{E}(1+\pi)^{\theta}\bigg]\hat{y}_{t}\\+ \alpha \beta^{2}\bigg(\alpha \mathbb{E}(1+\pi)^{\theta-1}\mathbb{E}(1+\pi)^{\theta} + \frac{(1-\alpha (1+\pi)^{\theta-1})}{(1+\pi)^{\theta-2}} \times \\  \bigg[\frac{\theta \mathbb{E}(1+\pi)^{\theta-1}\mathbb{E}(1+\pi)^{\theta-1}\nu'(\Delta Y/A)Y/A(1-\alpha \beta \mathbb{E}(1+\pi)^{\theta-1})}{\nu'(\Delta Y/A)Y/A + \alpha \beta \mathbb{E}(1+\pi)^{\theta}\nu'(\Delta Y/A)Y/A(1-\alpha \beta \mathbb{E}(1+\pi)^{\theta})}\\ -(\theta-1)
\frac{\mathbb{E}(1+\pi)^{\theta}\mathbb{E}(1+\pi)^{\theta-2}\psi u'(Y)Y/(1-\alpha \beta \mathbb{E}(1+\pi)^{\theta-1})}{\psi u'(Y)Y + \alpha \beta \mathbb{E}(1+\pi)^{\theta-1}\psi u'(Y)Y/(1-\alpha  \beta \mathbb{E}(1+\pi)^{\theta-1})}\bigg]\bigg)\mathbb{E}_{t}\hat{\pi}_{t+1} \\ + v_{4}(\hat{u}_{t-1}, \, \hat{u}_{t})
\end{multline}
The expression can be compressed further, by removing the two lagged terms $\hat{\Delta}_{t-1}$ and $\hat{y}_{t-1}$. To do so it is necessary to log-linearize the Euler and policy rules, whilst adjusting the output gap measure. 
\begin{multline}
\hat{\psi}_{t}-\sigma \hat{y}_{t}=\beta\frac{\mathbb{E}\psi u'(Y)/(1+\pi)}{\psi u'(Y)}(a_{\pi}\hat{\pi}_{t}+a_{y}\hat{y}_{t})+ \beta a_{y}\frac{(1 + \eta)}{(\sigma+\eta)}\frac{\mathbb{E}\psi u'(Y)/(1+\pi)}{\psi u'(Y)}\hat{a}_{t} \\ -\frac{\mathbb{E}\psi u'(Y)/(1+\pi)^{2}}{\mathbb{E}\psi u'(Y)/(1+\pi)}\mathbb{E}_{t}\hat{\pi}_{t+1}-\frac{u'(Y)}{u''(Y)}\frac{\mathbb{E}\psi u''(Y)/(1+\pi)}{\mathbb{E}\psi u'(Y)/(1+\pi)}\sigma \mathbb{E}_{t}\hat{y}_{t+1}
\end{multline}
Lagging the relationship and including the later period error term yields 
\begin{multline}
\hat{y}_{t-1}=\frac{u'(Y)}{u''(Y)}\frac{\mathbb{E}\psi u''(Y)/(1+\pi)}{\mathbb{E}\psi u'(Y)/(1+\pi)}\sigma \bigg/ \bigg(\sigma + \frac{a_{y} \beta} {\psi u'(Y)} \mathbb{E}\psi u'(Y)/(1+\pi)\bigg)\hat{y}_{t}\\ -\frac{a_{\pi}\beta}{\psi u'(Y)} \mathbb{E}\psi u'(Y)/(1+\pi) \bigg/ \bigg(\sigma + \frac{a_{y} \beta} {\psi u'(Y)} \mathbb{E}\psi u'(Y)/(1+\pi)\bigg)\hat{\pi}_{t-1} \\ + \frac{\mathbb{E}\psi u'(Y)/(1+\pi)^{2}}{\mathbb{E}\psi u'(Y)/(1+\pi)} \bigg/ \bigg(\sigma + \frac{a_{y} \beta} {\psi u'(Y)} \mathbb{E}\psi u'(Y)/(1+\pi)\bigg)\hat{\pi}_{t}+v_{5}(\hat{u}_{t-1}, \, \hat{u}_{t})
\end{multline} 
Finally, the log-linearized price dispersion relationship and its lagged form are respectively where I have used the expression for the stochastic steady of $\Delta$ 
\begin{multline}
\mathbb{E}_{t}\hat{\Delta}_{t+1}=\alpha \theta  \bigg[\mathbb{E}(1+\pi)^{\theta-1}-\frac{\mathbb{E}(1+\pi)^{\theta-2}(1-\alpha (1+\pi)^{\theta-1})^{1/(\theta-1)}}{(1-\alpha)^{1/(\theta-1)}\Delta} \bigg]\mathbb{E}_{t}\hat{\pi}_{t+1}\\ +\alpha \mathbb{E}(1+\pi)^{\theta}\hat{\Delta}_{t}
\end{multline}
\begin{multline}
\hat{\Delta}_{t-1}=
\frac{1}{\alpha \mathbb{E}(1+\pi)^{\theta}}\hat{\Delta}_{t}- 
\theta \bigg( (1-\alpha)^{1/(\theta-1)}\Delta  \mathbb{E}(1+\pi)^{\theta-1}- \\ \mathbb{E}(1+\pi)^{\theta-2}(1-\alpha (1+\pi)^{\theta-1})^{1/(\theta-1)}\bigg) \bigg/ (1-\alpha)^{1/(\theta-1)}\Delta \mathbb{E}(1+\pi)^{\theta}\hat{\pi}_{t}
\end{multline}
Combining (112), (114) and (116) yields the expression for the Phillips curve equivalent to (2) 
\begin{equation}
\hat{\pi}_{t}=b_{0}\hat{\pi}_{t-1} +b_{1}\hat{y}_{t}+ b_{2}\hat{\Delta}_{t} + b_{3}\mathbb{E}_{t}\hat{\pi}_{t+1}+
v_{6} \end{equation}
By substituting the Phillips curve into the Euler, we have 
\begin{equation}
\hat{y}_{t}=c_{0}\hat{\pi}_{t-1} + c_{1}\hat{\pi}_{t} + c_{2}\hat{\Delta}_{t}+ c_{3}\mathbb{E}_{t}\hat{y}_{t+1} + v_{7}
\end{equation}
For the price dispersion recursion we have
\begin{equation}
\hat{\Delta}_{t}=d_{0}\hat{\pi}_{t-1}+d_{1}
\hat{\pi}_{t}+d_{2}\hat{y}_{t}+d_{3}\mathbb{E}_{t}\hat{\Delta}_{t+1} +  v_{8} 
\end{equation}
These are sufficient to describe the canonical form.  
\subsubsection{Econometrics and Errors}
This part briefly discusses useful forms for econometric investigation, their basic properties and challenges.
\par It is possible to remove price dispersion as follows\footnote{$$d'_{0}=\alpha \theta  \bigg[\mathbb{E}(1+\pi)^{\theta-1}-\frac{\mathbb{E}(1+\pi)^{\theta-2}(1-\alpha (1+\pi)^{\theta-1})^{1/(\theta-1)}}{(1-\alpha)^{1/(\theta-1)}\Delta} \bigg]$$
$$d'_{1}=\alpha \mathbb{E}(1+\pi)^{\theta}$$}
\begin{equation}\hat{\Delta}_{t}=d'_{0}\hat{\pi}_{t}/(1-d'_{1}
\mathbb{L}) 
\end{equation}
After taking lags and expanding some expectations, this forms the following Vector Autoregressive Moving Average (VARMA) $(p, \, q)$ 
\begin{equation}\bf{\hat{X}}_{t}=\bf{R}\sum^{p}_{i=0}\bf{\hat{X}}_{t-i}+K\sum^{q}_{j=0}
\bf{\hat{u}}_{t-j}
\end{equation}
where $\textbf{X}=(\pi, \, y)'$ and $\textbf{u}=(\psi, \, A)$ and $p=q=3$.
This result is of macroeconometric interest, since VARMA are popular data descriptors and data on price dispersion are not widely available. In the small noise limit, when using the efficient output gap measure $\hat{y}^{e}_{t}$ 
inference through standard VARMA theory is feasible (see for example \cite{hamilton1995time}). However, this formulation presents two new econometric challenges away from the small noise limit. The first is that the stochastic coefficients are not amenable to standard method of moments formulations. This will prevent the use of existing estimation and hypothesis testing. Secondly, when $\sigma \neq 1$ the presence of persistent technology shocks will prevent the use of benchmark VARMA analysis.
\par The first problem is a long term opportunity for econometrics and mathematical statistics. The second is a more pressing economic concern. It would arise in the benchmark model if one allowed for persistent Euler disturbances. Truncation strategies have been popular (see \cite{ascari2014macroeconomics}) but there may be scope for refinement.
\subsubsection{Slope Coefficients}
This part details the crucial slope coefficients in the simplest form of the model, with log utility when impatience and noise have vanished, associated with $\sigma=1$, $\beta \rightarrow 1$ and $\vert{\varepsilon}\vert \rightarrow 0$. 
Each coefficient comprises a numerator indicated by a tilde superscript and an equation specific denominator, so $b_{0}=\tilde{b}_{0}/b$ and so forth.
\begin{equation} b= 1+ \alpha + \frac{(1-\alpha)^{2}}{\alpha}\frac{(1+\eta)}{1+a_{y}}\end{equation}
\begin{equation} \tilde{b}_{0}= 1+ a_{\pi}\frac{(1-\alpha)^{2}}{\alpha}\frac{(1+\eta)}{1+a_{y}}\end{equation}
\begin{equation} \tilde{b}_{1}=(1-\alpha)^{2}(1+\eta)\bigg\{\frac{\alpha (1+a_{y})-1}{\alpha(1+a_{y})}\bigg\} \end{equation}
\begin{equation} \tilde{b}_{2}=-\eta(1-\alpha)^{3}\frac{(1+\alpha)}{\alpha^{2}}
\end{equation}
\begin{equation} \tilde{b}_{3}=\alpha \end{equation}
\begin{equation} c = 1 + \frac{(1-\alpha)^{2}}{\alpha^{2}}\frac{(1+\eta)}{(1 + a_{y})^{2}}
\bigg(\alpha (1+a_{y})-1\bigg)\end{equation}
\begin{equation}  \tilde{c}_{0}=-\bigg( 1+ a_{\pi}\frac{(1-\alpha)^{2}}{\alpha}\frac{(1+\eta)}{1+a_{y}} \bigg) \frac{1}{\alpha(1+a_{y})} \end{equation}
\begin{equation} \tilde{c}_{1}= \bigg(1+ \alpha(1-a_{\pi}) + \frac{(1-\alpha)^{2}}{\alpha}\frac{(1+\eta)}{1+a_{y}}\bigg)\frac{1}{\alpha(1+a_{y})}  \end{equation}
\begin{equation} \tilde{c}_{2}= \eta(1-\alpha)^{3}\frac{(1+\alpha)}{\alpha^{3}(1+a_{y})}\end{equation}
\begin{equation} \tilde{c}_{3}=\frac{1}{1+a_{y}} \end{equation}
\begin{equation} d =\alpha \end{equation}
\begin{equation} \tilde{d}_{0}=\tilde{d}_{1}=\tilde{d}_{2}=0\end{equation}
\begin{equation} \tilde{d}_{3} =1 \end{equation}
Some of these expressions will be discussed and interpreted further in Section 10. 
\subsection{Error Coefficients}
This subsection focuses on the properties of the structural error terms, which are missing from existing approximations. The first part sets out the general expressions for each of the error terms $v_{i}$ listed. The second solves these out for the limiting expansion around ZINSS. 
\subsubsection{Coefficient Expressions}
The starting point is the infinite horizon solutions, which states how present deviations depend on present and past shocks.
\begin{equation}\hat{\pi}_{t}=\sum_{i=0}^{\infty} (e_{i}\hat{a}_{t-i} + f_{i}\hat{\psi}_{t-i})\end{equation}
\begin{equation}\hat{y}_{t}=\sum_{i=0}^{\infty} (g_{i}\hat{a}_{t-i} + h_{i}\hat{\psi}_{t-i})\end{equation}
Price dispersion works differently, away from ZINSS, it will be determined by (115). In the
$\vert{\varepsilon}\vert$ it will be first order and the magnitude will determined by the the coefficient on $\pi^{2}_{t}$ in its Taylor expansion, uncovered in the proof of Proposition 3 in Appendix B.3.2. Finally, it will vanish in the $\sqrt{\varepsilon}$ case.  
\begin{equation} \hat{\Delta}_{t}= C_{0}(\hat{\pi}_{t}+ \alpha \hat{\pi}_{t-1}+ \alpha^{2}\hat{\pi}_{t-2}+ \cdots )
\end{equation}
where 
\begin{equation}C_{0}=
\frac{\alpha \theta} {(1-\alpha)^{1/(\theta-1)}} \end{equation}
For the expectation variables it implies 
\begin{equation} \label{eq1}
\begin{split}
\mathbb{E}_{t}\hat{y}_{t+1} & = \mathbb{E}_{t} g_{0}\hat{a}_{t+1} + g_{1}\hat{a}_{t} + g_{2}\hat{a}_{t-1} + \cdots + h_{0}\hat{\psi}_{t+1} +h_{1}\hat{\psi}_{t} + h_{2}\hat{\psi}_{t-1} + \cdots \\
 & = (g_{1}+\rho g_{0}) \hat{a}_{t} + \sum_{i=1}^{\infty}g_{i+1}\hat{a}_{t-i} + \sum_{i=1}^{\infty}h_{i+1}\hat{\psi}_{t-i} 
\end{split}
\end{equation}
\begin{equation} \mathbb{E}_{t}\hat{\pi}_{t+1} = (e_{1}+\rho e_{0}) \hat{a}_{t} + \sum_{i=1}^{\infty}e_{i+1}\hat{a}_{t-i} + \sum_{i=0}^{\infty}f_{i+1}\hat{\psi}_{t-i} \end{equation}
\begin{equation}\begin{split} \hat{\Delta}_{t} & = C_{0}\big[e_{0}\hat{a}_{t} + (e_{1}+ \alpha e_{0} )\hat{a}_{t-1} + \cdots + f_{0}\hat{\psi}_{t} + (f_{1}+ \alpha f_{0})\hat{\psi}_{t-1} + \cdots  \big] \\
& = C_{0} \sum_{i=0}^{\infty}\bigg[\sum_{k=0}^{i}{\alpha}^{i-k}(e_{k}\hat{a}_{t-i} + f_{k}\hat{\psi}_{t-i}) \bigg] \end{split} \end{equation}
The first two steps remove expectations to focus on future variables. The third treats lag terms. The fourth is consolidation, the fifth deals with the Euler and the sixth puts them all together. Proceeding through the derivation in order, it is clear that
\begin{multline} v_{0}=
\frac{ \psi \nu'(\Delta Y/A)Y}{A \aleph}\hat{\psi}_{t} -\alpha \beta \bigg\{ \theta \frac{\mathbb{E}(1+\pi)^{\theta-1}\, \aleph}{\aleph}(\hat{\pi}_{t+1}-\mathbb{E}_{t}\hat{\pi}_{t+1}) + \\ \mathbb{E} (1+\pi)^{\theta}(\hat{\aleph}_{t+1}-\mathbb{E}_{t}\hat{\aleph}_{t+1})\bigg\} \end{multline}
 \begin{multline} v_{1}=
 \frac{\psi u'(Y)Y}{\beth}\hat{\psi}_{t}
- \alpha \beta \bigg\{  (\theta-1)\frac{\mathbb{E}(1+\pi)^{\theta-2}\, \beth}{\beth}
(\hat{\pi}_{t+1}-\mathbb{E}_{t}\hat{\pi}_{t+1}) + \\ \mathbb{E} (1+\pi)^{\theta-1}(\hat{\beth}_{t+1}-\mathbb{E}_{t}\hat{\beth}_{t+1})\bigg\} \end{multline}
\begin{multline} v_{2}= \frac{(1-\alpha(1+\pi)^{\theta-1})}{\alpha (1+\pi)^{\theta-2}}\Bigg\{(\mathbb{L}-\alpha \beta \mathbb{E}(1+\pi)^{\theta-1})\bigg[
\frac{ \psi \nu'(\Delta Y/A)Y}{A \aleph}\hat{\psi}_{t}
- \\ \alpha \beta \theta \frac{\mathbb{E}(1+\pi)^{\theta-1}\, \aleph}{\aleph}(\hat{\pi}_{t+1}-\mathbb{E}_{t}\hat{\pi}_{t+1})-   \alpha \beta \mathbb{E} (1+\pi)^{\theta}(\hat{\aleph}_{t+1} -\mathbb{E}_{t}\hat{\aleph}_{t+1})\bigg] - \\  (\mathbb{L}-\alpha \beta \mathbb{E}(1+\pi)^{\theta})\bigg[ \frac{\psi u'(Y)Y}{\beth}\hat{\psi}_{t}- \alpha \beta(\theta-1)\frac{\mathbb{E}(1+\pi)^{\theta-2}\, \beth}{\beth}(\hat{\pi}_{t+1}-\mathbb{E}_{t}\hat{\pi}_{t+1})  \\   -   \alpha \beta \mathbb{E} (1+\pi)^{\theta-1} (\hat{\beth}_{t+1} -\mathbb{E}_{t}\hat{\beth}_{t+1})\bigg]\Bigg\} \end{multline}
\begin{multline} 
v_{3} = \frac{(1-\alpha(1+\pi)^{\theta-1})}{\alpha (1+\pi)^{\theta-2}} 
\Bigg\{
\bigg[\frac{\psi \nu'(\Delta Y/A)Y}{A \aleph}\hat{\psi}_{t-1} 
- \\  \alpha \beta \theta \frac{\mathbb{E}(1+\pi)^{\theta-1}\, \aleph}{\aleph}(\hat{\pi}_{t}-\mathbb{E}_{t-1}\hat{\pi}_{t})     - \alpha \beta \mathbb{E} (1+\pi)^{\theta}(\hat{\aleph}_{t} -  \mathbb{E}_{t-1}\hat{\aleph}_{t}) - \\ \frac{\psi \nu'(\Delta Y/A)Y}{A \aleph}\alpha \beta \mathbb{E}(1+\pi)^{\theta-1}\hat{\psi}_{t}\bigg] - \\ 
\bigg[ \frac{\psi u'(Y)Y}{\beth}\hat{\psi}_{t-1} - \alpha \beta(\theta-1)\frac{\mathbb{E}(1+\pi)^{\theta-2}\, \beth}{\beth} (\hat{\pi}_{t}-\mathbb{E}_{t-1}\hat{\pi}_{t}) - \\ 
 \alpha \beta \mathbb{E} (1+\pi)^{\theta-1}(\hat{\beth}_{t} -\mathbb{E}_{t-1}\hat{\beth}_{t}) - \frac{\psi u'(Y)Y}{\beth}\alpha \beta \mathbb{E} (1+\pi)^{\theta}\hat{\psi}_{t}\bigg]\Bigg\} 
\end{multline}
which simplifies to 
\begin{multline}\frac{(1-\alpha(1+\pi)^{\theta-1})}{\alpha (1+\pi)^{\theta-2}} \Bigg[ \bigg( \frac{\psi \nu'(\Delta Y/A)Y}{A \aleph} - \frac{\psi u'(Y)Y}{\beth}\bigg)\hat{\psi}_{t-1} - \\ \bigg( \frac{\psi \nu'(\Delta Y/A)Y}{A \aleph}\alpha \beta \mathbb{E}(1+\pi)^{\theta-1}- \frac{\psi u'(Y)Y}{\beth}\alpha \beta \mathbb{E} (1+\pi)^{\theta}\bigg)\hat{\psi}_{t} - \\  
\alpha \beta \bigg(\theta \frac{\mathbb{E}(1+\pi)^{\theta-1}\, \aleph}{\aleph} -(\theta-1)\frac{\mathbb{E}(1+\pi)^{\theta-2}\, \beth}{\beth} + \\  \frac{\alpha (1+\pi)^{\theta-2}\, \mathbb{E} (1+\pi)^{\theta}}{(1-\alpha(1+\pi)^{\theta-1})}\bigg)(\hat{\pi}_{t}-\mathbb{E}_{t-1}\hat{\pi}_{t}) - \alpha \beta \mathbb{E} (1+\pi)^{\theta-1}\pi(\hat{\beth}_{t} -\mathbb{E}_{t-1}\hat{\beth}_{t})
\Bigg] \end{multline} 
The coefficients have been organised so that the bracketed terms will be positive when computed in the small noise limit at strictly positive rates of trend inflation, as this is a focal alternative to ZINSS.\footnote{For $\hat{\psi}_{t}$ I have assumed that $2\alpha \beta -1>0$, which is consistent with all calibrations.} To dig down to primitives it is necessary to expand out the weighting term as follows:
\begin{multline}
\hat{\beth}_{t}= \frac{\psi u'(Y)Y}{\beth}\hat{\psi}_{t} + \frac{\psi u'(Y)(1-\sigma)}{\beth}\hat{y}_{t} + \frac{\alpha \beta (\theta-1)}{\beth}\mathbb{E}(1+\pi)^{\theta-2}\, \beth \, \mathbb{E}_{t}\hat{\pi}_{t+1} + \\  \alpha \beta \mathbb{E}(1+\pi)^{\theta-1}\mathbb{E}_{t}\hat{\beth}_{t+1} 
\end{multline}
\begin{multline}
\hat{\beth}_{t}= \frac{\psi u'(Y)Y}{\beth}\hat{\psi}_{t} + \frac{\psi u'(Y)(1-\sigma)}{\beth}\hat{y}_{t} +  \\  \frac{\alpha \beta (\theta-1)}{\beth}\mathbb{E}(1+\pi)^{\theta-2}\, \beth \, 
\mathbb{E}_{t}\sum_{j=1}^{\infty} [\alpha \beta\mathbb{E}(1+\pi)^{\theta-1}]^{j-1} \, \hat{\pi}_{t+j}
+ \\
\frac{\psi u'(Y)(1-\sigma)}{\beth}\mathbb{E}_{t}\sum_{j=1}^{\infty}[\alpha \beta\mathbb{E}(1+\pi)^{\theta-1}]^{j} \, \hat{y}_{t+j} 
\end{multline}
Expanding forward using (139), (140) and the primitive properties of the errors yields 
\begin{multline} \hat{\beth}_{t}-\mathbb{E}_{t-1}\hat{\beth}_{t}=\frac{\psi u'(Y)Y}{\beth}\hat{\psi}_{t} + \frac{\psi u'(Y)(1-\sigma)}{\beth}\bigg(g_{0}(\hat{a}_{t}-\rho \hat{a}_{t-1}) + h_{0}\hat{\psi}_{t}\bigg) + 
\\ \frac{\alpha \beta (\theta-1)}{\beth}\mathbb{E}(1+\pi)^{\theta-2}\, \beth \, \mathbb{E}_{t}\sum_{j=1}^{\infty} [\alpha \beta\mathbb{E}(1+\pi)^{\theta-1}]^{j-1} \, \bigg[\sum_{i=0}^{j}\rho^{j-i}e_{i}(\hat{a}_{t}-\rho \hat{a}_{t-1}) +f_{i}\hat{\psi}_{t} \bigg]   \\ + \frac{\psi u'(Y)Y}{\beth}\sum_{j=1}^{\infty}[\alpha \beta\mathbb{E}(1+\pi)^{\theta-1}]^{j} \,\bigg[\sum_{i=0}^{j}\rho^{j-i}g_{i}(\hat{a}_{t}-\rho \hat{a}_{t-1}) +h_{i}\hat{\psi}_{t} \bigg]
\end{multline}
Thus, I can substitute back into 
(146) to solve for the third error in terms of the canonical expansions. 
The unwieldy expression is suppressed. An algorithm for a solution, in terms of the stochastic equilibrium expansion coefficients, would be useful, alongside a method of numerical approximation. However, these tasks lie beyond the boundaries of this investigation. 
\par The rest of the derivation is quite simple 
\begin{multline} v_{4}=
-\beta (1-\alpha(1+\pi)^{\theta-1})\frac{\mathbb{E}(1+\pi)^{\theta-1}}{(1+\pi)^{\theta-2}}\frac{(1+\eta) \nu'(\Delta Y/A)Y}{A\aleph}\hat{a}_{t} + \\ \frac{1-\alpha(1+\pi)^{\theta-1}}{\alpha (1+\pi)^{\theta-2}}\frac{(1+\eta ) \nu'(\Delta Y/A)Y}{A \aleph}\hat{a}_{t-1} -v_{3}
\end{multline}
\begin{multline} v_{5}= 
\frac{1}{\sigma} \hat{\psi}_{t-1} - \frac{\mathbb{E}\psi u'(Y)/(1+\pi)^{2}}{\mathbb{E}\psi u'(Y)/(1+\pi)} \bigg/ \bigg(\sigma + \frac{a_{y} \beta} {\psi u'(Y)} \mathbb{E}\frac{\psi u'(Y)}{(1+\pi)}\bigg)({\pi}_{t}-\mathbb{E}_{t-1}{\pi}_{t}) - \\ 
\frac{u'(Y)}{u''(Y)}\frac{\mathbb{E}\psi u''(Y)/(1+\pi)}{\mathbb{E}\psi u'(Y)/(1+\pi)}\sigma \bigg/ \bigg(\sigma + \frac{a_{y} \beta} {\psi u'(Y)} \mathbb{E}\psi u'(Y)/(1+\pi)\bigg)(\hat{y}_{t}-\mathbb{E}_{t-1}\hat{y}_{t}) + \\  \frac{u'(Y)}{u''(Y)}\frac{\mathbb{E}\psi u''(Y)/(1+\pi)}{\mathbb{E}\psi u'(Y)/(1+\pi)}\sigma \bigg/ \bigg(\sigma + \frac{a_{y} \beta} {\psi u'(Y)} \mathbb{E}\psi u'(Y)/(1+\pi)\bigg)\frac{1+\eta}{\sigma + \eta}\hat{a}_{t}- \\ \bigg\{1+\rho \frac{u'(Y)}{u''(Y)}\frac{\mathbb{E}\psi u''(Y)/(1+\pi)}{\mathbb{E}\psi u'(Y)/(1+\pi)}\sigma \bigg/ \bigg(\sigma + \frac{a_{y} \beta} {\psi u'(Y)} \mathbb{E}\psi u'(Y)/(1+\pi)\bigg)\bigg\}\frac{1+\eta}{\sigma + \eta}\hat{a}_{t-1}
\end{multline}
Combining (139), (140), (149)-(151) gives the following cumbersome expression for the Phillips curve errors 
\begin{multline} v_{6}= \frac{1}{b} \Bigg[ -\beta (1-\alpha(1+\pi)^{\theta-1})\frac{\mathbb{E}(1+\pi)^{\theta-1}}{(1+\pi)^{\theta-2}}\frac{(1+\eta) \nu'(\Delta Y/A)Y}{A\aleph}\hat{a}_{t} + \\ \frac{1-\alpha(1+\pi)^{\theta-1}}{\alpha (1+\pi)^{\theta-2}}\frac{(1+\eta ) \nu'(\Delta Y/A)Y}{A \aleph}\hat{a}_{t-1}- \\ \frac{(1-\alpha(1+\pi)^{\theta-1})}{\alpha (1+\pi)^{\theta-2}} \Bigg[ \bigg( \frac{\psi \nu'(\Delta Y/A)Y}{A \aleph} - \frac{\psi u'(Y)Y}{\beth}\bigg)\hat{\psi}_{t-1} - \\ \bigg( \frac{\psi \nu'(\Delta Y/A)Y}{A \aleph}\alpha \beta \mathbb{E}(1+\pi)^{\theta-1}- \frac{\psi u'(Y)Y}{\beth}\alpha \beta \mathbb{E} (1+\pi)^{\theta}\bigg)\hat{\psi}_{t} 
 - \\  
\alpha \beta \bigg(\theta \frac{\mathbb{E}(1+\pi)^{\theta-1}\, \aleph}{\aleph} -(\theta-1)\frac{\mathbb{E}(1+\pi)^{\theta-2}\, \beth}{\beth} + \\  \frac{\alpha (1+\pi)^{\theta-2}\, \mathbb{E} (1+\pi)^{\theta}}{(1-\alpha(1+\pi)^{\theta-1})}\bigg)\bigg(e_{0}(\hat{a}_{t}-\rho \hat{a}_{t-1}) + f_{0}\hat{\psi}_{t}\bigg) - \\ \alpha \beta \mathbb{E} (1+\pi)^{\theta-1}\pi\bigg\{\frac{\psi u'(Y)Y}{\beth}\hat{\psi}_{t} + \frac{\psi u'(Y)(1-\sigma)}{\beth}\bigg(g_{0}(\hat{a}_{t}-\rho \hat{a}_{t-1}) + h_{0}\hat{\psi}_{t}\bigg) + 
\\ \frac{\alpha \beta (\theta-1)}{\beth}\mathbb{E}(1+\pi)^{\theta-2}\, \beth \, 
\mathbb{E}_{t}\sum_{j=1}^{\infty} [\alpha \beta\mathbb{E}(1+\pi)^{\theta-1}]^{j-1} \, \bigg[\sum_{i=0}^{j}\rho^{j-i}e_{i}(\hat{a}_{t}-\rho \hat{a}_{t-1}) +f_{i}\hat{\psi}_{t} \bigg]   \\ + \frac{\psi u'(Y)Y}{\beth}\sum_{j=1}^{\infty}[\alpha \beta\mathbb{E}(1+\pi)^{\theta-1}]^{j} \,\bigg[\sum_{i=0}^{j}\rho^{j-i}g_{i}(\hat{a}_{t}-\rho \hat{a}_{t-1}) +h_{i}\hat{\psi}_{t} \bigg]\bigg\}
\Bigg]\\ -  \frac{(1-\alpha (1+\pi)^{\theta-1})\nu'(\Delta Y/A)Y/ A \alpha(1+\pi)^{\theta-2}}{\nu'(\Delta Y/A)Y/A + \alpha \beta \mathbb{E}(1+\pi)^{\theta}\nu'(\Delta Y/A)Y/A(1-\alpha \beta \mathbb{E}(1+\pi)^{\theta})}\times \\ \bigg[(1+\eta)\nu'(\Delta Y/A)\frac{Y}{A} -(1-\sigma)\psi u'(Y)Y\frac{(\theta-1)}{\theta} \frac{(1-\alpha)^{1/(\theta-1)}}{(1-\alpha (1+\pi)^{\theta-1})^{1/(\theta-1)}}\bigg] \times \\ \Bigg\{\frac{1}{\sigma} \hat{\psi}_{t-1} - \frac{\mathbb{E}\psi u'(Y)/(1+\pi)^{2}}{\mathbb{E}\psi u'(Y)/(1+\pi)} \bigg/ \bigg(\sigma + \frac{a_{y} \beta} {\psi u'(Y)} \mathbb{E}\frac{\psi u'(Y)}{(1+\pi)}\bigg)\bigg(e_{0}(\hat{a}_{t}-\rho \hat{a}_{t-1}) + f_{0}\hat{\psi}_{t}\bigg)  \\ -
\frac{u'(Y)}{u''(Y)}\frac{\mathbb{E}\psi u''(Y)/(1+\pi)}{\mathbb{E}\psi u'(Y)/(1+\pi)}\sigma \bigg/ \bigg(\sigma + \frac{a_{y} \beta} {\psi u'(Y)} \mathbb{E}\frac{\psi u'(Y)}{(1+\pi)}\bigg)\bigg(g_{0}(\hat{a}_{t}-\rho \hat{a}_{t-1}) + h_{0}\hat{\psi}_{t}\bigg)  \\ + \frac{u'(Y)}{u''(Y)}\frac{\mathbb{E}\psi u''(Y)/(1+\pi)}{\mathbb{E}\psi u'(Y)/(1+\pi)}\sigma \bigg/ \bigg(\sigma + \frac{a_{y} \beta} {\psi u'(Y)} \mathbb{E}\psi u'(Y)/(1+\pi)\bigg)\frac{1+\eta}{\sigma + \eta}\hat{a}_{t}- \\ \bigg\{1+\rho \frac{u'(Y)}{u''(Y)}\frac{\mathbb{E}\psi u''(Y)/(1+\pi)}{\mathbb{E}\psi u'(Y)/(1+\pi)}\sigma \bigg/ \bigg(\sigma + \frac{a_{y} \beta} {\psi u'(Y)} \frac{\mathbb{E}\psi u'(Y)}{(1+\pi)}\bigg)\bigg\}\frac{1+\eta}{\sigma + \eta}\hat{a}_{t-1}\Bigg\} \Bigg]\end{multline}
All subsequent error forms can be derived easily from this point. 
\subsubsection{Limiting Solution}
The main focus in this paper is on the limiting case around ZINSS with a Phillips curve specified in terms of $\hat{y}^{e}_{t}$, as $\beta \rightarrow 1$. This is formed by taking the limit as the shock size $\vert \varepsilon \vert \rightarrow 0$. It exploits the fact that the difference between the stochastic and non-stochastic steady states is second order and will therefore vanish, leaving only parametric terms in the coefficients. In this case, the error terms (distinguished by prime subscripts) simplify considerably and yield a closed form solution used in Section 3. The crucial point to remember 
is that the efficient output gap formulation causes the 
shocks emanating inside the Phillips curve to cancel. 
The derivation proceeds as follows 
\begin{equation} v'_{0}=\hat{\psi}_{t}
-\alpha \bigg\{ \theta(\pi_{t+1}-\mathbb{E}_{t}\pi_{t+1}) + (\hat{\aleph}_{t+1}-\mathbb{E}_{t}\hat{\aleph}_{t+1})\bigg\} \end{equation}
 \begin{equation} v'_{1}=\hat{\psi}_{t}
 -\alpha \bigg\{ (\theta-1)(\pi_{t+1}-\mathbb{E}_{t}\pi_{t+1}) + (\hat{\beth}_{t+1}-\mathbb{E}_{t}\hat{\beth}_{t+1})\bigg\} \end{equation}
\begin{multline} v'_{2}= \frac{(1-\alpha)}{\alpha}(\mathbb{L}-\alpha)\bigg[\hat{\psi}_{t} 
-\alpha \theta({\pi}_{t+1}-\mathbb{E}_{t}{\pi}_{t+1})-  \alpha(\hat{\aleph}_{t+1} -\mathbb{E}_{t}\hat{\aleph}_{t+1})\bigg] - \\ \frac{(1-\alpha)}{\alpha}(\mathbb{L}-\alpha)\bigg[ \hat{\psi}_{t} - \alpha(\theta-1)(\pi_{t+1}-\mathbb{E}_{t}\pi_{t+1})-  \alpha(\hat{\beth}_{t+1} -\mathbb{E}_{t}\hat{\beth}_{t+1})\bigg] \end{multline}
\begin{multline}  v'_{3}  = -\frac{(1-\alpha)}{\alpha}\bigg[  \alpha \theta({\pi}_{t}-\mathbb{E}_{t-1}{\pi}_{t}) + 
\alpha(\hat{\aleph}_{t} -\mathbb{E}_{t-1}\hat{\aleph}_{t})\bigg]  + \\ 
\frac{(1-\alpha)}{\alpha}\bigg[ 
 \alpha (\theta-1)(\pi_{t}-\mathbb{E}_{t-1}\pi_{t}) + 
\alpha(\hat{\beth}_{t} -\mathbb{E}_{t-1}\hat{\beth}_{t})\bigg] 
\end{multline}
\begin{equation} v'_{4} = {\pi}_{t}-\mathbb{E}_{t-1}{\pi}_{t}\end{equation} 
\begin{equation} v'_{5} = 
\hat{\psi}_{t-1}-({\pi}_{t}-\mathbb{E}_{t-1}{\pi}_{t})-(\hat{y}^{e}_{t}-\mathbb{E}_{t-1}\hat{y}^{e}_{t})\end{equation}
Now denoting the errors in (117) in the style of the slope coefficients 
\begin{equation} v'_{6}= b_{4}(\hat{\psi}_{t}-\hat{\psi}_{t-1}) \end{equation}
Hence, I am able to solve for the error coefficient using the output term in (112), the error expression (150) and the symmetry implied by (151) and (152) 
\begin{equation} \tilde{b}_{4}= 
\frac{(1-\alpha)^{2}}{\alpha}\frac{(1+\eta)}{1+a_{y}}\end{equation}
Finally, I am able to deduce the following relationship between the present responses of output and inflation to the demand shock from the equation in $\hat{\psi}_{t}$ 
\begin{equation} \bigg(1+\frac{(1-\alpha)^{2}}{\alpha}\frac{(1+\eta)}{1+a_{y}}\bigg)f_{0} + \frac{(1-\alpha)^{2}}{\alpha}\frac{(1+\eta)}{1+a_{y}} h_{0} = \frac{(1-\alpha)^{2}}{\alpha}\frac{(1+\eta)}{1+a_{y}} \end{equation}
\section{Existence Results}
This section is divided into three subsections: the first sets out the mathematical results central to the entire paper. The second introduces a concept key to the proof and then proves two major theorems. The third offers general discussion. 
\subsection{Main Results}
The first two theorems consist of existence and uniqueness results, for the main equilibrium concept here in the model, with appropriate generalization to a wider class of DSGE models.
\subsubsection{Main Theorems}
This part sets out the main theorems and discusses some of the primitive assumptions. 
\begin{theorem}
Consider a DSGE model characterized by the canonical form $\mathbb{E}_{t}X_{t+1}=f(X_{t},\, e_{t}, \, \gamma)$ and an infinite horizon optimization problem $$\mathcal{W}(X_{t}, \, e_{t}, \, \gamma)= \sum_{T=t}^{\infty}\beta^{T-t}u(X_{T}, \, e_{T}, \, \gamma)$$, with transversality condition $$\lim_{T \rightarrow \infty}\beta^{T-t}u(X_{T}, \, e_{T}, \, \gamma)=0$$, where the endogenous variables $X=X^{J} \times X^{S}$ can be partitioned into jump and states of dimensionality $j \times 1$ and $s \times 1$ respectively and $e_{t}$ is an exogenous error term. Let the following properties hold, $u$ is continuous, $f \in C^{1}$ is Lebesgue measurable with non-trivial interdependence
$\mu$ a.e., $\mathcal{X} \times \mathcal{E}  \in \mathcal{M}$ a Euclidean manifold whilst $\{e_{T}\}$ are continuously distributed and mixing processes with at least one moment. 
Suppose further that the system is not compact valued so that stochastic structural sections (measurable with respect to $\mu$) $$\lim_{X \rightarrow cl(\mathcal{X})}u$$ is unbounded. Consider the limit as $\beta \rightarrow 1$ ensuring $1-\beta << \delta$. There exists a recursive equilibrium if and only if, for any limiting sequence $\{ \mathbb{E}_{T}Z_{T}\}$, contained inside a $\mu$-measurable set, there is $T>T'$ and $\delta >0$ such that when the eigenvalues are gathered in ascending order $$\lvert \lambda_{s, \, T} \rvert < 1 - \delta $$
$$\lvert \lambda_{s+1, \, T} \rvert > 1 + \delta $$.
\end{theorem}
\begin{remark} Non-trivial interdependence rules out systems where one or more endogenous variables actually evolve independently, such that diagonalization will not work to reassign eigenvalues. It formalizes intuitive understanding of the difference between endogenous and exogenous variables. \end{remark}
\begin{remark}$f \in C^{1}$ implies the primitives $u$ and $v$ are $C^{2}$ ($\mu$ a.s.). $C^{1}$ and Lebesgue measurable could be replaced with $f \in C^{2}$ and an application of Sard's theorem, requiring $u$ and $v$ are $C^{3}$, which is not burdensome in a discrete time economic application. 
\end{remark}
\begin{remark} The stochastic section idea is important to rule out boundaries that are reached with zero probability because they require precise configurations among the exogenous shocks. From a topological standpoint, this allows me to avoid considering pieces at the boundary of an open $m$ dimensional manifold that are not regular, in the sense of being $m-1$ dimensional.\footnote{Unfamiliar readers may wish to consult a text like \cite{munkres2018analysis}.} This point will prove useful in application. \end{remark}
\begin{remark}
The eigenvalue condition is necessary and sufficient for a stochastic equilibrium (ergodic invariant measure) for the recursive equilibrium of $X$, even if there is no transversality or infinite horizon optimization condition. The other conditions are used to generate necessity of stochastic equilibrium. \end{remark}
\begin{remark} The sequence of eigenvalues $\{\lambda_{i, \, T}\}$ can be replaced with their value at any equilibrium where $f$ is $C^{1}$, which it will be with probability one. 
\end{remark}
\begin{theorem} The expected trajectory sequence $\{ f^{n}(X): n \geq 0\}$ is unique $\mu$ a.s. from any initial set of predetermined endogenous variables and errors $(X^{\mathfrak{P}}_{0}, \,  e_{1})$. \end{theorem} 
Predetermined variables are defined as $\mathbb{E}_{t}X^{\mathfrak{P}}_{t+1}=
X^{\mathfrak{P}}_{t+1}$, capital accumulation is typically modelled as a predetermined variable, set by last periods savings decision. The predetermined variables are a strict subset of the state variables. For Calvo, Proposition 4 in Section 4.8 implies that $X^{S}_{t}=(\Delta_{t}, \, \pi_{t-1})'$ whilst 
$(X^{\mathfrak{P}}_{t}, \, e_{t})=(\pi_{t-1},\, A_{t}, \, A_{t-1}, \, \psi_{t}, \, \psi_{t-1})'$. $\Delta$ is not predetermined because it depends on present as well as the past of inflation. 
\begin{remark} An expected trajectory corresponds with the loose conception of recursive equilibrium prevailing in the 
previous New Keynesian literature. The theorem justifies the association of $f$ with recursive equilibrium and clarifies that Classical and New Keynesian notions are in fact equivalent. It definitively demonstrates that indeterminacy is not in fact a standard feature of DSGE. \end{remark}
\subsubsection{Application}
This part confirms that the theorems apply to the two main models here. It also discusses some models which are not covered and completes the analysis for Rotemberg.  
\begin{proposition} Both Calvo and Rotemberg models meet the primitive conditions to apply Theorem 3.
\end{proposition}
The proof is split into two points starting with Rotemberg.
\begin{proof}(i) Rotemberg is characterized by the aggregate supply system (88) and (89) and demand system (36) and (55). There are two jump variables $(Y_{t}, \, \pi_{t})$ and no states.
Assessing differentiability term by term, it is clearly $C^{2}$.
The task is to prove blow up of an underlying objective function. To do so, I formulate the representative household's problem as a dynamic programming problem, where the Langrangian $\mathcal{L}_{t}$ has $\mathcal{W}_{t}$ from (6) as the objective function and a sequence of budget and marketing clear conditions as constraints- with $\psi_{T}u'(C_{T}) =\lambda_{T}$ the Lagrange multiplier.
It is clear that as $Y_{t} \rightarrow 0$, $C_{t} \rightarrow 0$, which sends $\lambda_{t} \rightarrow \infty$ under the Inada condition (11) and blows up the problem, as the budget constraint term will be unbounded.\footnote{This is because lifetime income will be strictly bigger than zero. Technically with Calvo, I will be calling in Proposition 13 to ensure losses do not swallow household's labor income. If this happened the blow-up would be automatic.} Informally, let us call this "breaking" or "violating" the Inada condition.
 On the other hand, when $Y_{t} \rightarrow \infty $ we must have $L_{t} \rightarrow \infty$ and the objective will explode under (12). The resource constraint  
 (89) implies that $\pi_{t} < \sqrt{2/c_{p}} $. This sends $C_{t} \rightarrow 0$ breaking the Inada condition. By construction, $\pi_{t} > -100 \%$ to balance the Euler $\mathbb{E}_{t}u'(C_{t+1}) \rightarrow \infty$ again violating the Inada condition.
 \par (ii) With Calvo, 
 (32) ensures $\pi_{t} < (1/\alpha)^{1/(\theta-1)}$. When it does so for some $T > t$,  $\mathbb{E}_{t}MC_{T}\rightarrow \infty$, breaching (12). This can happen either through $\mathbb{E}_{t}\Delta_{T} \rightarrow \infty$, if the explosion is a long run phenomenon, or $\mathbb{E}_{t}Y_{T} \rightarrow \infty$, if it is in finite time. This also covers the upper bound for $\Delta$. Finally, the lower bound is not full dimension and therefore measure zero according to Proposition 2, as it implies $\pi_{T}=\pi_{T-1}=0$. 
\end{proof}
\begin{remark}If $\sigma =1$ we could just consider the objective function $\mathcal{W}$ and not worry about the dynamic programming problem for the Rotemberg component. 
\end{remark}
\begin{remark} In both cases the relevant objective function diverges to $-\infty$, so the result is not a figment of measurability restrictions. The problem with indeterminacy is that it is based on conditional convergence of one or more forward solution variables in the linear approximation. However, with the underlying non-linear system there is divergence. \end{remark}
\begin{remark}There are models that do not meet the boundary explosion condition. This includes 
menu costs and 
similar models, where there is a fixed cost of adjustment 
(see \cite{stokey1989recursive}, 
and \cite{stokey2009economics}).\footnote{This is also true in classical models where there is some outside option, for example a firm could exit the industry at no cost or the household could subsist from social security benefits.} 
In the standard menu cost model if inflation is high enough all firms will choose to pay it and a flexible price system will arise. The suggestion is that the various compactness assumptions used 
elsewhere in the general equilibrium and mean field game literatures to prove unconditional existence (not depending on parameters or shock distributions) are necessary in a broad class of models. \end{remark}
\begin{remark}Dropping the boundary conditions can also represent a threat to uniqueness. Consider the canonical overlapping generations model of \cite{samuelson1958exact} and its many subsequent extensions, where multiple asset bubble equilibria can develop from intrinsically valueless assets.
\end{remark} 
I can proceed immediately to analyzing Rotemberg.
\begin{theorem}In the limit where 
$\beta \rightarrow 1$ and shock size $\vert \varepsilon \vert \rightarrow 0$, with $1-\beta << \vert \varepsilon \vert$, there exists a recursive equilibrium in the Rotemberg model of $4.7$ if and only $a_{\pi}>1$.
\end{theorem}
\begin{proof} Follows immediately from Proposition 16 (i), Corollary 1, Proposition 8 and Theorem 3. 
\end{proof}
For Calvo the best general result is 
\begin{proposition} There exists a recursive equilibrium if in the limit as $\beta \rightarrow 1$ the following inequality holds
\begin{multline} \bigg|  \frac{bc}{\tilde{b}_{3}\tilde{c}_{3}} + \bigg( \frac{b}{\tilde{b}_{3}}+ \frac{c}{\tilde{c}_{3}}\bigg)\frac{d}{\tilde{d}_{3}} - 
\frac{\tilde{b}_{1}\tilde{c}_{1}}{\tilde{b}_{3}\tilde{c}_{3}}
 + \frac{\tilde{b}_{0}}{\tilde{b}_{3}} \bigg| > 1 + \bigg| \frac{b}{\tilde{b}_{3}} + \frac{c}{\tilde{c}_{3}} + \frac{d}{\tilde{d}_{3}} \bigg| +  \\ \bigg| -\frac{bcd}{\tilde{b}_{3}\tilde{c}_{3}\tilde{d}_{3}} + 
 \frac{\tilde{b}_{1}\tilde{c}_{1}d}{\tilde{b}_{3}\tilde{c}_{3}\tilde{d}_{3}} 
- \frac{\tilde{b}_{1}\tilde{c}_{0}}{\tilde{b}_{3}\tilde{c}_{3}} -   \frac{\tilde{b}_{0}c}{\tilde{b}_{3}\tilde{c}_{3}} - \frac{\tilde{b}_{0}d}{\tilde{b}_{3}\tilde{d}_{3}} \bigg| + \bigg| \frac{\tilde{b}_{1}\tilde{c}_{0}d}{\tilde{b}_{3}\tilde{c}_{3}\tilde{d}_{3}}   + \frac{\tilde{b}_{0}cd}{\tilde{b}_{3}\tilde{c}_{3}\tilde{d}_{3}} \bigg| \end{multline}
\end{proposition}
\begin{proof} Proposition 16 (i), Proposition 4 and Theorem 3 serve to demonstrate that there have to be two roots inside and two roots outside the unit circle. The final result comes from applying the bound of Rouche's theorem\footnote{This is a standard result in complex analysis, consult \cite{stein2010complex}.} (for precisely two roots outside the unit circle) to the eigenvalue polynomial derived in Appendix E.3 and displayed later as (167).  
\end{proof}
This condition is not necessary for a solution. Bounds are often far from sharp and could be economically uninformative. Alternatively, there is a rather unwieldy quartic formula, which can be found in \cite{auckly2007solving}, that would yield necessary conditions. Appendix E.4.4 extends the analysis to the case with no price dispersion and looks at conditions that imply non-existence. The idea could naturally be applied across a wide range of DSGE. It could potentially be used alongside the techniques in Theorem 2 and developments thereof to analyze the function space of shocks that support equilibrium. This would take us outside the scope of an economics journal. 
\subsection{Key Proofs}
This part has three subdivisions. The first covers analytic prerequisites, the second proves Theorem 3, the main effort of this paper, whilst the third proves Theorem 4.
\subsubsection{Compactness}
Before the proofs appear, it is necessary to introduce the key compactness property. First, a preliminary definition is needed. It prevents the probability measure "escaping to infinity". 
\begin{definition}A sequence of probability measures $\{\mu_{s}: s \in \mathbb{T}\}$ is called \textbf{tight} when for every $\epsilon > 0$ there exists a compact set 
$C \subseteq X$ 
$$\lim_{s \rightarrow \infty } \inf \mu_{s}(C) \geq 1-\epsilon $$ \end{definition}
It has been widely applied in econometrics and microeconomics. The compactness property specializes tightness to expected paths. 
\begin{definition}A dynamical system $T$ will be called \textbf{Bounded in Probability on Average} if for each initial condition $X_{0} \in X$ the sequence of sample averages $\{\mathcal{A}_{s}(X_{0}): s \in \mathbb{T}\}$ is tight. \end{definition}
\par The proof has four strands. The first uses compactness and continuity to prove that stochastic equilibrium is necessary. The second demonstrates the desired eigenvalue configuration is consistent. The third uses a bounding by a suitable linear function to demonstrate that alternative configurations are not consistent with stochastic equilibrium. Finally, a suitable smoothness argument is used to rule out some pathological cases. 
\subsubsection{Proof of Theorem 3}
\begin{proof}The task at the outset is to establish boundedness in probability on average. From the primitives on the differentiability of
$f$ and the distribution, it is clear that $\mathbb{E}X_{T}$ will be continuously distributed, so the only way it might fail would be if the expectation escaped to the boundaries. This would cause the period payoff function $u$ to tend to blow up. Taking the patient limit ($\beta \rightarrow 1$) first then ensures the value of the infinite horizon objective function also explodes. Hence Theorem 12.0.1 of \cite{meyn2009markov} can be invoked to prove the existence of a unique ergodic invariant measure that mixes. This suffices to prove necessity of existence of stochastic equilibrium. 
\par Begin by assuming $f$ is 
$C^{1}$ at the stochastic equilibrium. Focus on the "if" part. First, I will cover the straightforward case, where the matrix is diagonalizable but not diagonal, before coming to two more awkward cases, where the matrix is non-diagonalizable or already diagonal. Writing the linearized system as 
\begin{equation} \mathbb{E}_{\bf{t}}\hat{\bf{X}}_{\bf{t+1}}=\bf{B}\hat{\bf{X}}_{t} + \boldsymbol\phi\hat{\bf{e}}_{t} \end{equation}
where $\bf{X}_{t}=(\bf{X}^{J}_{t}, \, \bf{X}^{S}_{t})'$ is split into jump and state variables. 
The pertinent diagonalization is 
\begin{equation} 
\bf{B}=\bf{P^{-1}}\boldsymbol \Lambda\bf{P}
\end{equation} 
where $\bf{\Lambda}$ 
contains all the eigenvalues $\lambda_{i}$ along its diagonal and zeros elsewhere. This Jordan decomposition is not unique, in particular the eigenvalues can be reordered as desired.\footnote{All the matrix algebra here is contained in \cite{abadir2005matrix} or any good undergraduate linear algebra text.}
The convenient form is to block by ascending order so 
\begin{equation} \bf{\Lambda}=\begin{bmatrix} \bf{\Lambda}_{1} & 0 \\ 0 & \bf{\Lambda}_{2}  \end{bmatrix}\end{equation}
where the number of eigenvalues in the upper block correspond with the number of jump variables and the lower block with the number of state variables. The parameter vector and matrix $\bf{P}$ are similarly partitioned so that $\boldsymbol{\phi}=(\boldsymbol{\phi_{1}}, \, \boldsymbol{\phi_{2}})'$ and 
\begin{equation} \bf{P}=\begin{bmatrix} \bf{P}_{11} & \bf{P}_{12} \\
\bf{P}_{21} & \bf{P}_{22}
\end{bmatrix} \end{equation}
The general solution is 
\begin{multline} 
\hat{\bf{X}}^{\bf{J}}_{\bf{t}}= (\bf{P}_{11}-\bf{P}_{12}\bf{P}_{22}^{-1}\bf{P}_{21})^{-1} \mathbb{E}_{t} \sum_{i=0}^{\infty} \Lambda_{1}^{-(i+1)}(\bf{P}_{11}\boldsymbol{\phi}_{1}+ \bf{P}_{12}\boldsymbol{\phi}_{2})\hat{\bf{e}}_{t+i}- \\ (\bf{P}_{11}-\bf{P}_{12}\bf{P}_{22}^{-1}\bf{P}_{21})^{-1}\bf{P}_{12}\bf{P}_{22}^{-1}  
\mathbb{E}_{t}\sum_{i=0}^{t}\Lambda_{1}^{i}(\bf{P}_{21}\boldsymbol{\phi}_{1}+ \bf{P}_{22}\boldsymbol{\phi}_{2}) \hat{\bf{e}}_{t +i} \end{multline}
\begin{multline}\hat{\bf{X}}^{\bf{S}}_{\bf{t}}= (\bf{P}_{11}-\bf{P}_{12}\bf{P}_{22}^{-1}\bf{P}_{21})^{-1}\bf{P}_{21}\bf{P}_{11}^{-1}\mathbb{E}_{t}\sum_{i=0}^{\infty} \Lambda_{1}^{-(i+1)}(\bf{P}_{11}\boldsymbol{\phi}_{1}+ \bf{P}_{12}\boldsymbol{\phi}_{2})\hat{\bf{e}}_{t+i} + \\ 
\bf{P}_{22}^{-1}(\bf{I} +\bf{P}_{21}(\bf{P}_{11}-\bf{P}_{12}\bf{P}_{22}^{-1}\bf{P}_{21})^{-1}\bf{P}_{12}\bf{P}_{22}^{-1})  \mathbb{E}_{t}\sum_{i=0}^{t}\Lambda_{1}^{i}(\bf{P}_{21}\boldsymbol{\phi}_{1}+ \bf{P}_{22}\boldsymbol{\phi}_{2}) \hat{\bf{e}}_{ t + i}
\end{multline}
Appendix E.3 breaks down the linear algebra. This part of the proof is completed by verifying that the infinite horizon dynamical system, implied by the spectral decomposition, is consistent with stochastic equilibrium. 
The decomposition yields 
$$\bf{\tilde{X}}^{J}_{t+1}= \bf{\Lambda}_{1}\bf{\tilde{X}}^{J}_{t}+ \boldsymbol{\tilde{\phi}_{1}}\bf{\hat{e}}_{t} $$
and 
$$\bf{\tilde{X}}^{S}_{t+1}= \bf{\Lambda}_{2}\bf{\tilde{X}}^{J}_{t}+ \boldsymbol{\tilde{\phi}_{2}}\bf{\hat{e}}_{t} $$ where these new coefficients are displayed in (279) and (280).
From the mixing property, we know that 
for every $\epsilon_{i}>0$ there exists $t'>0$, such that for 
all $T> t +t'$
$$\vert \mathbb{E}_{t}\bf{\tilde{X}}^{J}_{T+1}-
\mathbb{E}\bf{\tilde{X}}^{J} \vert < \epsilon_{1}$$ and by the assumed properties of the error term 
$$\mathbb{E}_{\bf{t}}\vert \bf{\hat{e}}_{T} \vert < \epsilon_{2}$$ 
Applying the triangle inequality component-wise on $\bf{\tilde{X}}^{J}$ yields the estimate 
$$\vert \mathbb{E}_{t}\boldsymbol{{\tilde{X}}^{J}_{T+1}}-\mathbb{E}\boldsymbol{\tilde{X}}^{\bf{J}}\vert \leq K \max\{ {\vert 1/\lambda^{1}_{i} \vert }\} \vert \epsilon \vert $$ 
where $\vert \epsilon \vert  \leq \vert \epsilon_{1} \vert + \vert \epsilon_{1} \vert $ and $K$ denotes a strictly positive constant, that may change from line to line. 
A similar estimate for $\bf{\tilde{X}}^{S}$ implies $$\vert \mathbb{E}_{\bf{t}}\boldsymbol{\tilde{X}}^{\bf{S}}_{\bf{T+1}}-\mathbb{E}\boldsymbol{\tilde{X}}^{\bf{S}}\vert \leq K \max\{\vert \lambda^{2}_{i} \vert \}\vert \epsilon \vert $$
Furthermore, this crosses over to the nonlinear model. The proof follows from \cite{coayla2007hartman} $$\mathbb{E}_{\bf{t}}\vert{\bf{X}}_{\bf{T+1}} -\mathbb{E}\boldsymbol{X} \vert < K \vert  \max\{\vert 1/ \lambda^{1}_{i}, \, \lambda^{2}_{i}\vert \}  \epsilon + {R}(\epsilon)\vert$$
$\bf{R}(\epsilon)$ is the remainder term. It has expectation zero and is locally bounded. Therefore, the triangle inequality applied on the right hand side reveals 
$$\mathbb{E}_{t}\vert{\bf{X}}_{\boldsymbol{T+1}} -\mathbb{E}\boldsymbol{X} \vert \leq  K   \max\{\vert 1/ \lambda^{1}_{i}, \, \lambda^{2}_{i}\vert \}  \vert \epsilon \vert  + \vert \bar{R}(\epsilon)\vert$$
Hence, taking the limit as $\vert \epsilon \vert \rightarrow 0$ completes the proof, that the dynamics here are consistent with mixing about $\mathbb{E}X$. $1-\beta << \vert \epsilon \vert $ maintains the dominance of the patient limit. 
\par There are three special cases: first where the matrix is already diagonal, second where it cannot be diagonalized and third where inverse matrices used in the solution (167) and (168) do not exist. In all instances these problems function as removable singularities (because of the primitives on $f$) that can be treated by considering suitable perturbations. 
\par Let us start with the case where the matrix is already diagonal, where we cannot permute the eigenvalues. Clearly, this can only arise when all the off diagonal elements are zero.  
Local non-trivial dependence and Lebesgue measurability ensure this is a zero probability event that can be ignored in stochastic equilibrium. 
\par Lastly let us turn to the more complicated non-diagonal possibility. Towards a contradiction, suppose there were a surface (of positive measure) in $\mathcal{X}$ where two or more eigenvalues were the same. This means there would be two (non-trivial) eigenfunctions $g_{1}(X_{t}, \, \gamma ) = g_{2}(X_{t}, \, \gamma )$. \cite{magnus1985differentiating} Theorem 3 demonstrates that, any eigenvalue function inherits the property of the ambient differential topology, in this case $C^{1}$. Therefore the implicit function theorem (see \cite{spivak2018calculus}) guarantees that there will be some $C^{1}$ function of the form $X^{i}_{t}=f(X_{t}^{-i}, \, \gamma)$ 
This contradicts the assumption we are on the Euclidean manifold $\mathcal{X}$. Therefore without loss of generality, we can proceed as though dynamic solutions are governed by (167) and (168), by using nearby limiting approximations.
\par The next steps of the proof treat the reverse implication, by ruling out alternative eigenvalue configurations. I proceed by contradiction, starting with instances where there are more or fewer eigenvalues inside the unit circle than jump variables but no eigenvalues on the unit circle, before finishing with these knife-edge cases. The previous arguments have established that around any candidate stochastic equilibrium
$$ \mathbb{E}_{\bf{t}}(\bf{X}^{*}_{T+1}- \mathbb{E} \bf{X}^{*})=\lim_{T \rightarrow \infty}O\big((\max \{ \vert 1/ \lambda^{1}_{i}\vert , \, \vert \lambda^{2}_{i}\vert \})^{T}\big)\vert \epsilon \vert$$
When there are too few eigenvalues in the unit circle, the system will blow up ($\mathbb{E}X^{*} \rightarrow \infty$ for any  $\vert \epsilon \vert>0$) under the influence of an inverse eigenvalue lying outside the unit circle from the first block. With too many the result will be the same but the culprit will be an eigenvalue from the second block. 
\par The final configuration to consider is where there are eigenvalues on the unit circle. I cannot use the Grobman-Hartman theorem, but we can use the geometric properties of stochastic equilibrium. By the continuity properties of $f$ and its eigenvalues (\cite{rahman2002analytic}), we can see that any equilibrium with eigenvalues on the unit circle would have to be arbitrarily close to one where the number of jump variables and eigenvalues outside the unit circle do not match up. Likewise, the objective function would also have to be arbitrarily close to its values in these equilibria but this is impossible since we have already shown that the objective function $u$ diverges in these cases.\footnote{The dominance of the patient limit is crucial because the blow-up of the linearized system would be non-uniform in $\beta$ because no 
$\beta <1$ could rule out eigenvalues on the unit circle.}
\par Lastly, it is necessary to consider the cases where desirable smoothness assumptions fail at the equilibrium point. If $f$ is not $C^{1}$ we can use the Lebesgue differentiation theorem to assign its local dynamics to one of the three bins, using the limiting convention $1-\beta << \delta << \vert \epsilon \vert $. It is clear when the stipulated condition is met, the dynamics will be governed by the stable system. Where there is no $\delta >0$, it will be governed by the non-hyperbolic or explosive system. \end{proof}
\subsubsection{Proof of Theorem 4}
\begin{proof}
Towards a contradiction, note that if there were multiple trajectories stemming from the same combination of predetermined and state variables then there would exist two functions 
$$g_{1}: \mathcal{X}^{\mathfrak{P}} \times \mathcal{E} \rightarrow \mathcal{X}^{\mathfrak{-P}} \times  \mathcal{X}$$
$$g_{2}: \mathcal{X}^{\mathfrak{P}} \times \mathcal{E} \rightarrow \mathcal{X}^{\mathfrak{-P}} \times  \mathcal{X}$$
where $g_{1} \neq g_{2}$ on a measurable set (without loss of generality we can have these inheriting the topology of $f$.) 
The idea is these functions represent two proposed alternative trajectories for the dynamical system. These are maps that send the current errors and predetermined variables to the present values of the non-predetermined and the future position of all the exogenous variables. 
To arrive at a contradiction, consider projections of the system, since $f$ is unique, I can define distinct maps 
$$\tilde{g}_{1}: \mathcal{X}^{\mathfrak{P}} \times \mathcal{E} \rightarrow \mathcal{X}^{\mathfrak{-P}}$$
$$\tilde{g}_{2}: \mathcal{X}^{\mathfrak{P}} \times \mathcal{E} \rightarrow \mathcal{X}^{\mathfrak{-P}}$$
Now take expectations with respect to the ergodic measure, if there is no measurable set 
$\Omega$ such that  
$$\mathbb{E}({g}_{1})(X^{-\mathfrak{P}} \vert X^{\mathfrak{P}} \in \Omega) \neq \mathbb{E}({g}_{2})(X^{-\mathfrak{P}} \vert X^{\mathfrak{P}} \in \Omega)$$ 
take a subsection 
$${g}^{*}_{1}: \mathcal{X}^{\mathfrak{P}}   \rightarrow \mathcal{X}^{\mathfrak{-P}}$$
$$\tilde{g}^{*}_{2}: \mathcal{X}^{\mathfrak{P}} \rightarrow \mathcal{X}^{\mathfrak{-P}}$$
since the two maps are probabilistic and distinct, by assumption, it must be the case that for some $\Omega$
$$\mathbb{E}({g}^{*}_{1})(X^{-\mathfrak{P}} \vert X^{\mathfrak{P}} \in \Omega) \neq \mathbb{E}({g}^{*}_{2})(X^{-\mathfrak{P}} \vert X^{\mathfrak{P}} \in \Omega)$$ 
This violates Birkhoff's ergodic theorem (Theorem 1). We have reached a contradiction.
\end{proof}
\subsection{Discussion}
This stanza commences by defending the only real assumption in the theorems. Subsequent paragraphs proceed under the premise it has been met. 
\par The only economic restriction, that the patient limit $\beta \rightarrow 1$ dominates, is very weak. It rules out non-ergodic solutions and is a requirement for standard econometric analysis. The discussion surrounding Corollary 2, in Section 6.4, suggests it should prove a reasonable approximation. 
It is rare in the empirical literature to see calibrations of $\beta$ below $0.99$ for developed nations. I am committed to testing all limiting assumptions under plausible model conditions in a future paper.\footnote{There is a longstanding tradition in microeconomics of using the patient limit to rule out non-stationary solutions, which can be multitudinous in game theory. This 
extends back to the folk theorem proven in \cite{friedman1971non}. Proposition 23 is the only case where I consider dynamics away from the patient limit but without treating existence.} 
\par The key 
message of Theorem 3 is that dynamic stochastic general equilibrium need not exist. 
Theorem 5 implies they will not exist under plausible parameters for the Rotemberg model.\footnote{In addition to the historical evidence discussed in the introduction, a body of contemporary evidence typically favors under-active response to contemporaneous inflation (see \cite{chortareas2008we}, \cite{taylor2010simple}, \cite{hansen2011model} and \cite{svitak2013empirical}). This is also reflected in the views of economic forecasters and the lack of public appreciation for the textbook policy stance (\cite{mitchell2010wall}, \cite{pierdzioch2012believes}, \cite{carvalho2014people} and \cite{drager2016survey}).} This surely puts paid to its claim to be a model of the business cycle. 
This explains a widespread pattern in the literature, where leading practitioners have struggled to compute non-linear New Keynesian models accurately. 
\cite{carlstrom2014fiscal}, \cite{herbst2016bayesian} and \cite{ascari2018welfare} explicitly discuss non-convergent simulations from popular New Keynesian models.\footnote{\cite{judd2017lower} demonstrates, via a formal statistical test, that existing approximations do not accurately simulate a typical large New Keynesian model but his method conflates problems with \emph{a priori} accuracy and \emph{ex post} existence.} 
\par When a solution does not exist, a package such as Dynare will return an error message, whilst a self-programmed routine will show successive iterations diverging. A key insight is that when a linearized model is indeterminate about its stochastic steady state, at least one of the underlying infinite horizon optimization problems has no solution- in the sense that its objective function is divergent.
In this paper, it means there is no way to define expected lifetime utility or expected profitability of the resetting firm. The amplitude of the business cycle will be arbitrarily large so, the boundary properties mean that, the supposedly infinitely lived representative household will expect to starve and expire from over work with probability one. Popular programming packages warn macroeconomists about indeterminacy for good reason.
\par Claims of indeterminacy ignore the fact that dynamics have to be consistent with the existence of the underlying objective function. Macroeconomists have traditionally ignored this preferring to focus on linear approximations. This approach offers a natural tractability. From this standpoint, when I compute a limiting solution, I am using the non-linear objective function as a refinement strategy, to rule out multiplicity and other unsustainable solutions implied by the linear model. This parallels similar developments in microeconomics and game theory. Moreover, it utilises and indeed justifies the first strand of the Lucas critique. Microfoundations serve a scientific purpose over and above their role in deriving common approximate solutions. 
\par Nevertheless, the purpose of this paper is not to close down debate or rule out plausible concepts \emph{a priori}. \cite{fernandez2023financial} defines a coherent notion of multiple stochastic steady states in continuous time. There is no guarantee uniqueness would carry over to an environment with a single large player (a government or central bank) facing a non-linear optimization problem. There should be sufficient flexibility. Features like hysteresis and structural change, that appear inconsistent with stochastic equilibrium, could be incorporated as limiting cases, as in \cite{bouchaud2023self}. In so far as this new rigorous approach were to impose restrictions on the scenarios that can be analyzed in DSGE, I would advocate greater plurality, including application of ideas and techniques from agent-based modelling, along with greater emphasis on temporary policy deviations, as in \cite{ascari2018high}. 
\par The phenomenon uncovered in Theorem 5 has three mathematical interpretations. The first is failure of local-in-time existence. Reliance on infinite horizon optimization means that a DSGE model exists for all time or none. This is considered pathological in mathematical physics, where systems are typically observed for some time before any blow up.
It makes it easier to interpret blow-ups and understand the properties of a system, from a computational perspective, before rigorous proofs can be supplied. This existence problem has prevented the division of labor between theoretical and mathematical physics being successfully applied to macroeconomics. 
\par Secondly, from a statistical physics perspective, there is a phase transition at the mean field limit. This means that the equilibrium, which we expect would exist when a large number of firms are pricing strategically against one another, breaks down in the limit when the firms start to compete against the aggregate population. However, this interpretation is less intuitive in business cycle analysis; it may make more sense in other economic applications.\footnote{This issue is discussed more formally and explained with commuting diagrams by \cite{carmona2013control}.} 
\par The third perspective comes from theoretical computer science. They would rather re-scale the exploding objective to make it zero and declare its ratio with the social optimum, the \emph{Price of Anarchy}, to have exploded. For background consult \cite{koutsoupias1999worst}, \cite{roughgarden2005selfish} and \cite{roughgarden2015intrinsic}.
\par These results confirm, 
the instrumentalist perspective, that models are meant to be useful not realistic, famously expounded by \cite{friedman1953methodology}. No central banker should lose sleep worrying about exploding price dispersion or worry about inflation hitting some fictitious upper bound.\footnote{Consult \cite{alvarez2018hyperinflation} who provides actual evidence of modest responses of real quantities to high inflation in Argentina.} However, it does reveal that a plausible model can breakdown, when fed unfavorable inputs. This means one cannot take its predictions too seriously away from sensible parameter values. Indeed, the insight is general to DSGE that the facility to make testable predictions is intimately connected with the possibility of non-existence. 
\par This gives the models less robustness. 
Consider the plausible extension where households' are uncertain about the state of the economy and form Bayesian beliefs concerning the parameters of the economy $\gamma$, including the shock processes. Typical conjugate priors have unbounded tails and hence will typically display values for which the model implodes and the households optimization problem along with it. 
The focus here is on the inflation reaction in the policy rule, where there is evidence of considerable dispersion in estimates and variance across econometric methodology, such that confidence intervals would include blow up values. This accords with our intuition that macroeconomics is a scientific endeavour but a less precise one than physics, where it is very difficult to find flaws in the behavior of well-known equations, although a greater emphasis on optimal policy and learning might help, particularly in monetary economics.  
\section{Bifurcation Analysis}
\par This section sets out to give precise mathematical explanation of what is taking place around the
ZINSS of the Calvo model. A number of concepts from algebra, topology and analysis are introduced. Every mathematical construction has intuitive economic appeal. There is discussion of the significance and salience of bifurcation in macroeconomics. 
\par The first part focuses on analysis. The goal is to review how the results here square up with basic real analysis and topology. There follow three subsections introducing ideas from algebra. The first two are topological, the last is geometric. I begin by introducing singularities and covers, with a view to aiding our understanding of market failure. Next a subsection on homology allows me to rigorously discuss how singularities give rise to  "holes" in the state space and how they destroy the dynamics of the underlying model. Finally, I introduce the algebraic machinery of schemes to rigorously study limiting approximations and dig into the root cause of the mathematical pathology. 
\par There are two main theorems. The first proves the singularity construction, foretold back in Section 3 and a little more besides. It buttresses the principal decomposition of the paper. The closing subsection ensures all abstract objects relate back to primitive economic phenomenon. The second provides a tight link between bifurcation, the lag polynomial and the infinite horizon solution. Supporting appendices on the dual cohomology theory, topological groups and categories can be found in Section F of the Appendix. There is also space to consider robustness and extensions. 
\subsection{Analytic Aspects}
The first result is to establish, by analytic means, that the constraint induces irregularities in the derivative. 
\begin{proposition} Defined as a function $Z_{t+1}=f$ everywhere, the system is non-differentiable at ZINSS. \end{proposition}
\begin{proof} In the proof of Proposition 4, from (215)-(216), the system can be expressed as $K/g(\pi_{t+1}, \, \pi_{t})$, such that $g \rightarrow 0$, as $(\pi_{t},\, \pi_{t+1}) \rightarrow 0$. (A.16)-(A.23) constitute a sequence of continuous transformations of the steady state conditions. Therefore the singularity has to be removable. It cannot be differentiable however, since  there will be a term in the derivative $O[(\pi_{t}-\pi_{t+1})^{-1}]$ whose magnitude will explode. 
\end{proof}
This proof only focuses on the wall of the crossing. It confirms the general idea that when a "rearrangement pattern"
fails a function will not be differentiable. 
\begin{remark} This rules out the use of the standard non-stochastic Grobman-Hartman theorem, as described, in for example, \cite{teschl2012ordinary}. However, away from $a_{\pi} =1$, the linearization can be viewed as an approximation to local dynamics about the singular measure, defined as where the appropriate generalization of (3)-(5) bind. \end{remark}
\begin{proposition} The Calvo New Keynesian model, defined by the recursive equilibrium $f$ $\mu$ a.e. in Proposition 4, is not continuously differentiable at ZINSS.\end{proposition}
\begin{proof} It follows immediately from Proposition 8, which proves that all lagged terms vanish so $\mathrm{d}f^{sing}/\mathrm{d}\pi_{t-1}=0$, whilst from (122) and (123) $b_{0} > 0$.
\end{proof}
\subsection{Algebraic Aspects (I) Singularities and Covers}
At the end of this subsection, I precisely describe the behavior of the non-linear system local to the ZINSS. Beforehand, I develop the requisite material from algebraic topology. 
\begin{definition} Let $Y$ be a topological space. A covering space of $Y$ is a topological space $X$ with a surjective (onto) map
$$C : X \rightarrow Y$$ 
such that for every $y \in Y$ there exists a neighborhood $N$ of $x$ such that $C^{-1}(N)$ (the inverse or pre-image of $N$ under $C$) is a union of disjoint open sets in $X$, each of which is mapped homeomorphically onto $N$ by $C$.
\end{definition}
The space $Y$ is often called the base space, whilst $X$ is known as the total space of the covering. For any point $y$ in the base, the inverse image in $X$ is necessarily a discrete space called the fiber over $x$. I will use $C^{X}_{Y}$ to mean $X$ covers $Y$. I will be working simultaneously with multiple covering maps. Therefore I will use the notation $$C^{X_{1}}_{Y} \times \cdots \times C^{X_{n}}_{Y}$$ to denote a product system of covering maps, where every $X_{i}$ covers $Y$ for all $x_{-i}$. When the cardinality of the fiber does not vary we say $Y$ is evenly covered.\footnote{This motivates the cover terminology. The homeomorphic copies in $C$ of an evenly covered neighborhood $N$ form sheets over $N$. One can think of $X$ as "hovering above" $Y$, with the horizontal sheets  piled up on top of $N$ and the cover $C$ pointing down to the base. Thus the fibers over $y$ consist of those points in $X$ that lie directly above $y$.}
Then the cardinality $n$ is called its degree. 
The focus in the paper will be on even covers. 
\par Every space trivially covers itself. More interesting examples include $\mathbb{R}$ covering $\mathbb{S}^{1}$. In general, every immersion from a compact manifold to a manifold of the same dimension is a covering of its image. Turning to the degree, the space consisting of the components $X_{n}=\{x^{2}+y^{2}=n: \, n \in \mathbb{Z}_{++}\}$ is an n-cover of $Y = \{x^{2}+y^{2}=1\} $. A cake with $n+1$ layers is an n-cover of any one layer. The principle interest here will be with covers of degree two.\footnote{There are many well-known geometric instances of two covers. For example, the n-sphere $\mathbb{S}^{n}$ covers the real projective plane (formed of lines through the origin in $\mathbb{R}^{n+1}$). There is one fiber for each of the two intersections.} Since we are working with the usual topology, cover and base spaces will always have the same dimension. \cite{munkres1974topology} is an excellent source for this topic.
\par The last definition is central to the observed mathematical misbehavior. 
\begin{definition} A ramification point is where pairs of branches of a covering map meet.
\end{definition}
The most obvious example is the vertex of a parabola, such as the origin $y=x^{2}$, where the two covering maps $\{y=x^{2}: \, x>0 \}$ and $\{y=x^{2}: \, x< 0 \}$ touch and break apart. In fact, all the coverings will take the form of local quadratic relationships. I will say that a system of covering maps $C^{X_{1}}_{Y} \times \cdots \times C^{X_{n}}_{Y}$ ramifies at a point where every individual cover ramifies.\footnote{In algebraic geometry and complex analysis ramification also carries the connotation of "branching out". This is clearest when we think of paths around the origin for $x=y^{2}$, moving down the vertical axis, we start on $y=\sqrt{x}$ and finish on a different branch $y=-\sqrt{x}$. This interpretation is probably less intuitive in our economic context. A cover with no ramification, such as those discussed previously, is called a universal cover.} 
\par Finally, it helps to have some shorthand notation for all the removable singularities $$R^{x_{2}}_{x_{1}}(g(X(0)))$$ read as "$x_{2}$ replaces $x_{1}$", means that there is a removable singularity $x_{2}=x_{1}$ in the function $g$ at $X(0)$, such that 
$$g(x_{1}, \, x_{1}, \,  X_{-\{1, \, 2\}}(0), \,  \cdot )=g(x_{2}, \, x_{2}, \,  X_{-\{1, \, 2\}}(0), \,  \cdot )$$
The last ingredient is \textbf{De Rham's Theorem} (\cite{tu2011introduction}), which extends differential calculus to manifolds. The idea is that the fundamental theorem of calculus and its immediate implications, hold when appropriately accounting for singular surfaces. All that is required is a differential structure, roughly a formulation of the overall system that is continuously differentiable. It is now possible to reveal the theorem of this subsection. 
\begin{theorem} In the limit as $\beta \rightarrow 1$ at 
ZINSS of the Calvo New Keynesian model, defined by $f_{\pi}$ from Proposition 4 and (97), the following system of covering maps are ramifying $C^{\pi}_{MC}\times C^{\psi}_{MC} \times C^{\pi}_{\Delta}$, whilst the following removable singularities occur $R^{\pi_{t}}_{{\pi}_{t-1}} \times R^{\psi_{t}}_{{\psi}_{t-1}}$
\end{theorem}
\begin{proof} For the first result around ZINSS, note that in non-stochastic steady state marginal costs can be expressed as a univariate function of inflation, see (271). From Proposition 6, we know that with lump sum transfers in place ZINSS uniquely implements the social optimum. When there is no discounting, this amounts to maximizing marginal costs (equal to the wage rate) subject to the technical constraints. Therefore, inflation must provide a local cover of marginal costs. The focus here is on the case without corrective subsidies, which forms a linear transform of efficient marginal costs $MC^{NSS}=(\theta-1)/\theta MC^{*}$, which will preserve the covering. Approaching from the trend inflation side, it is transparent from Proposition 3, that price dispersion vanishes as the cover ramifies. It is evident from Proposition 4, that marginal costs can be expressed as a function of inflation and its lag. Moreover, since $u$ and the elementary transformations forming the steady state are sufficiently differentiable,  where we can apply De Rham's Theorem.  
This implies that around ZINSS $$\left. \frac{\mathrm{d} MC_{t}}{\mathrm{d}\pi_{t}} + \frac{\mathrm{d} MC_{t}}{\mathrm{d}\pi_{t-1}}\right|_{ZINSS}=0$$ thus 
$\mathrm{d} MC_{t}/\mathrm{d}\pi_{t-1}=-\mathrm{d} MC_{t}/\mathrm{d}\pi_{t}$
Therefore the equation of the singularity, which eliminates $\pi_{t-1}$, can be derived from a first order expansion of marginal costs from ZINSS $$\left. \frac{\mathrm{d} MC_{t}}{\mathrm{d}\pi_{t}} \pi_{t} + \frac{\mathrm{d} MC_{t-1}}{\mathrm{d}\pi_{t-1}} \pi_{t-1}\right|_{ZINSS}=0$$ which yields $\pi_{t}=\pi_{t-1}$ as desired. \par The construction for the demand shocks is identical, except the cover comes from the non-linear expansion corresponding to (167) and (168) around ZINSS. Technically, it is the section where higher lags ($\hat{\psi}_{t-i}$, \, $i \geq 2$) have vanished. It arises out of the differential structure of the steady state around ZINSS, where error terms vanish. The last covering result is a simple consequence of Proposition 2.\end{proof}
The second part justifies the calculations made in 7.2.2. What is happening is that the second order approximations are dropping off, which ensures that the coefficients from past and present shocks will have opposite signs in the first order approximation. The singularities cause them to be conflated and cancel out. This is an exceptional mathematical pathology. This problem is avoided with price dispersion because its covering by inflation is not inter-temporal (there are no lags in its recursion (48)). 
\par This symmetry will prove a universal feature of dynamic responses to shocks affecting \emph{efficient} resource allocation around the non-stochastic steady state of models with only static market failure, including other Keynesian models. This contrasts with ad hoc shocks to expectations or mark-ups that should not move the efficient allocation and will lack dynamical structure. This symmetry will break down around non-degenerate stochastic equilibrium because there will be differential exposure to uncertainty in other variables. This promises to be a major theme in modern macroeconomics. 
\begin{remark} The assumption $\beta \rightarrow 1$ is a requirement for the analysis, to ensure there is an inter-temporal covering by a choice variable, as opposed to a shock. When discounting is introduced the covering interpretation breaks down but the singularity will remain and smoothly approach the covering case, as will be demonstrated in the next subsection. \end{remark}
\subsection{Algebraic Aspects (II) Homology}
The first part introduces the theoretical building blocks. The second applies these constructs to the problem at hand. Appendix F.1 provides algebraic extensions and F.3 the dual theory.   
\subsubsection{Basic Homology and Cohomology Theory}
Homological algebra studies holes in topological spaces. The original motivation was the observation that two shapes can be distinguished by examining their holes. For instance, a circle is not a disk because it has a hole whilst the disk is solid. Likewise, the ordinary sphere differs from a circle because it encloses a two rather than a one dimensional hole.
\par The building block of homology theory is called a \textbf{cycle} which forms a closed submanifold. They can be thought of as loops. For example, a line on a surface is a 1-cycle, a two dimensional surface like a disc cut out of a sphere is a 2 cycle. These cycles can be thought of as cuts or zippers that can be fastened or unfastened. A \textbf{boundary} is a cycle which is also the boundary of a submanifold. A \textbf{homology class}, which represents the hole or cut, is an equivalence class of cycles modulo boundaries. Therefore, a homology class is represented by a cycle which is not the boundary of any closed submanifold. Effectively, it is a "missing manifold". It is part of the space that would be added to form a Euclidean 1 cover. Here it represents economic effects and dynamic trade-offs, intrinsic to the model, that are absent from approximations taken from ZINSS.  
\par The principle tool required is the \textbf{homology group}. The $n^{th}$ homology group represents behavior in dimension $n$. The $0^{th}$ is called the fundamental group. It is trivial if the space is connected because every loop can be continuously deformed into a single point.\footnote{A zero dimensional hole is the homology class of a space with a missing point, such as the punctured plane $\mathbb{R}^{2}/ \{ (0, \, 0)\}$.} Appendix F.1 provides formal justification for these results. 
\par The dual \textbf{cohomology} theory, expounded in F.3, explains how these holes are filled in. 
The main result here is once again De Rham's theorem. It implies that holes are filled by singular surfaces with dimension equal to the difference between the ambient surface and the size of the hole. The intuition is that the only way that standard approximations of these models can break down is when a "rearrangement pattern" fails. 
\subsubsection{Application} 
Before proceeding to the bifurcation analysis around ZINSS, it is necessary to modify Theorem 6 to accommodate discounting.
\begin{proposition} Around ZINSS there is a singularity $\pi_{t-1}=\beta \pi_{t}$. \end{proposition}
\begin{proof} The optimization of household welfare problem amounts to maximizing the real wage and therefore the real marginal costs, with respect to the choice of present inflation and its lag. The first order condition sets the marginal effect today to the discounted marginal wage loss tomorrow so
$$\left. \beta \frac{\mathrm{d} MC_{t}}{\mathrm{d}\pi_{t}} + \frac{\mathrm{d} MC_{t}}{\mathrm{d}\pi_{t-1}}\right|_{ZINSS}=0$$ Thus, 
the rest of the arguments proceed as with Theorem 6. It follows that
$\mathrm{d} MC_{t}/\mathrm{d}\pi_{t-1}=-\beta \mathrm{d} MC_{t}/\mathrm{d}\pi_{t}$. Therefore the equation of the singularity, which eliminates $\pi_{t-1}$, can be derived from a first order expansion of marginal costs from ZINSS $$\left. \beta \frac{\mathrm{d} MC_{t}}{\mathrm{d}\pi_{t}} \pi_{t} + \frac{\mathrm{d} MC_{t-1}}{\mathrm{d}\pi_{t-1}} \pi_{t-1}\right|_{ZINSS}=0$$ which yields $\pi_{t-1}=\beta \pi_{t}$. 
\end{proof}
Appendix F.4 provides an elementary demonstration of the dynamic relationship between marginal costs and inflation.
\par
The main application concerns the family of linear approximations at low volatility. 
\begin{proposition} 
Consider a sequence of linear approximations around ZINSS with volatility $\varepsilon \in [0, \, \bar{\varepsilon}]$, ZINSS is the location of a three dimensional hole in the first derivative of the function space. Otherwise the homology is trivial. Formally, 
$$H^{n}(f^{1})= \begin{cases} \{\mathbb{Z}\}  & n=3 \\ \{0\} & \textnormal{otherwise}
\end{cases}$$
\end{proposition}
\begin{proof} 
The proof follows swiftly from Proposition 8, the derivation of the two Phillips curves in $5.1$, $7.1$ and the calculations in E.1.4, which verify that none of the terms in the Phillips curve drop out, when we fix $\beta < 1$. 
\end{proof}
\begin{remark}The same construction would work for a sequence of approximations at different rates of trend inflation. There would be no problem if I considered the $\sqrt{\varepsilon}$ case because the ramification of this cover in a single variable does not generate a hole. 
\end{remark}
\begin{remark} By contrast the homology is trivial when we exclude ZINSS ($\varepsilon \in (0, \, \bar{\varepsilon}]$).\end{remark}
This motivates the limiting concept and indeed the entire stochastic equilibrium concept from an \emph{a priori} New Keynesian standpoint. 
\par The following decomposition, derived from (112), serves to locate the components of the singularity and justifies the claims for the $\sqrt{\varepsilon}$ 
limit case made at the start. It starts by moving all the terms to the left hand side.
\begin{decomposition}
 \begin{multline} \tag{A}
  \beta \bigg(\alpha \mathbb{E}(1+\pi)^{\theta-1}(2+\pi) - \\ \frac{\alpha (1+\pi)^{\theta-1}}{(1+\pi)^{\theta-2}} \bigg[ \frac{\theta \mathbb{E}(1+\pi)^{\theta-1} \nu'(\Delta Y/A)Y/A(1-\alpha \beta \mathbb{E}(1+\pi)^{\theta})} {\nu'(\Delta Y/A)Y/A + \alpha \beta \mathbb{E}(1+\pi)^{\theta}\nu'(\Delta Y/A)Y/A(1-\alpha \beta \mathbb{E}(1+\pi)^{\theta})}\\ -(\theta-1)
\frac{\mathbb{E}(1+\pi)^{\theta-2}\psi u'(Y)Y/(1-\alpha \beta \mathbb{E}(1+\pi)^{\theta-1})}{\psi u'(Y)Y + \alpha \beta \mathbb{E}(1+\pi)^{\theta-1}\psi u'(Y)Y/(1-\alpha  \beta \mathbb{E}(1+\pi)^{\theta-1})}\bigg]\bigg)\hat{\pi}_{t}
\end{multline}
\begin{multline} \tag{B}
\frac{\beta}{(1+\pi)^{\theta-2}} \bigg( \frac{\theta \mathbb{E}(1+\pi)^{\theta-1} \nu'(\Delta Y/A)Y/A(1-\alpha \beta \mathbb{E}(1+\pi)^{\theta})} {\nu'(\Delta Y/A)Y/A + \alpha \beta \mathbb{E}(1+\pi)^{\theta}\nu'(\Delta Y/A)Y/A(1-\alpha \beta \mathbb{E}(1+\pi)^{\theta})}\\ -(\theta-1)
\frac{\mathbb{E}(1+\pi)^{\theta-2}\psi u'(Y)Y/(1-\alpha \beta \mathbb{E}(1+\pi)^{\theta-1})}{\psi u'(Y)Y + \alpha \beta \mathbb{E}(1+\pi)^{\theta-1}\psi u'(Y)Y/(1-\alpha  \beta \mathbb{E}(1+\pi)^{\theta-1})}\bigg)\hat{\pi}_{t}
\end{multline}
\begin{multline} \tag{C}  \bigg( \frac{\theta \mathbb{E}(1+\pi)^{\theta-1} \nu'(\Delta Y/A)Y/A(1-\alpha \beta \mathbb{E}(1+\pi)^{\theta})} {\nu'(\Delta Y/A)Y/A + \alpha \beta \mathbb{E}(1+\pi)^{\theta}\nu'(\Delta Y/A)Y/A(1-\alpha \beta \mathbb{E}(1+\pi)^{\theta})} - \\ (\theta-1)
\frac{\mathbb{E}(1+\pi)^{\theta-2}\psi u'(Y)Y/(1-\alpha \beta \mathbb{E}(1+\pi)^{\theta-1})}{\psi u'(Y)Y + \alpha \beta \mathbb{E}(1+\pi)^{\theta-1}\psi u'(Y)Y/(1-\alpha  \beta \mathbb{E}(1+\pi)^{\theta-1})}\bigg) \times \\ \frac{\beta^{2}(1-\alpha(1+\pi)^{\theta-1})}{(1+\pi)^{2(\theta-2)}}\mathbb{E} (1+\pi)^{\theta-1}\pi(\hat{\beth}_{t} -\mathbb{E}_{t-1}\hat{\beth}_{t})-\hat{\pi}_{t-1}  \end{multline}
\begin{equation} \tag{D} \alpha \beta \mathbb{E}(1+\pi)^{\theta-1}(2+\pi)\hat{\pi}_{t} +(\alpha \beta)^{2}\mathbb{E}(1+\pi)^{\theta-1}\mathbb{E}(1+\pi)^{\theta}\mathbb{E}_{t}\hat{\pi}_{t+1}\end{equation}
\begin{multline} \tag{E} - \bigg(\frac{1-\alpha(1+\pi)^{\theta-1}}{\alpha (1+\pi)^{\theta-2}}\frac{\eta \nu'(\Delta Y/A)Y}{A \aleph}\hat{\Delta}_{t-1} \\ 
+ \frac{1-\alpha(1+\pi)^{\theta-1}}{\alpha (1+\pi)^{\theta-2}}\frac{(1+\eta )\nu'(\Delta Y/A)Y}{A \aleph}\hat{y}_{t-1} \\ -\frac{1-\alpha(1+\pi)^{\theta-1}}{\alpha (1+\pi)^{\theta-2}}\frac{(1+\eta ) \nu'(\Delta Y/A)Y}{A \aleph}\hat{a}_{t-1}
+ \\ \beta \theta(1-\alpha (1+\pi)^{\theta-1})\frac{\mathbb{E}(1+\pi)^{\theta-1}\aleph}{(1+\pi)^{\theta-2}\aleph} \hat{\pi}_{t} + v^{*}_{t}\bigg) \end{multline}
\begin{multline} \tag{F} \beta  (1-\alpha(1+\pi)^{\theta-1})\frac{\mathbb{E}(1+\pi)^{\theta-1}}{(1+\pi)^{\theta-2}}\frac{\eta \nu'(\Delta Y/A)Y}{A\aleph}\hat{\Delta}_{t} + \\ \beta (1-\alpha(1+\pi)^{\theta-1})\frac{\mathbb{E}(1+\pi)^{\theta-1}}{(1+\pi)^{\theta-2}}\frac{(1+\eta)\nu'(\Delta Y/A)Y}{A\aleph}\hat{y}_{t}-\\ \beta (1-\alpha(1+\pi)^{\theta-1})\frac{\mathbb{E}(1+\pi)^{\theta-1}}{(1+\pi)^{\theta-2}}\frac{(1+\eta) \nu'(\Delta Y/A)Y}{A\aleph}\hat{a}_{t}\\ + \alpha \beta^{2} \theta (1-\alpha(1+\pi)^{\theta-1}) \mathbb{E}(1+\pi)^{\theta-1}\frac{\mathbb{E}(1+\pi)^{\theta-1}\aleph}{(1+\pi)^{\theta-2}\aleph}\mathbb{E}_{t}\hat{\pi}_{t+1}\end{multline}
\begin{multline}\tag{G} \frac{1-\alpha(1+\pi)^{\theta-1}}{\alpha (1+\pi)^{\theta-2}}\frac{(1-\sigma)\psi u'(Y)Y}{\beth}\hat{y}_{t-1}+\\ \beta (\theta-1)(1-\alpha (1+\pi)^{\theta-1})\frac{\mathbb{E}(1+\pi)^{\theta-2}\beth}{(1+\pi)^{\theta-2}\beth}\hat{\pi}_{t}\end{multline}
\begin{multline}\tag{H} -\bigg(\beta (1-\sigma)(1-\alpha (1+\pi)^{\theta-1})\frac{\psi u'(Y)Y}{\beth}\frac{\mathbb{E}(1+\pi)^{\theta}}{(1+\pi)^{\theta-2}}\hat{y}_{t}\\ + \alpha\beta^{2}(\theta-1)(1-\alpha (1+\pi)^{\theta-1})\mathbb{E}(1+\pi)^{\theta}\frac{\mathbb{E}(1+\pi)^{\theta-2}\beth}{(1+\pi)^{\theta-2}\beth}\mathbb{E}_{t}\hat{\pi}_{t+1}\bigg) \end{multline}
The relation with the equations in Section 2 is as follows: 
\begin{enumerate}[i]
\item (A) + (F)+(H) $=0 \equiv$ (1) 
\item (B) + (C) $=0 \equiv$ (3)
\item (D) + (E) + (G) $=0 \equiv$ (5)
\end{enumerate}
\end{decomposition}
The De Rham cohomology allows me to integrate each of these collections of terms to form surfaces. It is therefore clear that the homology / cohomology would extend to higher order approximations, as this would simply improve the approximation of each singular surface. Geometrically, the sum of all the terms represents the ambient space, whilst (ii), (iii) are holes that glue together to form a singular surface enclosing (i). The demand side disturbance term, which vanishes at ZINSS, reflects interactions between the shocks and output terms, uncovered back in Proposition 10.
\par The second surface (ii) represents the constraint that prices today equal prices tomorrow. This follows thanks to Proposition 1 that maps between reset prices and inflation. This step arises out of the non-cancellation of the lag operators in (108) and (109). It is therefore necessary to the whole bifurcation, representing the wall of the crossing between the two Phillips curves. It is now clear why dynamics on surface (i) are entirely forward-looking once re-optimization constraints have been removed. It is sub-optimal for firms to condition on past events, as it would be for the policymaker, which is made clear in Proposition 9. The error term here represents intertemporal distortions caused by stochasticity. It is of order $\pi$ around ZINSS and therefore vanishes to leave a non-stochastic limiting relationship.
\par The third surface is the "residual" surface. It arises automatically from the need to use the lagged Euler (114) to substitute out lagged output. It cleanses the model of inter-temporal forces. In an exactly determined system it is not a problem if one of the surfaces is less intuitive; if the system has $N$ dimensions then there are $N-1$ degrees of freedom and the last adjusts to equilibriate the model. This confirms that it represents \emph{General Equilibrium Effects}. The expression for the approximation around ZINSS is housed in Appendix E.1.
\par  Returning to our prime approximation, the salient point is that ZINSS is the site of a "hole within a hole". Throughout the state space there is a two-dimensional hole with codimension one. This reflects the fact that re-optimization constraints necessarily induce a cost channel and an inter-temporal substitution motive. At ZINSS, in the first order expansion, there is an additional singularity "inside" the ambient singular surface given by (4). This acts to destroy the present shock term coming from the demand side necessitated by their cancellation from the other singular components. Elsewhere, this constraint drops out because in general there will be a value of inflation for which re-optimization constraints break down, even though there is a non-equilibrium realization of the main shock. We can always view the second constraint (5) as acting to remove this lagged error term, in the way that the first constraint (3) is removing lagged inflation. 
\par If the approximation is such that a supply shock enters, there must be another component of the singular surface to eliminate it. In general, surface (5) scales with the number of variables that are used to flesh out the economy's neoclassical skeleton. Therefore, this bifurcation analysis extends to the medium and large models used at central banks. 
The formalization is as follows: 
\newpage
\begin{proposition} The homology of the various versions of the non-linear Calvo Phillips curve are 
$$H^{n}_{\sigma =1} =\begin{cases} \mathbb{Z}, & \text{for } 
$k= 0, \, 2$ \\ 
\{0\}, & \text{otherwise} 
\\
\end{cases}$$ 
$$H^{n}_{\sigma \neq 1} =\begin{cases} \mathbb{Z}, & \text{for } 
$k= 0, \, 3$ \\
\{0\}, & \text{otherwise} 
 \\
\end{cases}$$ 
\end{proposition}
\begin{proof} The first augments Proposition 21,
by noting that Proposition 10 implies that on the singular surface there arise demand shock terms $\psi$, removing the need for singularity (5), and reducing the dimensionality of the hole by one. The second case employs the same logic but acknowledges, in the light of Appendix C.1.4, that away from $\sigma=1$, terms in the technology shock arise. 
This necessitates another dimension to the hole to remove the lagged technology term, that comes about under staggered optimization, as shown by Proposition 10. The mechanics of constructing these groups is discussed in F.3.2. 
\end{proof}
\begin{remark} This singularity would again take the symmetric form $\hat{a}_{t}=\hat{a}_{t-1}$ around ZINSS, reflecting that this fulfils the maximization of welfare at the efficient steady state, as argued for demand shocks in Theorem 6. 
\end{remark}
\par The following diagram offers insights into both the strands of bifurcation analysis throughout the paper 
\\ \\
\begin{center}
\textbf{Figure 1: Well-Behaved Approximations}
\end{center}
\begin{center}
\begin{tikzcd}
Z_{t} \arrow[r, "f^{*}"] \arrow[d, "h"] &  \mathbb{E}_{t}Z_{t+1} \arrow[d, "h^{loc}(0)"]\\
\hat{Z}_{t} \arrow[r, "\hat{f}"] &  \mathbb{E}_{t}\hat{Z}_{t+1}
\end{tikzcd}
\end{center}
$f^{*}$ is the primitive function created by the DSGE, taken to include both $f$ of Proposition 4 and $f^{sing}$ of Proposition 10.
$h$ is an isomorphism typically linear. It is a mechanical relationship, that represents the state of the business cycle, for example computing the percentage deviation from steady state. $\hat{f}$ is the approximating model, used for econometric or theoretical purposes. Taylor expansions would be most common.
$h^{loc}(0)$ is a local homeomorphism that preserves the origin.\footnote{The focus on local homeomorphism is motivated by the application of perturbation techniques. It would be too demanding to seek global relationships. After all, economists do not expect an approximation to be globally accurate.} It represents the mapping between the predictions of the approximate model and the mappings of the underlying nonlinear model. \par  \emph{A priori} the role of the stochastic Grobman-Hartman theorem used in the proof of Theorem 3 is to prove that this local homeomorphism is possible, with sequences of linear approximations derived around the stochastic steady state. The point of the bifurcation analysis is that, this breaks down for approximations taken from the non-stochastic steady state, which only approximate the singular surface $f^{sing}$. Appendix F.3  covers formal background to the diagrams in this paper.
There can also be a similar break down \emph{ex post} around the critical parametric threshold implied by Theorem 3, such as $a_{\pi}=1$ in Theorem 5. 
\subsection{Algebraic Aspects (III) Schemes}
This subsection is split in two. 
The first part is mathematical, where I introduce the central technical concepts of schemes, rings, ideals and the Zariski topology. The second part applies these techniques to prove necessary and sufficient conditions for bifurcation between the underlying non-linear DSGE and the dynamics of the linear approximation about an equilibrium point. 
\subsubsection{Preliminaries Algebraic Geometry}
 To appreciate the final argument it is necessary to develop some competence with the methods of algebraic geometry. The underlying principle is that one can associate a system of simultaneous equations with their roots.\footnote{This is an extension of the fundamental theorem of algebra, which states that an $n^{th}$ order polynomial is determined by the $n$ complex roots it factorizes into (allowing multiplicity).} This is formalized by a local ring system in the case of a linear system and algebraic variety for polynomials.\footnote{Recall that a ring $R$ is a set equipped with two binary operators $+$ and $\times$ satisfying the following three ring axioms 
\begin{enumerate}
\item $R$ is an Abelian group under addition meaning 
\begin{itemize}
\item $(a + b) + c = a + (b + c)$ for all $a, \, b, \, c$ in $R$ (that is, $+$ is associative).
\item $a + b = b + a$ for all $a, \, b$ in $R$  (that is, $+$ is commutative).
\item There is a zero element in $R$, such that $a + 0 = a$ for all $a$ in $R$ (that is, $0$ is the additive identity).
\item For each $a$ in $R$ there exists $-a$ in $R$, such that $a + (-a) = 0$ (that is, $-a$ is the additive inverse of $a$).
\end{itemize}
\item $R$ is a monoid under multiplication 
\begin{itemize}
\item $(a \times b) \circ c = a \times (b \times c)$ for all $a$, $b$, $c$ in $R$  (that is, $\circ$ is associative).
\item There is a unit element in $R$, such that $a \times 1 = a$ and $1 \times a = a$ for all $a$ in $R$  (that is, $1$ is the multiplicative identity).
\end{itemize}
\item Multiplication is distributive with respect to addition implying
\end{enumerate}
\begin{itemize}
\item $a \times (b + c) = (a \times b) + (a \times c)$ for all $a$, $b$, $c$ in $R$  (left distributivity)
\item $(b + c) \times a = (b \times a) + (c \times a)$ for all $a$, $b$, $c$ in $R$ (right distributivity).
\end{itemize}
The family of square matrices with the usual operations of addition and matrix multiplication is a ring, as is the set of all continuous functions on the real line.}
\par The particular concept we need is an affine variety. A preliminary definition is helpful. 
\begin{definition}
An affine space $\mathbb{A}$ comprises a set $A$, together with a vector space $\overrightarrow {A}$ and a transitive free action of the additive group of $\overrightarrow{A}$, on the set $A$. The elements of the affine space A are called points. The vector space $\overrightarrow{A}$ is said to be associated to the affine space and its elements are called vectors, translations or sometimes free vectors. Unpacking the definition reveals the following primitive properties of the mapping represented by the action 
$$A \times \overrightarrow {A} \rightarrow A$$
$$(a, \, v)\mapsto a + v$$
\begin{enumerate}
\item Right identity: \\ 
$\forall a \in A$, $a+0=a$, where $0$ is the zero vector in $\overrightarrow {A}$
\item Associativity: \\
$\forall v, \, w \in \overrightarrow {A}, \forall a \in A,\;(a+v)+w=a+(v+w)$ (here the last $+$ is the addition in $\overrightarrow{A}$)
\item Free and transitive action: \\
For every $a \in A$, the mapping ${\overrightarrow {A}}\to A \colon v \mapsto a+v$ is a bijection.
\end{enumerate}
\end{definition}
 Intuitively, an affine space is the minimal structure required to determine changes in the multiplicity of solutions for a system of simultaneous linear equations. It starts from the familiar notion of a vector space and then dispenses with angles and distances; as such there is no need for an origin. Points represent translations or displacements rather than positions. One can think of the affine subspace as resulting from translating the linear subspace (away from the origin) by addition of the translation vector. In finite dimensions, such an affine subspace is the solution set of an inhomogeneous linear system. 
 The displacement vectors for that affine space are the solutions of the corresponding homogeneous linear system, which is a linear subspace and therefore must contain the origin of the vector space.\footnote{Textbooks sometimes use alternative properties for axiomization:
\begin{itemize}
\item Existence of one-to-one translations:
\\ For all $v \in \overrightarrow {A}$, the mapping $A \to A \colon a \mapsto a+v $ is a bijection.
\item Subtraction:
\\ 
For every $a, \, b$ in $A$, there exists a unique $v \in \overrightarrow {A}$ denoted $b - a$, such that $b=a+v$
\end{itemize}
In fact, the first extra property follows from the first two axioms. The second is equivalent to axiom three. }
 \begin{definition} For an algebraically closed field $P$ \footnote{A set $P$ forms an algebraically closed field, if every non-constant polynomial in $P[x]$ (the univariate polynomial ring with coefficients in $P$) has a root in $P$. The fundamental theorem of algebra ensures that $\mathbb{C}^{n}$ is an example.} 
 and a natural number $n$, let $\mathbb{A}^{n}$ be the affine $n$-space over $P$. An affine variety is a set, where a collection of polynomials $S \in P$ simultaneously vanishes, that is to say
$$Z(S)=\{ x \in \mathbb{A}^{n} \; \vert \; p(x)=0 \, \text{for all} \; p \in S\} $$
\end{definition}
It is now possible to lay out the crucial sequence of 
topological objects. 
\begin{definition} The Zariski topology on the affine space $\mathbb{A}^{n}$ with respect to a subset $S \in P[x]$ is defined by its closed sets given by 
$$V(S)=\{x \in \mathbb {A} ^{n}\, \vert \; p(x)=0,\text{for all} p \in S\}$$ \end{definition}
In the Zariski topology the only closed sets are the solutions of the polynomial equations. This means Zariski is coarser, in the sense that it has many more open sets than the usual topology. A closed set in the Zariski topology generated by a particular polynomial family is often called an \textbf{ideal}. The topological invariant is called the \textbf{Krull dimension}. It measures the size of the solution set in the ambient space. Its definition is a little technical. The dimension of a polynomial ring over a field $p[x_{1}, \, \cdots , \, x_{n}]$ is the number of variables n. For a field it is zero. It will always correspond with the Euclidean dimension of the solution set in the present application.\footnote{The Krull dimension is defined as the dimension of the maximum prime ideal (by set inclusion). A prime ideal has the property that $ab \in R$ implies either $a \in R$, $b \in R$ or both. Suppose the solution of a system of equations is given by the ideal $y-p(x)=0$. Take two elements $a, \, b \in R$ $a=y-p_{1}(x)$ and $b=y-p_{2}(x)=0$. Suppose that their product were part of the ideal then $(y-p_{1}(x))(y-p_{2}(x))=0$. By the factor theorem, we know that either or both $y-p_{1}(x)=0$ in which case $a \in I$ else $y-p_{2}(x)=0$  then $b \in I$. This demonstrates that our solutions do conform with the definition of a prime ideal. Other examples include the even numbers over $\mathbb{Z}$ and $f(x)=0$ for some $x \in M$ (a manifold.)
The nominal inspiration comes from the fact that if a prime number $p$ divides $ab$ then either $p$ divides a or $p$ divides $b$, implying that a positive integer $n$ is a prime number if and only if $n \mathbb {Z}$ is a prime ideal in $\mathbb{Z}$.} 
\par The idea is that each of the surfaces (1)-(5) represent prime ideals corresponding to the distinct components of the solution.\footnote{An ideal would be any subset of these.} The Zariski topology will come into its own when we start associating systems of equations with their lag polynomial.
The next part is devoted to describing the machinery of these local constructs. Appendix F.2 offers help with category theory.
\begin{definition} Let $X$ be a topological space. A presheaf of sets $F$ on $X$ consists of the following data:
\begin{itemize}
\item For each open set $U$ of $X$ a set $F(U)$. This set is sometimes also denoted $\Gamma(U, \, F)$. The elements in this set are called the sections of $F$ over $U$.
\item For each inclusion of open sets $V \subseteq U$, a function ${res} _{V,\,U}\colon F(U)\rightarrow F(V)$ called a restriction morphism (maps from one sheaf to another). 
The restriction morphisms are required to satisfy two additional categorical properties:
\item For every open set $U$ of $X$, the restriction morphism $res_{U,\, U} F(U) \rightarrow F(U)$
\item If we have three open sets $W\subseteq V \subseteq U$, then the composite ${res}_{W,\,V}\circ {res}_{V,\,U}={res}_{W,\, U}$
\end{itemize}
\end{definition}
\begin{definition}A sheaf is a presheaf with two additional requirements 
\begin{itemize}
\item (Locality) If $\mathcal{U}$ is an open covering of an open set $U$, and if $s, \, t \in F(U)$ have the property $s|_{U_{i}}=t|_{U_{i}}$ for each set $U_{i}$ of the covering, then $s=t$ 
\item (Gluing) If $\mathcal{U}$ is an open cover of a set $U$ and if for each $i \in I$ a section $s_{i} \in F(U_{i})$ is given, such that for each pair $U_{i}$, $U_{j}$ belonging to the covering set system, the restrictions agree on the overlaps so $s_{i} \vert_{ U_{i} \cap U_{j}}=s_{j} \vert_{ U_{i} \cap U_{j}}$, then there exists a section $s \in F(U)$ where  $s|_{U_{i}}=s_{i}$ for all $i \in I$ 
\end{itemize}
\end{definition}
Sheaves are a tool to systematically assign local data to open sets of a topological space. A presheaf stipulates how data can be assigned to smaller sets, whilst a sheaf shows how they are glued together to move from local to global properties. Examples of sheaves include the space of $C^{k}$ differentiable functions on $\mathbb{R}^{n}$ or bounded real-valued functions on a closed manifold. Holomorphic functions on $\mathbb{C}^{n}$ and constant functions are presheaves that cannot be extended to full sheaves.\footnote{It is not possible to extend a holomorphic to the entire complex plane, since Liouville's theorem states that the only bounded constant function can be entire (holomorphic throughout the complex plane, as explained in \cite{stein2010complex}). It is not possible to glue two distinct constant functions together to make another constant function.} Last of all, to formalize discussion of local properties we need:
\begin{definition} The stalk of a sheaf $\mathcal{F}$ at $x \in X$ (a topological space), denoted $\mathcal{F}_{x}$, is given by the following direct limit\footnote{"Direct limit" is a term from category theory. Appendix Section E.2 explains how the direct limit coincides with the limit from common analysis when we consider classes of approximations.} indexed over all the open sets $U$ containing $x$:
$$\lim \mathcal{F}_{x} : = \lim_{\substack{\rightarrow \\ U \ni x}} \mathcal{F}(U)$$  
\end{definition}
 Note that there is an order relation induced by reverse inclusion  $U \leq V$ iff $U \subseteq V$.\footnote{An element of a stalk is an equivalence class of elements $x_{U} \in \mathcal{F}(U)$, such that where two such sections $x_{U}$ and $x_{V}$ are considered equivalent if the restrictions of the two sections coincide on some neighborhood of x.} The intuition is that sheaves permit us to rigorously speak about sequences of approximations with stalks, allowing us to specify a point around which to consider these limits. I will always choose ZINSS. 
\par Before I finish by looking at schemes, it is necessary to introduce two relatively straightforward terms. The \textbf{Spectrum} $R$ of a ring is the collection of all prime ideals. You can think of it as the set of all surfaces forming solutions of polynomial equations. A \textbf{Ringed Space} $(X, \, \mathcal{O}_{X})$ is a topological space $X$ together with a sheaf of rings $\mathcal{O}_{X}$ on $X$. The sheaf $\mathcal{O}_{X}$ is called the \textbf{Structure Sheaf} of $X$. A \textbf{Locally Ringed Space} is a ringed space such that all stalks are all local rings.\footnote{A local ring is defined by a unique maximal ideal $I$. A maximal ideal contains all proper ideals and is smaller than the ring itself.  All our ideals are maximal by the weak Nullstellensatz theorem, which 
informally carries over the idea from the Fundamental theorem of algebra, that polynomials can be uniquely determined by their roots to systems of equations.}
\par In our context, the spectrum of the ring is the set of all solutions describing the possible configurations of $Z_{t}$ and $Z_{t+1}$. Suitably localized, the direct product of (3)-(5) would be the spectrum of the singular boundary surface. 
The purpose of these local constructions is to accommodate the fact that the simultaneous equations of interest in this paper are not defined on the whole field like $\mathbb{R}^{n}$, as with $P[x_{1}, \, x_{2}, \,  \cdots, \, x_{n}]$, due to constraints on the state space, like $\Delta \geq 1$; also the singular surfaces could degenerate or otherwise misbehave away from ZINSS.
\par The theory develops by parametizing systems of solutions. As before, the foundation are sets of linear equations.
\begin{definition} An affine scheme is a locally ringed space isomorphic to the spectrum of a commutative ring, $Spec(R)$. A scheme is a locally ringed space X admitting a covering by open sets $U_{i}$, such that each $U_{i}$ (as a locally ringed space) is an affine scheme. \end{definition}
One can think of a scheme as being covered by "coordinate charts", which are affine schemes. The definition means that schemes are obtained by gluing together affine schemes, using the Zariski topology. In particular, $X$ comes with a sheaf $\mathcal{O}_{X}$, which assigns to every open subset $U$ a commutative ring $\mathcal{O}_{X}(U)$, called the ring of regular functions on $U$. 
Sheaves will be used to map between the parameter set $\gamma$ (more properly an appropriate subset) and the family of perturbation expansions of the DSGE model and associated statistical objects like goodness of fit metrics. 
\begin{definition} The localization at $x$ of an affine scheme is the affine local ring attached to that point by the sheaf $\mathcal{F}$ \end{definition}
This can be thought of as an approximation taken from the point $x$, whereas the stalk would be the sequence of approximations in the (deleted) neighborhood of $x$. It is now possible to provide a scheme theoretic formalization of the bifurcation seen in the Calvo model. First a generic bifurcation is represented by the breakdown of the following commuting diagram.\footnote{It can be viewed as a functor between categories of pointed topological spaces local to a certain point, for example ZINSS in Calvo.}
\begin{center}
\textbf{Figure 2: Well-Behaved Approximate Solutions } 
\end{center}
\begin{center}
\begin{tikzcd}
X 
\arrow[d, "\iota"] \arrow[r, "G"]&  X_{0} \arrow[d, "\iota"] \\ 
R^{loc}(X) \arrow[r, "\nabla"] &  R^{loc}(X_{0})
\end{tikzcd}
\end{center}
The idea is that movements through the system of rings reflect movements through the underlying space in a way that preserves the topological structure in both spaces, i.e. the fundamental group $G$ in the base space and the Zariski in the rings. Where this mapping breaks down, a bifurcation arises. The focus will be on spaces where $\iota$ is well-behaved, reflecting smoothness of the model primitives and the bifurcation arises because the ring becomes \textbf{reducible}. This is where the Krull dimension of the solution set (prime ideal at $x$) declines because of some cancellation in the equation system. 
\subsubsection{Detecting Bifurcations}
It is common for macroeconomists, at central banks in particular, to build large models to study the interplay of multiple frictions and sectors. Conducting a full bifurcation analysis would likely be infeasible or technically challenging. Fortunately, I am able to use the machinery of the previous subsection to construct a readily implementable fail-safe test for fixed points about which standard approximations will fail. 
\begin{theorem} Take any system satisfying the hypotheses of Theorem 3 that has a formulation $G(X_{t}, \, X_{t+1}, \, \cdot)=0$, which is continuously differentiable around some fixed point $X^{*}$. Consider a sequence of stochastic equilibrium $\{ X_{n} \}$ approaching the possibly degenerate equilibrium $X^{*}$ with sheaves of trajectories local to each fixed point $\{TX_{n}^{loc}\}$ and $\{T(RX^{loc}_{n})\}$. Suppose there is a sequence of local homeomorphisms $h^{loc}_{n}: X_{n}\rightarrow RX_{n}$. 
This sequence will extend to the limit $X^{*}$ if and only if there is no reduction in the scheme generated by the sequence of linear approximations of $G$ with respect to the Zariski topology. 
\end{theorem}
\begin{proof}
The "if" part is a consequence of the De Rham cohomology and our stochastic Grobman-Hartman theorem. De Rham's theorem ensures continuous differentiability away from singularities (where the dimension changes) whilst Theorem 
3 rules out non-hyperbolic dynamics, allowing us to apply Grobman-Hartman. The "only if" part follows from the definition of the Zariski topology, as the dimension of a linear system and the fact that homeomorphisms have a common dimension. 
\end{proof}
\begin{remark} This analysis is easily applied to Calvo, where $G$ consists of the aggregate demand curve with a Phillips curve built from the two recursive forms (30) and (31). In the $\sqrt{\varepsilon}$ case the
appropriate formulation is \newline $X_{t}=(\aleph_{t}, \, \beth_{t},  \, \pi_{t}, \, y^{e}_{t}, \, 
\psi_{t}\, \cdot )$ and the (Krull) dimensionality of the linear system would be given by the number of linearly independent lag polynomial functions. For those variables with dynamic recursions this amounts to the existence of distinct roots of the lag polynomial i.e. those that are not repeated in the other equation. For variables without dynamics ($\mathbb{L}=0$) they would also have to have the same marginal effect. At ZINSS there are two repeat roots, $\mathbb{L}=1/\alpha \beta $ stemming from $\aleph$ and $\beth$ and $\mathbb{L}=0$ corresponding to the constant function from $\psi$, inside the Phillips curve. The dimensionality of the singular Phillips curve corresponds to two $y^{e}$ and $\pi$, which do not have cancelling common lag functions in the two equations.
This rigorously confirms the claim back in Section 5 Error 1 that the cross equation cancellation step was the sole cause of the bifurcation. In the interests of universality, Appendix H shows that the same phenomenon reoccurs in the wage Phillips curve. 
\end{remark}
\begin{remark}
The result is a Grobman-Hartman style theorem for DSGEs without kinks. Borrowing constraints, an effective lower bound on nominal interest rates, or tax thresholds are economic phenomena that would generate kinks that would make $G$ non-differentiable and prevent us from applying the theorem. Nevertheless, these instances could be analyzed piece-wise. Thus the approach is general.
\end{remark}
Informally, you can check for bifurcations by ensuring there are no common roots in the lag polynomial of the first order conditions. This is because assuming a solution exists you can only have bifurcations where unrepresentative cross-equation cancellations arise. This should become a standard feature of the output of DSGE solution packages. 
\subsection{Wider Economic Interpretations} 
The final component of the section provides wider application and economic context to the mathematical objects and arguments developed here. 
\begin{enumerate}
\item 
\textbf{Invertibility} 
The idea of the Grobman-Hartman theorem for trajectories and inverse function theorems\footnote{Unlike Grobman-Hartman there are inverse function theorems for discontinuous derivatives but they presuppose the derivative is 
locally invertible, which is missing here (see https://terrytao.wordpress.com/2011/09/12/the-inverse-function-theorem-for-everywhere-differentiable-maps/).} for mappings is that linear approximations can be used to represent local behavior because the system is invertible. Invertibility breaks down at ZINSS because the singular surfaces restrict
the value of past variables, that otherwise determine the qualitative behavior of the cocycle, in the vicinity of ZINSS. This is most clear for (3) and (4) but as will become clear in the next section, it is also the case for (5). 
\item \textbf{Covers and Polydromy}
Whether price dispersion is first or second order around ZINSS depends on which limiting metric is used. This is a new idea to economists. The reason is that unlike the other two in Theorem 6, this cover is unramified by any singularity because around ZINSS it can be written in static form, going back to Proposition 3. 
The $\vert \varepsilon \vert$ limit can be viewed as the volatile regime, whilst $\sqrt{\varepsilon}$ is the stable regime where the effect of inflation volatility has vanished. 
It should prove useful to study the dynamic role of price dispersion absent its static effects. Results are likely to extend to wide class of models with real rigidity.
\par Furthermore, $\vert \varepsilon \vert $ limit is a natural way to incorporate volatility in trend inflation. The empirical evidence, considered in Appendix I.3, appears mixed on whether trend inflation shocks have first order dynamic effects. 
Therefore, I advise subsequent papers consider both until decisive evidence appears. 
\par Moreover, the result has immediate econometric and computational implications. 
Informally, the $\vert \varepsilon \vert$ small noise limit encompasses its counterpart $\sqrt{\varepsilon}$, the very small noise limit. This makes it the more precise approximation in computational terms and the robust model in an econometric sense. 
\par Alternatively, it introduces the possibility, albeit limited, for multiple equilibria back into DSGE. In fact, in Section 11, I show that this will always be the case because the equilibrium existence conditions will be the same for both. This result is general because price dispersion behaves as an error term around ZINSS. 
\item \textbf{Covers and Rigidity}
Two of the covers from Theorem 6 have special significance to a longstanding macroeconomic debate. \cite{ball1990real} decomposes the effect of monetary policy in a Keynesian model into two forces; nominal rigidity and real rigidity. Real rigidity is the effect of monetary non-neutrality on the behavior of flexible price firms, whilst nominal rigidity only pertains to those who have sticky prices. 
This dichotomy yields both theoretical and empirical implications. \par
\underline{Theory} \\
To link the distinction to Calvo, imagine there is a very small fringe of firms $v << \vert \varepsilon \vert $, then near ZINSS the effect on the aggregate would vanish. Now approach ZINSS keeping $\vert \varepsilon \vert << \vert \pi \vert $ (corresponding to the case where trend inflation dominates volatility) then the asymptotics around ZINSS are governed by the limit of the non-stochastic steady state as $\pi \rightarrow 0$, which means that the marginal cost covering system vanishes leaving only $C^{\pi}_{\Delta}$. Since $u$ is sufficiently smooth, we can think of this as representing the cost of monetary non-neutrality. Moreover, this cost will be common across firms with flexible and sticky prices. Therefore it represents real rigidity- the effect of monetary policy on prices in the flexible fringe. \par On the other hand, if $\vert \pi \vert << \vert \varepsilon \vert $ the real rigidity cover vanishes\footnote{This argument is a little more difficult to motivate; it would arise if the volatility of output were to dominate the volatility of inflation. Heuristically, imagine a static aggregate demand and supply model. This would correspond to instances where the supply curve is considerably steeper than the aggregate demand schedule. Alternatively, one could strike out price dispersion a priori with the motivation discussed earlier.} leaving the pricing decision of the flexible sector unaffected, with only the marginal cost covering system active. Therefore distortions will only be felt by the firms facing sticky pricing problems. Hence we can think of $C_{MC}^{\psi}$ as reflecting nominal rigidity. To summarize 
\begin{itemize}
    \item $C^{\psi}_{MC}$ represents \emph{Nominal Rigidity}
    \item $C^{\pi}_{\Delta}$ represents \emph{Real Rigidity}
\end{itemize}
\underline{Empirics} \\
The results speak to an old debate about the interplay between Classical and Keynesian distortions.
The weak relationship between price dispersion and inflation and the promising hybrid structure of the $\sqrt{\varepsilon}$ Phillips curve belie the claim in \cite{ball1990real} that real rigidity is required to fit the business cycle evidence and make the effects of monetary policy substantive. This underlines the importance of time as opposed to mere state dependence when modelling monetary policy, which was the basis for his claims.\footnote{An alternative less formal take on real rigidity has it flattening the Phillips curve. This will come up in the next section. The conclusions will not change.} A more complete analysis will appear in the empirical companion paper to follow. 
\item \textbf{Covers and Market Failures}
Moreover, the covering systems can be seen through a welfare economics lens, more akin to microeconomics. The nominal rigidity system could reflect individual failure on the part of the firms with rigid prices, in the terminology of \cite{barile2017individual} (see also \cite{bernheim2009behavioral} and \cite{bernheim2016good}). Otherwise, it could be institutional or governance failures; note perspectives from \cite{vives2000corporate} and \cite{tirole2010theory}.\footnote{Alternatively, it could be viewed as pro-social behavior on the part of the firm, as in \cite{rotemberg2011fair}. This is arguably a more significant avenue for future applied research.} On the other hand, real rigidity here reflects coordination failure, a traditional theme in macroeconomics (see \cite{cooper1988coordinating} and \cite{leijonhufvud1968keynesian}).
\item \textbf{Homology and Missing Equilibrium}
This explains how the limiting equilibrium Phillips curve $(\pi, \vert \varepsilon \vert) \rightarrow 0$ represents a limiting equilibrium that is "missing" from the tangent space, like a vein in a rock. 
\item \textbf{Discretization}
Small noise limiting equilibrium constructions are robust to discretization, in a 
certain sense. Suppose the continuous stochastic processes in Section 4.8 and employed throughout the paper were replaced by a non-degenerate discrete process. Now suppose the maximum distance between any two realizations of the shocks were $\varepsilon$. The limit $\vert \varepsilon \vert \rightarrow 0$ would recover our limiting equilibrium. Therefore, the results here can be seen as approximating regime switching frameworks, such as \cite{hamilton1989new} and \cite{hamilton2010regime}, which might be surprising.
\item \textbf{Lucas Critique}
Figure 1 represents "passing the Lucas critique" with respect to the microfoundations criterion. 
\item \textbf{Double Bifurcation}
Around ZINSS there is a double bifurcation in the local ring system, associated with gluing together all the linear approximations to stochastic and non-stochastic equilibria. There is a trend inflation bifurcation  $$\lim_{\substack{\pi \rightarrow 0 \\ \vert \varepsilon \vert =0 }}(\pi, \, \vert \varepsilon \vert )$$
that economists have been aware of since \cite{ascari2002staggered}. However, there is an additional stochastic bifurcation as the size of the error term drops to zero. 
$$\lim_{\substack{\pi =0 \\ \vert \varepsilon \vert \rightarrow 0 }}(\pi, \, \vert \varepsilon \vert )$$
It is this bifurcation that economists have been unaware of, which is causing all the approximations from the existing framework to give erroneous results. Some confusion may arise because a second order difference between lag polynomial roots is causing a first order bifurcation. This is certainly an unusual geometrical pathology. 
\item \textbf{Inter-temporal Trade-Offs}
The wall of the crossing represents the break down or emergence of inter-temporal pricing constraints. In Proposition 4 where the model is constructed in canonical form (217) represents the break down of \emph{forward-looking} nominal price constraints. In other words, (expected) inflation implies a fixed point for the (expected) reset price $p^{*}_{t}=p^{*}_{t+1}$. By contrast, where the Phillips curve is solved conventionally as an inflation equality, Proposition 19 shows us that it is past nominal price constraints that are non-binding because $p^{*}_{t}=p^{*}_{t-1}$. In Proposition A.7 (Appendix F. 5.4), I prove this is $\pi_{t}=\alpha \pi_{t+1}$. Once again the inter-temporal pricing problem is breaking down with dynamic trade-offs collapsing. 
\item 
\textbf{Codimensionality}
The ambient space has codimension one, in the sense that if you adjust one variable you move inside the singular surface (around ZINSS (3) implies this will be either current inflation or its lag). This ensures that the breakdown of inter-temporal pricing constraints "causes" the bifurcation. This would not increase however many other variables I added to flesh out the description of the supply side. 
 \par Arguably, the main interest for established economists is the codimension of the singular surface. This represents how many coefficients change when you move from the existing singular approximation (1) to the "correct" approximation (2). It is easy to see that this is equal to the dimension of the full space. One can think of the codimension of the singular surface less the codimension of the nonsingular surface as measuring the "size" of the bifurcation. It is a measure of how unrepresentative the ZINSS approximation is. 
 \par For our model this size is maximal. In some sense this is the worst possible pathology. It is impossible to learn anything from 
 the existing approximation because there is no component of the Phillips curve unaffected. Staggered optimization creates a whole new transmission mechanism for monetary policy analysis. This will enable me to overturn the existence and stabilization properties of the model
in Section 11, compared to Rotemberg back in Theorem 5. We can view the second dimension of the hole as representing the inter-temporal trade-offs, associated with the Euler equation and the cost channel, inherently, arising from the presence of lag terms.
It links the "hole within a hole" 
back to the error symmetry, appearing at a steady state free from inter-temporal distortions.
\item \textbf{Constraints and Efficiency}
The system of singularities are constraints imposed on the social planner or equivalently the representative firm of \cite{acemoglu2009introduction} by the economy's history of non-optimizing behavior. 
Formally, the representative firms problem takes the form 
$$\max_{\{p^{*}_{t}(i)\}}\mathcal{L}_{t}=V_{t}- \boldsymbol{\mu}_{t} \mathbf{m}_{t} - \sum_{j=1}^{\infty}{\lambda}_{j,\,  t}\, (\alpha_{j}p^{*}_{t-j}- 
p^{f}_{j}) $$
where $V_{t}$ is the present value of profits, $\mathbf{m}_{t}$ are the market clearing constraints, with multipliers contained in $\bf{\mu_{t}}$, whilst $\lambda_{j}$ is the Lagrange multiplier on the constraint that prices last set in period $j$ must remain rigid (at ${p}^{f}_{j}$). There needs to be lump-sum transfers in place to establish isomorphism with  \cite{acemoglu2009introduction}, where there is price-taking behavior, otherwise the firm would have an incentive to facilitate collusion. Note that the constraints interpretation itself is not sensitive to whether there are lump sum taxes. This is because as demonstrated back in Section 7, the linear approximation around ZINSS does not feature $\theta$, the parameter governing steady state mark-ups. 
\par At ZINSS, the Divine Coincidence arises because the set of constraint multipliers $\{\lambda_{j,\,  t}\}=\bf{0}$. This causes the non-differentiablity in the infinite dimensional structure of the model. This can alternatively be seen through partitioning the market constraints in the planner's problem. The Lagrange multiplier $\boldsymbol{\mu^{*}}>\bf{0}$ in the problem shown below governs the constraints $\bf{m}^{*}$ on efficient output, derived from technology and preferences, whilst $\boldsymbol{\mu}^{c}$ reflect inefficient constraints, derived from staggered optimization. At ZINSS bifurcation arises because $\boldsymbol{\mu}^{c}=0$.
$$\max_{\{\{\pi_{T}\}\}}\mathcal{L}_{t}=V_{t}- \boldsymbol{\mu}^{*}_{t} \mathbf{m}^{*}_{t} - \boldsymbol{\mu}^{c}\mathbf{m}^{c}_{t}$$ 
The break down of all these constraints \emph{simultaneously} is the "Coincidence" behind the "Divine Coincidence".
This completes the optimization theoretic account of the standard Calvo model around ZINSS. 
\par Divine Coincidence is intimately tied up with the infinite horizon of the Calvo optimization problem.
With heterogeniety in the pricing process of firms, it can be seen as of infinite codimension because just one measure of constrained firms would create market failure. 
This has practical implications, for example where price spells are truncated, as is common in empirical work.\footnote{Consider, for example, the Generalized Taylor formulation of \cite{dixon2012unified} and \cite{dixon2012generalised}, which approximates heterogeneous price adjustment with contracts of finite length that differ between firms. They show that these can arbitrarily well approximate the reset distribution under the standard Calvo here.}  Around ZINSS there would always be a positive constraint multiplier on the firms forced to reset their prices so there would be no Divine Coincidence. 
In general, heterogeneity can increase the size of the  bifurcation by raising codimension of the singular surface, without changing the dimension of the wall.\footnote{In fact, the bifurcation would theoretically be infinite dimensional if we used a non-parametric function to estimate the reset price probability.}
\item \textbf{Mathematical Economics} The results in this paper have demonstrated that the distinction between mathematics and physics, where physicists theorize and make conjectures, whilst mathematicians provide rigorous proofs, will not work for economics. DSGE and most other economic models are over-identified (possess negative degrees of freedom). This means that loose conjectures are liable to prove untrue and economists need to be aware of analytical pathologies. This should provide fertile ground for future collaboration between economists and mathematicians. 
\end{enumerate}
\section{Econometric and Theoretical Implications}
This section focuses on the most basic economic theory embedded in the coefficients of the Phillips curve. It sketches out the fundamental policy and econometric implications of this theory. I chart a new course for macroeconomic policy analysis and suggest new avenues for econometric research.
\subsection{Identification and Trade-offs}
The first task of this section is to prove that price rigidity implies first order trade-offs between stabilization objectives. The second is to show that this naturally resolves econometric problems with identification.
These conditions constitute natural well-posedness conditions for dynamic stochastic models.
\subsubsection{No Divine Coincidence}
Here, I establish the existence of trade-offs in monetary policy setting, previously thought not to exist in such benchmark setting. In particular, it overturns Proposition 6 for the Calvo model. I begin with an interesting preliminary result of interest throughout this subsection. 
\begin{proposition} As $\beta \rightarrow 1$ there exists no recursive equilibrium when $(a_{\pi}, \, a_{y})=(0, \, -1)$. 
Other settings that prevent the substitution of the Euler equation into the Phillips curve gives rise to persistence that breaks Divine Coincidence. \end{proposition}
The simple proof occupies Appendix G.3.1. It is instructive to discuss further this special case. The first viewpoint is topological, it complements the bifurcation analysis of Section 9. The second focus on theoretical interpretations. 
\begin{remark} When the Euler substitution breaks down, for example, when $\beta \rightarrow 1$ and $(a_{\pi}, \, a_{y})=(0, \, -1)$ there is an \emph{a priori}  \emph{non-stochastic} 
bifurcation characterized by a one dimensional hole. Clearly, it is of codimension two and the boundary surface has full codimension because every coefficient changes. It is associated with the infinite roots of the lag polynomial in the demand system. 
\end{remark}
\begin{remark}This special case amounts to the Central Bank engineering financial frictions, which move the economy back to an ad hoc demand curve. This was traditionally how Keynesian models were constructed. Microfounded demand curves were seen as a restrictive afterthought. I have overturned this view. They are a necessary stabilizing force. This 
symbiosis of Keynesian and Classical mechanisms is a subtle yet significant feature of modern Keynesian economics. 
\end{remark}
Nevertheless, the practical significance is limited. The policy would be inconsistent with central banks mandate to develop and stabilize financial markets.\footnote{Moreover, this is not how one would model credit constraints; even if some people were shut out of the financial system. There would be capitalists who owned the firms' and if they were credit constrained, they would choose to close down their business. Hence we should always expect there to be some unconstrained agents. A genuine heterogeneous agent model is considerably outside the scope of the paper.} 
\begin{theorem} \textbf{(No Divine Coincidence)} Under Calvo pricing, even if the government is able to correct static market failures, there will always arise a first order welfare loss from business cycle fluctuations.\end{theorem}
\begin{proof} The result is obvious away from the $\vert \varepsilon \vert$ limit because fluctuations do not vanish. Down the $\vert \varepsilon \vert$ limit, Divine Coincidence is ruled out by the cost of price dispersion. At $a_{y}=-1$, Proposition 23 suffices. This leaves the standard $(a_{y}\neq -1)$  $\sqrt \varepsilon$ neighborhood, where a cost channel emerges and a contradiction argument is in order. The Phillips curve is shown below. I display the form where $\beta \rightarrow 1$ and $\sigma=1$ because these simplify the expressions without changing signs.
\begin{multline} \pi_{t}=
\frac{1}{1+\alpha}\pi_{t-1} 
 -\frac{(1-\alpha) \tilde{b}^{\circ}_{4}}{1+\alpha}\hat{y}^{e}_{t} + \frac{\tilde{b}^{\circ}_{4}}{1+\alpha}\hat{i}_{t-1} + \\ 
\frac{\tilde{b}^{\circ}_{3}}{1+\alpha}\mathbb{E}_{t}\pi_{t+1} + \frac{\tilde{b}^{\circ}_{4}}{1+\alpha}\bigg(\hat{\psi}_{t} -\hat{\psi}_{t-1}\bigg)
\end{multline}
$b^{\circ}_{\cdot}$ is the slope coefficient evaluated at the inactive policy setting $(a_{\pi}, \, a_{y})=(0, \, 0)$. Without loss of generality, I can focus on stabilization based on present rules. The best the Central Bank can do is eliminate the lagged error term by following the policy rule, which has the added benefit of removing the (present) shock from the Euler equation, however this leaves the present demand shock in the Phillips curve which is 
non-zero with probability one because $b_{4} >0$ down the small noise limit.\footnote{In the case where $\sigma=1$ and $\beta \rightarrow 1$ all these coefficients were expressed back in 6.1.3, it is obvious the term in (160) is bigger than zero. 
This is also true elsewhere, as 
Appendix E.1.4 derives that 
$$b= \beta \bigg(1+ \alpha  + \frac{(1-\alpha)(1-\alpha \beta)}{\alpha}\frac{(\sigma+\eta)}{\sigma+\beta a_{y}}\bigg)$$
$$\tilde{b}_{4}= \frac{(1-\alpha)(1-\alpha \beta)}{\alpha}\frac{(\sigma+\eta)}{\sigma+\beta a_{y}}$$ hence $b_{4}>0$.} 
\end{proof}
\begin{remark}This trade-off confirms the constraint interpretation of the boundary surface, prevalent in the last section.
\end{remark}
\begin{remark} This morphs into a general stabilization trade-off away from ZINSS. This is because successfully stabilizing the economy to first order, around any stochastic equilibrium, would move the economy back to the neighborhood of the non-stochastic equilibrium, since the mixing property would ensure second order terms vanished. 
\end{remark}
The problem is there are two cost-push shocks in the Phillips curve and only one policy instrument. This is the simplest possible stabilization trade-off. Although a priority, it lies beyond the scope of this paper to analyze the policy prescriptions. It also some mathematical significance. 
\begin{remark} The demise of Divine Coincidence demonstrates that the cost of economic fluctuations has a non-degenerate Lipschitz bound as a function of the shock size. This is a basic regularity property of the optimal policy problem.\end{remark}
\subsubsection{Persistence and Identification}
Here I prove identification and persistence, stemming from structural shocks, without real rigidity, in opposition to Propositions 8 and 9.
\begin{proposition}
In the $\sqrt{\varepsilon}$ limit, the Phillips curve is persistent, 
unless the central bank pursues 
an interest rate rule  $\hat{i}_{t}=\hat{\psi}_{t}- (b^{\circ}_{0}/b^{\circ}_{4})\pi_{t} $.
\end{proposition}
\begin{proof}
For the "if" part substitute into the Phillips curve and note that all the lag terms are eliminated, leaving a system 
with just inflation and the efficient output gap, which 
 must be solved forward. No persistence then follows from the white noise error assumption. 
 \par The "only if" claim is established by contradiction. 
 Consider the well-behaved case where $a_{y} \neq -1$ and $\beta \rightarrow 1$.
 Note that there must be a lag term, in particular one of $\{\pi_{t-1}, \, \hat{\psi}_{t-1}\}$ in the first order expansion, after the policy rule has been substituted in. If it were the error the previous theorem would complete the proof. Therefore it must be the case that $\pi_{t-1}=0$, assuming $a_{y} \neq -1$, the only way to eliminate the lagged error would be with the interest rate, which would imply that $\hat{y}^{e}_{t}=0$ from the Euler. Thus present inflation would have to absorb the current cost push shock with $\pi_{t}=(b^{\circ}_{4}/b^{\circ}_{0})\hat{\psi}_{t}$ but then $\pi_{t-1}=(b^{\circ}_{4}/b^{\circ}_{0})\hat{\psi}_{t-1}$. A contradiction has appeared. Once again Proposition  23 treats the pathology. 
\end{proof}
 \begin{remark} This no persistence result breaks down in the large noise limit because of price dispersion but might arise if price dispersion were ignored. \end{remark}
 The no persistence policy rule is counter-intuitive. Central banks seek to raise the profile of interest rates in response to increases in inflation. However, 
 further analysis of the policy rule in Appendix G.3 is unable to rule it out. This means there is no direct counter to Proposition 8. Elsewhere in the text, I persist with non-negative reaction coefficients. 
 \begin{remark} Persistence is 
 necessary and sufficient for identification of the structural model (without ad hoc shocks). This can be seen from the solution in Appendix E.2, which theoretically implies an infinite number of over-identifying conditions through construction of a suitable auto-regression. 
 \end{remark}
  The (over)identification of structural parameters follows swiftly apart from $\theta$, which will be dealt with in a subsequent paper. It is surely a step forward to no longer rely on unexplained and poorly motivated shocks for identification, particularly when they are not compatible with rational expectations (Proposition 7).
 \par This pathology with persistence and the surprising non-existence results point towards optimal monetary policy. Optimal monetary policy should yield stronger regularity of solution, including the prospect of unconditional existence in many cases. It is likely to deliver clearer empirical predictions and may even simplify the solution or improve its fit. It should be a natural next step for central bank economists to include their own optimizing efforts in the regular process of monetary policy modelling. 
\subsection{Econometric Duality}
This crucial subsection uncovers the deep connection between the topology and goodness of fit of approximate solutions around ZINSS. I finish by discussing bias implications to stimulate future computational and econometric work. The key idea is that missing dynamics in the existing approximation can directly translate to poor econometric performance. 
\subsubsection{Theoretical Formulation}
This part sets out the mathematical foundation for the econometric analysis. 
\begin{principle} \textbf{(Econometric Duality)} Consider one sheaf consisting of a sequence of approximate constrained optimization problems denoted $\{\mathcal{L}_{X}\}$, 
taken in a neighborhood of ZINSS, 
where $\vert X \vert$ indexes the closeness to zero. In general
$\mathcal{L}_{x, \, t}=U(Z_{t})-\lambda_{x.\, t}c(Z_{t})$ and in particular the localization at $0$ indicates the approximation taken from ZINSS itself. Likewise, think of a second sheaf $\{\mathcal{M}_{X}\}$ consisting of goodness of fit metrics associated with each approximation, which would take the form 
$$m(x)\equiv \frac{\vert Z_{t+1}-\mathbb{E}_{x}Z_{t+1} \vert}{\vert Z_{t+1}-\mathbb{E}Z_{t+1} \vert }=\frac{\vert Z_{t+1}-(\mathbb{E}Z_{t+1}-\lambda^{m}_{x}c(Z_{t})\vert}{\vert Z_{t+1}-\mathbb{E}Z_{t+1} \vert}$$ where $\lambda^{m}_{x, \, t}$ is a constraint on the goodness of fit in the
prediction problem. There is \emph{econometric duality} because either $\lambda^{m}_{x, \, t}=0$ (from ZINSS) or $\lambda_{x.\, t}=0$ (elsewhere) but not both. 
Jensen's inequality ensures there will be a jump in goodness of fit, when moving from the singular approximations at ZINSS to nearby non-singular approximations.
\end{principle}
\begin{remark}
Different bifurcation experiments correspond to different sheaves. 
It is convenient to work with the following parametization $X=(\delta, \, n, \, m )$ where $\delta$ is the starting rate of trend inflation, $n$ is the order of the initial difference between stochastic and non-stochastic steady states and $m$ is the limiting order of perturbation. Thus the $\vert \varepsilon \vert$ limit would $(0, \, 2, \, 1 )$, $\sqrt \varepsilon$  would be  $(0, \, 2, \, 1/2 )$. The first order non-stochastic bifurcation would be $(\delta, \, 0, \,1)$, with $(\delta, \, 0, \,n)$ the corresponding bifurcation at a general order of approximation $n$.
\end{remark}
There are two informative visual representations. The first is the  commuting diagram below where $\nabla$ is the duality map and dependence on $X$ has been suppressed.
\begin{center}
\textbf{Figure 3: Duality Mapping}
\end{center}
\begin{center}
\begin{tikzcd}
\mathcal{L} \arrow[r] \arrow[d] &  m \arrow[d] \\
 \lambda  \arrow[r, "\nabla"] &  \lambda^{m}
\end{tikzcd}
\end{center}
$\nabla$ is anti-symmetric and it can be represented by the following correspondence. 
\begin{figure}
\begin{center}
    \textbf{Figure 4: Constraint Multiplier Graph} 
\end{center}
\begin{center}
\begin{tikzpicture}
    \begin{axis}[
    axis lines = middle,
    xlabel = {$\lambda$},
    ylabel = {$\lambda^{m}$},
    xmin=0, xmax=1,
    ymin=0, ymax=1]
    \addplot [name path = L1,
    -latex,
    domain = 0:4,
    samples = 250] {-2} 
    node [very near end, left] {};
    \addplot [name path = L3,
    -latex,
    domain = 0:4,
    samples = 250] {4*x-7} 
    node [very near end, left] {};
\draw[ultra thick, red](0.03,0.94)--(0.03,0.03);
\draw[ultra thick, teal](0.03,0.03)--(0.94,0.03);
\node (O) at (0.02, 0.02) {O}; 
\node(X) at (0.23,0.7) {Singular Surface};
\node(Y) at (0.60,0.2) {Non-Singular Surface};
\node(Z) at (0.20,0.10) {Missing Origin};
\end{axis}
\end{tikzpicture}
\end{center}
\end{figure}
It represents a morphism between the sheaf of local goodness of fit metrics around ZINSS and the sheaf of approximate optimization problems, through their constraints.\footnote{Or even a functor between the two categories of sheaves around ZINSS, as in Appendix F.2.} The correspondence can be represented graphically as shown above. The hole at the origin is a consequence of Divine Coincidence. There is an unacceptable trade-off between simplifying the model by cancelling lag terms and fitting the data.
\subsubsection{Discussion}
Duality justifies the terminology of the boundary surface. It represents constraints that are always binding on the representative firm / social planner or the econometrician. 
This principle demonstrates a direct connection between the microfoundations missing from the faulty approximations at ZINSS and its poor empirical performance. If we believe in our Keynesian microfoundations we should expect a substantial improvement in goodness of fit when we change to the correct approximation. This is a reassuring prospect. The general conclusion is surely that modern Keynesian economics is first and foremost about justifying the presence of lags in the Phillips Curve, operating through the constraints implied by staggered optimization. Previous approximations failed to represent the underlying microfoundations; as such they failed the Lucas critique. 
\par The constraint idea gives a clear explanation for why impulse response cannot be computed from ZINSS. Approximations from ZINSS are singular and only valid under wall-crossing constraint (3) that $\pi_{t}=\pi_{t-1}$ (when $\beta \rightarrow 1$). This induces a contradiction, by definition an impulse response function is where $\pi_{t-1}=0$ and $\pi_{t} \neq 0$. This is the non-stochastic bifurcation in action. 
\par Econometric duality is a theoretical result that holds under the null hypothesis the model is true. If the real data generating process is sufficiently close to the theory the associated improvement in goodness of fit will still arise. This is an econometric hypothesis. However, on the basis of the exercise in Section 3.1 and Appendix H, I will proceed as though it is true that (2) fits the data better than (1).
\par This can seriously bias inference, drawn from goodness of fit comparisons. Suppose an econometrician were trying to estimate the improvement in the goodness of fit associated with a change like modelling (non-zero) trend inflation $Z_{\pi}$ but were unaware of the bifurcation. They would use the estimator
\begin{equation} \Delta \hat{m}= \left\| \hat{Z}_{\pi}- Z\right\| - \left\| \hat{Z}_{sing}- Z\right\|\end{equation}
for the appropriate (convergent) goodness of fit metric $\left\| \cdot \right|\|$
but the true improvement would be 
\begin{equation} \Delta m= \left\| Z_{\pi}- Z\right\| - \left\| Z_{0}- Z\right\|\end{equation} 
This yields an upward bias in the inferred improvement in the goodness of fit. 
\begin{equation} \chi = \plim (\Delta \hat{m}- \Delta m) = \left\| Z_{0}- Z\right\|- \left\| Z_{sing}- Z\right\| >0 \end{equation}
It would distort hypothesis testing towards accepting ill-fitting alternatives to the benchmark Calvo model. This would extend to any alternative to the Calvo that did not suffer from the same singularity. This would include alternative pricing models, such as those with \cite{taylor1979staggered}
or menu costs.
\par In fact any bifurcation will cause overestimation of non-linearity nearby a faulty linearized benchmark. The theory is as follows:
an econometrician wishing to estimate the non-linearity compared to the singular solution would estimate 
\begin{equation} \hat{l'}=\lvert| \hat{f}(Z)-\hat{Z}_{sing}\rvert| \end{equation}
the true estimator would be 
\begin{equation} l'=\lvert| \hat{f}(Z)-\hat{Z}_{0}\rvert| \end{equation}
thus an asymptotic bias would emerge \begin{equation} \chi(l')=\plim (\hat{l'}-l)=\lvert| f(Z)-{Z}_{sing}\rvert| -\lvert| f(Z)-{Z}_{0}\rvert| \approx \lvert| {Z}_{0}-{Z}_{sing}\rvert|>0\end{equation}
It is strictly positive if the true non-linearity is small.
It remains to test whether this approximate bound holds in simulations.
Nevertheless, the \emph{a priori} case was typically avoided, now at the expense of irrelevance, by using Rotemberg pricing.\footnote{The motivation was typically to avoid the role of price dispersion, which was felt to be too strong under Calvo. I have suggested remedies already.} In theory for \emph{ex post} bifurcation the bias would be infinite in any neighborhood of the cut-off as calculated for Calvo in the next section. 
An underlying theme is that complicated singularities are not amenable to inferential statistical learning. 
The confusion should now be over. 
\subsection{Coefficient Properties}
This subsection considers the salient properties of the slope coefficients, with particular emphasis on the output. There are surprising connections with historical debates in macroeconomics and alternative models of price rigidity. The other coefficients embody subtle trade-offs or overturn existing intuition. The properties of the error structure deserve a comprehensive treatment elsewhere. 
\subsubsection{Output Neutrality}
At the outset I confirm that the output coefficient from Section 7 is strictly less than with the singular Phillips curve back in Section 5.  
\begin{proposition} In the limit where $\beta \rightarrow 1$ and $\sigma=1$, $b_{1} < \tilde{\omega}$  \end{proposition}
\begin{proof}The proof is accomplished with the following decomposition 
$$b_{1}=\frac{\alpha}{b}\tilde{\omega}-\frac{\tilde{\omega}}{b(1+a_{y})}
<\omega$$
the strict inequality arises because the underlying coefficients are all positive and $b> \alpha$.
\end{proof}
This is likely to improve the fit to the data, as the slope is typically too large in the positive direction. The crucial difference is that the slope need not be positive, in fact 
it inherits its sign from the term $\alpha(1+a_{y})-1$.

\par The classical dichotomy indicates the efficient value is zero. In fact, the term represents two competing inter-temporal distortions. $\alpha <1$ represents the discounting term in the price setting problem, additional to $\beta$, which we have taken towards unity. The lesson is that the uncertain prospect of re-optimization causes firms to overweight the present relative to the future in their pricing decision. This should give rise to staggered nominal adjustment in response to any shock. The theme if not the specific argument should be familiar to macroeconomists. 
\par On the other hand, the output response may be more surprising. When firms engage in staggered optimization it creates inter-temporal substitution motives. Therefore, anything that impacts financial conditions will directly influence pricing decisions. The central bank by committing to adjust monetary policy in response to shocks will effect how firms adjust to those shocks. This justifies the policy invariance aspect of Lucas' critique, whilst overturning the observational equivalence idea because these forces are missing from the classical model. 
The result is robust to a battery of real frictions that rescale $\tilde{\omega}$ by $1/(1+R)$.\footnote{Appendix I.2 contains more information on this argument.} This suggests that nominal rather than supply side factors determine the fundamental shape of the Phillips curve. 
\par Output neutrality plays into an old debate in macroeconomics featuring the Keynesian, Austrian and Classical schools. \cite{snowdon2005modern} is a suitable background source. The Austrian school, such as \cite{von2012prices}, emphasised that active monetary policy would distort inter-temporal production and investment decisions. It is unlikely that pricing was the foremost mechanism in their analysis.\footnote{They typically put more emphasis on financial market factors that sound like modern behavioural economics with alternative policy prescriptions.} Nevertheless, it is important to confirm that this mechanism is \emph{a priori} plausible and obviously should be investigated, in models with more complicated production and financial sectors. 
Mainstream economists argued correctly, that this was not possible, unless there were alternative sources of market failure. Keynesian economics at the time was not able to supply this but now this is possible. It should be achievable to supply objective content to this type of debate, even if not everyone will be convinced. With the credible prospect of two market failures cancelling out, this is a clear example of the theory of second best (\cite{lipsey1956general}), a mainstay of welfare analysis. 
\par Many will be surprised or even concerned about a Phillips curve without a positive feedback from output onto inflation. This is not necessarily a problem. The cursory analysis of Section 3 indicates my flat Phillips curve is a better fit than its very steep counterpart. It is an opportunity rather than a threat when a micro-founded solution fails to conform to a popular conjecture. There may be other frictions that favor a positive slope. The goal is a suite of models containing a plethora of well-understood dynamic market failures, that combined give a powerful insight into typical business cycle dynamics. 
\par A particular case of interest is where $a_{y}=0$ and the limit where $\alpha \rightarrow 1$. This can be viewed as the case where time periods become arbitrarily short, holding constant the overall frequency of adjustment.\footnote{In this case the Phillips curve would take the form $\pi_{t}=\pi_{t-\delta} + \pi_{t+\delta}$, where $\delta >0$ is the time increment. In the limit where $\delta \rightarrow 0$, this corresponds to the stochastic differential equation 
$$\frac{1}{2}\mathrm{d}^{2}\pi_{t} +\mathrm{d}e_{t}=0$$ 
Nevertheless, there may be better limiting arguments down which to address the model in continuous time, which allow consideration of $\alpha$ bounded away from one.}
Discounting distortions are absent, as opposed to just cancelling out. This effects how inflation is determined. It is a simple average of its past and (expected) present value. This agrees with the \cite{taylor1979staggered} model of staggered contracts, where every price is reset after a fixed time horizon. In particular, this pattern corresponds to the simplest possible form with contract lasting two periods. A feature specific to Calvo is that as the reset rate per period declines away the structural error terms vanish so an additional error $e_{t}$, such as the unfortunate mark-ups shocks would have to be added to make the Phillips curve identified. 
\par In fact output neutrality can be extended to Taylor pricing and may be universal across staggered optimization models. To see how consider the accelerationist Phillips curve 
\begin{multline} \pi_{t}=
\frac{1}{1+\alpha}\pi_{t-1} 
 -\frac{\alpha \tilde{b}^{\circ}_{4}}{1+\alpha}\hat{y}^{e}_{t} + \frac{\tilde{b}^{\circ}_{4}}{(1+\alpha)}\hat{y}_{t-1} + \\ 
\frac{\tilde{b}^{\circ}_{3}}{1+\alpha}\mathbb{E}_{t}\pi_{t+1} + 
\frac{\tilde{b}^{\circ}_{4}}{1+\alpha}\bigg(\hat{\psi}_{t} -\hat{\psi}_{t-1}\bigg)
\end{multline}
in the limit where $\alpha \rightarrow 1$. The coefficients on output and its lag are of common magnitude but opposite sign. I demonstrate in Appendix Section G that this symmetry characterizes the terms in real output or marginal costs in the Taylor pricing framework. It looks like price rigidity alone is an insufficient deviation from market efficiency to support an upward sloping Phillips curve. This will surprise many but should become a central tenant of modern Keynesian economics. 
Finally, subject to the other optimality conditions, the ideal represented by (5) can be interpreted as enforcing output neutrality. 
\subsubsection{Other Coefficients and Trade-Offs}
This final part discusses salient features contained in the remaining slope coefficients. It is supported by Appendix G.2 that reminds readers of the basic Taylor pricing format. 
\begin{proposition} In the limit where $\beta \rightarrow 1$ and $\vert \varepsilon \vert \rightarrow 0 $, $b_{0} > b_{3}>0$ and $b_{3} < 1/2$. 
\end{proposition}
\begin{proof} 
The first part follows from $\tilde{b}_{0} > 1 > \tilde{b}_{3} =\alpha >0$. The second is proved by noting that $$\tilde{b}_{3} < \frac{\alpha}{1+\alpha}$$ and then optimizing in $\alpha$. 
\end{proof}
This is the idea that in the patient limit, inflation is equally sensitive to announced stimulus policies, no matter how far into the future they are realised.\footnote{The idea was popularized by \cite{delnegro2023forward}. It comes about from the alternative forward expansion
$$\pi_{t}=\tilde{\omega}\hat{y}^{e}_{t} + \tilde{\omega}\mathbb{E}_{t}\sum^{\infty}_{T=t+1}\hat{y}^{e}_{T} + \mathbb{E}_{t}\sum^{\infty}_{T=t+1}\pi_{T}=\lim_{\beta \rightarrow 1} \tilde{\omega}\hat{y}^{e}_{t} + \tilde{\omega}\mathbb{E}_{t}\sum^{\infty}_{T=t+1}\beta^{T-t}\hat{y}^{e}_{T} + \mathbb{E}_{t}\sum^{\infty}_{T=t+1}\beta^{T-t}\pi_{T}$$
It is roundly rejected on the data for a variety of tests (see \cite{coibion2023forward}, \cite{swanson2021measuring} and \cite{bhattarai2022analysis}).} 
This is a figment of the lack of inter-temporal distortions on the singular surface of Calvo and the lack of staggered optimization under Rotemberg. 
\par The over-weighting of past relative to future inflation contrasts with Taylor pricing, where they are always given equal weighting. In fact there is an interesting trade-off here. Calvo contains only one lag but gives it extra importance in the pricing decision, whereas Taylor contains the same or more lags but current inflation determination must be balanced between past lags and future expectations. This is an interesting point of empirical departure between the two prominent inflation frameworks. 
\par The lagged inflation coefficient holds the key trade-off of monetary policy design. A more active effort to stabilize inflation, reflected in a higher $a_{\pi}$, leads to an inflation process that is more persistent and therefore more difficult to control. The Phillips curve is not policy invariant, as Lucas suspected. This has challenging econometric implications that I leave for future work.
\par Lastly, turning to the price dispersion effect, the first surprise is that its coefficient changes between the singular $\vert \varepsilon \vert$ and correct approximations. Thus, this slow moving distortion term helps to reduce the positive persistence and surely makes the inflation process more forward-looking.  
Moreover, it is also governed by our neutrality result because in the limit $\alpha \rightarrow 1$ the $\vert \varepsilon \vert $
approximation merges into the $\sqrt{\varepsilon}$. With Taylor pricing again there is symmetry. 
\par Finally, we arrive at the deepest generality of our results; there is \emph{Real Neutrality}. 
Suppose that staggered nominal adjustment is added into an otherwise business cycle environment, such that there are no inter-temporal distortions. Current inflation is determined by a weighted average of past and future expected inflation, whilst the coefficients on real shocks net out. This is modern Keynesian economics' answer to the Classical Dichotomy that inflation is determined by the money supply rather than real output. Verily this is a firm basis for a New Neo-Classical Synthesis.
\subsection{Microeconomic Interpretation}
The stability property of the model can be repurposed to give a new window on the traditional microeconomic theory of general equilibrium. One of the central claims of this paper is that when there is no trend inflation and small noise, the economy does not need systematic central bank intervention to maintain economic stability. This may be surprising to modern Keynesian macroeconomists but not to those specialising in general equilibrium theory, where there has been a great deal of work on equilibrium stability.
\par The model has to be re-imagined with idiosyncratic rather than aggregate risk (common noise). The aggregate economy will be in non-stochastic equilibrium thanks to any suitable law of large numbers. Therefore interest rates will be independent of individual shocks, which is isomorphic to having an inactive policy setting $(a_{y}=a_{\pi}=0)$. In the small noise $\sqrt{\varepsilon}$ limit, the aggregate dynamics will be repeated at the idiosyncratic level so long as there is firm-specific labor.\footnote{For simplicity throughout the main text, I have assumed there is perfect competition in the labor market. This is probably unrealistic but alternative flexible wage arrangements, like bargaining solutions or additional mark-downs, could be accommodated into $\theta$ and would therefore not appear.}  
\par The result assumes market incompleteness; with complete markets firms would fully insure against idiosyncratic risk. The focus on small noise may seem unusual given that idiosyncratic shocks are typically large relative to aggregate shocks. Nevertheless, it seems plausible in the case of nominal rigidity where there a substantial body of evidence favouring flexible adjustment in response to large shocks, consistent with the predictions of state-dependent pricing models (see for example \cite{boivin2009sticky}). 
Finally, an intriguing alternative explanation is that producers facing large shocks insure on a near competitive market, selecting allocations that are $\sqrt{\varepsilon}$ away from full insurance.\footnote{It would be natural to think of $\sqrt{\varepsilon}$ as the approximate size of the excess. It is possible to have $\vert{\varepsilon}\vert$ equilibria with $\Delta$ adjusted to reflect demand heterogeniety. I will leave this to others. However, it would not impact first order dynamics, under the assumption here of firm-specific labor (recalling Footnote 9). } 
\par Thus, I have demonstrated that a natural benchmark free market equilibrium is stable under staggered optimization, in response to small noise. I feel this is a great improvement upon previous attempts to incorporate dynamics into general equilibrium microeconomics. These focused on tatonnement processes, based on unrealistically flexible out of equilibrium adjustment.\footnote{The idea originated with \cite{walras2014leon}. The framework struggles to generate clear or intuitive empirical predictions (see \cite{sonnenschein1972market}, \cite{sonnenschein1973walras}, \cite{mantel1974characterization} and \cite{debreu1974excess}). For further perspective consult \cite{weintraub1993general}.} Staggered optimization should feature more prominently in microeconomic theory in future. 
\section{Policy Rule}
Here I prove the anti-Taylor policy rule and briefly outline its significance. 
\begin{theorem} There exists polydromy if and only if $a_{\pi}<1$, otherwise there is no solution at all. \end{theorem}
\begin{proof}
Start with $\sqrt{\varepsilon}$ where the eigenvalue polynomial takes the form
\begin{equation} \lambda^{3}-\bigg(\frac{b}{\tilde{b}_{3}} +\frac{c}{\tilde{c}_{3}}\bigg)\lambda^{2} + \bigg( \frac{\tilde{b}_{0}}{\tilde{b}_{3}} + \frac{bc}{\tilde{b}_{3}\tilde{c}_{3}}- \frac{\tilde{b}_{1}\tilde{c}_{1}}{\tilde{b}_{3}\tilde{c}_{3}} \bigg)\lambda -\bigg( \frac{\tilde{b}_{1}\tilde{c}_{0}}{\tilde{b}_{3}\tilde{c}_{3}} + \frac{\tilde{b}_{0}c}{\tilde{b}_{3}\tilde{c}_{3}}\bigg)=0\end{equation} this expands out to 
\begin{multline} 
   \lambda^{3}-\bigg\{\frac{1+ \alpha}{\alpha} + \frac{(1-\alpha)^{2}}{\alpha^{2}}\frac{(1+\eta)}{1+a_{y}} + 1+a_{y} + \\  \frac{(1-\alpha)^{2}}{\alpha^{2}}\frac{(1+\eta)}{(1 + a_{y})}
\big[\alpha (1+a_{y})-1\big]\bigg\} \lambda^{2}  + \bigg\{\frac{1}{\alpha}+ a_{\pi}\frac{(1-\alpha)^{2}}{\alpha^{2}}\frac{(1+\eta)}{1+a_{y}} + \\  \bigg( \frac{1+ \alpha}{\alpha} + \frac{(1-\alpha)^{2}}{\alpha^{2}}\frac{(1+\eta)}{1+a_{y}}\bigg)\bigg(1+a_{y} +   \frac{(1-\alpha)^{2}}{\alpha^{2}}\frac{(1+\eta)}{(1 + a_{y})}
\big[\alpha (1+a_{y})-1\big]\bigg)- \\ (1-\alpha)^{2}(1+\eta)\frac{\big[\alpha (1+a_{y})-1\big]}{\alpha^{3}(1+a_{y})} \bigg(1+ \alpha(1-a_{\pi}) + \frac{(1-\alpha)^{2}}{\alpha}\frac{(1+\eta)}{1+a_{y}}\bigg)
\bigg\}\lambda + \\ (1-\alpha)^{2}(1+\eta)\frac{\big[\alpha (1+a_{y})-1\big]}{\alpha^{3}(1+a_{y})}\bigg( 1+ a_{\pi}\frac{(1-\alpha)^{2}}{\alpha}\frac{(1+\eta)}{1+a_{y}} \bigg) 
- \\ \bigg(\frac{1}{\alpha}+ a_{\pi}\frac{(1-\alpha)^{2}}{\alpha^{2}}\frac{(1+\eta)}{1+a_{y}}\bigg)\bigg( 1+a_{y} +  \frac{(1-\alpha)^{2}}{\alpha^{2}}\frac{(1+\eta)}{(1 + a_{y})}
\big[\alpha (1+a_{y})-1\big]\bigg)=0
\end{multline}
which simplifies to 
\begin{multline} \lambda^{3}- \bigg\{ \frac{1 + \alpha}{\alpha} + \frac{(1-\alpha)^{2}}{\alpha}(1+\eta)+ 1+a_{y}\bigg\} \lambda^{2}+ \\
\bigg\{ \frac{1}{\alpha} + \frac{(1-\alpha)^{2}}{\alpha^{2}}(1+\eta) + a_{\pi}\frac{(1-\alpha)^{2}}{\alpha}(1+\eta) + \frac{(1 + \alpha)}{\alpha}(1+a_{y})\bigg\}\lambda - \\ \frac{(1+a_{y})}{\alpha}-a_{\pi}\frac{(1-\alpha)^{2}}{\alpha^{2}}(1+\eta)=0 \end{multline} 
It is clear that $\lambda=1/\alpha$ is a root- which will lie outside the unit circle. The equation factorizes to 
\begin{equation} \bigg( \lambda -\frac{1}{\alpha}\bigg)\bigg( \lambda^{2}- \bigg\{ 2+a_{y} + \frac{(1-\alpha)^{2}}{\alpha}(1+\eta)\bigg\} \lambda + 1+a_{y} + a_{\pi}\frac{(1-\alpha)^{2}}{\alpha}(1+\eta)\bigg)=0\end{equation}
Together Theorem 3 and Proposition 16 stipulate that a solution exists if and only if the quadratic contains precisely one root inside the unit circle. This rules out complex roots. The sign pattern rules out negative roots. Therefore, the cut-off must be associated with $\lambda=1$ which is clearly associated with $a_{\pi}=1$. Rouche's theorem implies that when $a_{\pi}$ is sufficiently large there will be no roots inside the unit circle. Hence the desired configuration must be associated with $a_{\pi}<1$.\footnote{In fact its roots are $$\lambda_{1}= 1+\frac{1}{2}a_{y} + \frac{(1-\alpha)^{2}}{2\alpha}(1+\eta) + \frac{1}{2}\sqrt{\bigg( a_{y} + \frac{(1-\alpha)^{2}}{\alpha}(1+\eta)\bigg)^{2} + 4(1-a_{\pi})\frac{(1-\alpha)^{2}}{\alpha}(1+\eta)} $$
$$\lambda_{2}= 1+\frac{1}{2}a_{y} + \frac{(1-\alpha)^{2}}{2\alpha}(1+\eta) - \frac{1}{2}\sqrt{\bigg( a_{y} + \frac{(1-\alpha)^{2}}{\alpha}(1+\eta)\bigg)^{2} + 4(1-a_{\pi})\frac{(1-\alpha)^{2}}{\alpha}(1+\eta)} $$
where the former always lies outside the unit circle, the latter is inside if $a_{\pi}<1$, on when $a_{\pi}=1$ and outside otherwise. Finally, the cut-off, above which there are complex roots is
$$\bar{a}_{\pi} =1 + \frac{1}{4}\bigg( a_{y} + \frac{(1-\alpha)^{2}}{\alpha}(1+\eta)\bigg)^{2}\frac{\alpha}{(1-\alpha)^{2}(1+\eta)}$$ } 
\par The extension to $\vert \varepsilon \vert$ is straightforward. Its eigenvalue equation is
\begin{multline} \lambda^{4}-\Bigg\{ \frac{b}{\tilde{b}_{3}} + \frac{c}{\tilde{c}_{3}} + \frac{d}{\tilde{d}_{3}} \Bigg\}\lambda^{3} + \Bigg\{\frac{bc}{\tilde{b}_{3}\tilde{c}_{3}} + \bigg( \frac{b}{\tilde{b}_{3}}+ \frac{c}{\tilde{c}_{3}}\bigg)\frac{d}{\tilde{d}_{3}} - \frac{\tilde{b}_{1}\tilde{c}_{1}}{\tilde{b}_{3}\tilde{c}_{3}}
 + \frac{\tilde{b}_{0}}{\tilde{b}_{3}}\Bigg\}\lambda^{2} 
+ \\  \Bigg\{ -\frac{bcd}{\tilde{b}_{3}\tilde{c}_{3}\tilde{d}_{3}} + \frac{\tilde{b}_{1}\tilde{c}_{1}d}{\tilde{b}_{3}\tilde{c}_{3}\tilde{d}_{3}} 
- \frac{\tilde{b}_{1}\tilde{c}_{0}}{\tilde{b}_{3}\tilde{c}_{3}} -   \frac{\tilde{b}_{0}c}{\tilde{b}_{3}\tilde{c}_{3}} - \frac{\tilde{b}_{0}d}{\tilde{b}_{3}\tilde{d}_{3}}
\Bigg\}\lambda + 
\frac{\tilde{b}_{1}\tilde{c}_{0}d}{\tilde{b}_{3}\tilde{c}_{3}\tilde{d}_{3}} + \frac{\tilde{b}_{0}cd}{\tilde{b}_{3}\tilde{c}_{3}\tilde{d}_{3}} =0
\end{multline}
\begin{multline} \lambda^{4}- \bigg\{ \frac{1 + \alpha + \alpha^{2}}{\alpha} + \frac{(1-\alpha)^{2}}{\alpha^{2}}\frac{1 + \eta}{1 + a_{y}} + 1+a_{y} + \\  \frac{(1-\alpha)^{2}}{\alpha^{2}}\frac{(1+\eta)}{(1 + a_{y})}
\bigg[\alpha(1+a_{y})-1\bigg]\bigg
\}\lambda^{3} + \\  \bigg\{ \bigg( \frac{1 + \alpha}{\alpha} + \frac{(1-\alpha)^{2}}{\alpha^{2}}\frac{(1+\eta)}{(1 + a_{y})}\bigg)\bigg( 1 +a_{y} + \frac{(1-\alpha)^{2}}{\alpha^{2}}\frac{(1+\eta)}{(1 + a_{y})} \bigg[\alpha (1+a_{y})-1\bigg]\bigg) \\ +  1 + \alpha + \frac{(1-\alpha)^{2}}{\alpha}\frac{(1+\eta)}{(1 + a_{y})} + \alpha (1 +a_{y})+ \frac{(1-\alpha)^{2}}{\alpha}\frac{(1+\eta)}{(1 + a_{y})} \bigg[\alpha (1+a_{y})-1\bigg] - \\ \frac{(1-\alpha)^{2}}{\alpha^{3}}\frac{(1+\eta)}{(1 + a_{y})} \bigg[\alpha (1+a_{y})-1\bigg]\bigg(1 + \alpha(1-a_{\pi}) + \frac{(1-\alpha)^{2}}{\alpha}\frac{(1+\eta)}{(1 + a_{y})}\bigg) + \\  \frac{1}{\alpha} + a_{\pi}\frac{(1-\alpha)^{2}}{\alpha^{2}}\frac{(1+\eta)}{(1 + a_{y})} \bigg\}\lambda^{2} - 
\\ \bigg\{ \bigg( 1+ \alpha+ \frac{(1-\alpha)^{2}}{\alpha}\frac{(1+\eta)}{1+a_{y}}\bigg)\bigg( 1 + a_{y} + \frac{(1-\alpha)^{2}}{\alpha^{2}}\frac{(1+\eta)}{(1 + a_{y})} \bigg[\alpha (1+a_{y})-1\bigg]\bigg) \\ -\frac{(1-\alpha)^{2}}{\alpha^{2}}\frac{(1+\eta)}{(1 + a_{y})} \bigg[\alpha (1+a_{y})-1\bigg]\bigg(1 + \alpha(1-a_{\pi}) + \frac{(1-\alpha)^{2}}{\alpha}\frac{(1+\eta)}{(1 + a_{y})}\bigg) - \\ \frac{(1-\alpha)^{2}}{\alpha^{3}}\frac{(1+\eta)}{(1 + a_{y})}\bigg[\alpha (1+a_{y})-1\bigg]\bigg(1 + a_{\pi}\frac{(1-\alpha)^{2}}{\alpha}\frac{(1+\eta)}{(1 + a_{y})}\bigg) + \\ \bigg( 1 +a_{y} + \frac{(1-\alpha)^{2}}{\alpha^{2}}\frac{(1+\eta)}{(1 + a_{y})}\bigg[\alpha (1+a_{y})-1\bigg] \bigg)\bigg( \frac{1}{\alpha} + a_{\pi}\frac{(1-\alpha)^{2}}{\alpha^{2}}\frac{(1+\eta)}{(1 + a_{y})}\bigg) \\ +  1 + a_{\pi}\frac{(1-\alpha)^{2}}{\alpha}\frac{(1+\eta)}{(1 + a_{y})}\bigg\}\lambda + \\
 \frac{(1-\alpha)^{2}}{\alpha^{2}}\frac{(1+\eta)}{(1 + a_{y})} \bigg[\alpha (1+a_{y})-1\bigg]\bigg(1 + a_{\pi}\frac{(1-\alpha)^{2}}{\alpha}\frac{(1+\eta)}{(1 + a_{y})}\bigg) + \\ \bigg(1 + a_{\pi}\frac{(1-\alpha)^{2}}{\alpha}\frac{(1+\eta)}{(1 + a_{y})}\bigg)\bigg( 1 +a_{y}+  \frac{(1-\alpha)^{2}}{\alpha^{2}}\frac{(1+\eta)}{(1 + a_{y})}
\bigg[\alpha (1+a_{y})-1\bigg] \bigg)
\end{multline}
This simplifies to 
\begin{multline}  \lambda^{4}- \bigg\{ \frac{1 + \alpha + \alpha^{2}}{\alpha} + \frac{(1-\alpha)^{2}}{\alpha}(1 + \eta) + 1+a_{y} \bigg\}\lambda^{3} +\\ 
 \bigg\{ \frac{1 + \alpha + \alpha^{2}}{\alpha} + \frac{1 + \alpha + \alpha^{2}}{\alpha}(1+a_{y})  + a_{\pi}\frac{(1-\alpha)^{2}}{\alpha}(1 + \eta) + \\  (1-\alpha)^{2}(1+\eta)\bigg( 1 + \frac{1}{\alpha^{2}}\bigg)+ 1+a_{y} \bigg\}\lambda^{2} -\\ 
\bigg\{ 1 +  \frac{1 + \alpha + \alpha^{2}}{\alpha}(1+a_{y})  + a_{\pi}(1-\alpha)^{2}(1 + \eta)\bigg( 1 + \frac{1}{\alpha^{2}}\bigg)  + \frac{(1-\alpha)^{2}}{\alpha}(1 + \eta) \bigg\}\lambda \\ + 1+a_{y} + a_{\pi}\frac{(1-\alpha)^{2}}{\alpha}(1 + \eta)=0
\end{multline}
Plugging in reveals that there is one root inside the unit circle at $\lambda=\alpha$ and another outside at $\lambda=1/\alpha$. When the polynomial is factorized, the other roots come from the same quadratic equation as in the
$\sqrt{\varepsilon}$ limit in (180). There have to be two roots inside the unit circle overall, so the parameter condition is identical. \end{proof}
There might be a wider lesson for general equilibrium theory and mathematical economics. Equilibrium existence is demanding a contraction on destabilizing general equilibrium effects. To see how, consider the following decomposition of the lagged inflation coefficient informed by Decomposition 1. 
\begin{decomposition}
$$b_{0}=\frac{1+a_{\pi}GE}{1+ \alpha + GE}$$
where $GE$ represents the general equilibrium impulse from the cost channel $$GE = \frac{(1-\alpha)^{2}(1+\eta)}{\alpha(1+a_{y})}$$ 
whilst the unitary terms come from the wall of the crossing $(A)$ and the $\alpha$ comes from the singular component $(D)$ and $(F)$ from Decomposition 1.
\end{decomposition}
It is clear that $\Delta b_{0}/\Delta GE \leq a_{\pi}$ where the bound is sharp because as $GE \rightarrow \infty$ (corresponding to $\alpha \rightarrow 0$ or $\eta \rightarrow \infty$) $\Delta b_{0}/\Delta GE \rightarrow a_{\pi}$. Hence $a_{\pi}< 1$ is necessary to ensure $\Delta b_{0}/\Delta GE <1$. This is the requirement to stop the cost channel overturning the demand reducing effect of increasing interest rates on inflation. 
\par Turning to the New Keynesian field in particular, the following observation should prove interesting for empirical specification design, as well as theoretical consideration of the property of structural disturbances. 
\begin{remark} Note that down the $\vert \varepsilon \vert$ limit $\Delta$ detaches from the rest of the system and behaves as an exogenous variable, like the cost-push shocks previously used to drive the Phillips curve.\end{remark}
I anticipate the policy rule results will generalize widely and may extend out to moderate rates of trend inflation. 
To reaffirm, in economic terms the blow up is unproblematic and even instructive. On the one hand, the economy is stable under inactive settings, when the Taylor rule represents a realistic shorthand description of monetary policy-making. On the other hand, the model explodes under active policy, precisely when following a simple rule is most unrealistic. Successful central banks employ macroeconomists and other skilled practitioners in order to implement a semblance of optimal monetary policy. Best practice appears to be inertial. In time, I expect the theory of monetary policy to catch-up.\footnote{There is nothing precluding 
active stabilization of the output gap. This would always act to moderate fluctuations in inflation because there are no supply shocks. For $a_{y}$ sufficiently large, empirically $a_{y}>1$ is sufficient, $b_{1}>0$ consistent with the traditional transmission from the output gap to inflation.}
\par Finally, there is a direct connection between \emph{ex post} constraints implied by the existence conditions and \emph{ex ante} constraints breaking Divine Coincidence. The policies that are ruled out \emph{ex post}, which seek to raise real interest rates immediately, are precisely those that fail to engage with the \emph{ex ante} restrictions that give the economy inertia. Elaborate mathematical machinery serves to rule out evidently \emph{unsustainable} macroeconomic policies. I anticipate this will become a universal lesson in New Keynesian economics. 
\section{Conclusions}
I have set out a New Keynesian model that should prove useful for policy. Its backbone, the Phillips Curve, compares favourably along all dimensions in preliminary data analysis; with a jump in predictive power guaranteed under the null it is the true model. 
My analysis shows nominal rigidity is central to understanding macroeconomic dynamics; whatever the inflation rate, no matter how small the shocks are. 
It hints at a surprising yet intuitive rationale for the inertial policies, ubiquitous at successful central banks. 
\par Natural next steps include a wage Phillips curve and optimal policy problems. I will theorize further about the error terms and there will be a renewed commitment to structural econometrics. No doubt alterations and extensions will arise but I am confident the salient features of the solution here will survive.
\par I have constructed a rigorous solution method for a class of previously intractable problems. Yet, I have only scratched the surface of what is conceivable with this mathematical framework. The techniques I have developed will have applications, elsewhere in economics and beyond. 
\par Numerous theories of real and nominal frictions have been set out with less intuition and empirical support than the one here. All were implicitly based on perceived failures of a faulty New Keynesian model. When solved correctly, a benchmark sticky price model implies clear and credible trade-offs. Together my results decisively answer the broadly fair criticisms of previous formulations, as levelled by \cite{chari2009new}.
\par New Keynesian economics now has striking results and firm foundations. In this paper, I have profoundly reinterpreted the Lucas critique and our understanding of the Phillips curve. By disproving observational equivalence between Keynesian and Classical models, I have surely ensured that: "We are all Keynesians now."\footnote{The phrase was supposedly uttered at different times, possibly with different intentions, by Milton Friedman and Richard Nixon. The respective attributions are here:
https://web.archive.org/web/20081023074323\\/http://www.time.com/time/magazine/article/0,9171,842353-3,00.html
https://www.nytimes.com/1971/01/07/archives/nixon-reportedly-says-he-is-now-a-keynesian.html
\newline It should not be surprising that a rigorous solutions ends up confirming policy consensus and economic intuition.}
\bibliographystyle{plainnat} 
\bibliography{seke.bib} 
\appendixpage
\appendix
\section{Proof of Theorem 2}
This appendix proves Theorem 2 providing stochastic comparative statics and quantitative estimates. It is subdivided into subsections each 
focusing on a particular part of the theorem.
\subsection{Proof of Part (i)}
This part focuses on the interest rates and an important intermediate result.
\begin{proof}
Rearranging the Euler gives the following expression for the equilibrium interest rate 
\begin{equation} i=\frac{1} {\beta}\frac{\psi u'(Y)}{\mathbb{E}\psi u'(Y)/(1+\pi)}-1\end{equation}
focus on the denominator applying Jensen's inequality reveals that 
\begin{equation} \mathbb{E}\psi u'(Y)/(1+\pi) \geq \mathbb{E}\psi \; \mathbb{E}u'(Y) \; \mathbb{E} 1/ (1+\pi) \end{equation}
applying Jensen's inequality once more implies 
\begin{equation}  \mathbb{E}\psi u'(Y)/(1+\pi) > \psi \; u'(Y) \;  1/ (1+\pi) \end{equation}
now substituting back in to (100) 
closes the proof. \end{proof}
\subsection{Behaviour of $\Delta$}
The following theorem accomplishes the proof of Theorem 2(ii) and Proposition 15.
\begin{theorem}
There exists $\underline{\underline{\pi}}<\underline{\pi}$, such that $\Delta$ is strictly increasing (decreasing), if and only if $\pi \gtrless \underline{\pi}$. Furthermore, $\Delta$ is convex (concave) above (below) if and only if $\pi \gtrless \underline{\underline{\pi}}$ and if $\theta \geq 2$ then $\underline{\pi}<0$.
\end{theorem} 
\begin{proof}
Focusing first on $\underline{\pi}$, we know by continuity of the first derivative that this must be a stationary point.
The derivative $\mathrm{d}\Delta/\mathrm{d}\pi$ is
\begin{multline}
\frac{\alpha \theta}{(1-\alpha)^{1/(\theta-1)}}\bigg[\mathbb{E}(1+\pi)^{\theta-1}\frac{\mathbb{E}(1-\alpha (1+\pi)^{\theta-1})^{\theta/(\theta-1)}}{(1-\alpha \mathbb{E}(1+\pi)^{\theta})^{2}}-\\  \frac{\mathbb{E}(1-\alpha (1+\pi)^{\theta-1})^{1/(\theta-1)}(1+\pi)^{\theta-2}}{1-\alpha \mathbb{E}(1+\pi)^{\theta}}\bigg]
\end{multline}
To prove existence of the bounds examine the limiting behavior of the derivatives around the poles. At the positive pole 
$$\bar{\pi}=\{\pi= d^{-1}(0):\, d=1-\alpha \mathbb{E}(1+\pi)^{\theta}\}$$ 
$$\lim_{\pi \rightarrow \bar{\pi} }\mathrm{d}\Delta/\mathrm{d}\pi =\infty$$
from the domination of the first term in the bracket which is $\mathbb{O}(x^{-2})$ term. Now there are more complications with the lower pole $\pi=\rightarrow -100 \%$ as the limit is zero for $\theta >2$. However, since it represents the non-negativity constraint on the price, we know that this pole is shared by both the whole model and the steady state. Therefore, whatever the other parameters, every sequence of (stochastic equilibrium) measures must converge on a sequence of non-stochastic steady state, as inflation approaches the pole. Therefore $$\lim_{\pi \rightarrow -100\%}\mathrm{d}\Delta/\mathrm{d}\pi \rightarrow \frac{\alpha \theta}{(1-\alpha)^{1/(\theta-1)}}\frac{(1+\pi)^{\theta-2}(1-\alpha(1+\pi)^{\theta-1})^{1/(\theta-1)}\pi}{(1-\alpha (1+\pi)^{\theta})^{2}}<0$$
in the neighborhood of the pole. The intermediate value theorem proves the existence of a stationary point. Since the derivative must move from negative to positive, uniqueness will prove the existence of the boundary value $\underline{\pi}$.
Now turning to the convexity claim,  direct computation reveals that 
\begin{multline}
\mathrm{d}^{2}\Delta/\mathrm{d}\pi^{2}=\frac{\alpha \theta (\theta-1)}{(1-\alpha)^{1/(\theta-1)}}\mathbb{E}(1+\pi)^{\theta-2}\frac{\mathbb{E}(1-\alpha(1+\pi)^{\theta-1})^{\theta/(\theta-1)}}{(1-\alpha \mathbb{E}(1+\pi)^{\theta})^{2}}- \\ \frac{\alpha^{2}\theta^{2}}{(1-\alpha)^{1/(\theta-1)}}\mathbb{E}(1+\pi)^{\theta-1}\frac{\mathbb{E}(1-\alpha(1+\pi)^{\theta-1})^{1/(\theta-1)}(1+\pi)^{\theta-2}}{(1-\alpha \mathbb{E}(1+\pi)^{\theta})^{2}}+\\ \frac{2 \alpha^{2}\theta^{2}}{(1-\alpha)^{1/(\theta-1)}}(\mathbb{E}(1+\pi)^{\theta-1})^{2}\frac{\mathbb{E}(1-\alpha(1+\pi)^{\theta-1})^{\theta/(\theta-1)}}{(1-\alpha \mathbb{E}(1+\pi)^{\theta})^{3}} -\\ \frac{\alpha \theta(\theta-2)}{(1-\alpha)^{1/(\theta-1)}}\frac{\mathbb{E}(1-\alpha(1+\pi)^{\theta-1})^{1/(\theta-1)}(1+\pi)^{\theta-3}}{(1-\alpha \mathbb{E}(1+\pi)^{\theta})} + \\ \frac{\alpha^{2}\theta}{(1-\alpha)^{1/(\theta-1)}} \mathbb{E}\frac{(1+\pi)^{2(\theta-2)}}{(1-\alpha (1+\pi)^{\theta-1})^{(\theta-2)/(\theta-1)}}\frac{1}{1-\alpha \mathbb{E}(1+\pi)^{\theta}} \\ -\frac{\alpha^{2}\theta^{2}}{(1-\alpha)^{1/(\theta-1)}}\mathbb{E}(1+\pi)^{\theta-1}\frac{\mathbb{E}(1-\alpha (1+\pi)^{\theta-1})^{1/(\theta-1)}(1+\pi)^{\theta-2}}{(1-\alpha \mathbb{E}(1+\pi)^{\theta})^{2}}
\end{multline}
As with the first derivative we can establish a candidate for the lower bound, in this case $\underline{\underline{\pi}}$. By taking limits around the poles 
we can see that 
$$\lim_{\pi \rightarrow \bar{\pi} }\mathrm{d}^{2}\Delta/\mathrm{d}\pi^{2} =\infty$$
from the domination of the third term that is $\mathbb{O}(x^{-3})$. 
Now in the limit as we approach the lower pole, the fourth term is dominant since $\theta>2$ so we know that 
$$\lim_{\pi \rightarrow \bar{\pi} }\mathrm{d}^{2}\Delta/\mathrm{d}\pi^{2}
\rightarrow \mathrm{d}^{2}\Delta^{NSS}/\mathrm{d}\pi^{2}<0$$
In this case the intermediate value theorem, applied to the second derivative, provides a candidate stationary point ,which will be equal to $\underline{\underline{\pi}}$, if we can establish uniqueness.
\par Next is the crucial step connecting convexity in stochastic equilibrium with the level of the stochastic steady state, relative to its non-stochastic counterpart. By applying the Jensen-Cheyshev inequality from Proposition 2 to the stochastic steady state $\Delta$, along with iterated expectations, we discover that $\Delta \gtrless \Delta^{NSS}$ for $\mathrm{d}^{2}{\Delta}/\mathrm{d}\pi^{2}\gtrless 0$ and therefore by continuity $\Delta=\Delta^{NSS}$ if and only if $\mathrm{d}^{2}{\Delta}/\mathrm{d}\pi^{2}= 0$
\par It is easiest to establish first the non-positivity of the bounds, starting with the first derivative.
The following sequence of inequalities apply when $\theta \geq 2$. 
\begin{align*}
\frac{d \Delta}{d \pi}&>\frac{\alpha \theta }{1-\alpha (1+\pi)^{\theta}}\mathbb{E}\bigg[(1+\pi)^{\theta-2}\bigg((1+\pi)\Delta -\frac{(1-\alpha (1+\pi)^{\theta-1})^{1/(\theta-1)}}{(1-\alpha)^{1/(\theta-1)}}\bigg)\bigg]\\ 
&>\frac{\alpha \theta }{1-\alpha (1+\pi)^{\theta}}\mathbb{E}\bigg[(1+\pi)^{\theta-2}\bigg((1+\pi) -\frac{(1-\alpha (1+\pi)^{\theta-1})^{1/(\theta-1)}}{(1-\alpha)^{1/(\theta-1)}}\bigg)\bigg]\\ 
& \geq \frac{\alpha \theta }{1-\alpha (1+\pi)^{\theta}}\mathbb{E}(1+\pi)^{\theta-2}\bigg((1+\pi) -\frac{\mathbb{E}(1-\alpha (1+\pi)^{\theta-1})^{1/(\theta-1)}}{(1-\alpha)^{1/(\theta-1)}}\bigg)\\ 
& \geq \frac{\alpha \theta }{1-\alpha (1+\pi)^{\theta}}\mathbb{E}(1+\pi)^{\theta-2}\bigg((1+\pi) -\frac{(1-\alpha (1+\pi)^{\theta-1})^{1/(\theta-1)}}{(1-\alpha)^{1/(\theta-1)}}\bigg)\\ 
\end{align*}
The first, second and fourth follow from Proposition 2. The first and fourth are Jensen-Chebyshev, the second uses the properties of $\Delta$ and monotonicity of the Lebesgue integral. The fourth requires $\theta \geq 2$ as the relevant second derivative changes sign (becomes convex) as it approaches the negative pole as
$$\lim_{\substack{\theta <2 \\ \pi \rightarrow -100\%}}\frac{\alpha (2-\theta)(1+\pi)^{\theta-3}}{(1-\alpha (1+\pi)^{\theta-1})^{1/(\theta-1)}}-\frac{\alpha^{2}(1+\pi)^{2(\theta-2)}}{(1-\alpha (1+\pi)^{\theta-1})^{(\theta-2)/(\theta-1)}}=\infty $$
The third inequality follows from Lemma 1, since with $\theta >2$, both $(1+\pi)^{\theta-2}$ and the bracketed term inside the expectation are strictly increasing, as the first derivative of the brackets is $$1+\alpha(1+\pi)^{\theta-2}/(1-\alpha(1+\pi)^{\theta-1})^{(\theta-2)/(\theta-1)}(1-\alpha)^{1/(\theta-1)}>0$$ The weak inequality incorporates the case $\theta=2$ where $(1+\pi)^{\theta-2}$ is a constant rather than strictly increasing. Therefore the root $\pi=0$ is unique and the strict inequality implies that any $\underline{\pi}<0$.
\par Turning to the convexity bound in search of a contradiction, it is necessary to work first with the non-stochastic steady state $\Delta^{NSS}$. Its second derivative is simply (188) with the expectations removed. It can be written as follows 
\begin{multline}
(1+\pi)^{-1}\bigg(\frac{(\theta-1)+\alpha (\theta+1)(1+\pi)^{\theta}}{1-\alpha (1+\pi)^{\theta}}\bigg)\frac{\mathrm{d}\Delta}{\mathrm{d}\pi} +\\  \frac{\alpha \theta}{(1-\alpha)^{1/(\theta-1)}}\frac{(1-\alpha(1+\pi)^{\theta-1})^{1/(\theta-1)}}{1-\alpha (1+\pi)^{\theta}}(1+\pi)^{\theta-3} +\\  \frac{\alpha^{2}\theta}{(1-\alpha)^{1/(\theta-1)}} \frac{(1+\pi)^{2(\theta-2)}}{(1-\alpha (1+\pi)^{\theta-1})^{(\theta-2)/(\theta-1)}}\frac{1}{1-\alpha (1+\pi)^{\theta}}
\end{multline}
where the first term comes from the first and fourth terms of the previous expression, the second comes from combining terms two, three and six whilst the final term is term five. By taking the non-stochastic limit of the first derivative (187)
\begin{equation}\frac{\mathrm{d}\Delta}{\mathrm{d}\pi}= \alpha \theta \frac{(1+\pi)^{\theta-2}\Delta}{(1-\alpha (1+\pi)^{\theta}}\pi\end{equation}
it is clear that $sgn(\mathrm{d}\Delta/\mathrm{d}\pi)=sgn(\pi)$. As the first derivative is (strictly) positive for (strictly) positive inflation, we know that is strictly convex for non-negative inflation.
\par Furthermore, we know that $\Delta$ must be convex at the origin, since it always lies strictly above $\Delta^{NSS}=1$. This rules out $\underline{\underline{\pi}}>0$ since, by the intermediate value theorem, there must be some $\pi$ in the neighborhood of the origin where $\Delta$ is still convex. However, to prove we can still have $\underline{\underline{\pi}}<0$, it is necessary to rule out the possibility that $\Delta$ is convex at any positive $\pi$. Suppose this were the case, there would have to be $0<\tilde{\pi}<x$ such that $\Delta$ is strictly convex for $0<\pi<\tilde{\pi}$. Therefore, $\Delta^{NSS}$ must cut $\Delta$ from below. Hence at $\tilde{\pi}$ the non-stochastic steady state profile must be steeper $$\frac{\mathrm{d} \Delta^{NSS}}{\mathrm{d} \pi}>\frac{d \Delta}{d \pi}$$ since this inequality is strict. Continuity ensures there is some region to the right of $\tilde{\pi}$ such that $\Delta^{NSS}> \Delta$. Since we have established $\Delta$ must become convex, eventually there must be some point $\tilde{\tilde {\pi}}$ where the curve again becomes convex such that $\Delta$ is convex on $(\tilde{\pi},\tilde{\tilde{\pi}})$.
However, since $\Delta^{NSS}$ is still convex while $\Delta$ is concave and both cases are strict near $\tilde{\pi}$,
Gronwall's differential inequality (see \cite{evans2022partial}) tells us that 
$$\Delta^{NSS}(\tilde{\tilde{\pi}})>\Delta({\tilde{\pi}})+(\tilde{\tilde{\pi}}-\tilde{\pi})\frac{\mathrm{d}\Delta^{NSS}}{\mathrm{d}\pi}\vert_{\pi=\tilde{\pi}}$$
$$\Delta(\tilde{\tilde{\pi}})<\Delta({\tilde{\pi}})+(\tilde{\tilde{\pi}}-\tilde{\pi})\frac{\mathrm{d}\Delta}{\mathrm{d}\pi}\vert_{\pi=\tilde{\pi}}$$
Hence $$\Delta^{NSS}(\tilde{\tilde{\pi}})>\Delta(\tilde{\tilde{\pi}})$$
\par However by Jensen's inequality $$\Delta^{NSS}(\tilde{\tilde{\pi}})\leq \Delta(\tilde{\tilde{\pi}})$$ and we have reached a contradiction. Therefore $\Delta$ must be strictly convex in $\pi$ when inflation is non-negative so any $\underline{\underline{\pi}}<0$ will suffice. The bulk of the work for the case of negative inflation is achieved by the following powerful single crossing result.
\par Suppose towards a contradiction, that there is an intersection $\Delta(\tilde{\pi})=\Delta^{NSS}(\tilde{\pi})$ where $\Delta$ is increasing but $\Delta^{NSS}$ is strictly decreasing. In search of a contradiction, consider the boundary of the concave region, containing $\tilde{\pi}$ which can be denoted by $\tilde{\tilde{\pi}}$. This is either the negative pole or another solution to $\mathrm{d}^{2}\Delta/\mathrm{d} pi^{2}=0$. An application of Gronwall's inequality reveals that 
$$\lim_{\pi \rightarrow \tilde{\tilde{\pi} }}\;\Delta^{NSS}>\Delta({\tilde{\pi}})$$
whilst $$\lim_{\pi \rightarrow \tilde{\tilde{\pi}}}\; \Delta< \Delta({\tilde{\pi}})-(\tilde{\pi}-\tilde{\tilde{\pi}})\frac{\mathrm{d}\Delta}{\mathrm{d}\pi}\bigg|_{\pi}=\tilde{\pi}$$
Hence $$\lim_{\pi \rightarrow \tilde{\tilde{\pi} }}\; \Delta^{NSS}>\lim_{\pi \rightarrow \tilde{\tilde{\pi}}}\Delta$$
which contradicts either Jensen's equality under continuity assumptions or convergence in distribution established at the negative pole. This establishes that any $\underline{\underline{\pi}}<\underline{\pi}$. It proves that there is a rate of inflation below which $\Delta$ is weakly concave. Therefore, using the fact that inflation is strictly decreasing in the limit approaching the negative pole completes the proof of the existence of $\underline{\pi}$.
\par Finally, turning to the sharpness of the parameter condition required by Theorem 2, consider the limit case where $$\rho \rightarrow 1$$
$$\vert \varepsilon_{\psi} \vert \rightarrow 0$$
suppose further that 
$$\mu(A) \rightarrow p(\bar{A})\bar{A} + (1-p(\bar{A}))\underbar{A}$$
so the distribution of $A$ approaches a two point discrete distribution. The persistent limit means we can transfer this to the distribution of inflation 
$$\mu(\pi) \rightarrow p(\bar{A})\pi(\bar{A}) +(1-p(\bar{A}))\pi(\underbar{A})$$
Furthermore let the upper bound of the shock be arbitrarily increased  
$$\bar{A} \rightarrow \infty$$
so by taking the non-stochastic limit of the steady state (detailed in Appendix D) it is clear that $$\pi(\bar{A}) \rightarrow -100\%$$
whilst the probability is scaled down to preserve the boundedness of the expectation by
$$p=O(A^{1+\kappa_{1}})$$
where $k_{1}>0$. Finally, suppose that 
$$\theta = 2-\kappa_{2}$$ where $k_{2}>0$. Now it is clear that, the only way to overturn Theorem 2 (ii) is to make the fourth and only negative term of the second derivative (188) dominate. This can only happen at the upper bound where inflation moves towards its lower bound
$$\mathrm{d}^{2}\Delta/\mathrm{d}\pi^{2}=O(p(\underbar{A})(1+\pi)^{\theta-3})=O(1+\pi)^{\kappa_{1}-\kappa_{2}}$$
selecting $\kappa_{2} < \kappa_{1}$ shows $\mathrm{d}^{2}\Delta/\mathrm{d}\pi^{2} \rightarrow \infty$, which means that price dispersion must become concave- reversing the direction of Jensen's inequality used in the proof. Taking the limit $\kappa_{1} \rightarrow 0$ completes the proof. 
\end{proof}
\subsection{Proof Part (iii)}
This part proves the claim concerning output.
\begin{proof}Take the optimal reset price condition with the object of interest the right hand side $\aleph(Z)/\beth(Z)$. The first step is to show this is a convex function of $\pi$. The easiest way to do so is to differentiate the left-hand side which is a function only of $\pi$. The task is accomplished by computing the second derivative and concluding that it is strictly positive using the parameter restriction for $\theta$ $$\alpha(1-\alpha)^{1/(\theta-1)}(1+\pi)^{\theta-3}\bigg[\frac{(\theta-1)(\theta-2)(1-\alpha (1+\pi)^{\theta-1}) + \alpha \theta (1+\pi)^{\theta-1}}{(1-\alpha(1+\pi)^{\theta-1})^{(2 \theta-1)/(\theta-1)}}\bigg]$$ 
Note that the condition is sharp in $\theta$, by considering the limits respectively as $\alpha \rightarrow 0$ and $\pi \rightarrow -100\%$. Now the Jensen-Chebyshev inequality implies that the left-hand side strictly exceeds the right-hand side, evaluated at the non-stochastic steady state. This implies that $\Delta$ or $Y$ must differ from their non-stochastic steady state values to restore equilibrium. The right-hand side is clearly increasing in $\Delta$. Therefore to correct the imbalance $\Delta$ would have to fall. However, this induces a contradiction since $\Delta$ is convex in $\pi$ for the relevant parametization. Hence $Y$ must change to equilibriate the system. To see that it must fall take the derivative of the right-hand side 
\begin{multline}
\bigg(\frac{\nu'(\Delta Y/A)}{A} + \frac{Y\nu''(\Delta Y/A)}{A}+ \\ \alpha \beta \frac{\mathbb{E}\{\nu'(\Delta Y /A)+Y\nu''(\Delta Y/A)\}(1+\pi)^{\theta}/A}{1-\alpha \beta \mathbb{E}(1+\pi)^{\theta}}\bigg)\bigg/ \\ 
\bigg(\psi u'(Y)Y+ \alpha \beta \frac{\mathbb{E}\psi u'(Y)Y(1+\pi)^{\theta-1}}{1-\alpha \beta \mathbb{E}(1+\pi)^{\theta-1}}\bigg)+\\(\sigma -1)\bigg(u'(Y)+\alpha \beta \frac{\mathbb{E}\psi u'(Y)(1+\pi)^{\theta-1}}{1-\alpha \beta \mathbb{E}(1+\pi)^{\theta-1}}\bigg)\times \\ \bigg(\frac{Y\nu'(\Delta Y/A)}{A}+\alpha \beta \frac{\mathbb{E}Y\nu'(\Delta Y/A)(1+\pi)^{\theta}/A}{1-\alpha \beta \mathbb{E}(1+\pi)^{\theta}}\bigg)\bigg/ \\ \bigg(\psi u'(Y)Y+\alpha \beta \frac{\mathbb{E}\psi u'(Y)Y(1+\pi)^{\theta-1}}{1-\alpha \beta \mathbb{E}(1+\pi)^{\theta-1}}\bigg)^{2}
\end{multline}
since when $\sigma \geq 1$ the second term is non-negative and therefore unable to cancel the strictly positive first term. 
\par Sharpness depends on parametric assumptions. Suppose $0<\sigma <1$ and consider the joint limit, where the economy approaches any non-stochastic steady state and $A \rightarrow 0$. There are two competing terms. The first represented by the top two lines above is always positive and the second is negative by assumption.
\par The idea is to show that sufficiently close to these limiting cases the second term will dominate. To achieve this consider the ratio of the negative term to the positive term. The proof will be complete if its limit is greater than unity. The expression is the following product \begin{multline}
\lim_{\substack{A \rightarrow 0 \\ \pi \rightarrow \bar{\pi}}}\Bigg\{\bigg(\psi u'(Y)Y+\alpha \beta \frac{\mathbb{E}\psi u'(Y)Y(1+\pi)^{\theta-1}}{1-\alpha \beta \mathbb{E}(1+\pi)^{\theta-1}}\bigg)\bigg/ \\ \bigg(\psi u'(Y)Y+\alpha \beta \frac{\mathbb{E}\psi u'(Y)Y(1+\pi)^{\theta-1}}{1-\alpha \beta \mathbb{E}(1+\pi)^{\theta-1}}\bigg)\Bigg\} \\ \times \bigg(\frac{Y\nu'(\Delta Y/A)}{A}+\alpha \beta \frac{\mathbb{E}Y\nu'(\Delta Y/A)(1+\pi)^{\theta}/A}{1-\alpha \beta \mathbb{E}(1+\pi)^{\theta}}\bigg)\bigg/ \\ \bigg(\frac{\nu'(\Delta Y/A)}{A} + \frac{Y\nu''(\Delta Y/A)}{A}+ \\ \alpha \beta \frac{\mathbb{E}\{\nu'(\Delta Y /A)+Y\nu''(\Delta Y/A)\}(1+\pi)^{\theta}/A}{1-\alpha \beta \mathbb{E}(1+\pi)^{\theta}}\bigg) 
\end{multline}
With functional forms in place, this can be computed with big $\mathcal{O}$ notation. The first ratio (top line) is 
\\$\mathcal{O}(Y^{-1})=\mathcal{O}(A^{-(1+\gamma)/(\sigma+\gamma)})$ The second term is more complicated, the numerator of the second line is $\mathcal{O}(\Delta Y/A)/A=\mathcal{O}(A^{-\sigma(1+\gamma)/(\sigma + \gamma)})$.
\par Now turning to the denominator line three, there are terms in this order and others in $$\mathcal{O}(Y \nu''(\Delta Y/A)/A)=\mathcal{O}(A^{-(1+\sigma \gamma)/(\sigma + \gamma)})$$ since $\sigma <1$, this term will dominate and the total order of the second term will be $\mathcal{O}(A^{(1-\sigma)/(\sigma + \gamma)})$. Hence, the order of the whole limit will be $\mathcal{O}(A^{-(1+\sigma \gamma)/(\sigma + \gamma)} +{(1-\sigma)/(\sigma + \gamma)})=\mathcal{O}(A^{-1})$. Thus, the limit will be infinite. This ensures that for productivity sufficiently low and inflation sufficiently stable the derivative will be negative and the result overturned for any $0< \sigma < 1$. This proves that $\sigma \geq 1$ is a sharp restriction. 
\end{proof}
\section{Proofs from Section 4}
This part contains three proofs from the framework section. The first from Subsection 2 completes a standard derivation of product demand under monopolistic competition, associated with \cite{armington1969theory} and \cite{dixit1977monopolistic}. \subsection{Individual Product Demand (4.2)}
The purpose of this part is to derive the demand for an individual variety. The firms are monopolistically competitive. This ensures that firms face a meaningful pricing decision, avoiding the case of unbounded sales possible under perfect or simple Bertrand competition. Under monopolistic competition firms produce differentiated products that are imperfect substitutes and consumers prefer variety. This implies that aggregate consumption has to be a non-linear function of the underlying individual varieties. This contrasts with labor where, if we broke up the representative household into constituents $j$, the aggregation would be linear with $l=\sum_{j=1}^{J} l_{j}$.
\par The consumption objective (utility from consumption) is
\begin{equation}
u=\bigg[\int_0^1 {c_{t}(i)}^{(\theta -1)/\theta} \,{d}i\bigg]^{\theta/(\theta-1)}\end{equation}
where $\theta > 1$ is required for a well-behaved problem. The household's problem is to maximize consumption utility subject to an expenditure constraint
\begin{equation}P_{t}C_{t}=\int_0^1 p_{t}(i)\, c_{t}(i)\,{d}i\end{equation}
For any two varieties $i$ and $i'$ this yields relative demand. 
\begin{equation}\frac{{c}_{t}(i)}{c_{t}(i')}=\bigg(\frac{p_{t}(i)}{p_{t}(i')}\bigg)^{-\theta} \end{equation}
Now a little manipulation and then integration with respect to $i'$ yields 
\begin{equation} \int_0^1 p_{t}(i')\, c_{t}(i')\,{d}i'=\int_0^1 c_{t}(i)\, p_{t}(i)^{\theta}\, p_{t}^{1-\theta}(i')\,{d}i' \end{equation} 
Guessing and verifying yields the demand system price level pair 
\begin{equation}
P_{t}=\bigg[\int_0^1 {p_{t}(i)}^{1-\theta}{d}i\bigg]^{1/(1-\theta)}
\end{equation}
\begin{equation}c_{t}(i)=\bigg(\frac{p_{t}(i)}{P_{t}}\bigg)^{-\theta} C_{t}\end{equation}
combining with the individual market clearing $c_{t}(i)=y_{t}(i)$ yields the stated expression (17)
\qed
\subsection{Flexible Price Equilibrium and ZINSS (4.4)}
This subsection clarifies remarks concerning the relationship between the optimal reset price and the flexible price equilibrium, including their coincidence when there is zero inflation.
To achieve this I introduce a little mathematical machinery, that I understand to be original. Under monopolistic competition, I find that the optimal price is a mark-up $m=1/(\theta-1)$ over marginal costs, so in real terms
\begin{equation}\frac{p^{f}_{t}}{P_{t}}=\frac{\theta}{\theta-1}MC_{t}\bigg(\frac{p^{f}_{t}}{P_{t}}\bigg)\end{equation}
Particular to the flexible price equilibrium is that whatever has happened in the past all firms have the same price $p^{f}_{t}=P_{t}$. Furthermore, markups $m$ are common across firms and real marginal costs $MC$ remain in steady state with $MC_{t}=M C=(\theta -1)/\theta=1/(1+m)$. The connection between the optimal flexible price $p^{f}_{t}$ and the optimal reset price $p^{*}_{t}$ underpins the equivalence between ZINSS and the flexible price equilibrium. It takes the form of a \emph{Generalized Stochastic Mean} (GSM) defined as follows:
\begin{definition}
A GSM $\mathfrak{M}$ over a collection $X$ ordered by $i \in \mathbb{N}$, whose elements take on values $x$ in a reflexive space\footnote{A reflexive space has the property that the set of possible ensemble averages and the set of possible realizations coincide.}  $\mathcal{X}$ is a function stochastic process pair $(\mathcal{M}(X), \, \mathcal{S}(X))$ such that: \begin{enumerate} \item $\mathcal{M}(\bar{x},\, \bar{x}, \, \cdots ,\, \bar{x})=\bar{x}, \: $ for any  sequences $\langle \bar{x}, \, \bar{x}, \, \cdots \rangle$.
\item $\mathcal{M}$ is strictly increasing and continuous on $X^{\infty}$, the space of all possible convergent sequences, and is weakly measurable with respect to the stochastic process $\mathcal{S}(X)$.
\end{enumerate}\end{definition}
It is worth briefly motivating this definition. Continuity and increasing requirements support the idea of the mean as an average and rule out alternative measures of location, such as median and mode.\footnote{Consult \cite{bullen2003handbook} for a review of mean concepts and their inequalities.} 
The function space assumption allows us to include expectations of future variables, as part of our average ensemble. It is this averaging property that underpins ergodicity and mixing in the model.
\begin{proposition}The optimal reset price $p^{*}_{t}/P_{t}$ is a GSM of the optimal flexible price series $\langle p^{f}_{t}/P_{t}, \, \mathbb{E}_{t}p^{f}_{t+1}/P_{t+1}\, \cdots 
\rangle $  for the duration of the contract $T \geq t$.\end{proposition}
\begin{proof}
Weak measurability comes from the structure of the optimization problem. Monotonicity and continuity follow from the following implicit function argument, where the left-hand side is (27), combined with (199) and the gross inflation definition (23) on the right-hand side 
$$\mathcal{B}_{t}=\mathbb{E}_{t}\sum_{T=t}^{\infty}(1-\alpha)^{T-t}Q_{t,\, T}\bigg(\frac{p^{*}_{t}}{P_{t}}\bigg)^{-\theta}(\Pi_{t,\, T})^{-\theta}Y_{T}\bigg[ \frac{p^{*}_{t}}{P_{t}}\frac{1}{\Pi_{t,T}}-\frac{p^{f}_{T}}{P_{T}}\bigg]=0$$
$$\frac{\partial \mathcal{B}_{t}}{\partial (p^{*}_{t}/P_{t})}=\mathbb{E}_{t}\sum_{T=t}^{\infty}(1-\alpha)^{T-t}Q_{t,\, T}\bigg(\frac{p^{*}_{t}}{P_{t}}\bigg)^{-\theta }(\Pi_{t,\, T})^{-(\theta+1)}Y_{T} > 0 $$
$$\frac{\partial \mathcal{B}_{t}}{\partial (p^{f}_{T}/P_{T})}=-\mathbb{E}_{t}\sum_{T=t}^{\infty}(1-\alpha)^{T-t}Q_{t,\, T}\bigg(\frac{p^{*}_{t}}{P_{t}}\bigg)^{-\theta}(\Pi_{t,\, T})^{-\theta}Y_{T}<0$$
Hence $$\frac{\partial (p^{*}_{t}/P_{t}) }{\partial (p^{f}_{T}/P_{T})}=-\frac{\partial \mathcal{B}_{t}}{\partial (p^{f}_{T}/P_{T})}\bigg/\frac{\partial \mathcal{B}_{t}}{\partial (p^{*}_{t}/P_{t})}>0$$
This fulfils the monotonicity and continuity conditions. It is also sufficient to invoke the inverse function theorem of \cite{lang2001fundamentals} for the Banach space induced by the norm $\langle p^{f}_{t}/P_{t},\,  \mathbb{E}_{t}p^{f}_{t+1}/P_{t+1}, \,  \cdots \rangle $. Finally, the averaging property is obvious from (24) and (25). 
\end{proof}
\begin{corollary}
ZINSS is equivalent to a flexible price equilibrium.
\end{corollary}
\begin{proof}
From condition [ii] we have $p^{*}_{t}=\mathcal{M}(\langle p^{f}_{NSS},\, p^{f}_{NSS}, \, \cdots \rangle )=p^{f}_{NSS}$
\end{proof}
Therefore uniquely at ZINSS the reset price constraint is not binding. 
This confirms the breakdown of \emph{nominal rigidity} as the root cause 
of the bifurcation. The other weighted measures feature expectations of non-linear terms and are therefore not amenable to our definition. 
\begin{remark}These results carry over easily to alternative settings with finite price lives. This includes Taylor pricing and models where the price spell is truncated, such as \cite{wolman1999sticky} and \cite{dixon2012generalised}.\end{remark}
The final result justifies a comment in Footnote 25 in Section 4.4. It explains the distinction between the reset price and the flexible price. 
Moreover, it presages the idea of a limiting fringe of flexible price firms, used in Section 9.5 to make distinctions between real and nominal rigidity. 
\begin{proposition}The magnitude of the change in the reset price weakly exceeds the change in the price of a firm that always had flexible prices, with equality only when neither price changes. \end{proposition}
\begin{proof}Since price re-setters are randomly assigned here, by a standard law of large numbers argument, the reset price inflation is given by 
\begin{equation}1+\pi^{r}_{t}=\frac{p^{*}_{t}}{P_{t-1}}=\frac{(1-\alpha)^{1/(\theta-1)}(1+\pi_{t})}{(1-\alpha(1+\pi_{t})^{\theta-1})^{1/(\theta-1)}} \end{equation}
A flexible price firm will always set its price equal to the general price level
\begin{equation}1+\pi^{f}_{t}=\frac{p^{f}_{t}}{P_{t-1}}=1+\pi_{t}
\end{equation}
Here it is possible to think of a very small fringe $\eta$ of flexible price firms too small to influence aggregate quantities. The claimed result requires us to prove three cases. The first that $\pi_{t}=\pi^{r}_{t}=\pi^{f}_{t}=0$ is trivial. The second that $\pi^{r}_{t}>\pi^{f}_{t}$ when $\pi_{t}>0$ and conversely $\pi^{r}_{t}<\pi^{f}_{t}$ when $\pi_{t}<0$. Focusing on the second which corresponds to the inequality 
\begin{equation}\frac{(1-\alpha)^{1/(\theta-1)}}{(1-\alpha(1+\pi_{t})^{\theta-1}}>1\end{equation}
when $\pi_{t}>0$, we can see this is true because of the equality case and the fact that the left hand side is clearly increasing in $\pi_{t}$. The third case follows likewise. 
\end{proof}
(202) corresponds to Proposition 1, the economic idea that the relative price of the re-setter is increasing in inflation. Intuitively, with rigid prices to generate inflation or deflation, re-setters have to adjust more aggressively than in the flexible price world, in order to make up for the portion of the price level that does not change. This is why we cannot naively compare observed price changes with a classical benchmark to obtain the spillover from rigid prices to flexible prices. 
\subsection{Price Dispersion (4.5)}
This item proves that the price aggregator is convex and therefore represents price dispersion. It is a standard exercise in real analysis. 
\subsubsection{Proof of Proposition 2}
\begin{proof}The proof that $\Delta \geq 1$ is an application of Jensen's inequality. First define two functions $$g(p_{i})=p_{i}*\bigg(\frac{p_{i}}{P}\bigg)^{-\theta}$$ $$\phi(p_{i})=\bigg(\frac{p_{i}}{P}\bigg)^{\theta/(\theta-1)}$$
It is clear that $\phi$ is strictly convex because
$$\frac{d^{2}\phi}{d{p_{i}}^{2}}=\frac{\theta}{(\theta-1)^{2}}\frac{\phi(p_{i})}{p_{i}^2}>0, \: \forall \, p_{i} > 0$$
Although, in the first part of the proof, I will only use the weak convexity property. Note that $P=\int_{\Omega} \; g \; \mathrm{d}\mu$. Now since $\phi$ is a convex function defined on a metric space, it follows from Rockafeller's theorem that I may select $a$ and $b$ such that
$$ap^{*}+b \leq \phi(p^{*})$$
for all possible reset prices $p*$ and for the particular value $p^{*}=P$ $$aP+b=\phi(P)$$
This object is called a sub-derivative of $\phi$ at $P$. It follows that for all $p^{*}$ $$\phi \circ g(p^{*}) \geq ag(p^{*})+b$$ 
since we have a probability measure the integral is monotone with $\mu(\Omega)=1$. 
This implies that: $$\Delta= \int_{\Omega}\phi \circ g \; \mathrm{d}\mu$$ $$\geq \int_{\Omega} (ag+b) \; \mathrm{d}\mu$$
$$=a\int_{\Omega} g \; \mathrm{d}\mu + \int_{\Omega} b \; \mathrm{d}\mu$$ $$=aP+b$$ $$=\phi(P)$$ $$=1$$
The "only if" claim follows because under strict convexity the sub-derivative will be a tangent. Thus, away from price equality, I can proceed replacing weak with strict inequalities. \end{proof}
\subsubsection{Proof of Proposition 3}
\begin{proof}
To begin with, express price dispersion in non-stochastic steady state to reveal a function of trend inflation. The precise expression is (274) available along with the other steady state terms in Appendix D. Differentiate with respect to inflation to reveal the first derivative
$$\alpha \theta \pi \frac{(1+\pi)^{\theta-2}({1-\alpha(1+\pi)^{\theta-1})}^{1/(\theta-1)}}{(1-\alpha)^{1/(\theta-1)}(1-\alpha (1+\pi)^{\theta})}$$
as expected there is a stationary point at ZINSS. It is profitable to rewrite this as follows 
$$ \frac{\alpha \theta \pi(1+\pi)^{\theta-2}\Delta}{(1-\alpha(1+\pi)^{\theta-1})^{1/(\theta-1)}}$$
It is clear that when we I use the product rule to find the second derivative all the terms will vanish except for the derivative with respect to the function $\pi$, leaving 
$$ \frac{\alpha \theta (1+\pi)^{\theta-2}\Delta}{(1-\alpha(1+\pi)^{\theta-1})^{1/(\theta-1)}}>0$$ which justifies the claim.
\end{proof}
\subsection{Cost Minimization (4.6) and (10.4)}
For simplicity, there is only one factor of production: labor purchased on a competitive market. I will focus on the general case from (10.4) where labor is firm-specific.\footnote{ In this context it would be more natural to think of firms having monopsony power. Nevertheless, this causes no problem around ZINSS, where a constant mark-down would be isomorphic to an increase in $\theta$, which would cancel from the first order dynamics. \cite{nash1950bargaining} supplies bargaining foundations.} The production function takes the form \begin{equation}y_{t}(i)=f(l_{t}(i))\end{equation}
where $f$ must always be weakly concave ($f_{ll} \leq 0$) to ensure standard first order conditions apply. The problem is as follows:
\begin{equation}\min_{l_{t}(i)}\: W_{t}(i)l_{t}(i) \end{equation}
subject to the production constraint
\begin{equation}f(l_{t}(i))=\bar{y}_{t}(i)
\end{equation}
The Lagrange multiplier gives the real marginal cost of production paid by the firm. Hence we can solve for real marginal cost
\begin{equation}MC_{t}(i) = \frac{W_{t}(i)}{f'(l_{t}(i))}\end{equation}
In the main analysis, I will work with a linear production technology with an economy-wide labor market. This ensures all firms will have the same marginal costs, simplifying analysis considerably \begin{equation}
MC_{t}(i)=MC_{t}=\frac{W_{t}}{A_{t}}
\end{equation}
In particular, the real marginal cost is the ratio of the real wage $W_{t}$ to aggregate technical efficiency term $A_{t}$, from the RBC model. If I allowed for decreasing returns to scale, it would be equivalent to rescaling $\eta$, so key results would go through. This is one example of the forces interpreted by \cite{gopinath2011search} as real rigidities. These are discussed in Section 9.5 and Appendix I.2.
\subsection{Consolidation (4.8)}
Here, I derive the state space form of the non-linear model. The proof uses only techniques from basic analysis. The main complexity arises from the non-linear Phillips curve. 
\subsubsection{Proof of Proposition 4}
\begin{proof}
Start with the non-linear Phillips curve, formed from substituting the relationship between the optimal relative reset price and inflation derived in Proposition 1 into the left-hand side of (27). Now put (30) and (31) into the right-hand side. Finally, lag the relationship one period to reveal that 
\begin{multline} \frac{(1-\alpha)^{1/(\theta-1)}}{(1-\alpha(1+\pi_{t-1})^{\theta-1})^{1/(\theta-1)}}  = \\ \frac{\theta}{\theta-1}\frac{\psi_{t-1}u'(Y_{t-1})Y_{t-1}MC_{t-1}+
\alpha \beta \mathbb{E}_{t-1}(1+\pi_{t})^{\theta}\aleph_{t}}{\psi_{t-1}u'(Y_{t-1})Y_{t-1}+\alpha \beta \mathbb{E}_{t-1}(1+\pi_{t})^{\theta-1}\beth_{t}} \end{multline}
Rearranging for $\beth_{t}$ yields 
\begin{multline}\beth_{t}=\frac{(1-\alpha(1+\pi_{t-1})^{\theta-1})^{1/(\theta-1)}}{(1-\alpha)^{1/(\theta-1)}\alpha \beta (1+\pi_{t})^{\theta-1}}\bigg[\\ \frac{\theta}{\theta-1} \bigg( \psi_{t-1}u'(Y_{t-1})Y_{t-1}MC_{t-1}+
\alpha \beta (1+\pi_{t})^{\theta}\aleph_{t}+ W^{0}_{t}(u^{J}_{t})\bigg)-\\ \frac{(1-\alpha)^{1/(\theta-1)}}{(1-\alpha(1+\pi_{t-1})^{\theta-1})^{1/(\theta-1)}} \bigg(\psi_{t-1}u'(Y_{t-1})Y_{t-1}+W^{1}_{t}(u^{J}_{t})\bigg)\bigg]\end{multline}
where $W^{.}_{t}$ are error terms reflecting the difference between expected and true values. They are white noise thanks to the rational expectation hypothesis. Subsequently, dependence on the current shocks will be suppressed.\footnote{The transversality condition prevents additional non-fundamental terms entering.}
\begin{equation}
 \aleph_{t}=\frac{\theta-1}{\theta} \frac{(1-\alpha)^{1/(\theta-1)}}{(1-\alpha(1+\pi_{t})^{\theta-1})^{1/(\theta-1)}}\beth_{t} 
\end{equation}
I am able to solve for $\aleph_{t}$ as follows 
\begin{multline}\aleph_{t}=\frac{\theta-1}{\theta}\Bigg[ \frac{\theta}{\theta-1}(1-\alpha(1+\pi_{t-1})^{\theta-1})^{1/(\theta-1)} \times 
\\ \bigg( \psi_{t-1}u'(Y_{t-1})Y_{t-1}MC_{t-1}+
\alpha \beta (1+\pi_{t})^{\theta}\aleph_{t}+ W^{0}_{t}(u^{J}_{t})\bigg)- \\ (1-\alpha)^{1/(\theta-1)}\bigg(\psi_{t-1}u'(Y_{t-1})Y_{t-1}+W^{1}_{t}(u^{J}_{t})\bigg)
\Bigg] \Bigg/ \\ (1+\pi_{t})^{\theta-1}\bigg\{ (1-\alpha(1+\pi_{t})^{\theta-1})^{1/(\theta-1)}-(1-\alpha(1+\pi_{t-1})^{\theta-1})^{1/(\theta-1)}(1+\pi_{t})\bigg\}
\end{multline}
where the formula is valid whenever the denominator is non-zero. Now rearranging the future price equation 
\begin{equation}
\beth_{t+1}=\frac{\theta}{\theta-1}\frac{(1-\alpha(1+\pi_{t})^{\theta-1})^{1/(\theta-1)}} {(1-\alpha)^{1/(\theta-1)}}\aleph_{t+1}
\end{equation}
then substituting into the right-hand side and simplifying yields 
\begin{multline}\beth_{t}=\psi_{t}u'(Y_{t})Y_{t}+ \\ \alpha \beta \frac{\theta}{\theta-1} (1+\pi_{t+1})^{\theta-1}\frac{(1-\alpha(1+\pi_{t+1})^{\theta-1})^{1/(\theta-1)}} {(1-\alpha)^{1/(\theta-1)}}\aleph_{t+1} + W^{2}_{t+1}\end{multline}
Moving (207) forward one period and substituting in reveals
\begin{equation}\beth_{t}=\tilde{f}_{0} + \tilde{f}_{1}\tilde{f}_{2}(\pi_{t+1})\end{equation}
where $\tilde{f}_{0}$ and $\tilde{f}_{1}$ do not depend on $\pi_{t+1}$. In fact, $\tilde{f}_{2}$ is the only expression in the whole Phillips curve (205) determined by future inflation. Therefore, decisively signing its derivative will be sufficient to solve for an appropriately smooth Phillips curve, via the implicit function theorem (see \cite{fritzsche2002holomorphic}). Hence the result follows from the following calculations
\begin{equation}\tilde{f}_{2}=\frac{(1-\alpha(1+\pi_{t+1})^{\theta-1})^{1/(\theta-1)}}{(1-\alpha(1+\pi_{t+1})^{\theta-1})^{1/(\theta-1)}-(1-\alpha(1+\pi_{t})^{\theta-1})^{1/(\theta-1)}(1+\pi_{t+1})}\end{equation}
\begin{multline}\frac{\partial \tilde{f}_{2}}{\partial \pi_{t+1}}=(1-\alpha(1+\pi_{t+1})^{\theta-1})^{1/(\theta-1)}+ \\ \alpha (\theta-1)(1+\pi_{t+1})^{\theta-1}(1-\alpha(1+\pi_{t})^{\theta-1})^{1/(\theta-1)} 
\\ \Bigg/ 
\bigg((1-\alpha(1+\pi_{t+1})^{\theta-1})^{1/(\theta-1)}-(1-\alpha(1+\pi_{t})^{\theta-1})^{1/(\theta-1)}(1+\pi_{t+1})\bigg)^{2}>0\end{multline}
To prove this holds with probability one consider 
$$(1-\alpha(1+\pi_{t+1})^{\theta-1})^{1/(\theta-1)}-(1-\alpha(1+\pi_{t})^{\theta-1})^{1/(\theta-1)}(1+\pi_{t+1})$$ 
which appears in the denominator of (207) and (210) through (212), setting the expression equal to zero gives the equation of the singularity 
\begin{equation} \pi_{t+1}=\bigg(\frac{1}{(\alpha + 1-\alpha(1+\pi_{t})^{\theta-1})}\bigg)^{1/(\theta-1)}-1\end{equation} We can see that this function is surjective, by evaluating its derivative, which shows that it is strictly increasing. 
\begin{equation}\frac{\mathrm{d}\pi_{t+1}}{\mathrm{d}\pi_{t}}=-\frac{\alpha (1+\pi_{t})^{\theta-2}}{{(\alpha + 1-\alpha(1+\pi_{t})^{\theta-1})}^{\theta/(\theta-1)}}<0\end{equation}
This means that the singularity can only occur at a point in time $t+1$, if inflation is equal to one particular value, determined by inflation in the previous period $t$. We know this must occur with probability zero as inflation is continuously distributed, since it is determined by the reset price that depends non-trivially on the continuously distributed shock processes. 
\par This can be converted into a relationship in expected inflation, by noting that equation (27) and the underlying optimization problems bound inflation above and below as follows
\begin{equation}-1< \pi_{t}<\bigg(\frac{1}{\alpha}\bigg)^{1/(\theta-1)}\end{equation} 
and then noting continuity properties are sufficient for the existence of expectation. Note that the singularities in $\aleph_{t}$ and $\beth_{t}$ are both removable since the system is continuous with respect to $Y$, $\pi$ and the increments of the summation of $\aleph_{t}$ and $\beth_{t}$ in (28) and (29).\footnote{ZINSS corresponds to a particular case where both singularities arise simultaneously and "cancel out" from the right hand side of (205). The nature of the cocycle at the singularity is treated in the next section.} 
Therefore we can write the Phillips curve as \begin{equation}\mathbb{E}_{t}\pi_{t+1}=f_{\pi}(\pi_{t},\, Y_{t}, \, \pi_{t-1},\, \Delta_{t},\, Y_{t-1},\, MC_{t},\, MC_{t-1}, \, A_{t},\, \psi_{t},\, A_{t-1},\, \psi_{t-1})\end{equation}
To complete this section of the derivation all that is required is to substitute out terms in marginal costs and lagged income. Present marginal costs can be removed using (19) and (37) as follows 
\begin{equation}MC_{t}=\frac{\nu(\Delta_{t}Y_{t}/A_{t})}{\psi_{t}u'(Y_{t})} 
\end{equation}
leaving only lagged income which can be substituted out via the Euler equation. 
To this end let us define the following function 
\begin{multline}\tilde{f}_{3}=u'(Y_{t-1})-\beta(1+ i_{t-1}(Y_{t-1},\, \pi_{t-1}))\bigg\{\frac{u'(Y_{t})}{1+\pi_{t}}\frac{\psi_{t}}{\psi_{t-1}}+ W_{3}\bigg\}= \\ u'(Y_{t})-\beta(1+ i_{t}(Y_{t},\, \pi_{t}))\mathbb{E}_{t}\frac{u'(Y_{t+1})}{1+\pi_{t+1}}\frac{\psi_{t+1}}{\psi_{t}}= 0\end{multline}
Calling in the policy rule (35), including the output smoothing restriction gives
\begin{equation}\frac{\partial \tilde{f}_{3}}{\partial Y_{t-1}}=u''(Y_{t-1})-\beta\frac{a_{y}}{y_{t}}(1+ i_{t-1}(Y_{t-1},\, \pi_{t-1}))\bigg\{\frac{u'(Y_{t})}{1+\pi_{t}}\frac{\psi_{t}}{\psi_{t-1}}+ W_{3}\bigg\}<0\end{equation}
which allows us to form a function \begin{equation}Y_{t-1}=\tilde{f}_{4}(\pi_{t},\, Y_{t}, \, \pi_{t-1},\,  A_{t},\,  \psi_{t},\, \psi_{t-1})\end{equation} substituting into (217) gives the appropriate form for the Phillips curve 
\begin{equation}\mathbb{E}_{t}\pi_{t+1}=f_{\pi}(\pi_{t},\, Y_{t},\, \pi_{t-1},\, \Delta_{t}, \, A_{t}, \, \psi_{t},\, A_{t-1},\, \psi_{t-1})\end{equation}
Finally, we can form the canonical Euler by inverting the Euler using 
\begin{equation}\frac{\partial \tilde{f}_{3}}{\partial Y_{t+1}}=-\beta(1+ i_{t}(Y_{t},\pi_{t}))\mathbb{E}_{t}\frac{u''(Y_{t+1})}{1+\pi_{t+1}}\frac{\psi_{t+1}}{\psi_{t}}>0\end{equation}
then substituting in the Phillips curve gives the final section 
\begin{equation}\mathbb{E}_{t}Y_{t+1}=f_{y}(\pi_{t},\, Y_{t},\, \pi_{t-1},\, \Delta_{t},\, A_{t},\, \psi_{t},\, A_{t-1},\, \psi_{t-1})\end{equation}
Lastly, by inverting (34), moving one period into the future and substituting in the Phillips curve and noting the expectation must exist to prevent transversality condition blowing again, which means that 
\begin{equation}\mathbb{E}_{t}\Delta_{t+1}=f_{\Delta}(\pi_{t},\, Y_{t},\, \pi_{t-1}, \,\Delta_{t}, \, A_{t},\, \psi_{t},\, A_{t-1},\, \psi_{t-1})\end{equation}
Existence of expectations follows from mean field game arguments. The only way for the cocycle (225) to not exist would be if $\mathbb{E}_{t}\Delta_{t+1}$ were infinity. However, this would force the household to violate its $t+1$ Inada condition (6) causing its optimization problem to blow up. Likewise, $\mathbb{E}_{t}Y_{t+1} = \infty$ in (224) would make the firms' reset problem (24) 
and that of the household (6) ill-defined. Combining (222), (224) and (225) forms the claimed recursion, thus completing the proof.
\end{proof}
\section{Proofs from Section 5}
This section contains proof of the remaining propositions in Section 5, as well as additional supporting material and an extension of arguments concerning persistence. 
\subsection{Puzzles, Policy and Persistence}
This subsection focuses on proving the main dynamics and policy results
pertinent to first order dynamics. \subsubsection{Proof of Lemma A.1}
 Here is the proof of Chebyshev's correlation inequality which is used to complete the proof of Proposition 5.
 \begin{proof}
\begin{equation}
\mathbb{E}UV=\int U(V-\mathbb{E}V)\mathrm{d}\mu + \mathbb{E}V\mathbb{E}U\end{equation}
By assumptions placed on the measure, the first term can be decomposed into the (non-zero) contribution of terms above and below the expected value $\mathbb{E}V$
\begin{equation}
\int U(V-\mathbb{E}V)\; \mathrm{d}\mu=\int_{V>\mathbb{E}V} U(V-\mathbb{E}V)\; \mathrm{d}\mu +\int_{V<\mathbb{E}V} U(V-\mathbb{E}V)\; \mathrm{d}\mu\end{equation}
(Strict) Monotonicity of $U$, $V$ and the Lebesgue integral allow us to conclude that 
\begin{equation}
\int_{V>\mathbb{E}V} U(V-\mathbb{E}V)\; \mathrm{d}\mu> U(\mathbb{E}V)\int_{V>\mathbb{E}V}(V-\mathbb{E}V)\; \mathrm{d}\mu
\end{equation}
Similarly when $V$ is below average by the same argument 
$$\int_{V<\mathbb{E}V} U(\mathbb{E}V-V)\; \mathrm{d}\mu < U(\mathbb{E}V)\int_{V<\mathbb{E}V}(\mathbb{E}V-V)\; \mathrm{d}\mu$$
Now reversing yields
\begin{equation}
\int_{V<\mathbb{E}V} U(V-\mathbb{E}V)\; \mathrm{d}\mu > U(\mathbb{E}V)\int_{V<\mathbb{E}V}(V-\mathbb{E}V)\; \mathrm{d}\mu
\end{equation}
combining the numbered equations completes the proof.
\end{proof}
\subsubsection{Proof of Proposition 6}
 \begin{proof} The planner can set interest rates to offset the demand shock $\hat{i}_{t}=\hat{\psi}_{t}$. This implements a non-stochastic steady state.
 Therefore the only remaining distortion is the markup of prices over marginal costs associated with imperfect competition. This can be remedied by a subsidy set to cancel the desired mark up as follows
 \begin{equation}\frac{p^{*}}{P}=\frac{\theta}{\theta-1}(1-\tau^{*})MC\end{equation}
 with the optimal level
 \begin{equation}\tau^{*}=\frac{1}{\theta}\end{equation}
 This creates an equivalence with the allocations of a standard competitive equilibrium. Therefore, these transfers can be funded by lump transfers thanks to Ricardian equivalence (see \cite{barro1974government}). 
 \par To apply this version of the welfare theorem, it is necessary to write the household's problem as an infinite budget constraint, subject to the standard objective function (7). To do so consider the Lagrangian form of the problem from which the Euler is derived
 \begin{multline}\mathcal{L}_{t}= {\mathbb{E}_{t}\sum_{T=t}^{\infty}\beta^{T-t}\,\bigg\{\bigg[u(C_{T})-\varphi \nu(1-N_{T})\bigg]\,\psi_{T}}-\\ \lambda_{T}\bigg(P_{T}C_{T}+W_{T}N_{T}-(1+i_{T})B_{T}-W_{T}-\int_{0}^{1}\Pi_{T}\, \mathrm{d}i\bigg)\bigg\}\end{multline}
  Note that in this context it is easier to work with $N_{T}$, the good leisure. Here, $\{\lambda_{T}\}_{t}^{\infty}>0$ is the sequence of Lagrange multipliers on the budget constraint. It is strictly positive because of non-satiation. \cite{acemoglu2009introduction} considers the non-stochastic case; therefore implicitly maximization takes place with respect to the future plans  $\{C_{T}\}_{t}^{\infty}>0$ and $\{N_{T}\}_{t}^{\infty}>0$. His analysis is adapted from a finite good setting, so to apply his proof I have to construct a single budget constraint. To this end, I repeatedly substitute out Lagrange multipliers, using the first order optimality conditions
\begin{equation}\frac{\mathrm{d}\mathcal{L}_{t}}{\mathrm{d}C_{T}}=
 \beta^{T-t}\mathbb{E}_{t}\Bigg[\psi_{T}u'(C_{T})-\lambda_{T}P_{T}
 \Bigg]=0\end{equation}
 In fact, the budget constraint can be given a particular compact form
 \begin{equation} \sum_{T=t}^{\infty}M_{t,T}C_{T} + \sum_{T=t}^{\infty}Q_{t,T}\bigg[(1+i_{T})B_{T}+W_{T}N_{T}\bigg]=\sum_{T=t}^{\infty}Q_{t,T}\bigg[W_{T}+\int_{0}^{1}\Pi_{T}\,\mathrm{d}i\bigg]
 \end{equation}
where I have used the expression for the nominal discount factor 
\begin{equation}M_{t, t+k}=\beta ^k \frac{\psi_{t+k}u'(C_{t+k})}{\psi_{t}u'(C_{t})}\end{equation}
familiar from asset pricing literature.
The final adjustment is to the measure of firms. He works with a finite number and therefore implicitly with the discrete measure, whereas here with a mean field of monopolistic competition in the background, we need to work with the standard Lebesgue measure.  
\par His proof can now be applied to demonstrate that the proposed scheme is Pareto efficient, in the set of all non-stochastic allocations. He is working in $\mathbb{R}^{\infty}$. All that is left is to show that this holds for policies that do not fully offset the shock. This is achieved via a simple convexity argument. From the primitives, we know that the period utility is strictly concave in $(C,\, L)$. This extends to the objective function via the entire sequence $\{C_{T},\, L_{T}\}$. Then Jensen's inequality tells us that any equilibrium that eliminates uncertainty in each period will be strictly preferred but still affordable. Hence I have constructed a Pareto optimal policy. 
\end{proof}
\begin{remark}\cite{stokey1989recursive} provide an alternative treatment of welfare theorems, using production sets and allowing directly for stochasticity, from which an alternative proof might be developed. \end{remark}
\begin{remark}From the infinite horizon budget constraint, we can see that nominal rigidity creates inefficiency, through a wedge between the effective future real wage $Q_{t,\, T}W_{T}$ and the future real wage prevailing in an efficient economy $M_{t,\, T}W_{T}$. Non-zero inflation creates an inter-temporal labor supply distortion. \end{remark}
Labor supply distortions have previously been a focus of business cycle accounting popular in the neoclassical tradition 
(see for example \cite{chari2007business} and \cite{shimer2009convergence}). Although, inter-temporal labor substitution is somewhat controversial (see \cite{bianchi2001iceland} and \cite{martinez2021intertemporal}), it is likely some similar inter-temporal distortions would arise with any alternative micro-foundations.
\subsubsection{Proof of Proposition 8}
\begin{proof}
 This is a linear system of expectational equations, with two jump variables and no state variables.
 There are two cases to consider, one where there is a unique solution, the other where there is a multiplicity of solutions. When there is indeterminacy it will not be possible to rule out persistent fluctuations, arising from expectational bubbles. Later, when I connect the non-bifurcated approximation to the non-linear model, I will be able to rule out indeterminacy. The proof will be completed by showing that whenever there is a determinate solution it is non-persistent and that determinacy corresponds to the expression given in the proposition. To do so, I express the system in canonical form by substituting the Phillips curve and policy rule to form the canonical Euler. It can be expressed in the following matrix form
 \begin{equation} \mathbb{E}_{t}\bf{X}_{t+1}=\bf{B}\bf{X}_{t}+
 \bf{e}_{t}\end{equation}
 \begin{equation} \bf{B}^{-1}=\begin{bmatrix} 1/ \beta & -\kappa / \beta \\ (\beta a_{\pi} -1 )/\beta \sigma & \sigma \kappa + \beta(\sigma + a_{y})/ \beta \sigma \end{bmatrix} \end{equation}
 The condition for a unique solution is that we can solve the system forward. This requires both eigenvalues be outside the unit circle, as in \cite{blanchard1980solution}. It is convenient to work with the inverse eigenvalue polynomial 
 \begin{equation} x^{2} +q_{0}x +q_{1}=0\end{equation}
 where $x =1/\lambda$ 
  \begin{equation} q_{0}= -\bigg( \frac{\kappa + (1+\beta)\sigma + \beta a_{y}}{\sigma + a_{y}+ \kappa a_{\pi}}\bigg) \end{equation}
 \begin{equation} q_{1}=\frac{\beta \sigma}{\sigma + a_{y}+ \kappa a_{\pi}} \end{equation}
 \cite{lasalle2012stability} shows that there will be two inverse eigenvalues inside the unit circle if and only if
 \begin{equation} \vert q_{0} \vert < 1+q_{1}\end{equation}
 \begin{equation} \vert q_{1} \vert <1\end{equation}
  The second condition amounts to $$\frac{\kappa + (1+\beta)\sigma + \beta a_{y}}{\sigma + a_{y}+ \kappa a_{\pi}}< 1+ \frac{\beta \sigma}{\sigma + a_{y}+ \kappa a_{\pi}} $$
 which rearranges to give the desired expression. The second condition is redundant 
 because $$(1-\beta)\sigma + a_{y}+ \kappa a_{\pi}>0$$
 \end{proof}
 \subsection{Extending No Persistence}
 This subsection generalizes lack of persistence in two directions. The first is to show that in the absence of price dispersion, it extends to all approximations from ZINSS. Similar results are well-known for Rotemberg. The second shows that even when the output gap is adjusted to allow technology shocks, inflation still responds to demand shocks only on impact.
 \subsubsection{Higher Order Dynamics}
This part demonstrates that when $\sigma \neq 1$ persistence arises, through the technology shock, but it is only second order. It is clear that the second stream of terms in (57) coming from (56) will remain unchanged, whilst those in (55) will become 
\begin{multline} \kappa 
(1+ \psi A^{1 -\sigma}\hat{\psi}_{t} )\bigg(\sum^{\infty}_{j=1}\prod^{\infty}_{j=1}(1-\sigma) \cdots (1-\sigma -\{ j-1\})\hat{a}^{j}_{t}\bigg) \times \\ \bigg(\sum^{\infty}_{k=1}\prod^{\infty}_{k=1} (1+\eta) \cdots (1+\eta -\{k-1\})(\hat{y}^{e}_{t})^{k}\bigg) \end{multline}
Since the system is forward-looking by assumption, it will be possible to expand out $(\hat{y}^{e}_{t}, \, \pi_{t})$ in terms of present and expected future terms in the errors $(\hat{\psi}_{t}, \, \hat{a}_{t})$, which can then be expressed in terms of just the contemporaneous shocks. Cov$(\hat{y}^{e}_{t}, \, \hat{a}_{t}) \neq 0$, which will spread to the rest of the system so persistence will be $O(\hat{\varepsilon}^{2})$. This justifies the claim that persistence will be second order, on the singular surface around ZINSS, away from $\sigma =1$.
\subsubsection{Dynamics with Unobserved Natural Rate}
This part considers the case where the central bank cannot observe the natural rate so the intercept ${i}^{*}_{t}$ is removed from the policy rule (35) that aggregate demand takes the form
\begin{equation}\hat{y}_{t}=-\frac{1}{\sigma}(a_{\pi}\pi_{t}+ a_{y}\hat{y}^{e}_{t})+ \mathbb{E}_{t+1}\hat{y}_{t}-\frac{1}{\sigma}\mathbb{E}_{t}\pi_{t+1}-\frac{1}{\sigma}\hat{\psi}_{t}
\end{equation}
Although no persistence breaks down, propagation problems still arise, in particular, with respect to inflation and aggregate demand shocks. The destruction of endogenous persistence when shock persistence takes limiting values is a theme common to the whole paper. This 
analysis has appeared before. The task at hand is to diagonalize (240) and substitute into the forward solution of
(239) and then compute the persistence. First solve (241)-(243), this yields  
\begin{equation} x_{1}= \frac{1}{2} \Bigg\{\frac{\kappa + (1+\beta)\sigma + \beta a_{y}}{\sigma + a_{y}+ \kappa a_{\pi}} + \sqrt{\bigg( \frac{\kappa + (1+\beta)\sigma + \beta a_{y}}{\sigma + a_{y}+ \kappa a_{\pi}}\bigg)^{2}-\frac{4\beta \sigma}{\sigma + a_{y}+ \kappa a_{\pi}}}\Bigg\} \end{equation}
\begin{equation} x_{2}= \frac{1}{2} \Bigg\{\frac{\kappa + (1+\beta)\sigma + \beta a_{y}}{\sigma + a_{y}+ \kappa a_{\pi}} - \sqrt{\bigg( \frac{\kappa + (1+\beta)\sigma + \beta a_{y}}{\sigma + a_{y}+ \kappa a_{\pi}}\bigg)^{2}-\frac{4\beta \sigma}{\sigma + a_{y}+ \kappa a_{\pi}}}\Bigg\} \end{equation}
The corresponding eigenvectors are $[1 \, -(1-\beta x_{1})/\kappa]'$ and $[1 \, -(1-\beta x_{2})/\kappa]'$
\begin{equation} \bf{e}_{t}= \begin{bmatrix} 0 & 0  \\ -(1-\rho_{a})(\sigma +\eta)/\sigma  & 1/\sigma \end{bmatrix} \begin{bmatrix} \hat{a}_{t} \\ \hat{\psi}_{t} \end{bmatrix}
\end{equation}
The relevant inverse matrix is 
\begin{equation} \bf{P}^{-1}=\frac{\kappa}{\beta(x_{1}-x_{2})}\begin{bmatrix}(1- \beta x_{2})/ \kappa & 1 \\ -(1- \beta x_{1})/ \kappa  & -1 \end{bmatrix}\end{equation}
using the forward solution and several rounds of substitution yields the following expression for inflation 
\begin{equation} \pi_{t}=\frac{\kappa}{\beta\sigma(x_{1}-x_{2})}\mathbb{E}_{t}\sum^{\infty}_{T=t+1}(x^{T-t}_{1}-x^{T-t}_{2})\hat{\psi}_{t} -(1-\rho_{a})(\sigma + \eta)(x^{T-t}_{1}-x^{T-t}_{2})\hat{a}_{t}  \end{equation}
Deploying the persistent properties of the errors, communicated in Subsection 4.7, this yields 
\begin{equation} \pi_{t}=\frac{\kappa}{\beta \sigma(x_{1}-x_{2})}\hat{\psi}_{t}-\frac{\kappa \rho_{a}(1-\rho_{a})}{\beta(x_{1}-x_{2})}\frac{(\sigma + \eta)}{\sigma}\bigg(\frac{\rho_{a} x_{1}}{1-\rho_{a} x_{1}}-\frac{\rho_{a} x_{2}}{1-\rho_{a} x_{2}}\bigg)\hat{a}_{t}  \end{equation}
It is intuitive that a positive supply shock drives down inflation. Away from the two limiting cases where it vanishes, 
as with Divine Coincidence, there is positive persistence, measured by the product moment correlation coefficient
\begin{equation} Corr (\pi_{t}, \, \pi_{t-1})= \sqrt{\rho_{a}}\end{equation}
It is still pathological to Keynesian intuition that inflation does not respond to demand shocks. For output the expression is 
\begin{multline} \hat{y}^{e}_{t}= \frac{1}{\sigma}\hat{\psi}_{t} - \\  (1-\rho_{a})\frac{(\sigma + \eta)}{\sigma}\bigg(1-\frac{\rho_{a}}{\beta(x_{1}-x_{2})}\bigg[ \frac{(1-\beta x_{1})}{(1-\rho_{a}x_{1})}x_{1}- \frac{(1-\beta x_{2})}{(1-\rho_{a}x_{2})}x_{2}\bigg]\bigg)\hat{a}_{t} \end{multline}
\begin{equation} 
Corr(y^{e}_{t}, \, y^{e}_{t-1})=\sqrt{\frac{K\sigma^{2}_{a}}{K\sigma^{2}_{a} + \sigma^{2}_{\psi}/\sigma^{2} }}\end{equation}
where 
\begin{equation} K=\rho_{a}(1-\rho_{a})\frac{(\sigma + \eta)}{\sigma}\bigg(1-\frac{\rho_{a}}{\beta(x_{1}-x_{2})}\bigg[ \frac{(1-\beta x_{1})}{(1-\rho_{a}x_{1})}x_{1}- \frac{(1-\beta x_{2})}{(1-\rho_{a}x_{2})}x_{2}\bigg]\bigg) \end{equation} 
Again, the desired limiting cases arise. There can be overshooting or under-shooting with respect to the efficient benchmark.\footnote{When $\beta \rightarrow 1$, $x_{1}>0$ whilst $x_{2}\rightarrow 0$ the overall technology coefficient is positive. On the other hand it can also go negative because the internal bracket term simplifies to 
$$1-\rho_{a}/\beta + \rho_{a}(x_{1}+x_{2})-\rho^{2}_{a}x_{1}x_{2}$$ which has to be positive when $\rho_{a} \rightarrow 0$, making the coefficient negative.} However, the underlying message is that this system cannot generate persistent responses from inflation to demand shocks.
\section{Stochastic Equilibrium}
 This section aligns with Section 6 of the main text. It begins by setting out the non-stochastic equilibrium conditions, before moving onto to the three omitted proofs and addressing the claims made in Remark 8.
 \subsection{Non-Stochastic Equilibrium}
 This subsection divided in two the first part details the closed form of the solution in terms of primitive parameters. The second explains why we must restrict the model to allow for economic growth and de-trending. 
 \subsubsection{Exact Form}
 With functional forms in, the non-stochastic steady state is as follows
 \begin{equation}
MC^{NSS}=\frac{\theta-1}{\theta}\bigg(\frac{1-\alpha \beta (1+\pi)^{\theta}}{1-\alpha \beta (1+\pi)^{\theta-1}}\bigg)\bigg(\frac{1-\alpha}{1-\alpha (1+\pi)^{\theta-1}}\bigg)^{1/(\theta-1)}
\end{equation}
where I have used the price-setting relation. 
Therefore, thanks to constant returns the wage rate is 
\begin{equation}
W^{NSS}=\frac{\theta-1}{\theta}\bigg(\frac{1-\alpha \beta (1+\pi)^{\theta}}{1-\alpha \beta (1+\pi)^{\theta-1}}\bigg)\bigg(\frac{1-\alpha}{1-\alpha (1+\pi)^{\theta-1}}\bigg)A
\end{equation}
Price dispersion is 
\begin{equation}
\Delta^{NSS}= \frac{(1-\alpha (1+\pi)^{\theta-1})^{\theta/(\theta-1)}}{(1-\alpha)^{1/(\theta-1)}(1-\alpha (1+\pi)^{\theta})}
\end{equation}
Inverting the marginal cost function and substituting in the price dispersion relationship yields 
\begin{multline}
Y^{NSS}=\bigg(\frac{\theta-1}{\theta}\bigg)^{1/(\sigma + \eta)}\frac{(1-\alpha)^{(1+\eta)/(\theta-1)}(1-\alpha (1+\pi)^{\theta})^{\eta /(\sigma + \eta)}}{(1-\alpha \beta (1+\pi)^{\theta-1})^{(1+\eta \theta)/(\theta-1)}}\times\\ \bigg( \frac{1-\alpha \beta (1+\pi)^{\theta}}{1-\alpha \beta (1+\pi)^{\theta-1}}\bigg)^{1/(\sigma + \eta)}{A}^{(1+\eta)/(\sigma +\eta)}
\end{multline}
The resource constraint yields the solution for labor supply
\begin{multline}
L^{NSS}=\bigg(\frac{\theta-1}{\theta}\bigg)^{1/(\sigma + \eta)}\frac{(1-\alpha)^{(1-\sigma)/(\theta-1)(\sigma+\eta)}}{(1-\alpha (1+\pi)^{\theta})^{\sigma/(\sigma + \eta)}}\times \\
(1-\alpha (1+\pi)^{\theta-1})^{(\sigma \theta-1)/(\theta-1)(\sigma +\eta)}\times \\ \bigg( \frac{1-\alpha \beta (1+\pi)^{\theta}}{1-\alpha \beta (1+\pi)^{\theta-1}}\bigg)^{1/(\sigma + \eta)}A^{(1-\sigma)/(\sigma + \eta)}
\end{multline}
The aggregate profit rate is  
\begin{equation}\frac{\Pi^{NSS}}{Y^{NSS}}=\bigg(1-\bigg(\frac{\theta-1}{\theta}\bigg)\bigg(\frac{1-\alpha \beta(1+\pi)^{\theta}}{1-\alpha \beta (1+\pi)^{\theta-1}}\bigg) \bigg(\frac{1-\alpha}{1-\alpha (1+\pi)^{\theta-1}}\bigg)^{1/(\theta-1)} \bigg)
\end{equation}
substituting into the utility function reveals that 
\begin{multline}
u^{NSS}=\frac{1}{1-\beta}\bigg[\frac{1}{1-\sigma} \bigg(\ \frac{\theta-1}{\theta}\bigg)^{(1-\sigma)/(\sigma +\eta)}(1-\alpha)^{(1-\sigma)(1+\eta)/(\theta-1)(\sigma +\eta)}\times \\ \frac{(1-\alpha (1+\pi)^{\theta})^{(1-\sigma)\eta/(\sigma +\eta)}}{(1-\alpha (1+\pi)^{\theta-1})^{(1-\sigma)(\eta \theta +1)/(\theta-1)(\sigma +\eta)}}\bigg(\frac{1-\alpha \beta (1+\pi)^{\theta}}{1-\alpha \beta (1+\pi)^{\theta-1}}\bigg)^{(1-\sigma)/(\sigma + \eta)}\times \\ A^{(1+\eta)(1-\sigma)/(\sigma +\eta)}-\frac{1}{1+\eta}\bigg( \frac{\theta-1}{\theta} \bigg)^{(1+\eta)/(\sigma +\eta)}(1-\alpha)^{(1-\sigma)(1+\eta)/(\theta-1)(\sigma +\eta)}\times \\\frac{(1-\alpha (1+\pi)^{\theta-1})^{(1+\eta)(\sigma \theta-1)/(\theta-1)(\sigma +\eta)}}{(1-\alpha (1+\pi)^{\theta-1})^{\sigma(1+\eta)/(\sigma +\eta)}}\bigg( \frac{1-\alpha \beta (1+\pi)^{\theta}}{1-\alpha \beta (1+\pi)^{\theta-1}}\bigg)^{(1+\eta)/(\sigma + \eta)} \times \\ A^{(1-\sigma)(1+\eta)/(\theta-1)(\sigma +\eta)}\bigg]
\end{multline}
Finally, present discounted welfare is 
\begin{equation}\mathcal{W}^{NSS}= \frac{1}{1-\beta}u^{NSS} \end{equation}
In line with the discussion in the text about profitability with trend inflation, 
firms will make different profits depending on when they last reset their price. Denoting the time since the last reset by the age of its price $a$
\begin{equation}
\Pi_{a}=\bigg((1+\pi)^{-a}-MC\bigg)(1+\pi)^{-a\theta}Y
\end{equation}
Indeed with positive trend inflation there will be some firms making negative profits. To understand this fix $\pi>0$ and consider the infinite limit $\lim_{a \rightarrow \infty} \Pi_{a} =-\infty$ and then continuity of profit with respect to $a$ implies a cutoff, such that all firms with a current price spell $a\geq \bar{a}$ will make negative profits.
The bracketed term is the steady state markup.
\subsubsection{Economic Growth and De-trending}
It is useful to examine 
the consequences of allowing for economic growth. To do so suppose there is exogenous technological progress at rate $g$ so 
\begin{equation}A_{t}=(1+g)A_{t-1}\end{equation}
To justify the practice of linear de-trending standard in empirical work, a balanced growth path is required. However, the following result demonstrates that this necessitates a specific parametric restriction. 
\begin{proposition}
There exists a balanced growth path if and only if $\sigma =1$.   
\end{proposition}
\begin{proof} There can be a balanced growth path if and only if we can solve for de-trended output $Y/A$ that does not depend on the growing variable $A$. Repeating the steps used to derive the steady state with no growth reveals that $$\frac{Y}{A}= kA^{(1-\sigma)/ (\sigma + \eta)}$$
where $k$ is a function of only the behavioural parameters that could be calculated directly from (261). It is clear independence only arises when the exponent is zero which corresponds to $\sigma =1$.
\end{proof}
Therefore, I select $\sigma =1$ for quantitative work here, since I am working with small noise limits and desire consistency with standard econometric practice. Although, this value is not ubiquitous in the literature, in the calibration section I demonstrate that it is consistent with a body of empirical evidence and prior work. In fact, Theorem 3 and all the subsequent analysis would go through in this case. Expected utility on the balanced growth path (BGP) would be augmented by a term reflecting the value of future growth.\footnote{The derivation is quite simple consider the non-stochastic case $$u^{BGP}= \ln(Y_{0}) + \beta \ln((1+g)Y_{0}) + \beta^{2}\ln((1+g)^{2}Y_{0}) + \cdots + $$
The growth effect is transmitted through 
$$\beta \ln(1+g)\bigg[ 1 + 2\beta + 3\beta^{2} + \cdots + \bigg]$$
It is clear that $\beta \rightarrow 1$ 
mandates stationary of de-trended output $y^{e}_{T}=Y_{T}/A_{0}(1+g)^{T}$
Noting that the bracketed term the derivative of $1/(1-x)$ yields the desired result $$u^{BGP}= \frac{\ln(Y_{0})}{(1-\beta)} + \beta \frac{\ln(1+g)}{(1-\beta^{2})}$$
This shows the versatility of the stochastic Grobman-Hartman in particular and Stochastic Equilibrium theory in general.} Adding capital would avoid this knife-edge condition, at the expense of complicating the rest of the analysis. 
\subsection{Profits and Long-Run Growth}
The first two subsubsections house the proofs of the two main propositions concerning long-run profits. The final subsubsection contains a final proof and fortifies empirical arguments concerning labour supply made in the profit section.
\subsubsection{Proof of Proposition 13} 
\begin{proof}The proof starts with the lower bound where there are three cases and then finishes with the upper bound. For the lower bound the order of argument is (ii), (iii) then (i). Aggregate profit can be written as 
\begin{equation} \Pi = (1-\alpha) \Pi_{0}+ \alpha(1-\alpha)\Pi_{1} + \alpha^{2}(1-\alpha)\Pi_{2} + \cdots +  \end{equation}
where the subscript indicates the time since the last reset. Cross-multiplying, taking expectations and using the non-stochastic limit, to cancel out terms in the preference shock $\psi$ from both sides, reveals that 
\begin{equation}\mathbb{E}u'(Y)\Pi = (1-\alpha)\mathbb{E}\sum_{T=t}^{\infty} (\alpha)^{T-t}u'(Y)\Pi_{T}\end{equation}
in the patient limit; with the weight coming from the terms in (25) and the context of the resetting firms problem \begin{equation} \mathbb{E}u'(Y)\Pi \rightarrow (1-\alpha) u'(Y)\mathbb{E} \sum_{T=t}^{\infty} w(AR_{T} -AC_{T}) \end{equation}
because the demand curve slopes down and marginal costs are constant, we can see that 
\begin{equation} \mathbb{E}u'(Y)\Pi > (1-\alpha) u'(Y) \mathbb{E} \sum_{T=t}^{\infty} w_{T}(MR_{T} -MC_{T})=0 \end{equation}
The claim based on Assumption 2 (ii) follows swiftly, as $\vert \varepsilon \vert \rightarrow 0$ then $\mathbb{E}u'(Y)\Pi \rightarrow u'(Y)\Pi$, which implies $\Pi =\mathbb{E} \Pi >0$.
\par For Assumption 2 (iii), the proof comes from an analysis of covariance and (strict) stochastic monotonicity properties. By application of Theorem 1.A.3 and 1.A.8 of \cite{shaked2007stochastic} to the cases of above and below average profit, we know that the covariance 
\begin{equation} \mathbb{E}(u'(Y)-\mathbb{E}u'(Y))(\Pi-\mathbb{E}\Pi)>0 \end{equation}
is strictly positive. Applying the definition of covariance. \begin{equation}\mathbb{E}u'(Y)\mathbb{E}\Pi  > \mathbb{E}u'(Y)\Pi >0 \end{equation}
hence $E\Pi >0$, as required. 
For the third part, modify the three equations at the start (268)-(270),
by dividing by $u'(Y_{t})$, then taking expectations from $T=t-1$ allows us to have profits discounted by the stochastic discount factor kernel on the left hand side. Since the inequalities and limits still going through, it follows that $\mathbb{E}M \Pi >0$ and then a basic covariance argument yields $\Pi >0$, as desired. 
\par Finally concentrating on the upper bound, recall the labor supply condition (19) and simplify the budget constraint to $C =WL+ \Pi$.
To analyze the move from a non-stochastic to a stochastic environment, I consider a modified Slutsky decomposition.
Workers will choose their labor supply, subject to the wage rate and the level of profits. In stochastic equilibrium we can view the world as if it is static problem, with standard labor-leisure preferences (see \cite{varian1992microeconomic}). 
\par Therefore there will be three channels. A substitution effect from (lower) wages onto labor supply and two income effects, one from the change in wages and the other from the change in profits. Theorem 2 tells us that labor supply rises and wages fall. The strategy is to show that the first two effects net out to reduce labor supply. This means that profits must decrease, to generate a wealth effect that increases labor supply. To demonstrate this point, it is sufficient to show that the labor supply curve is positively sloped. Totally differentiating with respect to $L$
\begin{equation} \frac{\mathrm{d}W}{\mathrm{d}L}u'(Y)+ 
u''(Y)\bigg[ W + L\frac{\mathrm{d}W}{\mathrm{d}
L}\bigg]=\nu''(L)
\end{equation}
Simplifying yields 
\begin{equation} \frac{\mathrm{d}W}{\mathrm{d}L}\bigg(1-\sigma \frac{L}{Y}\bigg)=\frac{\nu''}{u'} + \sigma \frac{W}{Y}\end{equation}
Since the right hand side is always positive, the sign of the derivative is determined by the bracket. Now substitute the production function in to create the determinant \begin{equation} 1- \sigma \frac{\Delta}{A} \end{equation}
Taking the limit $\mathbb{E}A \rightarrow \infty $ and then passing expectations ensures $\mathbb{E} \mathrm{d}W / \mathrm{d}L>0$, giving the desired bound on $\mathbb{E}\Pi$.
\end{proof}
\subsubsection{Proof of Proposition 14}
Let $\mathcal{V}^{*}_{t}(i)$ be the present value of a firm, which has reset its price today and $\mathbb{E}_{t}\mathcal{V}^{*}_{T}(i)$ denote the expectation of that firm's present value, given that it does not get to re-optimize until period $T$.
\begin{proof} The result comes from a chain of inequalities. From the principle of optimality, it is clear that the firm is weakly better off when it can change prices. Furthermore, since we know that profit function is strictly concave, the optimization problem at any time will have a unique solution. Therefore, we can upgrade to strict inequalities $$\mathcal{V}_{t}(i) > \mathbb{E}_{t}\mathcal{V}^{*}_{t+1}(i)> \mathbb{E}_{t}\mathcal{V}^{*}_{t+2} > \cdots $$
Consider the limit where $\beta \rightarrow 1$ and $T \rightarrow \infty$. From the first limit, we know that the reset price will always dominate any initial spell of price rigidity. The second ensures that $\mathcal{V}^{*}_{T}\rightarrow \mathbb{E}\mathcal{V}^{*}(i)$ hence 
\begin{equation} \mathcal{V}_{t}(i) > \mathbb{E} \max_{p_{t}^{*}(i)}\sum_{T=t}^{\infty}\alpha^{T-t}Q_{t, \, T}\bigg[\frac{p_{t}(i)}{P_{T}}y_{T}(i)-C (y_{T})(i)\bigg]  \end{equation}
where I am using (25) (the profit function). Since the demand curve slopes down and there are no increasing returns to scale, we know that 
\begin{multline} \mathcal{V}_{t}(i) > \mathbb{E} \max_{p_{t}^{*}(i)}\sum_{T=t}^{\infty}\alpha^{T-t}Q_{t, \, T}\bigg[\frac{p_{t}(i)}{P_{T}}y_{T}(i)-C (y_{T})(i)\bigg] > \\ \mathbb{E}_{t}\sum_{T=t}^{\infty}w_{t, \, T}\bigg[MR_{T}(y_{T}(i))-MC_{T}(y_{T}(i))\bigg]=0
\end{multline}
\end{proof}
\subsubsection{Additional Arguments}
The evidence on long-run labor supply elasticity is mixed. The consensus estimate appears to be zero associated with \cite{doran2014long} and \cite{berger2018drivers}, in which case the main result goes through. \cite{ashenfelter2010shred} argues that the likely range is between -0.2 and 0.2 with their study at the lower range. The concern with unemployment and underemployment, in \cite{mas2019labor}, favours a positive figure. There is more discussion in the parametization Appendix I.
\begin{proposition} The normalized share of profits $\mathbb{E} \Pi/ A$ is strictly lower under stochastic balanced growth than non-stochastic balanced growth, under the maintained assumptions from Proposition 13.
\end{proposition}
\begin{proof}
With balanced growth it has been established that $\sigma =1$, whilst the labor supply and profit conditions take the form 
\begin{equation} \frac{W}{A}\bigg(\frac{C}{A}\bigg)^{-1}=\nu'(L) \end{equation}
\begin{equation} \frac{C}{A} = \frac{W}{A}L + \frac{\Pi}{A}\end{equation}
Via the substitution $\tilde{W}=W/A$, 
$\tilde{C}=C/A=Y/A=\tilde{Y}$
and the arguments in Proposition 13,
(275) and (276) in particular, it suffices to prove that 
\begin{equation} \frac{\mathrm{d}\tilde{W}}{\mathrm{d}L}\bigg(1-\sigma \frac{\Delta}{A}\bigg)=\nu''(\tilde{C})
+ \sigma \frac{\tilde{W}}{\tilde{Y}}
\end{equation}
and the proof will be complete if we can argue that the left hand bracket is strictly positive with probability one. To do so suppose the converse, then it follows from the definition of balanced growth in stochastic equilibrium that
\begin{equation}\mathbb{P}(\Delta > A_{0}\mathbb{E}(1+g)^{T})= \epsilon>0)\end{equation}
where $\epsilon$ does not depend on $T$, taking the limit as $T \rightarrow \infty$ means that $\mathbb{E}\Delta \rightarrow \infty$, which is a contradiction of the Inada condition.
\end{proof}
\section{Slopes and Eigenvalues}
This section contains details involving eigenvalues and slope coefficients, omitted from the main text. The first subsection gives the general expressions for the slope coefficients. The second finishes off the derivation of the general linearized solution of a DSGE.
The third constructs the characteristic equations. The final part covers the general existence conditions implied by Rouche's theorem.
\begin{subsection}{Slope Coefficients}
This subsection begins by gathering together the expression for the non-linear Phillips curve coefficients.
The expressions are grouped into three subsubsections dealing in turn with the Phillips curve, the Euler and the price dispersion relation. The fourth covers the Phillips curve in the general small noise limit case. The final part solves for the singular surfaces around ZINSS. 
\subsubsection{Phillips Curve (116)}
\begin{multline}
b= \alpha \beta \mathbb{E}(1+\pi)^{\theta-1}(2+\pi) \\   + \frac{1-\alpha (1+\pi)^{\theta-1}}{\alpha (1+\pi)^{\theta-2}}\bigg/ \bigg(\nu'(\Delta Y/A)Y/A + \alpha \beta \frac{\mathbb{E}(1+\pi)^{\theta}\nu'(\Delta Y/A)Y/A}{1-\alpha \beta \mathbb{E}(1+\pi)^{\theta}} \bigg) \\ \bigg[ \alpha \beta \bigg\{ \bigg( \theta \frac{\mathbb{E}(1+\pi)^{\theta-1}\nu'(\Delta Y/A)Y/A}{1-\alpha \beta \mathbb{E}(1+\pi)^{\theta}} \\ - \beta \frac{(\theta-1)^{2}}{\theta} \frac{(1-\alpha)^{1/(\theta-1)}}{(1-\alpha(1+\pi)^{\theta-1})^{1/(\theta-1)}}\frac{\mathbb{E}(1+\pi)^{\theta-2}\psi u'(Y)Y}{1-\alpha \beta \mathbb{E}(1+\pi)^{\theta-1}} \bigg) \\ + \bigg((1+\eta)\nu'(\Delta Y/A)\frac{Y}{A}\mathbb{E}(1+\pi)^{\theta-1} \\ -(1-\sigma)\psi u'(Y) \frac{\theta-1}{\theta} \frac{(1-\alpha)^{1/(\theta-1)}}{(1-\alpha (1+\pi)^{\theta-1})^{1/(\theta-1)}}\mathbb{E}(1+\pi)^{\theta} \bigg)\\ \times 
\frac{1}{\alpha}
\frac{\mathbb{E}\psi u'(Y)/(1+\pi)^{2}}{\mathbb{E}\psi u'(Y)/(1+\pi)} \bigg/ \bigg(\sigma + \frac{a_{y} \beta} {\psi u'(Y)} \mathbb{E}\psi u'(Y)/(1+\pi)\bigg)   \bigg\} \\ - \eta \theta \nu'(\Delta Y/A)Y/A \bigg(\bigg[ \mathbb{E}(1-\alpha (1+\pi)^{\theta-1})^{\theta/(\theta-1)}\mathbb{E}(1+\pi)^{\theta-1}\\ -\mathbb{E}(1+\pi)^{\theta-2}(1-\alpha (1+\pi)^{\theta-1})^{1/(\theta-1)}(1-\alpha \mathbb{E}(1+\pi)^{\theta})  \bigg] \\ \bigg/ \mathbb{E}(1-\alpha (1+\pi)^{\theta-1})^{\theta/(\theta-1)}\mathbb{E}(1+\pi)^{\theta}(1-\alpha \mathbb{E}(1+\pi)^{\theta}) \bigg)\bigg] 
\end{multline}
\begin{multline}
\tilde{b}_{0}=1+ \frac{1-\alpha (1+\pi)^{\theta-1}}{(1+\pi)^{\theta-2}}\bigg/ \\ \bigg(\nu'(\Delta Y/A)Y/A + \alpha \beta \frac{\mathbb{E}(1+\pi)^{\theta}\nu'(\Delta Y/A)Y/A}{1-\alpha \beta \mathbb{E}(1+\pi)^{\theta}} \bigg)\\ \times \bigg\{\frac{1}{\alpha}
\frac{a_{\pi} \beta} {\psi u'(Y)} \mathbb{E}\psi u'(Y)/(1+\pi) \bigg/ \bigg(\sigma + \frac{a_{y} \beta} {\psi u'(Y)} \mathbb{E}\psi u'(Y)/(1+\pi)\bigg)  \\ \times \bigg((1+\eta)\nu'(\Delta Y/A)\frac{Y}{A}\mathbb{E}(1+\pi)^{\theta-1} \\ -(1-\sigma)\psi u'(Y) \frac{\theta-1}{\theta} \frac{(1-\alpha)^{1/(\theta-1)}}{(1-\alpha (1+\pi)^{\theta-1})^{1/(\theta-1)}}\mathbb{E}(1+\pi)^{\theta}\bigg)\bigg\} 
\end{multline}
\begin{multline}
\tilde{b}_{1}=\frac{1-\alpha (1+\pi)^{\theta-1}}{(1+\pi)^{\theta-2}}\bigg/ \bigg(\nu'\bigg( \frac{\Delta Y}{A}\bigg)\frac{Y}{A} + \alpha \beta \frac{\mathbb{E}(1+\pi)^{\theta}\nu'(\Delta Y/A)Y/A}{1-\alpha \beta \mathbb{E}(1+\pi)^{\theta}} \bigg)\\ \times \bigg\{ \beta \bigg[(1+\eta)\nu'(\Delta Y/A)\frac{Y}{A}\mathbb{E}(1+\pi)^{\theta-1}\\ -(1-\sigma)\psi u'(Y) \frac{\theta-1}{\theta} \frac{(1-\alpha)^{1/(\theta-1)}}{(1-\alpha (1+\pi)^{\theta-1})^{1/(\theta-1)}}\mathbb{E}(1+\pi)^{\theta}\bigg]\\-
\frac{1}{\alpha}
\frac{u'(Y)}{u''(Y)}\frac{\mathbb{E}\psi u''(Y)/(1+\pi)}{\mathbb{E}\psi u'(Y)/(1+\pi)}\sigma \bigg/ \bigg(\sigma + \frac{a_{y} \beta} {\psi u'(Y)} \mathbb{E}\psi u'(Y)/(1+\pi)\bigg)\\ \times \bigg( (1+\eta)\nu'\bigg(\frac{\Delta Y}{A}\bigg)\frac{Y}{A} - \\(1-\sigma)\psi u'(Y)\frac{(\theta-1)}{\theta} \frac{(1-\alpha)^{1/(\theta-1)}}{(1-\alpha (1+\pi)^{\theta-1})^{1/(\theta-1)}}\bigg)\bigg\}
\end{multline}
\begin{multline}
\tilde{b}_{2}=\frac{\eta(1-\alpha (1+\pi)^{\theta-1})}{\alpha^{2}(1+\pi)^{\theta-2}}\bigg(\alpha^{2} \beta \mathbb{E}(1+\pi)^{\theta-1}\nu'\bigg(\frac{\Delta Y}{A}\bigg)\frac{Y}{A}-\frac{\nu'(\Delta Y/A)Y/A}{\mathbb{E}(1+\pi)^{\theta}}\bigg) \\ \bigg/ \bigg(\nu'(\Delta Y/A)Y/A + \alpha \beta \frac{\mathbb{E}(1+\pi)^{\theta}\nu'(\Delta Y/A)Y/A}{1-\alpha \beta \mathbb{E}(1+\pi)^{\theta}} \bigg) 
\end{multline}
\begin{multline}
\tilde{b}_{3}=\alpha \beta^{2}\bigg(\alpha \mathbb{E}(1+\pi)^{\theta-1}\mathbb{E}(1+\pi)^{\theta} + \frac{(1-\alpha (1+\pi)^{\theta-1})}{(1+\pi)^{\theta-2}} \times \\  \bigg[\frac{\theta \mathbb{E}(1+\pi)^{\theta-1}\mathbb{E}(1+\pi)^{\theta-1}\nu'(\Delta Y/A)Y/A(1-\alpha \beta \mathbb{E}(1+\pi)^{\theta-1})}{\nu'(\Delta Y/A)Y/A + \alpha \beta \mathbb{E}(1+\pi)^{\theta}\nu'(\Delta Y/A)Y/A(1-\alpha \beta \mathbb{E}(1+\pi)^{\theta})}\\ -(\theta-1)
\frac{\mathbb{E}(1+\pi)^{\theta}\mathbb{E}(1+\pi)^{\theta-2}\psi u'(Y)Y/(1-\alpha \beta \mathbb{E}(1+\pi)^{\theta-1})}{\psi u'(Y)Y + \alpha \beta \mathbb{E}(1+\pi)^{\theta-1}\psi u'(Y)Y/(1-\alpha  \beta \mathbb{E}(1+\pi)^{\theta-1})}\bigg]\bigg)
\end{multline}
\subsubsection{Euler Coefficients (117)}
\begin{multline}
c= \beta^{2}\bigg(\alpha \mathbb{E}(1+\pi)^{\theta-1}\mathbb{E}(1+\pi)^{\theta} + \frac{(1-\alpha (1+\pi)^{\theta-1})}{(1+\pi)^{\theta-2}} \times \\  \bigg[\frac{\theta \mathbb{E}(1+\pi)^{\theta-1}\mathbb{E}(1+\pi)^{\theta-1}\nu'(\Delta Y/A)Y/A(1-\alpha \beta \mathbb{E}(1+\pi)^{\theta})}{\nu'(\Delta Y/A)Y/A + \alpha \beta \mathbb{E}(1+\pi)^{\theta}\nu'(\Delta Y/A)Y/A(1-\alpha \beta \mathbb{E}(1+\pi)^{\theta})}\\ -(\theta-1)
\frac{\mathbb{E}(1+\pi)^{\theta}\mathbb{E}(1+\pi)^{\theta-2}\psi u'(Y)Y/(1-\alpha \beta \mathbb{E}(1+\pi)^{\theta-1})}{\psi u'(Y)Y + \alpha \beta \mathbb{E}(1+\pi)^{\theta-1}\psi u'(Y)Y/(1-\alpha  \beta \mathbb{E}(1+\pi)^{\theta-1})}\bigg]\bigg)\\ + \frac{1-\alpha (1+\pi)^{\theta-1}}{(1+\pi)^{\theta-2}}\bigg/ \bigg(\nu'(\Delta Y/A)Y/A + \alpha \beta \frac{\mathbb{E}(1+\pi)^{\theta}\nu'(\Delta Y/A)Y/A}{1-\alpha \beta \mathbb{E}(1+\pi)^{\theta}} \bigg)\\ \times \bigg\{ \beta \bigg[(1+\eta)\nu'(\Delta Y/A)\frac{Y}{A}\mathbb{E}(1+\pi)^{\theta-1}\\ -(1-\sigma)\psi u'(Y) \frac{\theta-1}{\theta} \frac{(1-\alpha)^{1/(\theta-1)}}{(1-\alpha (1+\pi)^{\theta-1})^{1/(\theta-1)}}\mathbb{E}(1+\pi)^{\theta}\bigg]\\-
\frac{1}{\alpha}
\frac{u'(Y)}{u''(Y)}\frac{\mathbb{E}\psi u''(Y)/(1+\pi)}{\mathbb{E}\psi u'(Y)/(1+\pi)}\sigma \bigg/ \bigg(\sigma + \frac{a_{y} \beta} {\psi u'(Y)} \mathbb{E}\psi u'(Y)/(1+\pi)\bigg)\\ \times \bigg((1+\eta)\nu'(\Delta Y/A)\frac{Y}{A} -(1-\sigma)\psi u'(Y)\frac{(\theta-1)}{\theta} \frac{(1-\alpha)^{1/(\theta-1)}}{(1-\alpha (1+\pi)^{\theta-1})^{1/(\theta-1)}}\bigg)\bigg\} \\ \times 
\frac{1}{\alpha} 
\frac{\mathbb{E}\psi u'(Y)/(1+\pi)^{2}}{\mathbb{E}\psi u'(Y)/(1+\pi)} \bigg/ \bigg(\sigma + \frac{a_{y} \beta} {\psi u'(Y)}\end{multline}
\begin{multline}
\tilde{c}_{0}=-\bigg\{1+ \frac{1-\alpha (1+\pi)^{\theta-1}}{(1+\pi)^{\theta-2}}\bigg/ \\ \bigg(\nu'(\Delta Y/A)\frac{Y}{A} + \alpha \beta \frac{\mathbb{E}(1+\pi)^{\theta}\nu'(\Delta Y/A)Y/A}{1-\alpha \beta \mathbb{E}(1+\pi)^{\theta}} \bigg)\\ \times \bigg[\frac{1}{\alpha}
\frac{a_{\pi} \beta} {\psi u'(Y)} \mathbb{E}\psi u'(Y)/(1+\pi) \bigg/ \bigg(\sigma + \frac{a_{y} \beta} {\psi u'(Y)} \mathbb{E}\psi u'(Y)/(1+\pi)\bigg)  \\ \times \bigg((1+\eta)\nu'(\Delta Y/A)\frac{Y}{A}\mathbb{E}(1+\pi)^{\theta-1} \\ -(1-\sigma)\psi u'(Y) \frac{\theta-1}{\theta} \frac{(1-\alpha)^{1/(\theta-1)}}{(1-\alpha (1+\pi)^{\theta-1})^{1/(\theta-1)}}\mathbb{E}(1+\pi)^{\theta}\bigg)\bigg]\bigg\}  \\ 
\times \frac{1}{\alpha}
\frac{\mathbb{E}\psi u'(Y)/(1+\pi)^{2}}{\mathbb{E}\psi u'(Y)/(1+\pi)} \bigg/ \bigg(\sigma + \frac{a_{y} \beta} {\psi u'(Y)} \mathbb{E}\psi u'(Y)/(1+\pi)\bigg)
\end{multline}
\begin{multline}
\tilde{c}_{1}=-
\beta^{2}\bigg(\alpha \mathbb{E}(1+\pi)^{\theta-1}\mathbb{E}(1+\pi)^{\theta} + \frac{(1-\alpha (1+\pi)^{\theta-1})}{(1+\pi)^{\theta-2}} \times \\  \bigg[\frac{\theta \mathbb{E}(1+\pi)^{\theta-1}\mathbb{E}(1+\pi)^{\theta-1}\nu'(\Delta Y/A)Y/A(1-\alpha \beta \mathbb{E}(1+\pi)
^{\theta})}{\nu'(\Delta Y/A)Y/A + \alpha \beta \mathbb{E}(1+\pi)^{\theta}\nu'(\Delta Y/A)Y/A(1-\alpha \beta \mathbb{E}(1+\pi)^{\theta})}\\ -(\theta-1)
\frac{\mathbb{E}(1+\pi)^{\theta}\mathbb{E}(1+\pi)^{\theta-2}\psi u'(Y)Y/(1-\alpha \beta \mathbb{E}(1+\pi)^{\theta-1})}{\psi u'(Y)Y + \alpha \beta \mathbb{E}(1+\pi)^{\theta-1}\psi u'(Y)Y/(1-\alpha  \beta \mathbb{E}(1+\pi)^{\theta-1})}\bigg]\bigg)\\ \times 
\frac{a_{\pi}\beta}{\psi u'(Y)} \mathbb{E}\psi u'(Y)/(1+\pi) \bigg/ \bigg(\sigma + \frac{a_{y} \beta} {\psi u'(Y)} \mathbb{E}\psi u'(Y)/(1+\pi)\bigg)\\ + \frac{1}{\alpha}
\frac{\mathbb{E}\psi u'(Y)/(1+\pi)^{2}}{\mathbb{E}\psi u'(Y)/(1+\pi)} \bigg/ \bigg(\sigma + \frac{a_{y} \beta} {\psi u'(Y)} \mathbb{E}\psi u'(Y)/(1+\pi)\bigg)\\ 
\Bigg\{ \alpha \beta \mathbb{E}(1+\pi)^{\theta-1}(2+\pi) \\   + \frac{1-\alpha (1+\pi)^{\theta-1}}{\alpha (1+\pi)^{\theta-2}}\bigg/ \bigg(\nu'(\Delta Y/A)Y/A + \alpha \beta \frac{\mathbb{E}(1+\pi)^{\theta}\nu'(\Delta Y/A)Y/A}{1-\alpha \beta \mathbb{E}(1+\pi)^{\theta}} \bigg) \\ \bigg[ \alpha \beta \bigg\{ \bigg( \theta \frac{\mathbb{E}(1+\pi)^{\theta-1}\nu'(\Delta Y/A)Y/A}{1-\alpha \beta \mathbb{E}(1+\pi)^{\theta}} \\ - \beta \frac{(\theta-1)^{2}}{\theta} \frac{(1-\alpha)^{1/(\theta-1)}}{(1-\alpha(1+\pi)^{\theta-1})^{1/(\theta-1)}}\frac{\mathbb{E}(1+\pi)^{\theta-2}\psi u'(Y)Y}{1-\alpha \beta \mathbb{E}(1+\pi)^{\theta-1}} \bigg) \\ + \bigg((1+\eta)\nu'(\Delta Y/A)\frac{Y}{A}\mathbb{E}(1+\pi)^{\theta-1} \\ -(1-\sigma)\psi u'(Y) \frac{\theta-1}{\theta} \frac{(1-\alpha)^{1/(\theta-1)}}{(1-\alpha (1+\pi)^{\theta-1})^{1/(\theta-1)}}\mathbb{E}(1+\pi)^{\theta} \bigg)\\ \times \frac{1}{\alpha} 
\frac{\mathbb{E}\psi u'(Y)/(1+\pi)^{2}}{\mathbb{E}\psi u'(Y)/(1+\pi)} \bigg/ \bigg(\sigma + \frac{a_{y} \beta} {\psi u'(Y)} \mathbb{E}\psi u'(Y)/(1+\pi)\bigg)   \bigg\} \\ - \eta \theta \nu'(\Delta Y/A)Y/A \bigg(\bigg[ \mathbb{E}(1-\alpha (1+\pi)^{\theta-1})^{\theta/(\theta-1)}\mathbb{E}(1+\pi)^{\theta-1}\\ -\mathbb{E}(1+\pi)^{\theta-2}(1-\alpha (1+\pi)^{\theta-1})^{1/(\theta-1)}(1-\alpha \mathbb{E}(1+\pi)^{\theta})  \bigg] \\ / \mathbb{E}(1-\alpha (1+\pi)^{\theta-1})^{\theta/(\theta-1)}\mathbb{E}(1+\pi)^{\theta}(1-\alpha \mathbb{E}(1+\pi)^{\theta}) \bigg)\bigg]\Bigg\} 
\end{multline}
\begin{multline}
\tilde{c}_{2}=\frac{\eta(1-\alpha (1+\pi)^{\theta-1})}{\alpha^{2}(1+\pi)^{\theta-2}}\bigg(\frac{\nu'(\Delta Y/A)Y/A}{\mathbb{E}(1+\pi)^{\theta}}-\alpha^{2} \beta \mathbb{E}(1+\pi)^{\theta-1}\nu'\bigg( \frac{\Delta Y}{A}\bigg)\frac{Y}{A} \bigg) \\ \bigg/ \bigg(\nu'(\Delta Y/A)Y/A + \alpha \beta \frac{\mathbb{E}(1+\pi)^{\theta}\nu'(\Delta Y/A)Y/A}{1-\alpha \beta \mathbb{E}(1+\pi)^{\theta}} \bigg) \\ \times \frac{1}{\alpha} 
\frac{\mathbb{E}\psi u'(Y)/(1+\pi)^{2}}{\mathbb{E}\psi u'(Y)/(1+\pi)} \bigg/ \bigg(\sigma + \frac{a_{y} \beta} {\psi u'(Y)} \mathbb{E}\psi u'(Y)/(1+\pi)\bigg)
\end{multline}
\begin{multline}
\tilde{c}_{3}=\frac{1}{\alpha}
\frac{\mathbb{E}\psi u'(Y)/(1+\pi)^{2}}{\mathbb{E}\psi u'(Y)/(1+\pi)} \bigg/ \bigg(\sigma + \frac{a_{y} \beta} {\psi u'(Y)} \mathbb{E}\psi u'(Y)/(1+\pi)\bigg)\times \\  \bigg[\frac{\theta \mathbb{E}(1+\pi)^{\theta-1}\mathbb{E}(1+\pi)^{\theta-1}\nu'(\Delta Y/A)Y/A(1-\alpha \beta \mathbb{E}(1+\pi)^{\theta-1})}{\nu'(\Delta Y/A)Y/A + \alpha \beta \mathbb{E}(1+\pi)^{\theta}\nu'(\Delta Y/A)Y/A(1-\alpha \beta \mathbb{E}(1+\pi)^{\theta})}-\\ (\theta-1)
\frac{\mathbb{E}(1+\pi)^{\theta}\mathbb{E}(1+\pi)^{\theta-2}\psi u'(Y)Y/(1-\alpha \beta \mathbb{E}(1+\pi)^{\theta-1})}{\psi u'(Y)Y + \alpha \beta \mathbb{E}(1+\pi)^{\theta-1}\psi u'(Y)Y/(1-\alpha  \beta \mathbb{E}(1+\pi)^{\theta-1})}\bigg] \\ \times \alpha \beta^{2}\bigg(\alpha \mathbb{E}(1+\pi)^{\theta-1}\mathbb{E}(1+\pi)^{\theta} + \frac{(1-\alpha (1+\pi)^{\theta-1})}{(1+\pi)^{\theta-2}}\bigg)
\end{multline}
\subsubsection{Price Dispersion} 
\begin{multline}
d=\alpha \beta^{2}\bigg(\alpha \mathbb{E}(1+\pi)^{\theta-1}\mathbb{E}(1+\pi)^{\theta} + \frac{(1-\alpha (1+\pi)^{\theta-1})}{(1+\pi)^{\theta-2}} \times \\  \bigg[\frac{\theta \mathbb{E}(1+\pi)^{\theta-1}\mathbb{E}(1+\pi)^{\theta-1}\nu'(\Delta Y/A)Y/A(1-\alpha \beta \mathbb{E}(1+\pi)^{\theta-1})}{\nu'(\Delta Y/A)Y/A + \alpha \beta \mathbb{E}(1+\pi)^{\theta}\nu'(\Delta Y/A)Y/A(1-\alpha \beta \mathbb{E}(1+\pi)^{\theta})}\\ -(\theta-1)
\frac{\mathbb{E}(1+\pi)^{\theta}\mathbb{E}(1+\pi)^{\theta-2}\psi u'(Y)Y/(1-\alpha \beta \mathbb{E}(1+\pi)^{\theta-1})}{\psi u'(Y)Y + \alpha \beta \mathbb{E}(1+\pi)^{\theta-1}\psi u'(Y)Y/(1-\alpha  \beta \mathbb{E}(1+\pi)^{\theta-1})}\bigg]\bigg)\\
+\theta \bigg(\bigg[ \mathbb{E}(1-\alpha (1+\pi)^{\theta-1})^{\theta/(\theta-1)}\mathbb{E}(1+\pi)^{\theta-1}\\ -\mathbb{E}(1+\pi)^{\theta-2}(1-\alpha (1+\pi)^{\theta-1})^{1/(\theta-1)}(1-\alpha \mathbb{E}(1+\pi)^{\theta})  \bigg] \\ \bigg/ \mathbb{E}(1-\alpha (1+\pi)^{\theta-1})^{\theta/(\theta-1)}\mathbb{E}(1+\pi)^{\theta}(1-\alpha \mathbb{E}(1+\pi)^{\theta}) \bigg) \\ \times \frac{\eta(1-\alpha (1+\pi)^{\theta-1})}{\alpha(1+\pi)^{\theta-2}}\bigg(
\nu'(\Delta Y/A)Y/A 
-\beta \frac{\mathbb{E}(1+\pi)^{\theta-1}\nu'(\Delta Y/A)/A}{\mathbb{E}(1+\pi)^{\theta}}\bigg) \\ \bigg/ \bigg(\nu'(\Delta Y/A)Y/A + \alpha \beta \frac{\mathbb{E}(1+\pi)^{\theta}\nu'(\Delta Y/A)Y/A}{1-\alpha \beta \mathbb{E}(1+\pi)^{\theta}} \bigg)  
\end{multline}
\begin{multline}
\tilde{d}_{0}=\bigg\{1+\frac{1-\alpha (1+\pi)^{\theta-1}}{(1+\pi)^{\theta-2}}\bigg/ \\ \bigg(\nu'(\Delta Y/A)\frac{Y}{A} + \alpha \beta \frac{\mathbb{E}(1+\pi)^{\theta}\nu'(\Delta Y/A)Y/A}{1-\alpha \beta \mathbb{E}(1+\pi)^{\theta}} \bigg)\\ \times \bigg[ 
\frac{1}{\alpha} \frac{a_{\pi} \beta} {\psi u'(Y)} \mathbb{E}\psi u'(Y)/(1+\pi) \bigg/ \bigg(\sigma + \frac{a_{y} \beta} {\psi u'(Y)} \mathbb{E}\psi u'(Y)/(1+\pi)\bigg)  \\ \times \bigg((1+\eta)\nu'(\Delta Y/A)\frac{Y}{A}\mathbb{E}(1+\pi)^{\theta-1} \\ -(1-\sigma)\psi u'(Y) \frac{\theta-1}{\theta} \frac{(1-\alpha)^{1/(\theta-1)}}{(1-\alpha (1+\pi)^{\theta-1})^{1/(\theta-1)}}\mathbb{E}(1+\pi)^{\theta}\bigg)\bigg] \bigg\} \\ \times \theta \bigg(\bigg[ \mathbb{E}(1-\alpha (1+\pi)^{\theta-1})^{\theta/(\theta-1)}\mathbb{E}(1+\pi)^{\theta-1}\\ -\mathbb{E}(1+\pi)^{\theta-2}(1-\alpha (1+\pi)^{\theta-1})^{1/(\theta-1)}(1-\alpha \mathbb{E}(1+\pi)^{\theta})  \bigg] \\ \bigg/ \mathbb{E}(1-\alpha (1+\pi)^{\theta-1})^{\theta/(\theta-1)}\mathbb{E}(1+\pi)^{\theta}(1-\alpha \mathbb{E}(1+\pi)^{\theta}) \bigg)
\end{multline}
\begin{multline}\tilde{d}_{1}=-\theta \bigg(\bigg[ \mathbb{E}(1-\alpha (1+\pi)^{\theta-1})^{\theta/(\theta-1)}\mathbb{E}(1+\pi)^{\theta-1}\\ -\mathbb{E}(1+\pi)^{\theta-2}(1-\alpha (1+\pi)^{\theta-1})^{1/(\theta-1)}(1-\alpha \mathbb{E}(1+\pi)^{\theta})  \bigg] \\ \bigg/ \mathbb{E}(1-\alpha (1+\pi)^{\theta-1})^{\theta/(\theta-1)}\mathbb{E}(1+\pi)^{\theta}(1-\alpha \mathbb{E}(1+\pi)^{\theta}) \bigg)
\end{multline}
\begin{multline}
\tilde{d}_{2}= \frac{1-\alpha (1+\pi)^{\theta-1}}{(1+\pi)^{\theta-2}}\bigg/  \bigg(\nu'(\Delta Y/A)Y/A + \alpha \beta \frac{\mathbb{E}(1+\pi)^{\theta}\nu'(\Delta Y/A)Y/A}{1-\alpha \beta \mathbb{E}(1+\pi)^{\theta}} \bigg)\\ 
 \times \bigg\{ \beta \bigg[(1+\eta)\nu'(\Delta Y/A)\frac{Y}{A}\mathbb{E}(1+\pi)^{\theta-1}\\ -(1-\sigma)\psi u'(Y) \frac{\theta-1}{\theta} \frac{(1-\alpha)^{1/(\theta-1)}}{(1-\alpha (1+\pi)^{\theta-1})^{1/(\theta-1)}}\mathbb{E}(1+\pi)^{\theta}\bigg]\\- 
 \frac{1}{\alpha} \frac{u'(Y)}{u''(Y)}\frac{\mathbb{E}\psi u''(Y)/(1+\pi)}{\mathbb{E}\psi u'(Y)/(1+\pi)}\sigma \bigg/ \bigg(\sigma + \frac{a_{y} \beta} {\psi u'(Y)} \mathbb{E}\psi u'(Y)/(1+\pi)\bigg)\times \\ \bigg( (1+\eta)\nu'(\Delta Y/A)\frac{Y}{A} -(1-\sigma)\psi u'(Y)\frac{(\theta-1)}{\theta} \frac{(1-\alpha)^{1/(\theta-1)}}{(1-\alpha (1+\pi)^{\theta-1})^{1/(\theta-1)}}\bigg)\bigg\} 
 \times \\ \theta \bigg(\bigg[ \mathbb{E}(1-\alpha (1+\pi)^{\theta-1})^{\theta/(\theta-1)}\mathbb{E}(1+\pi)^{\theta-1}\\ -\mathbb{E}(1+\pi)^{\theta-2}(1-\alpha (1+\pi)^{\theta-1})^{1/(\theta-1)}(1-\alpha \mathbb{E}(1+\pi)^{\theta})  \bigg] \\ \bigg/ \mathbb{E}(1-\alpha (1+\pi)^{\theta-1})^{\theta/(\theta-1)}\mathbb{E}(1+\pi)^{\theta}(1-\alpha \mathbb{E}(1+\pi)^{\theta}) \bigg) 
\end{multline}
\begin{multline}
\tilde{d}_{3}=\frac{\beta^{2}}{\mathbb{E}(1+\pi)^{\theta}}\bigg(\alpha \mathbb{E}(1+\pi)^{\theta-1}\mathbb{E}(1+\pi)^{\theta} + \frac{(1-\alpha (1+\pi)^{\theta-1})}{(1+\pi)^{\theta-2}} \times \\  \bigg[\frac{\theta \mathbb{E}(1+\pi)^{\theta-1}\mathbb{E}(1+\pi)^{\theta-1}\nu'(\Delta Y/A)Y/A(1-\alpha \beta \mathbb{E}(1+\pi)^{\theta-1})}{\nu'(\Delta Y/A)Y/A + \alpha \beta \mathbb{E}(1+\pi)^{\theta}\nu'(\Delta Y/A)Y/A(1-\alpha \beta \mathbb{E}(1+\pi)^{\theta})}\\ -(\theta-1)
\frac{\mathbb{E}(1+\pi)^{\theta}\mathbb{E}(1+\pi)^{\theta-2}\psi u'(Y)Y/(1-\alpha \beta \mathbb{E}(1+\pi)^{\theta-1})}{\psi u'(Y)Y + \alpha \beta \mathbb{E}(1+\pi)^{\theta-1}\psi u'(Y)Y/(1-\alpha  \beta \mathbb{E}(1+\pi)^{\theta-1})}\bigg]\bigg)
\end{multline}
\begin{subsubsection}{Trend Inflation Phillips Curve} 
This final part lays out the general Phillips curve in the small noise limit with arbitrary time preference and trend inflation. It compliments discussions of trend inflation and allows me to complete the proofs of Theorem 8 and Proposition 23. The justification of Footnote 92 is immediate.
\begin{multline} b = \beta \bigg(\alpha(1+\pi)^{\theta-1}(2+\pi) + (1-\alpha)^{1/(\theta-1)}\frac{(1-\alpha(1+\pi)^{\theta-1})^{(\theta-2)/(\theta-1)}}{(1+\pi)^{\theta-2}} + \\ \frac{(1-\alpha(1+\pi)^{\theta-1})}{\alpha(1+\pi)^{\theta-2}}(1-\alpha \beta (1+\pi)^{\theta})\frac{(\sigma + \eta)}{\sigma + \beta a_{y}}\bigg)- 
\frac{\eta \theta \pi (1-\alpha)^{1/(\theta-1)}} {(1-\alpha(1+\pi)^{\theta-1})^{2/(\theta-1)}}
\end{multline}
\begin{equation} \tilde{b}_{0}= 1 + a_{\pi}\frac{(1-\alpha(1+\pi)^{\theta-1})}{\alpha
(1+\pi)^{\theta-2}}(1-\alpha \beta (1+\pi)^{\theta})\frac{(\sigma + \eta)}{\sigma + \beta a_{y}}\end{equation}
\begin{multline} \tilde{b}_{1}=\frac{(1-\alpha(1+\pi)^{\theta-1})}{(1+\pi)^{\theta-2}}(1-\alpha \beta (1+\pi)^{\theta})(\sigma + \eta)\bigg[\beta(1-\alpha\pi(1+\pi)^{\theta-1})- \\ \frac{1}{\alpha(\sigma + \beta a_{y})}\bigg]\end{multline}
\begin{equation} \tilde{b}_{2}= -\eta(1+\pi) \frac{(1-\alpha(1+\pi)^{\theta-1})}{\alpha^{2}}
(1-\alpha \beta (1+\pi)^{\theta})(\beta-\alpha^{2}(1+\pi)^{\theta-1})\end{equation}
\begin{equation} \tilde{b}_{3}=\alpha \beta^{2}\bigg( \alpha (1+\pi)^{2 \theta-1} + \frac{(1-\alpha(1+\pi)^{\theta-1})^{\theta/(\theta-1)}}{(1-\alpha)^{1/(\theta-1)}(1+\pi)^{\theta-2}}\bigg)\end{equation}
\begin{remark} Symmetry arises in the error terms under trend inflation as under ZINSS. This reflects the inefficiency of aggregate fluctuations discovered in Theorem 6, which holds at any rate of trend inflation. However, it cannot be solved for without finding an exact form for the entire model which is beyond the perimeter of this study.\end{remark}
\begin{remark}Symmetry breaks down in non-degenerate stochastic steady state, where non-vanishing uncertainty creates additional inter-temporal distortions, hence, the more complicated expressions in Section 6.2. \end{remark}
\end{subsubsection}
\begin{subsubsection}{Singular Surfaces}
This part completes the solution of the boundary surface. It justifies (5) and  extends it to the $\vert {\varepsilon} \vert $ limit, whilst generalizing it to the whole parameter space. Parametize the surface as follows 
\begin{equation} \pi_{t}= g_{0}\pi_{t-1} +g_{1}\hat{y}^{e}_{t} + g_{2} \hat{\Delta}_{t}\end{equation}
The bifurcated Phillips curve takes the form 
\begin{equation}\pi_{t}=\tilde{\omega} \hat{y}_{t}^{e} + \eta \kappa \hat{\Delta}_{t} + \beta \mathbb{E}_{t}\pi_{t+1}\end{equation}
where $\kappa$ was defined back in (44). Substitute into the main Phillips curve (116) and suppressing the errors reveals \begin{equation} \pi_{t}=\frac{g_{0}(1-b_{0})}{g_{0}+1}\pi_{t-1} +\frac{g_{1}-g_{0}b_{1}}{g_{0}+ 1} \hat{y}^{e}_{t} + \frac{g_{2}-g_{0}b_{2}}{g_{0}+1} \hat{\Delta}_{t} + \frac{g_{0}b_{3}}{g_{0}+1}\mathbb{E}_{t}\pi_{t+1}
\end{equation}
Substituting in the wall-crossing constraint from Proposition 20 \begin{multline} \pi_{t}=\frac{g_{0}b_{1}-g_{1}}{g_{0}(1+\beta -\beta b_{0})+1}\hat{y}^{e}_{t} + \frac{g_{0}b_{2}-g_{2}}{g_{0}(1+\beta -\beta b_{0}) +1}\hat{\Delta}_{t}+ \\
\frac{g_{0}b_{3}}{g_{0}(1+\beta - \beta b_{0})+1}\mathbb{E}_{t}\pi_{t+1}
\end{multline}
Equating the coefficients between
(302) with (304)
\begin{equation} g_{0}=-\frac{\beta}{\beta \{1+\beta-\beta b_{0}\} -b_{3}}=-\frac{\beta b}{\beta         \{(1+\beta)b-\beta \tilde{b}_{0}\} -\tilde{b}_{3}}\end{equation}
\begin{equation} g_{1}=\frac{b_{3}\tilde{\omega}- \beta b_{1}}{\beta \{1+\beta-\beta b_{0}\} -b_{3}}=\frac{\tilde{b}_{3}\tilde{\omega}- \beta \tilde{b}_{1}}{\beta         \{(1+\beta)b-\beta \tilde{b}_{0}\} -\tilde{b}_{3}}\end{equation}
\begin{equation}\tilde{g}_{2}=\frac{b_{3}\eta \kappa- \beta b_{2}}{\beta \{1+\beta-\beta b_{0}\} -b_{3}}=\frac{\tilde{b}_{3}\eta\tilde{\omega}- \beta \tilde{b}_{2}}{\beta    \{(1+\beta)b-\beta \tilde{b}_{0}\} -\tilde{b}_{3}}\end{equation}
substituting in the primitives and splitting numerators from denominators, as before, yields
\begin{multline} g = 
\bigg(\beta^{2}(1+\alpha) -(1- \beta) + \big\{\beta(1+\beta)-a_{\pi}\big\}\frac{(1-\alpha)(1-\alpha \beta)}{\alpha}\frac{(\sigma+\eta)}{\sigma+\beta a_{y}} \bigg)\end{multline}
\begin{equation} \tilde{g}_{0}=-
\beta \bigg(1+ \alpha  + \frac{(1-\alpha)(1-\alpha \beta)}{\alpha}\frac{(\sigma+\eta)}{\sigma+\beta a_{y}}\bigg)\end{equation}
\begin{equation} \tilde{g}_{1}=
\frac{(1-\alpha)(1-\alpha \beta)}{\alpha}\frac{(\sigma+\eta) } {(\sigma+\beta a_{y})}\end{equation}
\begin{equation} \tilde{g}_{2}=
\eta \frac{(1-\alpha)(1-\alpha \beta)}{\alpha^{2}}\bigg( \beta -\alpha^{2}(1-\beta)\bigg)\end{equation}
Finally, down the limit where $\beta \rightarrow 1$ and $\sigma=1$.
\begin{equation} g = \bigg(1+ \alpha + (2-a_{\pi})\frac{(1-\alpha)^{2}}{\alpha}\frac{(1+\eta)}{1+a_{y}}\bigg)\end{equation}
\begin{equation} \tilde{g}_{0}= -b= -\bigg(1+ \alpha + \frac{(1-\alpha)^{2}}{\alpha}\frac{(1+\eta)}{1+a_{y}}\bigg)\end{equation}
\begin{equation} \tilde{g}_{1}=\frac{(1-\alpha)^{2}}{\alpha}\frac{(1+\eta) } {(1 + a_{y})} \end{equation}
\begin{equation} \tilde{g}_{2}= \eta \frac{(1-\alpha)^{3}}{\alpha^{2}}(1+\alpha)\end{equation}
Note that these coefficients have clear predicted signs.
\end{subsubsection}
\end{subsection}
\subsection{Linearized DSGE Solution}
This part is devoted to the step by step derivation of equations (160) and (161), from the proof of Theorem 3. Start off with the substitution 
\begin{equation} 
\bf{\tilde{X}}=\begin{bmatrix} \bf{\tilde{X}^{J}} \\ \bf{\tilde{X}^{S}} \end{bmatrix}=\bf{P} \begin{bmatrix} \bf{\hat{X}^{J}}\\ \bf{\hat{X}^{S}} \end{bmatrix}
\end{equation}
This generates the diagonal system 
\begin{equation} \mathbb{E}_{\bf{t}}\bf{\tilde{X}}_{t+1}=\bf{\Lambda} \bf{\tilde{X}}_{t} + \bf{P}\boldsymbol{\phi} \hat{\bf{e}}_{t}\end{equation}
comprised of the non-predetermined system
\begin{equation} \mathbb{E}_{\bf{t}}\bf{\tilde{X}}^{J}_{t+1}=\bf{\Lambda_{1}}
\bf{\tilde{X}^{J}}_{t} + (\bf{P}_{11}\boldsymbol{\phi}_{1}+ \bf{P}_{12}\boldsymbol{\phi}_{2})\hat{\bf{e}}_{t}
\end{equation}
and its predetermined counterpart
\begin{equation} \mathbb{E}_{\bf{t}}\bf{\tilde{X}}^{S}_{t+1}=\bf{\Lambda_{2}}\bf{\tilde{X}^{S}}_{t} + (\bf{P}_{21}\boldsymbol{\phi}_{1}+ \bf{P}_{22}\boldsymbol{\phi}_{2})\hat{\bf{e}}_{t}
\end{equation}
The non-linear system stipulates that the first equation be solved forwards and the second backwards so 
\begin{equation} \bf{\tilde{X}}^{J}_{t}= \mathbb{E}_{\bf{t}}\sum_{i=0}^{\infty} \Lambda_{1}^{-(i+1)}(\bf{P}_{11}\boldsymbol{\phi}_{1}+ \bf{P}_{12}\boldsymbol{\phi}_{2})\hat{\bf{e}}_{t+i}
\end{equation}
\begin{equation} \bf{\tilde{X}^{S}}_{t}=\sum_{i=0}^{t}\Lambda_{1}^{i}(\bf{P}_{21}\boldsymbol{\phi}_{1}+ \bf{P}_{22}\boldsymbol{\phi}_{2}) \hat{\bf{e}}_{i}  \end{equation}
Using the formula for a partitioned inverse 
\begin{equation}\begin{bmatrix} \bf{P}_{11} & \bf{P}_{12} \\ 
\bf{P}_{21} & \bf{P}_{22}
\end{bmatrix}^{-1}= \begin{bmatrix} \bf{\tilde{P}}_{11} & \bf{\tilde{P}}_{12} \\ 
\bf{\tilde{P}}_{21} & \bf{\tilde{P}}_{22}
\end{bmatrix} \end{equation}
where 
\begin{equation} \bf{\tilde{P}}_{11} =(\bf{P}_{11}-\bf{P}_{12}\bf{P}_{22}^{-1}\bf{P}_{21})^{-1} \end{equation}
\begin{equation} \bf{\tilde{P}}_{12} =-(\bf{P}_{11}-\bf{P}_{12}\bf{P}_{22}^{-1}\bf{P}_{21})^{-1}\bf{P}_{12}\bf{P}_{22}^{-1}
\end{equation}
\begin{equation}
\bf{\tilde{P}}_{21} =(\bf{P}_{11}-\bf{P}_{12}\bf{P}_{22}^{-1}\bf{P}_{21})^{-1}\bf{P}_{21}\bf{P}_{11}^{-1}
\end{equation}
\begin{equation}
\bf{\tilde{P}}_{22} =\bf{P}_{22}^{-1}(\bf{I} +\bf{P}_{21}(\bf{P}_{11}-\bf{P}_{12}\bf{P}_{22}^{-1}\bf{P}_{21})^{-1}\bf{P}_{12}\bf{P}_{22}^{-1})
\end{equation}
Using (318) and (324) yields
\begin{equation}\bf{X}^{J}_{t}=\tilde{P}_{11} \bf{\tilde{X}}^{J}_{t} + \tilde{P}_{12} \bf{\tilde{X}}^{S}_{t}  \end{equation}
\begin{equation}\bf{X}^{S}_{t}=\tilde{P}_{21} \bf{\tilde{X}}^{J}_{t} + \tilde{P}_{22} \bf{\tilde{X}}^{S}_{t}  \end{equation}
substituting in (325)-(328) yields the desired expressions in the text. There is an alternative form appropriate when $\bf{P}_{22}$ or $\bf{P}_{11}-\bf{P}_{12}\bf{P}_{22}^{-1}
\bf{P}_{21}$ are not invertible but $\bf{P}_{11}$ and $\bf{P}_{22}-\bf{P}_{21}\bf{P}_{11}^{-1}$ are
\begin{equation} \bf{\tilde{P}}_{11} =\bf{P}_{11}^{-1}(\bf{I} + \bf{P}_{12}(\bf{P}_{22}-\bf{P}_{21}\bf{P}_{11}^{-1}\bf{P}_{12})^{-1}\bf{P}_{21}\bf{P}_{11}^{-1})
\end{equation}
\begin{equation} \bf{\tilde{P}}_{12} =-\bf{P}_{11}^{-1}\bf{P}_{12}(\bf{P}_{22}-\bf{P}_{21}\bf{P}_{11}^{-1}\bf{P}_{12})^{-1}
\end{equation}
\begin{equation} \bf{\tilde{P}}_{21} =- (\bf{P}_{22}-\bf{P}_{21}\bf{P}_{11}^{-1}\bf{P}_{12})^{-1}P_{21}P_{11}^{-1}\end{equation}
\begin{equation}\bf{\tilde{P}}_{22} = (\bf{P}_{22}-\bf{P}_{21}\bf{P}_{11}^{-1}\bf{P}_{12})^{-1} \end{equation}
This gives rise to the final forms 
\begin{multline} \bf{X}^{J}_{t}= \bf{P}_{11}^{-1}(\bf{I} + \bf{P}_{12}(\bf{P}_{22}-\bf{P}_{21}\bf{P}_{11}^{-1}\bf{P}_{12})^{-1}\bf{P}_{21}\bf{P}_{11}^{-1}) \times \\ \mathbb{E}_{\bf{t}}\sum_{\bf{i=0}}^{\infty} \Lambda_{1}^{-(i+1)}(\bf{P}_{11}\boldsymbol{\phi}_{1}+ \bf{P}_{12}\boldsymbol{\phi}_{2})\hat{\bf{e}}_{t+i}  -  \bf{P}_{11}^{-1}\bf{P}_{12}(\bf{P}_{22}-\bf{P}_{21}\bf{P}_{11}^{-1}\bf{P}_{12})^{-1} \\ \times 
\sum_{\bf{i=0}}^{\bf{t}}\Lambda_{1}^{i}(\bf{P}_{21}\boldsymbol{\phi}_{1}+ \bf{P}_{22}\boldsymbol{\phi}_{2}) \hat{\bf{e}}_{i} \end{multline}
\begin{multline}\bf{X}^{S}_{t}= (\bf{P}_{22}-\bf{P}_{21}\bf{P}_{11}^{-1}
\bf{P}_{12})^{-1} \sum_{i=0}^{t}\Lambda_{1}^{i}(\bf{P}_{21}\boldsymbol{\phi}_{1}+ \boldsymbol{P}_{22}\bf{\phi}_{2}) \hat{\bf{e}}_{i}- \\ (\bf{P}_{22}-\bf{P}_{21}\bf{P}_{11}^{-1}\bf{P}_{12})^{-1}P_{21}P_{11}^{-1}\mathbb{E}_{t}\sum_{i=0}^{\infty} \Lambda_{1}^{-(i+1)}(\bf{P}_{11}\boldsymbol{\phi}_{1}+ \bf{P}_{12}\boldsymbol{\phi}_{2})\hat{\bf{e}}_{t+i} 
\end{multline}
\subsection{Eigenvalue Conditions}
This last part contains derivations of the two eigenvalue polynomials in the text, starting with the simplest limiting case and moving onto the general.
\subsubsection{$\sqrt{\varepsilon}$ Characteristic Equation}
Abstracting from the errors, the system can be written as follows
\begin{equation} \begin{pmatrix} \mathbb{E}_{t}\pi_{t+1} \\  \mathbb{E}_{t}\hat{y}^{e}_{t+1} \\ \pi_{t} \end{pmatrix} = \begin{bmatrix} b/\tilde{b}_{3} & -\tilde{b}_{1}/\tilde{b}_{3} & -\tilde{b}_{0}/\tilde{b}_{3} \\ -\tilde{c}_{1}/\tilde{c}_{3} & c/\tilde{c}_{3} & -\tilde{c}_{0}/\tilde{c}_{3}  \\ 1 & 0 & 0\end{bmatrix}  \begin{pmatrix} \pi_{t} \\ \hat{y}^{e}_{t} \\ \pi_{t-1} \end{pmatrix} \end{equation}
The characteristic polynomial comes from the following determinant equation 
\begin{equation}\begin{vmatrix} b/\tilde{b}_{3}  -\lambda & -\tilde{b}_{1}/\tilde{b}_{3} & -\tilde{b}_{0}/\tilde{b}_{3} \\ \tilde{c}_{1}/\tilde{c}_{3} & c/\tilde{c}_{3}-\lambda & -\tilde{c}_{0}/\tilde{c}_{3}  \\ 1 & 0 & -\lambda \end{vmatrix}=0\end{equation}
expressing in terms of minors gives 
\begin{multline} \bigg(\frac{b}{\tilde{b}_{3}} -\lambda\bigg)
\begin{bmatrix} c/\tilde{c}_{3}-\lambda & -\tilde{c}_{0}/\tilde{c}_{3}  \\ 0 & -\lambda \end{bmatrix} + \frac{\tilde{b}_{1}}{\tilde{b}_{3}}\begin{bmatrix} -\tilde{c}_{1}/\tilde{c}_{3} & -\tilde{c}_{0}/\tilde{c}_{3}  \\ 1 & -\lambda \end{bmatrix} - \\ \frac{\tilde{b}_{0}}{\tilde{b}_{3}}\begin{bmatrix} -\tilde{c}_{1}/\tilde{c}_{3} & c/\tilde{c}_{3}-\lambda \\ 1 & 0 \end{bmatrix}=0\end{multline}
expanding out yields 
$$-\bigg(\frac{b}{\tilde{b}_{3}} -\lambda \bigg)\bigg(\frac{c}{\tilde{c}_{3}}-\lambda \bigg)\lambda + \frac{\tilde{b}_{1}\tilde{c}_{1}}{\tilde{b}_{3}\tilde{c}_{3}}\lambda + \frac{\tilde{b}_{1}\tilde{c}_{0}}{\tilde{b}_{3}\tilde{c}_{3}} + \frac{\tilde{b}_{0}}{\tilde{b}_{3}}
\bigg(\frac{c}{\tilde{c}_{3}}-\lambda \bigg)=0$$
then a simple rearrangement gives the expression in the text. 
\subsubsection{General Characteristic Equation} 
Start from the characteristic matrix 
\begin{equation}\begin{bmatrix}
b/\tilde{b}_{3}-\lambda & -\tilde{b}_{1}/\tilde{b}_{3} & -\tilde{b}_{2}/\tilde{b}_{3} & -\tilde{b}_{0}/\tilde{b}_{3} \\
-\tilde{c}_{1}/\tilde{c}_{3} & c/\tilde{c}_{3}-\lambda & -\tilde{c}_{2}/\tilde{c}_{3} & -\tilde{c}_{0}/\tilde{c}_{3} \\ -\tilde{d}_{1}/\tilde{d}_{3} & -\tilde{d}_{2}/\tilde{d}_{3} & d/\tilde{d}_{3}-\lambda & -\tilde{d}_{0}/\tilde{d}_{3} 
\\ 1 & 0 & 0 & -\lambda
\end{bmatrix}\end{equation}
where I have reordered the variables $(\hat{\pi}_{t}, \, \hat{y}_{t}, \, \hat{\Delta}_{t}, \, \hat{\pi}_{t-1})'$ for convenience, substitution yields 
\begin{multline}
\bigg(\frac{b}{\tilde{b}_{3}}-\lambda \bigg)\bigg(\frac{c}{\tilde{c}_{3}}-\lambda \bigg)
\begin{vmatrix}  d/\tilde{d}_{3}-\lambda & -\tilde{d}_{0}/\tilde{d}_{3} \\
  0 & -\lambda \end{vmatrix} +
  \bigg(\frac{b}{\tilde{b}_{3}}-\lambda \bigg)\frac{\tilde{c}_{2}}{\tilde{c}_{3}}\begin{vmatrix}  -\tilde{d}_{2}/\tilde{d}_{3}  & -\tilde{d}_{0}/\tilde{d}_{3} \\
 0 & -\lambda \end{vmatrix} 
 \\  - \bigg(\frac{b}{\tilde{b}_{3}}-\lambda \bigg)\frac{\tilde{c}_{0}}{\tilde{c}_{3}}\begin{vmatrix}  -\tilde{d}_{2}/\tilde{d}_{3}  & d/\tilde{d}_{3}-\lambda  \\
 0 & 0  \end{vmatrix}
  - \frac{\tilde{b}_{1}\tilde{c}_{1}}{\tilde{b}_{3} \tilde{c}_{3}}
  \begin{vmatrix} d/\tilde{d}_{3}-\lambda & -\tilde{d}_{0}/\tilde{d}_{3} \\
0 & -\lambda \end{vmatrix} + \\ 
\frac{\tilde{b}_{1}\tilde{c}_{2}}{\tilde{b}_{3} \tilde{c}_{3}}
\begin{vmatrix}  -\tilde{d}_{1}/\tilde{d}_{3}  &  -\tilde{d}_{0}/\tilde{d}_{3} \\
1 & -\lambda \end{vmatrix}- 
\frac{\tilde{b}_{1}\tilde{c}_{0}}{\tilde{b}_{3} \tilde{c}_{3}}\begin{vmatrix} -\tilde{d}_{1}/\tilde{d}_{3}  & d/\tilde{d}_{3} -\lambda  \\
1 & 0  \end{vmatrix} + \\
\frac{\tilde{b}_{2}\tilde{c}_{1}}{\tilde{b}_{3}\tilde{c}_{3}}\begin{vmatrix}   -\tilde{d}_{2}/\tilde{d}_{3}  & -\tilde{d}_{0}/\tilde{d}_{3}  \\
0 & -\lambda \end{vmatrix} +  
\bigg(\frac{\tilde{c}}{\tilde{c}_{3}}-\lambda \bigg)\frac{\tilde{b}_{2}}{\tilde{b}_{3}}\begin{vmatrix}  -\tilde{d}_{1}/\tilde{d}_{3}  & -\tilde{d}_{0}/\tilde{d}_{3}\\
1  & -\lambda  \end{vmatrix} + \\ 
\frac{\tilde{b}_{2}\tilde{c}_{0}}{\tilde{b}_{3}\tilde{c}_{3}}\begin{vmatrix}  -\tilde{d}_{1}/\tilde{d}_{3} & -\tilde{d}_{2}/\tilde{d}_{3} \\
1 & 0  \end{vmatrix}-  
\frac{\tilde{b}_{0}\tilde{c}_{1}}{\tilde{b}_{3}\tilde{c}_{3}}
\begin{vmatrix} -\tilde{d}_{2}/\tilde{d}_{3} & d/\tilde{d}_{3}-\lambda \\
 0 & 0  \end{vmatrix} - \\ 
 \bigg(\frac{c}{\tilde{c}_{3}}-\lambda \bigg)\frac{\tilde{b}_{0}}{\tilde{b}_{3}}\begin{vmatrix} -\tilde{d}_{1}/\tilde{d}_{3}  & d/\tilde{d}_{3}-\lambda   \\
1 &  0  \end{vmatrix} - 
\frac{\tilde{b}_{0}\tilde{c}_{2}}{\tilde{b}_{3}\tilde{c}_{3}}\begin{vmatrix}  -\tilde{d}_{1}/\tilde{d}_{3}  &  -\tilde{d}_{2}/\tilde{d}_{3} \\
1 & 0  \end{vmatrix}=0 \end{multline} 
expanding out the matrices gives 
\begin{multline}
- \bigg(\frac{b}{\tilde{b}_{3}}-\lambda \bigg)\bigg(\frac{c}{\tilde{c}_{3}}-\lambda \bigg)\bigg(\frac{d}{\tilde{d}_{3}}-\lambda \bigg)\lambda +   \bigg(\frac{b}{\tilde{b}_{3}}-\lambda \bigg)\frac{\tilde{c}_{2}\tilde{d}_{2}}{\tilde{c}_{3}\tilde{d}_{3}} \lambda  +  \frac{\tilde{b}_{1}\tilde{c}_{1}}{\tilde{b}_{3} \tilde{c}_{3}} \bigg(\frac{d}{\tilde{d}_{3}}-\lambda \bigg) \lambda + \\  \frac{\tilde{b}_{1}\tilde{c}_{2}\tilde{d}_{1
}}{\tilde{b}_{3} \tilde{c}_{3}\tilde{d}_{3}}\lambda + \frac{\tilde{b}_{1}\tilde{c}_{2}\tilde{d}_{0}}{\tilde{b}_{3} \tilde{c}_{3}\tilde{d}_{3}} + \frac{\tilde{b}_{1}\tilde{c}_{0}}{\tilde{b}_{3} \tilde{c}_{3}}\bigg(\frac{d}{\tilde{d}_{3}}-\lambda \bigg) +   \frac{\tilde{b}_{2}\tilde{c}_{1}\tilde{d}_{2}}{\tilde{b}_{3} \tilde{c}_{3}\tilde{d}_{3}}\lambda +  \frac{\tilde{b}_{2}\tilde{d}_{1}}{\tilde{b}_{3}\tilde{d}_{3}}\bigg(\frac{c}{\tilde{c}_{3}}-\lambda \bigg) \lambda + \\  \frac{\tilde{b}_{2}\tilde{d}_{0}}{\tilde{b}_{3}\tilde{d}_{3}}\bigg(\frac{c}{\tilde{c}_{3}}-\lambda \bigg)+ \frac{\tilde{b}_{2}\tilde{c}_{0}\tilde{d}_{2}}{\tilde{b}_{3}\tilde{c}_{3}\tilde{d}_{3}} + \frac{\tilde{b}_{0}}{\tilde{b}_{3}}\bigg(\frac{c}{\tilde{c}_{3}}-\lambda \bigg)\bigg(\frac{d}{\tilde{d}_{3}}-\lambda \bigg) - \frac{\tilde{b}_{0}\tilde{c}_{2}\tilde{d}_{2}}{\tilde{b}_{3}\tilde{c}_{3}\tilde{d}_{3}}=0 \end{multline}
collecting terms implies the polynomial
\begin{multline} \lambda^{4}-\Bigg\{
\frac{b}{\tilde{b}_{3}} + \frac{c}{\tilde{c}_{3}} + \frac{d}{\tilde{d}_{3}} \Bigg\}\lambda^{3} + \\
\Bigg\{\frac{bc}{\tilde{b}_{3}\tilde{c}_{3}} + \bigg( \frac{b}{\tilde{b}_{3}}+ \frac{c}{\tilde{c}_{3}}\bigg)\frac{d}{\tilde{d}_{3}} - \frac{\tilde{c}_{2}\tilde{d}_{2}}{\tilde{c}_{3}\tilde{d}_{3}}- 
\frac{\tilde{b}_{1}\tilde{c}_{1}}{\tilde{b}_{3}\tilde{c}_{3}}
- \frac{\tilde{b}_{2} \tilde{d}_{1}}{\tilde{b}_{3}\tilde{d}_{3}} +  \frac{\tilde{b}_{0}}{\tilde{b}_{3}}\Bigg\}\lambda^{2} 
+ \\ \Bigg\{ 
\frac{b\tilde{c}_{2}\tilde{d}_{2}}{\tilde{b}_{3}\tilde{c}_{3}\tilde{d}_{3}} - \frac{bcd}{\tilde{b}_{3}\tilde{c}_{3}\tilde{d}_{3}} + 
 \frac{\tilde{b}_{1}\tilde{c}_{1}d}{\tilde{b}_{3}\tilde{c}_{3}\tilde{d}_{3}} + \frac{\tilde{b}_{1}\tilde{c}_{2}\tilde{d}_{1}}{\tilde{b}_{3}\tilde{c}_{3}\tilde{d}_{3}} 
- \frac{\tilde{b}_{1}\tilde{c}_{0}}{\tilde{b}_{3}\tilde{c}_{3}} + \frac{\tilde{b}_{2}\tilde{c}_{1}\tilde{d}_{2}}{\tilde{b}_{3}\tilde{c}_{3}\tilde{d}_{3}}+ \frac{\tilde{b}_{2}c\tilde{d}_{1}}{\tilde{b}_{3}\tilde{c}_{3}\tilde{d}_{3}}-
 \frac{\tilde{b}_{2}\tilde{d}_{0}}{\tilde{b}_{3}\tilde{d}_{3}} \\ - \frac{\tilde{b}_{0}c}{\tilde{b}_{3}\tilde{c}_{3}} - \frac{\tilde{b}_{0}d}{\tilde{b}_{3}\tilde{d}_{3}}
\Bigg\}\lambda + 
\frac{\tilde{b}_{1}\tilde{c}_{2}\tilde{d}_{0}}{\tilde{b}_{3}\tilde{c}_{3}\tilde{d}_{3}} +
\frac{\tilde{b}_{1}\tilde{c}_{0}d}{\tilde{b}_{3}\tilde{c}_{3}\tilde{d}_{3}} + \frac{\tilde{b}_{2}c \tilde{d}_{0}}{\tilde{b}_{3}\tilde{c}_{3}\tilde{d}_{3}} + \frac{\tilde{b}_{2}\tilde{c}_{0} \tilde{d}_{2}}{\tilde{b}_{3}\tilde{c}_{3}\tilde{d}_{3}} + \frac{\tilde{b}_{0}cd}{\tilde{b}_{3}\tilde{c}_{3}\tilde{d}_{3}} - \frac{\tilde{b}_{0}
\tilde{c}_{2}\tilde{d}_{2}}{\tilde{b}_{3}\tilde{c}_{3}\tilde{d}_{3}}
\end{multline}
where perhaps the most challenging step is the expansion 
 \begin{multline}-\bigg(\frac{b}{\tilde{b}_{3}}-\lambda \bigg)\bigg(\frac{c}{\tilde{c}_{3}}-\lambda \bigg)\bigg(\frac{d}{\tilde{d}_{3}}-\lambda \bigg)\lambda \equiv -\frac{bcd}{\tilde{b}_{3}\tilde{c}_{3}\tilde{d}_{3}}\lambda + \Bigg\{\frac{bc}{\tilde{b}_{3}\tilde{c}_{3}} + \bigg( \frac{b}{\tilde{b}_{3}}+ \frac{c}{\tilde{c}_{3}}\bigg)\frac{d}{\tilde{d}_{3}}\Bigg\} \lambda^{2} \\ -\Bigg\{ \frac{b}{\tilde{b}_{3}} + \frac{c}{\tilde{c}_{3}} + \frac{d}{\tilde{d}_{3}}\Bigg\} \lambda^{3} + \lambda^{4} \end{multline}
 substituting in the ZINSS vanishing condition (138) yields the expression in Theorem 9.
 \subsection{Rouche's Theorem Conditions}
 This part augments Proposition 17 with more cases derived from Rouche's theorem.
 \begin{proposition} Any of the following conditions would prevent the existence of a recursive equilibrium.\end{proposition} 
 \begin{multline} \tag{I} 1  > \bigg| \frac{b}{\tilde{b}_{3}} + \frac{c}{\tilde{c}_{3}} + \frac{d}{\tilde{d}_{3}}\bigg| + \bigg| \frac{bc}{\tilde{b}_{3}\tilde{c}_{3}} + \bigg( \frac{b}{\tilde{b}_{3}}+ \frac{c}{\tilde{c}_{3}}\bigg)\frac{d}{\tilde{d}_{3}} - \frac{\tilde{c}_{2}\tilde{d}_{2}}{\tilde{c}_{3}\tilde{d}_{3}}-
\frac{\tilde{b}_{1}\tilde{c}_{1}}{\tilde{b}_{3}\tilde{c}_{3}}
- \frac{\tilde{b}_{2} \tilde{d}_{1}}{\tilde{b}_{3}\tilde{d}_{3}} +  \frac{\tilde{b}_{0}}{\tilde{b}_{3}}  \bigg| + \\ 
\bigg| \frac{b\tilde{c}_{2}\tilde{d}_{2}}{\tilde{b}_{3}\tilde{c}_{3}\tilde{d}_{3}}-\frac{bcd}{\tilde{b}_{3}\tilde{c}_{3}\tilde{d}_{3}}  + 
 \frac{\tilde{b}_{1}\tilde{c}_{1}d}{\tilde{b}_{3}\tilde{c}_{3}\tilde{d}_{3}} + \frac{\tilde{b}_{1}\tilde{c}_{2}\tilde{d}_{1}}{\tilde{b}_{3}\tilde{c}_{3}\tilde{d}_{3}} 
- \frac{\tilde{b}_{1}\tilde{c}_{0}}{\tilde{b}_{3}\tilde{c}_{3}} + \frac{\tilde{b}_{2}\tilde{c}_{1}\tilde{d}_{2}}{\tilde{b}_{3}\tilde{c}_{3}\tilde{d}_{3}}+ \frac{\tilde{b}_{2}c\tilde{d}_{1}}{\tilde{b}_{3}\tilde{c}_{3}\tilde{d}_{3}}-
\\ \frac{\tilde{b}_{2}\tilde{d}_{0}}{\tilde{b}_{3}\tilde{d}_{3}}-   \frac{\tilde{b}_{0}c}{\tilde{b}_{3}\tilde{c}_{3}} - \frac{\tilde{b}_{0}d}{\tilde{b}_{3}\tilde{d}_{3}} \bigg| 
+  \bigg| \frac{\tilde{b}_{1}\tilde{c}_{2}\tilde{d}_{0}}{\tilde{b}_{3}\tilde{c}_{3}\tilde{d}_{3}} +
\frac{\tilde{b}_{1}\tilde{c}_{0}d}{\tilde{b}_{3}\tilde{c}_{3}\tilde{d}_{3}} + \frac{\tilde{b}_{2}c \tilde{d}_{0}}{\tilde{b}_{3}\tilde{c}_{3}\tilde{d}_{3}} + \frac{\tilde{b}_{2}\tilde{c}_{0} \tilde{d}_{2}}{\tilde{b}_{3}\tilde{c}_{3}\tilde{d}_{3}} + \frac{\tilde{b}_{0}cd}{\tilde{b}_{3}\tilde{c}_{3}\tilde{d}_{3}} - \frac{\tilde{b}_{0}
\tilde{c}_{2}\tilde{d}_{2}}{\tilde{b}_{3}\tilde{c}_{3}\tilde{d}_{3}} \bigg|
\end{multline}
\begin{multline} \tag{II} \bigg| \frac{b}{\tilde{b}_{3}} + \frac{c}{\tilde{c}_{3}} + \frac{d}{\tilde{d}_{3}}\bigg| > 1 + \bigg| \frac{bc}{\tilde{b}_{3}\tilde{c}_{3}} + \bigg( \frac{b}{\tilde{b}_{3}}+ \frac{c}{\tilde{c}_{3}}\bigg)\frac{d}{\tilde{d}_{3}} - \frac{\tilde{c}_{2}\tilde{d}_{2}}{\tilde{c}_{3}\tilde{d}_{3}}- 
\frac{\tilde{b}_{1}\tilde{c}_{1}}{\tilde{b}_{3}\tilde{c}_{3}}
- \frac{\tilde{b}_{2} \tilde{d}_{1}}{\tilde{b}_{3}\tilde{d}_{3}} +  \frac{\tilde{b}_{0}}{\tilde{b}_{3}}  \bigg|  \\ +
\bigg| \frac{b\tilde{c}_{2}\tilde{d}_{2}}{\tilde{b}_{3}\tilde{c}_{3}\tilde{d}_{3}} -\frac{bcd}{\tilde{b}_{3}\tilde{c}_{3}\tilde{d}_{3}} + 
 \frac{\tilde{b}_{1}\tilde{c}_{1}d}{\tilde{b}_{3}\tilde{c}_{3}\tilde{d}_{3}} +
 \frac{\tilde{b}_{1}\tilde{c}_{2}\tilde{d}_{1}}{\tilde{b}_{3}\tilde{c}_{3}\tilde{d}_{3}} 
- \frac{\tilde{b}_{1}\tilde{c}_{0}}{\tilde{b}_{3}\tilde{c}_{3}} + \frac{\tilde{b}_{2}\tilde{c}_{1}\tilde{d}_{2}}{\tilde{b}_{3}\tilde{c}_{3}\tilde{d}_{3}}+ \frac{\tilde{b}_{2}c\tilde{d}_{1}}{\tilde{b}_{3}\tilde{c}_{3}\tilde{d}_{3}}-
\frac{\tilde{b}_{2}\tilde{d}_{0}}{\tilde{b}_{3}\tilde{d}_{3}} \\ -  \frac{\tilde{b}_{0}c}{\tilde{b}_{3}\tilde{c}_{3}} - \frac{\tilde{b}_{0}d}{\tilde{b}_{3}\tilde{d}_{3}} \bigg| 
+ \bigg| \frac{\tilde{b}_{1}\tilde{c}_{2}\tilde{d}_{0}}{\tilde{b}_{3}\tilde{c}_{3}\tilde{d}_{3}} +
\frac{\tilde{b}_{1}\tilde{c}_{0}d}{\tilde{b}_{3}\tilde{c}_{3}\tilde{d}_{3}} + \frac{\tilde{b}_{2}c \tilde{d}_{0}}{\tilde{b}_{3}\tilde{c}_{3}\tilde{d}_{3}} + \frac{\tilde{b}_{2}\tilde{c}_{0} \tilde{d}_{2}}{\tilde{b}_{3}\tilde{c}_{3}\tilde{d}_{3}} + \frac{\tilde{b}_{0}cd}{\tilde{b}_{3}\tilde{c}_{3}\tilde{d}_{3}} - \frac{\tilde{b}_{0}
\tilde{c}_{2}\tilde{d}_{2}}{\tilde{b}_{3}\tilde{c}_{3}\tilde{d}_{3}} \bigg|
\end{multline}
\begin{multline} \tag{III} \bigg| \frac{bcd}{\tilde{b}_{3}\tilde{c}_{3}\tilde{d}_{3}} - \frac{b\tilde{c}_{2}\tilde{d}_{2}}{\tilde{b}_{3}\tilde{c}_{3}\tilde{d}_{3}} -
 \frac{\tilde{b}_{1}\tilde{c}_{1}d}{\tilde{b}_{3}\tilde{c}_{3}\tilde{d}_{3}} -
 \frac{\tilde{b}_{1}\tilde{c}_{2}\tilde{d}_{1}}{\tilde{b}_{3}\tilde{c}_{3}\tilde{d}_{3}} 
+ \frac{\tilde{b}_{1}\tilde{c}_{0}}{\tilde{b}_{3}\tilde{c}_{3}} - \frac{\tilde{b}_{2}\tilde{c}_{1}\tilde{d}_{2}}{\tilde{b}_{3}\tilde{c}_{3}\tilde{d}_{3}}+ \frac{\tilde{b}_{2}c\tilde{d}_{1}}{\tilde{b}_{3}\tilde{c}_{3}\tilde{d}_{3}}+
\frac{\tilde{b}_{2}\tilde{d}_{0}}{\tilde{b}_{3}\tilde{d}_{3}} +  \\ \frac{\tilde{b}_{0}c}{\tilde{b}_{3}\tilde{c}_{3}} + \frac{\tilde{b}_{0}d}{\tilde{b}_{3}\tilde{d}_{3}} \bigg|   > 1 +  \bigg| \frac{b}{\tilde{b}_{3}} + \frac{c}{\tilde{c}_{3}} + \frac{d}{\tilde{d}_{3}}\bigg| + \bigg| \frac{bc}{\tilde{b}_{3}\tilde{c}_{3}} + \bigg( \frac{b}{\tilde{b}_{3}}+ \frac{c}{\tilde{c}_{3}}\bigg)\frac{d}{\tilde{d}_{3}} - \frac{\tilde{c}_{2}\tilde{d}_{2}}{\tilde{c}_{3}\tilde{d}_{3}}- 
\frac{\tilde{b}_{1}\tilde{c}_{1}}{\tilde{b}_{3}\tilde{c}_{3}}
- \\  \frac{\tilde{b}_{2} \tilde{d}_{1}}{\tilde{b}_{3}\tilde{d}_{3}} +  \frac{\tilde{b}_{0}}{\tilde{b}_{3}}  \bigg| 
+  \bigg| \frac{\tilde{b}_{1}\tilde{c}_{2}\tilde{d}_{0}}{\tilde{b}_{3}\tilde{c}_{3}\tilde{d}_{3}} +
\frac{\tilde{b}_{1}\tilde{c}_{0}d}{\tilde{b}_{3}\tilde{c}_{3}\tilde{d}_{3}} + \frac{\tilde{b}_{2}c \tilde{d}_{0}}{\tilde{b}_{3}\tilde{c}_{3}\tilde{d}_{3}} + \frac{\tilde{b}_{2}\tilde{c}_{0} \tilde{d}_{2}}{\tilde{b}_{3}\tilde{c}_{3}\tilde{d}_{3}} + \frac{\tilde{b}_{0}cd}{\tilde{b}_{3}\tilde{c}_{3}\tilde{d}_{3}} - \frac{\tilde{b}_{0}
\tilde{c}_{2}\tilde{d}_{2}}{\tilde{b}_{3}\tilde{c}_{3}\tilde{d}_{3}} \bigg| 
\end{multline}
\begin{multline} \tag{IV}  \bigg| \frac{\tilde{b}_{1}\tilde{c}_{2}\tilde{d}_{0}}{\tilde{b}_{3}\tilde{c}_{3}\tilde{d}_{3}} +
\frac{\tilde{b}_{1}\tilde{c}_{0}d}{\tilde{b}_{3}\tilde{c}_{3}\tilde{d}_{3}} + \frac{\tilde{b}_{2}c \tilde{d}_{0}}{\tilde{b}_{3}\tilde{c}_{3}\tilde{d}_{3}} + \frac{\tilde{b}_{2}\tilde{c}_{0} \tilde{d}_{2}}{\tilde{b}_{3}\tilde{c}_{3}\tilde{d}_{3}} + \frac{\tilde{b}_{0}cd}{\tilde{b}_{3}\tilde{c}_{3}\tilde{d}_{3}} - \frac{\tilde{b}_{0}
\tilde{c}_{2}\tilde{d}_{2}}{\tilde{b}_{3}\tilde{c}_{3}\tilde{d}_{3}} \bigg| 
> 1 +   \\ \bigg| \frac{b}{\tilde{b}_{3}} + \frac{c}{\tilde{c}_{3}} + \frac{d}{\tilde{d}_{3}}\bigg|  +  \bigg| \frac{bc}{\tilde{b}_{3}\tilde{c}_{3}} + \bigg( \frac{b}{\tilde{b}_{3}}+ \frac{c}{\tilde{c}_{3}}\bigg)\frac{d}{\tilde{d}_{3}} - \frac{\tilde{c}_{2}\tilde{d}_{2}}{\tilde{c}_{3}\tilde{d}_{3}}- 
\frac{\tilde{b}_{1}\tilde{c}_{1}}{\tilde{b}_{3}\tilde{c}_{3}}
- \frac{\tilde{b}_{2} \tilde{d}_{1}}{\tilde{b}_{3}\tilde{d}_{3}} +  \frac{\tilde{b}_{0}}{\tilde{b}_{3}}  \bigg| + \\ 
\bigg| \frac{bcd}{\tilde{b}_{3}\tilde{c}_{3}\tilde{d}_{3}} - \frac{b\tilde{c}_{2}\tilde{d}_{2}}{\tilde{b}_{3}\tilde{c}_{3}\tilde{d}_{3}} -
 \frac{\tilde{b}_{1}\tilde{c}_{1}d}{\tilde{b}_{3}\tilde{c}_{3}\tilde{d}_{3}} -
 \frac{\tilde{b}_{1}\tilde{c}_{2}\tilde{d}_{1}}{\tilde{b}_{3}\tilde{c}_{3}\tilde{d}_{3}} 
+ \frac{\tilde{b}_{1}\tilde{c}_{0}}{\tilde{b}_{3}\tilde{c}_{3}} - \frac{\tilde{b}_{2}\tilde{c}_{1}\tilde{d}_{2}}{\tilde{b}_{3}\tilde{c}_{3}\tilde{d}_{3}}+ \frac{\tilde{b}_{2}c\tilde{d}_{1}}{\tilde{b}_{3}\tilde{c}_{3}\tilde{d}_{3}}+
\frac{\tilde{b}_{2}\tilde{d}_{0}}{\tilde{b}_{3}\tilde{d}_{3}} + \\ \frac{\tilde{b}_{0}c}{\tilde{b}_{3}\tilde{c}_{3}} +  \frac{\tilde{b}_{0}d}{\tilde{b}_{3}\tilde{d}_{3}} \bigg|
\end{multline}
\begin{proof} Following Proposition 
17, it is sufficient to note that the cases (I), (II), (III) and (IV) correspond respectively to the conditions guaranteeing zero, one, three and four roots outside the unit circle.
\end{proof}
If price dispersion were jettisoned the system would be governed by (163), in which case the sufficient condition would be 
\begin{equation}  \bigg| \frac{\tilde{b}_{0}}{\tilde{b}_{3}} + \frac{bc}{\tilde{b}_{3}\tilde{c}_{3}}- \frac{\tilde{b}_{1}\tilde{c}_{1}}{\tilde{b}_{3}\tilde{c}_{3}} \bigg| > 1 + \bigg| \frac{b}{\tilde{b}_{3}} +\frac{c}{\tilde{c}_{3}}\bigg| + \bigg| \frac{\tilde{b}_{1}\tilde{c}_{0}}{\tilde{b}_{3}\tilde{c}_{3}} + \frac{\tilde{b}_{0}c}{\tilde{b}_{3}\tilde{c}_{3}}\bigg|\end{equation} 
whilst the following would imply non-existence
\begin{equation} \tag{I} 1  > \bigg| \frac{\tilde{b}_{0}}{\tilde{b}_{3}} + \frac{bc}{\tilde{b}_{3}\tilde{c}_{3}}- \frac{\tilde{b}_{1}\tilde{c}_{1}}{\tilde{b}_{3}\tilde{c}_{3}} \bigg| + \bigg| \frac{b}{\tilde{b}_{3}} +\frac{c}{\tilde{c}_{3}}\bigg| + \bigg| \frac{\tilde{b}_{1}\tilde{c}_{0}}{\tilde{b}_{3}\tilde{c}_{3}} + \frac{\tilde{b}_{0}c}{\tilde{b}_{3}\tilde{c}_{3}}\bigg|\end{equation} 
\begin{equation} \tag{II} \bigg| \frac{b}{\tilde{b}_{3}} +\frac{c}{\tilde{c}_{3}}\bigg| > 1 + \bigg| \frac{\tilde{b}_{0}}{\tilde{b}_{3}} + \frac{bc}{\tilde{b}_{3}\tilde{c}_{3}}- \frac{\tilde{b}_{1}\tilde{c}_{1}}{\tilde{b}_{3}\tilde{c}_{3}} \bigg|   + \bigg| \frac{\tilde{b}_{1}\tilde{c}_{0}}{\tilde{b}_{3}\tilde{c}_{3}} + \frac{\tilde{b}_{0}c}{\tilde{b}_{3}\tilde{c}_{3}}\bigg|\end{equation} 
\begin{equation} \tag{III} \bigg| \frac{\tilde{b}_{1}\tilde{c}_{0}}{\tilde{b}_{3}\tilde{c}_{3}} + \frac{\tilde{b}_{0}c}{\tilde{b}_{3}\tilde{c}_{3}}\bigg|  > 1 + \bigg| \frac{\tilde{b}_{0}}{\tilde{b}_{3}} + \frac{bc}{\tilde{b}_{3}\tilde{c}_{3}}- \frac{\tilde{b}_{1}\tilde{c}_{1}}{\tilde{b}_{3}\tilde{c}_{3}} \bigg|   + \bigg| \frac{b}{\tilde{b}_{3}} +\frac{c}{\tilde{c}_{3}}\bigg| \end{equation}
\section{Abstract Algebra}
This section supplies an introductory primer on homological algebra, De Rham cohomology and category theory, complimentary to the informal discussion in the text. Each occupies a subsection. At the end, a short part is devoted to additional details of a calculation from Proposition 20, which applies the methods here to the ultimate goal of the paper. \subsection{Homology Groups}
 This part develops the theory of singular homology to support the claims and analysis in the main text. Other popular homology constructions are discussed in \cite{nathan1980basic}. Informally, the singular homology is built by a succession of mappings between $n$ dimensional blocks and the paths around their boundaries. 
 \subsubsection{Singular Simplex} The standard $n$-simplex $\Delta^{n}$ is comprised of $$\{ x \in \mathbb{R}^{k}: x _{0} + \cdots + x_{k-1} = 1, \, x_{i} \geq 0 \; \textrm{for} \; i=0, \, \cdots , \, k-1 \}$$
 its vertices are the $k$ standard unit vectors and the origin.\footnote{In other contexts these are sometimes referred to as the probability simplex, such as in \cite{boyd2004convex}.} Familiar examples include: 
 \begin{itemize}
 \item $\Delta^{0}$ is the point one in $\mathbb{R}$
 \item $\Delta^{1}$ is the line segment joining $(1,\, 0)$ and $(0,\, 1)$ in $\mathbb{R}^{2}$.
 \item $\Delta^{2}$ is the equilateral triangle with vertices $(1, \,  0, \, 0)$, $(0, \, 1, \, 0)$ and $(0, \,  0, \, 1)$ in $\mathbb{R}^{3}$.
 \end{itemize}
A singular $n$-simplex, in a topological space $X$, is any continuous function from the standard $n$-simplex $\Delta^{n}$ to $X$, written $\sigma: \Delta^{n} \rightarrow X$.
\par The boundary of $\sigma$ denoted $\partial \sigma$ is defined as the formal sum of the singular $(n-1)$-simplices created by the restriction of $\sigma$ to the faces of the standard $n$-simplex, with an alternating sign pattern to account for orientation.\footnote{A formal sum is an element of the free abelian group on the simplices. The basis for this group is the infinite set of all possible singular simplices. It should be clear from basic topological properties of manifolds that these form a valid abelian group action $+$ since $a+b$ represents another singular simplex, $a+a=2a$ is $2 \sigma $ and $-a$ is $-\sigma$. For more background on this topic, consult \cite{jacobson2009basic}.} Let me designate $\sigma$ 
by its vertices 
$$ [p_{0}, \, p_{1} , \, \cdots , \, p_{n}]=[\sigma(e_{0}), \, \sigma(e_{1}), \, \cdots , \, \sigma(e_{n})] $$
corresponding to the vertices $e_{k}$ of the standard $n$-simplex. Naturally, this is not sufficient to specify a particular singular complex $\sigma$. Then $$\partial_{n}\sigma = \sum_{k=0}^{n}(-1)^{k}\sigma \vert _{[p_{0}, \, p_{1}, \, \cdots , \, p_{n}]}$$
 is a formal sum of the faces, such that each is a restriction of $\sigma$ to $\Delta^{n}$, according to the listing of the vertices. For example, the boundary of $\sigma = [p_{0}, \, p_{1}]$ is a curve going from $p_{0}$ to $p_{1}$, called either the formal sum or the formal difference (labelled $[p_{1}]- [p_{0}]$).
\subsubsection{Singular Chain Complex}
The construction of the singular homology involves defining formal sums of simplices, which form the elements of a free abelian group, and then using the boundary operators to derive the homology group of the topological space. First, consider the set of all possible singular $n$-simplices $\sigma_{n}(X)$ on a topological space $X$. This set can be used as the basis of a free abelian group, such that each singular $n$-simplex is the generator of the group.\footnote{It does not matter that this basis is uncountable, since there are $\aleph_{1}$ (the cardinality of $\mathbb{R}$) continuous mappings $\mathbb{R} \rightarrow \mathbb{R}$.} This free abelian group generated by this basis is written $C_{n}(X)$. Elements of $C_{n}(X)$ are referred to as singular chains - the formal sum of singular simplices with integer coefficients $(\mathbb{Z})$. This explains how the groups were constructed in the main text, a result called the universal coefficient theorem shows that, this is without loss of generality (see \cite{hatcher2002algebraic}).
\par The boundary $\partial$ readily extends from singular complexes to singular $n$-chains, where it is known as the boundary operator
 $$\partial_{n}: C_{n} \rightarrow C_{n-1}$$
\begin{center}
\textbf{Figure 4: Singular Homology Chain Complex}
\begin{tikzcd}
\cdots \arrow{r}{\partial_{n+1}}
& C_{n}\arrow{r}{\partial_{n}}
& C_{n-1}  \arrow{r}{\partial_{n-1}} &
\cdots \arrow{r}{\partial_{2}} & C_{1} \arrow{r}{\partial_{1}} & C_{0} \arrow{r}{\partial_{0}} & 0
\end{tikzcd}
\end{center}
It forms a group homomorphism.\footnote{Recall that this is a mapping between two groups $(G, \, \ast )$ and $(H, \, \cdot)$, a group homomorphism from $(G, \, \ast)$ to $(H, \, \cdot)$ is a function $h : G \rightarrow H$, such that for all $u$ and $v$ in $G$, it holds that $h(u \ast v)=h(u)\cdot h(v)$, where the group operation on the left side of the equation is that of $G$ and on the right side that of $H$. From this property, one can deduce that $h$ maps the identity element $eG$ of $G$ to the identity element $eH$ of $H$, $h(e_{G})=e_{H}$ and it also maps inverses to inverses in the sense that
$h(u^{-1})=h(u)^{-1}$. Hence one can say that $h$ "is compatible with the group structure". The application is natural from the geometry of the problem.} The boundary operator together with the family of $C_{n}$ are called the singular complex $(C_{\cdot}(X), \, \partial_{\cdot})$, which is often shortened to $C_{\cdot}(X)$.
\par The kernel of the boundary operator is called $Z_{n}(X)=\ker(\partial_{n})$ and is known as the group of singular $n$-cycles. Its image is $B_{n}(X)=\newcommand{\Ima}{\text{im}}\Ima(\partial_{n+1})$.
It is called the group of singular $n$-boundaries. By geometric arguments, it is clear that $\partial_{n} \circ \partial_{n+1}=0$.
The $n^{th}$ homology group of $X$ is then defined as the quotient group\footnote{A quotient group is obtained by aggregating similar elements of a larger group, using an equivalence relation that preserves some of the group structure- the rest of the structure is "factored" out- hence it is also called a factor group. The best known example is addition modulo $n$. A quotient group is formally defined in terms of two intermediate objects, a normal subgroup and a coset. A normal subgroup is a subgroup that is invariant under conjugation, by members of the parent group. In other words, a subgroup $N$ of is normal in $G$ if and only if $gng^{-1} \in N$ for all $g \in G$ and $n \in N$; clearly this is true with abelian groups. The usual notation for this relation is $ N \vartriangleright G$. The cosets of $N$ in $G$ are those formed from $n + g$, for all $n \in N$ and $g \in G$. The quotient group of $N$ in $G$ is the set of cosets of $N$ in $G$. Here it represents represents the idea of cycles modulo boundaries, in that elements from the boundary are all sent to zero.} $$H_{n}(X)=Z_{n}(X)/ B_{n}(X)$$
The elements of $H_{n}(X)$ are called homology classes. 
\subsubsection{Homotopy Invariance}
The final task is to specialize results from $\mathbb{R}^{n}$ to general manifolds. In fact, a more wide-ranging argument can be demonstrated. If $X$ and $Y$ are topological spaces that are homotopy equivalent\footnote{Recall that two spaces are homotopically equivalent if they can be continuously deformed into one another, a space is called contractible (or homotopically trivial) if it can be deformed to a point. \cite{munkres1974topology} is a place to review the formal definition.} then they have equivalent homology 
$$H_{n}(X) \cong H_{n}(Y)$$
for all $n \geq 0$. This means that homology groups are topologically invariants. In particular, if $X$ is a connected contractible space, then all its homology groups are $0$ except $$H_{0}(X) \cong \mathbb{Z}$$. 
\par I finish with a sketch of the proof of homotopy invariance of singular homology. A continuous map $f: X \rightarrow Y$ induces a homomorphism 
$$f_{\sharp}: C_{n}(X) \rightarrow C_{n}(Y)$$
It is easy to see that 
$$\partial{f}_{\sharp}=f_{\sharp}\partial $$
where $f_{\sharp}$ is a chain map which descends to homomorphisms on homology 
$$f_{*}: H_{n}(X) \rightarrow H_{n}(Y)$$
It is left to show that, if $f$ and $g$ are homotopically equivalent, then $f_{*}=g_{*}$. This implies that if $f$ is a homotopy equivalence then $f_{*}$ is an isomorphism.
\par Let $F: X \times [0,1] \rightarrow Y$ be a homotopy, that takes $f$ to $g$ 
on the level of chains. It defines a homomorphism 
$$P : C_{n}(X) \rightarrow C_{n+1}(Y)$$
that from a geometric standpoint sends a basis element $\sigma: \Delta^{n} \rightarrow X$ of $C_{n}(X)$ to the "prism" $P_{n} \times I \rightarrow Y $. The boundary of $P(\sigma)$ can be expressed as 
$$\partial{\sigma}= f_{\sharp}(\sigma)-g_{\sharp}(\sigma)-P(\partial{\sigma})$$
so if $\alpha$ in $C_{n}(X)$ is an $n$-cycle, then $f_{\sharp}(\alpha)$ and $g_{\sharp}(\alpha)$ differ by a boundary 
$$f_{\sharp}(\alpha)- g_{\sharp}(\alpha) = \partial P (\alpha)$$ 
Therefore, they are homologous, which settles the claim. 
\subsection{Basic Categories}
This part helps with understanding the concepts underpinning Definition 13. It also develops the theory behind Figure 1 in Section 9 and draws links with other items in this appendix. Category theory uses formal diagrams to study underlying mathematical relations. A \textbf{Category} is a labelled directed graph whose nodes are called \textbf{Objects} and labelled directed edges are called morphisms or \textbf{Arrows}. Categories have two basic properties: arrows can be composed associatively and for every object there is an identity map. \cite{vakil2017rising} has an excellent chapter expositing category theory, with a view to algebraic geometry. \cite{leinster2014basic} is a general introduction.
\subsubsection{Discussion and Examples} Categories represent abstractions of other mathematical concepts. Many familiar concepts and structures can be viewed as categories. It simplifies the process of stating and proving theorems and helps with understanding similarities between seemingly disparate areas of mathematics. 
\par Perhaps the most basic example of a category is the category of sets, where the objects are sets and the arrows represents functions between these sets. The strength of the framework is that any system, that can be built from associative mappings, can be analyzed in this way. Arrows are often said to represent \textbf{Structure-Preserving Transformations} between objects.
\par This is most apparent with the family of concrete categories, informally, those that can be viewed as part of the category of sets but with additional restrictions on the arrows. 
First is \textbf{Top}, whose objects are topological spaces with continuous functions as morphisms. From basic analysis and topology these preserve compactness, connectedness and separability properties. In differential topology the category \textbf{Smooth} has smooth manifolds as objects and maps that preserve this property (smooth maps) as arrows. In \textbf{Homotopy} maps preserve homotopy between pointed topological spaces.
\par The insight is more general. The class \textbf{Grp} consists of groups and homomorphisms that retain the group structure of the object in the target. Categories have their own structure-preserving processes called \textbf{Functor}.
\par A functor associates to every object in one category an object in another and every morphism one in the other. Categories allow for significant abstraction, for instance, it is possible to define the category of \textbf{Categories and Functors} with categories as objects and functors as arrows. Finally, \textbf{Natural Transformations} are mappings between functors that obey certain commutativity conditions. Functors and natural transformations underpin the "diagram chasing" method, exemplified in Figure 1 in the text.
\subsubsection{Categories and Morphisms}
A category $C$ comprises the following three mathematical entities
 \begin{itemize}
\item A family $ob(C)$, whose elements are called objects.
\item A class $hom(C)$, whose elements are known as morphisms, maps or arrows. $f: a \rightarrow b$ denotes a morphism from source $a$ to target $b$.
\item $hom(a,\, b)$ denotes the hom-class of all morphisms from $a$ to $b$, for example, all the smooth functions between sets $a$ and $b$ in \textbf{Smooth}.
\item A binary operation $\circ$, called \emph{composition of morphisms}, such that for any three objects $a$, $b$ and $c$ $\circ : hom(a, \, b) \times hom (b,\, c) \rightarrow hom(a,\, c)$. The composition of $f:a \rightarrow b$ with $g: b \rightarrow c$ rendered $g \circ f$ or $gf$ is governed by two familiar axioms 
\end{itemize}
\begin{enumerate} [i]
\item Associativity: If $f:a \rightarrow b$, $g: b \rightarrow c$ and $h: c \rightarrow d$ then $h \circ (g \circ f)= (h \circ g) \circ f$
\item Identity: For every object $x$, there exists a morphism $1_{x}: x \rightarrow x$, called the \emph{identity morphism}, such that for every morphism $f: a \rightarrow b$, it is the case that $1_{b} \circ f = f \circ 1_{a}$
\end{enumerate}
Intuitively, the first condition is essential to define paths around diagrams. It is common to notions of multiplication and addition. The second, is a requirement that no matter how layered the algebraic structure, one can drill down to the primitive level typically points, sets or maps.
\par Relations between morphisms, like $fg=h$, are often depicted in commutative diagrams, with vertices representing objects and arrows representing morphism.
A morphism $f: a \rightarrow b$ can possess the following properties:
\begin{itemize}
\item \textbf{Monomorphism} (or monic) if $f \circ g_{1}= f \circ g_{2}$ implies $g_{1}=g_{2}$, for all morphisms $g_{1}, \, g_{2} : x \rightarrow a$
\item \textbf{Epimorphism} (or epic) if $g_{1} \circ f = g_{2} \circ f$ for any $g_{1}, \, g_{2} : b \rightarrow x$
\item \textbf{Bimorphism} if $f$ is both monic and epic. 
\item An \textbf{Isomorphism} if there is another morphism $g: b \rightarrow a$ where $f \circ g =1_{b}$ and $g \circ f=1_{a}$
\end{itemize}
A category is \textbf{Abelian} if it has the subsequent attributes: 
\begin{itemize}
\item It is pre-additive, so for any three morphisms $f$, $g$ and $h$ and the group operator $+$
$$f \circ(g +h) = f \circ g + f \circ h$$
$$(f+g) \circ h = f \circ h + g \circ h$$
Thus every set $Hom(a,\, b)$ has an abelian group structure and function composition is distributive over the group operator. 
\item It has a zero object $I$, such that for any other object $Z$, the object is initial, so there is exactly one morphism $I \rightarrow Z$ and final, in the sense that there is only one mapping $Z \rightarrow I$.
\item All bi-products exist- this is equivalent to the existence of direct sums of elements (recall $\mathbb{R} \bigoplus \mathbb{R}= \mathbb{R}^{2}$)
\item It has all kernels and cokernels. 
\item All morphisms and epimorphisms are normal.This means the category behaves as though there is a set which is the destination for all mappings.
\end{itemize} 
\begin{itemize}
\item  An object $F(x)$ in D for any object $x$ in $C$ and
\item A morphism $F(f): F(x) \rightarrow F(y)$ for each morphism $f: x \rightarrow y$ in $C$.
\end{itemize}
with the properties that
\begin{itemize}
\item  for every object $x$ in $C$, $F(1_{x})=1_{F(x)}$ 
\item all morphisms compose associatively so $f: x \rightarrow y$ and $g: y \rightarrow z$ imply $F(g \circ f)=F(g) \circ F(f)$
\end{itemize}
A \textbf{Contravariant} functor $F: C \rightarrow D$ behaves like a covariant functor, except with all the morphisms turned around. More specifically, every morphism $f: x \rightarrow y$ in $C$ is assigned to a morphism $F(f): F(y) \rightarrow F(x)$ in $D$. Equivalently, a contravariant functor acts as a covariant functor from the opposite category $C^{op}$ to D, where the opposite category $C^{op}$ is $C$ with all its 
arrows reversed. 
\par Formally, a \textbf{Natural Transformation} is a relation between two functors. Functors often describe so called natural constructions and natural transformations represent natural homomorphisms between these constructions. When two different constructions yield the same result an isomorphism arises between the two functors.  
\par If $F$ and $G$ are covariant functors between the categories $C$ and $D$, then a natural transformation $\eta$ from $F$ to $G$ associates with every object $x$ in $C$ a morphism $\eta_{x} : F(x) \rightarrow G(x)$ in $D$, such that for every morphism $f: x \rightarrow y$, the relationship $\eta_{y} \circ F(f) =G(f) \circ \eta_{x}$. Thus, I construct a commuting diagram below. The two functors $F$ and $G$ are referred to as \emph{naturally isomorphic} because there exists a natural transformation from $F$ to $G$, such that $\eta_{x}$ is an isomorphism for every $x$ in $C$.
\begin{center}
\begin{tikzcd}
F(x)  \arrow[r, "F(f)"]  \arrow[d, "\eta_{x}"] &  F(y)   \arrow[d, "\eta_{y}"]\\
G(x) \arrow[r, "G(f)"] &  G(y) 
\end{tikzcd}
\end{center}
Next I can define a \textbf{Concrete Category} as a pair $(C, \, U)$, such that
$C$ is a category, and $U : C \rightarrow$  \textbf{Set} (the category of sets and functions) is a faithful functor- one which is one-to-one when restricted to morphisms with the same target and source. The functor $U$ is to be thought of as a so called forgetful functor, which assigns to every object of C its "underlying set", and to every morphism in C its "underlying function". Thus a concrete category can be viewed as the category of sets but with a restriction on the set of possible morphisms, like differentiability or requiring uniform continuity (often denoted \textbf{Uni}).
\par An \textbf{Exact Functor} is one which preserves short exact sequences (those that end in zero after three or fewer steps). Formally, let $P$ and $Q$ be abelian categories, and let $F: P \rightarrow Q$ be a covariant additive functor (so that, in particular, $F(0)=0$). We say that $F$ is an exact functor if, whenever
\begin{center}
\begin{tikzcd}
0 \arrow{r}
& A \arrow{r}{f}
& B \arrow{r}{g} & C \arrow{r}
& 0
\end{tikzcd}
\end{center}
is a short exact sequence in $P$, then
\begin{center}
\begin{tikzcd}
0 \arrow{r}
& F(A) \arrow{r}{F(f)}
& F(B) \arrow{r}{F(g)} & F(C) \arrow{r}
& 0
\end{tikzcd}
\end{center}
is a short exact sequence in $Q$. It is often easier to work with the weaker notion of a \textbf{Left Exact} sequence defined below 
\begin{center}
\begin{tikzcd}
0 \arrow{r}
& A \arrow{r}{f}
& B \arrow{r}{g} & C \arrow{r}
& 0
\end{tikzcd}
\end{center}
is a short exact sequence in $P$, then
 \begin{center}
 \begin{tikzcd}
0 \arrow{r}
& F(A) \arrow{r}{F(f)}
& F(B) \arrow{r}{F(g)} & F(C)
\end{tikzcd}
\end{center}
Finally, I will talk  briefly later about \textbf{Derived Functors}. This advanced topic is probably best appreciated by studying \cite{nathan1980basic} or \cite{vakil2017rising}. I will provide a more heuristic presentation here. Consider a covariant left exact functor $F: \bf{A} \rightarrow \bf{B}$ between two abelian categories $\bf{A}$ and $\bf{B}$. If $0 \rightarrow A \rightarrow B \rightarrow C \rightarrow 0$ is a short exact sequence in $\bf{A}$ then applying $F$ yields the exact sequence $0 \rightarrow F(A) \rightarrow F(B) \rightarrow F(C) $ and the sequence could continue to the right to form a long exact sequence.\footnote{A long exact sequence is simply one with more than two arrows not pointing to or from zero.} A right derived functor $R^{i}F: \bf{A} \rightarrow \bf{B}$ defined for each $i \geq 1$ continues the sequence as follows: 
\begin{figure}[hb]
\centering
\label{fig:enter-label}
\begin{tikzcd}
0 \arrow[r] & F(A) \arrow[r] & F(B) \arrow[r] & F(C) \arrow[llld] \\
R^{1}F(A) \arrow[r] 
& R^{1}F(B) \arrow[r] & R^{1}F(C) \arrow[lld] \\ 
R^{2}F(A) \arrow[r] & R^{2}F(B) \arrow[r] & R^{2}F(C) \arrow[r] & \cdots 
\end{tikzcd}
\end{figure}
\newline
It is clear that, $F$ is an exact functor if and only if $R^{1}F=0$, in this sense, the right derived functor measures how far $F$ is from being exact.
\par If the objects in the sequence are injective, then the sequence splits, which corresponds to the case where you can "skip links" in the chain, %like how%
for example, if there are no holes of dimensions $k< d \leq n$ from the $n$ dimensional surface to its $k$-dimensional building blocks. Any right additive functor sends split sequences to split sequences,\footnote{An exact sequence of an abelian category is called split exact if 
\begin{center}
\begin{tikzcd}[ampersand replacement=\&]
0 \arrow{r}
\& A \arrow{r}{a}
\& B \arrow{r}{b} \& C \arrow{r} \& 0
\end{tikzcd}
\end{center}
is isomorphic to the sequence where the middle term is the direct sum of the outer ones. 
\begin{center}
\begin{tikzcd}[ampersand replacement=\&]
0 \arrow{r}
\& A \arrow{r}{i}
\& A \bigoplus
C \arrow{r}{p} \& C \arrow{r} \& 0
\end{tikzcd} 
\end{center}} meaning $R^{1}F(A)=0$. Right derived functors are zero on injectives (for $i >0$). 
\subsubsection{Applications}
 I supply one mathematical and one economic interpretation, focusing on Figure 1 in the text.
 The mapping idea of the Lucas critique can be viewed categorically, as the requirement for a natural transformation between the structural model and its econometric approximation in the category of sets or concrete categories, such as categories of local approximations. 
 \par Homological algebra can also be viewed through a category theoretic lens.
 Chain complexes form a category: A morphism from the chain complex  $\partial_{n}: A_{n} \rightarrow A_{n-1}$
 to the chain complex $e_{n}:B_{n} \rightarrow B_{n-1}$ is a sequence of homomorphisms $f_{n}: A_{n} \rightarrow B_{n}$, such that $f_{n-1} \circ \partial_{n}=e_{n} \circ f_{n}$ for all $n$. The $n$-th homology $H_{n}$ is a covariant functor from the category of chain complexes to the category of abelian groups (or modules).
If the chain complex depends on the object $X$ in a covariant manner (meaning that any morphism $X \rightarrow Y$ induces a morphism from the chain complex of X to the chain complex of Y), then the $H_{n}$ are covariant functors from the category that $X$ belongs to into the category of abelian groups (or modules). In fact, homology groups can be seen as derived functors on appropriate abelian categories, measuring the failure of a functor to be exact; since the break down of a short exact sequence, from a higher dimension to a lower dimension, is the defining feature of a hole or tear.
\par This links back to the Lucas critique. We can think of the structural mapping as breaking down, when the trivial homology fails. In the next part, we will explore a particular cohomology, which is contravariant to homology, which is central to Econometric Duality. This (Principle 1) represents the deepest connection between micro-foundations, econometrics and the Lucas critique. From this standpoint, category theoretic reasoning is a cornerstone of the paper.
\subsection{De Rham Cohomology}
This last mathematical stanza
seeks to explain the power of the linear approximation in the most general setting, with a view to justifying major results in the text. This includes the linearization step from (3)-(5) and Decomposition 1, specific to Calvo, and 
the general bifurcation result Theorem 7. The first subsubsection sets out preliminaries of differential topology and cohomology. The second states the main result. The best textbook here is probably \cite{tu2011introduction}; some readers may find the informal presentation in \cite{stone2009mathematics} helpful. \subsubsection{Preliminaries}
A \textbf{Differential Structure} on a topological manifold forms a \textbf{Differential Manifold}. It allows us to do calculus on a manifold consistent with its topology. An $n$-dimensional $C^{k}$ manifold is defined by a $C^{k}$-atlas, comprised of a set of bijections, called charts, between a collection of subsets (whose union is $M$) and a family of open subsets in $\mathbb{R}^{n}$
$$\varphi_{i}:M \supset W_{i} \rightarrow U_{i} \subset \mathbb{R}^{n}$$
which are $C^{k}$-compatible.
This means that they can be pieced together in an appropriately smooth fashion.
Formally, consider two charts 
$$\varphi_{i}: W_{i} \rightarrow U_{i} $$
$$\varphi_{j}: W_{j} \rightarrow U_{j} $$
denote the intersection of domains by 
$$W_{ij}=W_{i} \cap W_{j}$$
and its map by the two chart maps to the two images 
$$U_{ij}=\varphi_{i}(W_{ij})$$
$$U_{ji}=\varphi_{j}(W_{ij})$$
It follows that the transition map between the two charts is the map between the two images of this intersection under the two chart maps. 
$$\varphi_{ij}: U_{ij} \rightarrow U_{ji} $$
$$\varphi_{ji}(x)= \varphi_{j}(\varphi^{-1}_{i}(x))$$
Two charts $\varphi_{i}$, $\varphi_{j}$ are $C^{k}$-compatible if and only if $U_{ij}$ and $U_{ji}$ are open and the transition maps $\varphi_{ij}$ and $\varphi_{ji}$ have continuous partial derivatives of order $k$. 
\par A \textbf{Differential Form} is a coordinate-independent approach to calculus. It has its roots in physics (see \cite{stone2009mathematics}). It differs from the approach to integration in the rest of the paper because it is orientation-dependent, so integrals can be zero or negative for non-zero functions. In general, for a differential form $\omega$ integrated over a manifold $M$ and the same manifold but with opposite orientation $M'$ then 
$$\int_{M} \omega = -\int_{M'} \omega$$
A $k$ form is an object that can be integrated over a $k$ dimensional oriented manifold and is homogeneous of degree $k$ in the coordinate differentials. 
\par A one form can be integrated over an oriented interval $[a, \, b]$ in the domain of $f$:
$$\int_{a}^{b}f(x)\, \mathrm{d}x$$
Similarly, the expression $$f(x, \, y, \, z) \,  \mathrm{d}x \wedge \mathrm{d}y + g(x, \, y, \, z) \,  \mathrm{d}z \wedge \mathrm{d}x + h(x,\, y,\, z) \, \mathrm{d}y \wedge \mathrm{d}z$$ is a 2-form, that has the following integral over an oriented surface $S$
$$\int_{S}(f(x, \, y, \,  z) \,  \mathrm{d}x \wedge \mathrm{d}y + g(x, \, y, \, z) \, \mathrm{d}z \wedge \mathrm{d}x + h(x, \, y, \, z) \, \mathrm{d}y \wedge \mathrm{d}z) $$
likewise, the 3-form $f(x, \, y, \, z) \, \mathrm{d}x \wedge \mathrm{d}y \wedge \mathrm{d}z$ is called a volume element, integrated over an oriented region of three dimensional space. $\wedge$ is called the wedge or \textbf{Exterior Product}. It is an alternating product, in the sense that 
$$\mathrm{d}x^{1}\wedge \mathrm{d}x^{2}=- \mathrm{d}x^{2} \wedge \mathrm{d}x^{1} $$ 
because an integral over a square, whose first side is $\mathrm{d}x^{1}$ and second side is $\mathrm{d}x^{2}$, has the opposite orientation to one starting with $\mathrm{d}x^{2}$ and finishing with $\mathrm{d}x^{1}$. From the alternating property, it follows that $\mathrm{d}x^{i} \wedge \mathrm{d}x^{i} =0$, in the same way that the cross product of parallel vectors, whose magnitude is the area of the parallelogram spanned by those vectors, is zero. In higher dimensions 
$$\mathrm{d}x^{i_{1}}\wedge \cdots \wedge \mathrm{d}x^{i_{m}}=0$$
if any two of the indices $i_{1}, \, \cdots , \, i_{m}$ are equal, in the same way that the volume enclosed by a parallelotope, whose edge vectors are linearly dependent. We can lean on our knowledge of the Lebesgue measure. $\mathrm{d}x^{i_{1}}\wedge \cdots \wedge \mathrm{d}x^{i_{m}}$ defines an integral in $m$-dimensional space. If two coordinate sections are dependent then you have an $m-1$ dimensional surface, which has measure zero in the ambient space. 
\par The exterior derivative operator $d$ sends $k$ forms to $k+1$ forms, so perimeters go to areas, surface areas to enclosed volumes etc.\footnote{The axiomization is as follows.
The exterior derivative is the unique $\mathbb{R}$-linear mapping from $k$-forms to $(k + 1)$-forms with these properties:
\begin{enumerate}
\item $\mathrm{d}f$ is the differential of $f$ for a $0$-form $f$.
\item $\mathrm{d}(\mathrm{d}f) = d^{2}= 0$ for a $0$-form $f$.
\item $\mathrm{d}(\alpha \wedge \beta) = \mathrm{d}\alpha \wedge \beta  + (-1)^{p}(\alpha \wedge \mathrm{d} \beta) $ where $\alpha$ is a $p$-form.
\end{enumerate}
Intuitively, the first condition ensures that the exterior derivative is a generalization of differential calculus. The second has been discussed already. The last is a version of the chain rule respecting orientation.}
The generalized Stokes theorem states that the integral of a differential form 
$\omega$ over the boundary of some orientable manifold $\Omega$ equals the integral of its exterior derivative. 
$$\int_{\partial \Omega} \omega = \int_{\Omega} \mathrm{d} \omega $$
This means that the volume of a sphere is equal to the surface area of the family of spheres covering the circle, as their diameters become arbitrarily small. In general, we can approximate the integral over $k+1$- form arbitrarily well, by summing over the boundary of its $k$ form. This constitutes a deep extension of the fundamental theorem of calculus. 
\par Lastly, returning to algebraic topology, it is necessary to introduce \textbf{Cohomology} in general, specifically the \textbf{Singular Cohomology}. Cohomology is the dual construction to homology. It arises from \textbf{Cochains}, which are functions defined on the group of chains in homology theory. 
\par Cohomology develops from the geometry of functions and their pullbacks. Taking two spaces $X$ and $Y$ and a function $F$ on $Y$, any mapping $f: X \rightarrow Y$ will lead to another $F \circ f$ called the pullback of $F$ on $X$ under $f$. Typically, there is interest in the relation between continuity and differentiability properties of the functions. Leading cohomology theories have a product called the \textbf{Cup Product}, which gives them a ring structure. For this reason cohomology can provide richer algebraic invariants.
\par The singular cohomology is a powerful topological invariant, associating a graded-commutative ring to any topological space.\footnote{A graded commutative ring is a ring with a direct sum structure $$R=\bigoplus^{\infty}_{n=0}R_{0} \bigoplus R_{1} \bigoplus R_{2} \bigoplus \cdots $$ of additive groups such that for all non-negative $n, \, m$ it holds that $R_{n}R_{m} \subset R_{n+m}$.} Decomposition 1 exemplifies this structure. Every continuous map $f: X \rightarrow Y$ creates a homomorphism from the cohomology ring of $Y$ to that of $X$; this puts strong restrictions on the behavior of maps between the two spaces. 
\par Recall the chain complex underlying the singular homology described by Figure 4. Now fix an abelian group $A$, and replace each $C_{i}$ by its dual group $C^{*}=Hom(C_{i}, \, A)$ and $\partial_{i}$ by its dual homomorphism.\footnote{Let $U$ and $V$ be abelian groups and $T: U \rightarrow V$ a homomorphism between them (think of it as a linear map). Denote the corresponding dual spaces by $U^{*}$ and $V^{*}$, the dual homomorphism $T^{*}: V^{*} \rightarrow U^{*}$ is the linear mapping with action from  
$$\alpha \rightarrow \alpha \circ T$$ for $\alpha \in V^{*}$.
Since we are interested in cases where $U$ and $V$ are finite dimensional, the dualizing operator can be viewed as a composition of canonical isomorphism
\begin{center}
\begin{tikzcd}[ampersand replacement=\&]
Hom(U, \, V) \arrow{r}{\backsimeq}
\& U^{*} \bigoplus V \arrow{r}{\backsimeq}
\& (V^{*})^{*} \bigoplus U^{*} \arrow{r}{\backsimeq} \&
Hom(U^{*},\, V^{*})
\end{tikzcd}
\end{center}
There are also links with the adjoint operator, an extension of the transpose from basic linear algebra. \cite{aliprantis2007infinite} presents these results in the context of vector spaces. The dual group $C^{*}$ encompasses the idea of all possible movements around the relevant space, in a particular dimension.}
$$d_{i-1}: C^{*}_{i-1}\rightarrow C^{*}_{i}$$
This reverses all the arrows 
\begin{center}
\textbf{Figure 5: Singular Cohomology Cochain Complex}
 \begin{tikzcd}
\cdots 
& \arrow{l} C^{*}_{i+1} 
& \arrow{l} {d_{i}} C^{*}_{i}  &
\arrow{l} {d_{i-1}}  C^{*}_{i-1}  & \cdots  \arrow{l}    
\end{tikzcd}
\end{center}
For an integer $i$ the $i^{th}$ \textbf{Cohomology Group} of $X$ with coefficients in $A$, denoted $H^{i}(X, \, A)$, is defined by $ker(d_{i})/ im(d_{i-1})$. Elements of $C^{*}_{i}$ are referred to as \textbf{Singular i-Cochains} with coefficients in $A$. Each $i$- cochain on $X$ is identified with a function from the family of singular simplices in $X$ to $A$. Elements of $ker(d)$ and $im(d)$ are known respectively as \textbf{Cocycles} and \textbf{Coboundaries}. Members of $ker(d_{i})/ im(d_{i-1})$ are equivalence classes of cocycles and are therefore called \textbf{Cohomology Classes}.
\par Many properties of cohomology are only minor variants of homology: 
\begin{itemize}
\item A continuous function $f: X \rightarrow Y$ determines what is known as a pushforward homomorphism $f_{*}: H_{i}(X) \rightarrow H_{i}(Y)$ on homology  and a pullback homomorphism $f^{*}: H^{i}(Y) \rightarrow H^{i}(X)$ for cohomology. This makes cohomology a contravariant functor from topological spaces to abelian groups. 
\item Two homotopic maps from $X$ to $Y$ induce the same homomorphism on cohomology (as they do on homology.)
\item If a space $X$ is the union of open subsets $U$ and $V$, then there is a long exact sequence 
$$ \cdots \rightarrow H^{i}(X) \rightarrow H^{i}(U) \bigoplus H^{i}(V)  \rightarrow H^{i}(U \cap V) \rightarrow H^{i+1}(X) \rightarrow \cdots $$
\item There are relative cohomology groups $H^{i}(X, \, Y ; A)$, which are related to the usual cohomology group by a long exact sequence 
$$ \cdots \rightarrow H^{i}(X, Y) \rightarrow H^{i}(X) \rightarrow H^{i}(Y) \rightarrow H^{i+1}(X) \rightarrow \cdots $$
Informally, these properties allow cohomology calculations to be broken down over suitable partitions. 
\item The cohomology group over a field $\mathbb{F}$ $H^{i}(X, \, \mathbb{F})$ is the dual space of its homology group $H_{i}(X, \, \mathbb{F})$.
\item The cohomology groups $H^{i}(X, \, A) $ are zero for $i$ greater than the dimension of $X$.
\end{itemize}
\par On the other hand, cohomology has a special structure absent in homology. For any topological space $X$ and commutative ring $R$, there is a bilinear\footnote{A bilinear mapping is a function from two vector spaces to a third which is linear in both arguments. Formally, consider $V$, $W$ and $X$, suppose they are vector spaces over a field $\mathbb{F}$. It would work with weaker structures, where the relevant operators are still defined. A bilinear map is a function 
$$B: V \times W \rightarrow X$$ such that for all $w \in W$, the map $B_{w}$
$$v \mapsto B(v, \, w)$$
is a linear map from $V$ to $X$, likewise for all $v \in V$, the map $B_{v}$
$$w \mapsto B(v,\, w)$$ is a linear map from $W$ to $X$. In other words, holding the first entry constant and letting the second entry vary yields a linear operator and vice versa. Key properties are homogeniety and additivity detailed below 
\begin{itemize}
\item For any $\lambda \in \mathbb{F}$, $B(\lambda v, \, w)= B(v, \, \lambda w)=\lambda B(v,\, w)$
\item If $v_{1}, \, v_{2} \in V$ and $w_{1}, \, w_{2} \in W$, then $B(v_{1}+v_{2}, \, w)=B(v_{1}, \, w) + B(v_{2},\, w)$ and $B(v, \, w_{1}+w_{2})=B(v, \, w_{1}) +B(v, \, w_{2})$
\end{itemize}
Matrix multiplication $M(m,\, n) \times M(n,\, p) \rightarrow M(m,\, p)$ is a familiar example.} map called the \textbf{Cup Product}.  $$H^{i}(X,\, R) \times H^{j}(X,\, R) \rightarrow H^{i+j}(X, \, R)$$
calculated explicitly from the singular cochains. The product of cohomology classes $u$ and $v$ is written $u \cup v$ or even $uv$. This implies that the direct sum 
$$H^{*}(X,\, R)=\bigoplus_{i}H^{i}(X,\, R)$$
forms a graded ring, named the \textbf{Cohomology Ring} of $X$. It has the intrinsic feature of graded commutativity, that for any $u \in H^{i}(X,\, R)$ and $v \in H^{j}(X,\, R)$. 
$$uv=(-1)^{ij}vu$$
For any continuous function $f: X \rightarrow Y$, the pullback $f^{*}: H^{*}(Y,\, R) \rightarrow H^{*}(X,\, R)$ is a homomorphism of graded $R$-algebras. This proves that two homotopy equivalent spaces have isomorphic cohomology rings. 
\par To close, the focus moves to geometric intuition. It is helpful to clarify a few topological terms at the outset. A closed manifold will mean a compact manifold without boundary, whereas a closed submanifold $N$ of $M$ will refer to a submanifold that is a closed subset of $M$ but not necessarily compact (although obviously it will be if $N$ itself is compact).
\par Let $X$ be a closed oriented manifold of dimension $n$. Poincar\'e duality gives an isomorphism $H^{i}(X) \cong H_{n-i}(X)$. As a consequence, a closed oriented submanifold $S$ of codimension $i$ in $X$ determines a cohomology class in $H^{i}(X)$ labelled $[S]$. The cup product describes the intersection of submanifolds, in the sense that, if $S$ and $T$ are submanifolds of codimension $i$ and $j$ that have a transverse intersection\footnote{A transversal intersection is a well-behaved meeting between two surfaces, where there is no singularity. Formally, two submanifolds $L_{1}$ and $L_{2}$ of a larger manifold $M$, form a transversal intersection $L_{1} \pitchfork L_{2}$ if the direct sum of the two tangent spaces forms the tangent space of the ambient space, formally $$\forall p \in L_{1} \cap L_{2}, \, T_{p}M=T_{p}L_{1}+T_{p}L_{2}$$ Another way of looking at transversality, is that the codimension of the intersection must be equal to the sum of the codimension of the two manifolds.}, then $$[S][T]=[S \cap T] \in H^{i+j}(X)$$
where the intersection is of codimension $i+j$, with an orientation determined by the orientation of $S$, $T$ and $X$. In the case of a smooth manifold, this process can still be applied if $S$ and $T$ do not intersect transversely by perturbing $S$ or $T$ to make the intersection transverse. The strategy later will be to effectively shrink the ambient space to fill the hole in the singular surface.
\par Intuitively, for any topological space $X$, elements of $H^{i}(X)$ can be thought of as co-dimension-$i$ subspaces of $X$, that can be moved freely around $X$. For example, members of $H^{i}(X)$ can be defined by a continuous function $f$ from $X$ to a manifold $M$ and a closed codimension-$i$ submanifold $N$ of $M$, with an orientation on the normal bundle. Informally, one can think of the resulting class $f^{*}([N]) \in H^{i}(X)$ as lying on the subspace $f^{-1}(N)$ of $X$; this is justified since the class $f^{*}([N])$ is zero on the cohomology of the open subsets of $X-f^{-1}(N)$. The cohomology class $f^{*}([N])$ can move freely on $X$, in the sense that, $N$ can be replaced by any continuous transformation of itself that stays inside $M$. Poincar\'e duality implies that for a continuous map $\mathbb{E}_{t}X_{t+1}=f(X_{t}, \, \cdot)$ on a state space forming $n$-dimensional manifold, an $n-i$ dimensional manifold, where dynamics behavior differs qualitatively from the rest of the system, is the result of an $i$-dimensional singularity.
\subsubsection{Formalization and Application}
The De Rham Cohomology is a technique common to both algebraic and differential topology. It is a powerful tool capable of expressing basic topological information about smooth manifolds, making it easy to compute concrete representations of cohomology classes. It is a cohomology theory based on the behavior of differential forms. 
\par The key distinction is between exact and closed forms. Here, an \textbf{Exact Form} is is a differential form $\alpha$, that is the exterior derivative of another differential form $\beta$, whilst a closed form is a differential form $\alpha$ whose exterior derivative is zero ($\mathrm{d}\alpha = 0$). Thus an exact form is an image of $\mathrm{d}$ and a closed form is a kernel of $\mathrm{d}$. Every exact form is closed but a closed form need not be exact.\footnote{An intuitive geometric example is that of a circle with the 1-form $\mathrm{d}\theta$ corresponding to the derivative of the angle from a reference point at its center. On the entire circle, there is no function $\theta$ that has $\mathrm{d}\theta$ as its derivative. The increase of $2 \pi$ in going around the circle implies a multi-valued map. Removing one point would alleviate this problem but would create a gap that would change the homology of the manifold. Every circle contains singularities- where it is not differentiable- associated with two vertical tangency in $\mathbb{R}^{2}$.} Poincar\'e's lemma ensures it is true for contractible spaces. However, there is a critical connection between the existence of exactness and the existence of holes. From this standpoint, the De Rham cohomology measures how the Fundamental Theorem of Calculus\footnote{Recall at this point how we argued that the Stokes theorem could be viewed as the generalization of this theorem to higher dimensions.} fails on general manifolds in higher dimensions. \par The De Rham complex is the cochain complex of differential forms on a smooth manifold $M$ with the exterior as its differential: 
\begin{center}
 \begin{tikzcd}
0 \arrow{r}
    & \Omega^{0}(M) \arrow{r}{d}
        & \Omega^{1}(M) \arrow{r}{d} & \Omega^{2}(M) \arrow{r}
                 & \cdots 
\end{tikzcd}
\end{center}
where $\Omega^{0}(M)$ is the space of smooth functions on $M$, $\Omega^{0}(M)$ is the space of 1-forms on $M$ and so on. The De Rham cohomology works by creating equivalence classes of closed forms. One classifies two closed forms $\alpha, \, \beta \in \Omega^{k}(M)$ as cohomologous, if they differ by an exact form that is if $\alpha -\beta $ is exact. This classification induces an equivalence relation on the family of closed forms $\Omega^{k}(M)$. The next step is to define the $k$-th De Rham cohomology $H^{k}_{dR}(M)$ to be the set of equivalence classes, containing closed forms modulo exact forms. Note that we can apply this to functions that are $C^{k}$, $k< \infty$, by setting $\Omega^{k+1}(M)=0$ and replacing some long exact sequences by short ones. 
\par The De Rham cohomology is an extension of Stokes' theorem. Stokes' theorem implies a duality between the De Rham cohomology $H^{k}_{dR}(M)$ and the singular cohomology $H^{k}(M; \mathbb{R})$. De Rham's theorem proves this is in actually an isomorphism. More precisely, consider the map $$I : H^{p}_{dR}(M) \rightarrow H^{p}(M; \mathbb{R})$$
with the following definition: for any $[\omega] \in H^{p}_{dR}(M) $, let $I(\omega)$ be the element of $\newcommand{\Homo}{\text{Hom}}\Homo(H_{p}(M)) \cong H^{p}(M;\mathbb{R})$ which acts as follows: 
$$H_{p}(M) \owns [c] \mapsto \int_{c} \omega $$
The exterior product endows the direct sum of groups with a ring structure. A further consequence of the theorem is that two cohomology rings are isomorphic, when viewed as graded rings, where the analagous product for the singular cohomology is the cup product. 
\par At last, we come to the application, the cohomology computation for our main model. 
\begin{proposition} The Calvo Phillips curve model has the following De Rham cohomology representations 
$$H^{k}_{dR}=\begin{cases} \mathbb{R}, & \text{for } 
$k=0, \,
5, \, 8$ \\ \{0\}, & \text{otherwise } \end{cases}$$
$$H^{k}_{dR}\vert_{\Delta'}=\begin{cases} \mathbb{R}, & \text{for } 
$k=0, \, 
4, \,  7$ \\ \{0\}, & \text{otherwise } \end{cases}$$
$$H^{k}_{dR}\vert_{\sigma =1}=\begin{cases} \mathbb{R}, & \text{for } 
$k=0, \,
4, \, 6$ \\ \{0\}, & \text{otherwise } \end{cases}$$
$$H^{k}_{dR}\vert_{\sqrt{\varepsilon}\rightarrow 0}
=\begin{cases} \mathbb{R}, & \text{for } 
$k=0, \, 2, \,
5$ \\ \{0\}, & \text{otherwise } \end{cases}$$
\end{proposition}
\begin{proof} From Proposition 4 and Proposition 10, it follows that the model in general, away from $\sigma=1$, is comprised of an eight dimensional surface \newline $f=(\pi_{t-1}, \, \pi_{t}, \, y_{t}, \, \Delta_{t}, \, a_{t}, \, a_{t-1}, \, \psi_{t}, \, \psi_{t-1})$ \newline with a one dimensional passage leading to a five dimensional singular surface \newline $f^{sing}(\pi_{t}, \, y_{t}, \, \Delta_{t}, \, a_{t}, \, \psi_{t})$. Focusing on the outer surface first, since the codimension of the bifurcation 
(seven) exceeds its dimension (one) there is always a passage around it. Thus this surface can be homotopically mapped to $\mathbb{R}$. This makes the cohomology in dimensions six and seven trivial. The closed forms in $\mathbb{R}$ are the maps from $\mathbb{R} \rightarrow \mathbb{R}$ such that $df=0$, these are the constant functions justifying the claim in dimension seven. The singular surface is simply connected so the previous argument can repeated in dimension one to
five. The case of dimension zero is the result of connectedness which is established by the continuity of the entire cocycle $\{ f, \, f^{sing}\}$.  
\par The smallest surface relates to dynamics around ZINSS and parallels the analysis of homology. The other rows deal with the rest of the cases in Proposition 10. In particular, $\Delta'$ represents the case where there is no price dispersion. \end{proof}
In conjunction with Theorem 9, this formally justifies the use of the linear approximations (3)-(5) to represent the non-binding constraint on the representative firm or social planner.
\subsection{Marginal Costs and Inflation}
Here, I derive the equation of the wall of the singularity. First, I lay out the steps of the proof of the singularity from Proposition 20.
\begin{proof}
Start from the efficient steady state expression (80) expressed as a function of $\pi_{t}$ and $\pi_{t-1}$ 
\begin{equation} MC^{*}= \bigg(\frac{1-\alpha \beta (1+\pi_{t})^{\theta}}{1-\alpha \beta (1+\pi_{t})^{\theta-1}}\bigg)\bigg(\frac{1-\alpha}{1-\alpha (1+\pi_{t-1})^{\theta-1}}\bigg)^{1/(\theta-1)}\end{equation}
Differentiating with respect to $\pi_{t}$ and $\pi_{t-1}$ and employing the optimality, as before, reveals that
\begin{equation} \mathrm{d} MC^{*}\equiv \frac{\partial MC}{\partial \pi_{t}}\mathrm{d}\pi_{t} +\frac{\partial MC}{\partial \pi_{t-1}}(k\mathrm{d}\pi_{t-1})=0 \end{equation}
\begin{equation} \mathrm{d} MC^{*}\equiv \frac{\partial MC}{\partial \pi_{t}}\mathrm{d}\pi_{t} +\frac{\partial MC}{\partial \pi_{t-1}}(k\mathrm{d}\pi_{t-1})=0 \end{equation}
where $\mathrm{d}\pi_{t}=k\mathrm{d}\pi_{t-1}$ represents the equation of the infinitesimal singularity.
Substituting in and simplifying gives
\begin{multline} 0 = -\beta \bigg(\frac{1-\alpha \beta (1+\pi_{t})^{\theta}}{1-\alpha \beta (1+\pi_{t})^{\theta-1}}\bigg)\bigg(\frac{(1-\alpha)^{1/(\theta-2)}}{1-\alpha (1+\pi_{t-1})^{\theta-1}}\bigg)^{(\theta-2)/(\theta-1)}\mathrm{d}\pi_{t} + \\ \bigg(\frac{1-\alpha \beta (1+\pi_{t})^{\theta}}{1-\alpha \beta (1+\pi_{t})^{\theta-1}}\bigg)\bigg(\frac{(1-\alpha)^{1/(\theta-2)}}{1-\alpha (1+\pi_{t-1})^{\theta-1}}\bigg)^{(\theta-2)/(\theta-1)}(k\mathrm{d}\pi_{t-1})\end{multline}
Cancelling out, it is clear that this implies that $k=\beta$ and $\mathrm{d}\pi_{t}=\beta \mathrm{d}\pi_{t-1}$ which integrates to $\pi_{t}=\beta \pi_{t-1}$, as desired. 
\end{proof}
\begin{remark} In the differential structure, where we set $\mathrm{d}\pi_{t}= \mathrm{d}\pi_{t-1}$ in accordance with the fundamental theorem of calculus, this corresponds to the optimization condition 
$$ \frac{\partial MC_{t}}{\partial \pi_{t}} =  \beta \frac{\partial MC_{t}}{\partial \pi_{t-1}}$$
that is a feature of the first best solution. It is the first order condition for inflation to minimize discounted (real) marginal costs. \end{remark}
The forward-looking form (217) implements the alternative optimality condition 
$$ \frac{\partial MC_{t}}{\partial \pi_{t+1}} =  \alpha  \frac{\partial MC_{t}}{\partial \pi_{t}} $$
This states that the costs to society, in terms of the increase in present marginal costs, from expected inflation must balance out the effect from present inflation, discounted by the share of firms whose prices are expected to stay fixed.\footnote{The reason there is no time discounting is that these effects work through the (forward-looking) choice of the present price instrument so the relationship should read 
$$ \beta \frac{\partial \pi_{t+1}}{\partial p^{*}_{t}} \frac{\partial MC_{t}}{\beta \partial \pi_{t+1}} =  \alpha \frac{\partial \pi_{t}}{\partial p^{*}_{t}}  \frac{\partial MC_{t}}{\partial \pi_{t}} $$
with the time preference cancelling out at ZINSS, where $\frac{\partial \pi_{t}}{\partial p^{*}_{t}}=\frac{\partial \pi_{t+1}}{\partial p^{*}_{t}}$. Each form of the wall-crossing singularities represent a different first best condition true at ZINSS.} It is the condition for minimizing long-run marginal costs, using present and announced future reset pricing rules. The constraint interpretation of the wall of the singular surface is once again confirmed. 
\section{Further Keynesian Models}
The first two subsections lay out two Keynesian models that come up in the course of the discussion in Section 10. The aim is not to solve them out fully but to provide the reader with background. The last part verifies claims made in Section 10.1 concerning unusual policy settings. The idea is that firms re-optimize the price periodically. The difference with Calvo is that they know the length of contract for sure, in advance.  Contracts overlap so that in each period the same fraction come up for renewal. This set up seems intuitively appealing.
\subsection{Taylor Pricing}
The \cite{taylor1979staggered} model of staggered contracts is one of the most popular in applied macroeconomics. The idea is that firms re-optimize the price periodically. The difference with Calvo is that they know the length of contract for sure, in advance. Contracts overlap so that in each period the same fraction come up for renewal. This set up seems intuitively appealing.
\par The objective function for a resetting firm is just (24) summed over the length of the contract $M$. Assume for simplicity at the outset that $M=2$, as prominent in the text. The focus is on the small noise limit around ZINSS $O(\sqrt{\varepsilon})$. For clarity, I will focus on the non-stochastic expansion. It is clear that the price change will be the weighted average of the expected changes in marginal cost and aggregate price level throughout the contract. 
\begin{equation} \hat{p}^{*}_{t}= \frac{1}{2}\hat{P}_{t} + \frac{1}{2}\hat{P}_{t+1} + \frac{1}{2}\hat{mc}_{t} + \hat{mc}_{t+1} \end{equation}
\begin{equation} \hat{p}^{*}_{t-1}= \frac{1}{2}\hat{P}_{t-1} + \frac{1}{2}
\hat{P}_{t} + \frac{1}{2}\hat{mc}_{t-1} + \frac{1}{2}
\hat{mc}_{t} \end{equation}
The price level is simply
\begin{equation} \hat{P}_{t}=\frac{1}{2}\hat{p}^{*}_{t-1} + \frac{1}{2}\hat{p}^{*}_{t}\end{equation}
Lagging this relationship leads to 
\begin{equation} \hat{P}_{t-1}=\frac{1}{2}\hat{p}^{*}_{t-2} + \frac{1}{2}\hat{p}^{*}_{t-1}\end{equation}
Hence, (352)-(353) 
\begin{equation} \pi_{t}= \frac{1}{2}(\hat{p}^{*}_{t}-\hat{p}^{*}_{t-1}) + \frac{1}{2}(\hat{p}^{*}_{t-1}-\hat{p}^{*}_{t-2})\end{equation}
(350)-(351) evaluates to 
\begin{equation} \hat{p}^{*}_{t}-\hat{p}^{*}_{t-1} = \frac{1}{2} \pi_{t} + \frac{1}{2} \pi_{t+1} + \frac{1}{2} (\hat{mc}_{t} -\hat{mc}_{t-1}) + \frac{1}{2} (\hat{mc}_{t+1} -\hat{mc}_{t})
\end{equation}
lagging, substituting in, then simplifying yields
\begin{equation} \pi_{t}= \frac{1}{2}\pi_{t-1} + \frac{1}{2}\pi_{t+1} + \frac{1}{2} (\hat{mc}_{t-1} -\hat{mc}_{t-2}) + \frac{1}{2} (\hat{mc}_{t+1} -
\hat{mc}_{t}) + 
(\hat{mc}_{t} -\hat{mc}_{t-1}) 
\end{equation}
As argued in Section 9.3, inflation is determined half by its lag and half by its lead. Furthermore, the terms occur in difference pairs, so it is clear that the sum of the real coefficients on marginal costs is zero, in keeping with real neutrality. These results extend over to contracts of general length $M \geq 2$, where 
\begin{multline} \hat{p}^{*}_{t}= \frac{1}{M}\hat{P}_{t} + \frac{1}{M}\hat{P}_{t+1} + \cdots + \frac{1}{M}\hat{P}_{t+M-1}  + \\ \frac{1}{M}\hat{mc}_{t} + \frac{1}{M}\hat{mc}_{t+1} + \cdots + \frac{1}{M}\hat{mc}_{t+M-1}\end{multline}
\begin{equation} \pi_{t}= \frac{1}{M}(\hat{p}^{*}_{t}-\hat{p}^{*}_{t -(M-1)})\end{equation}
before a similar substitution process yields the Phillips curve \begin{multline} \pi_{t}= \frac{1}{M}\pi_{t-(M-1)} + \cdots + \frac{1}{M}\pi_{t + M-1} + \\ \frac{1}{M}\hat{mc}_{t+M -1} + \cdots + \frac{1}{M}\hat{mc}_{t} - \frac{1}{M}\hat{mc}_{t-1} - \cdots - \frac{1}{M}\hat{mc}_{t-M} 
\end{multline} 
where I have employed a more compact form than with the two period contracts. The symmetry would naturally extend to generalizations of Taylor pricing with differing contract lengths (\cite{dixon2012generalised}).
There is a powerful link back to generalized stochastic mean back in Appendix B.2 and crystallized in Proposition 27. The property that the optimal reset price is a weighted average of all the flexible prices extends to inflation determination, if uncertainty and discounting (efficient or otherwise) vanish. This is a universal result.
\subsection{Calvo Wage Phillips Curve}
This subsection shows a wage Phillips curve problem and how results of Theorem 
8 apply. First, I introduce imperfect competition into the labor market before showing that the associated resetting problem is analogous to that of the price Phillips curve (26)-(31).
\subsubsection{Labor Unions}
In the benchmark model of Section 4 the labor market is perfectly competitive. To generate wage rigidity, we need to introduce imperfect substitutability between wage setters. The easiest way to achieve this is to introduce unions, which combine different types of labor into a composite labor service, that 
is leased to firms at a wage rate $W$. 
The aggregation is as follows 
\begin{equation} L_{t}=\bigg(\int_{0}^{1}l_{t}^{(\theta_{w}-1)/\theta_{w}}(x) \; \mathrm{d}x\bigg)^{\theta_{w}/(\theta_{w}-1)}\end{equation}
where $x$ indexes the differentiated labor inputs and $\theta_{w}>1$ the elasticity of substitution among different varieties, which populate the unit interval. The profit maximization problem for a union in a competitive market reads 
\begin{equation} \max_{l_{t}(x)}\bigg(\int_{0}^{1}l_{t}^{(\theta_{w}-1)/\theta_{w}}(x) \; \mathrm{d}x\bigg)^{\theta_{w}/(\theta_{w}-1)}W_{t}-\int_{0}^{1}w_{t}(x)l_{t}(x)\; \mathrm{d}x
\end{equation}
This yields the labor demand system 
\begin{equation} l_{t}(x)=\bigg(\frac{w_{t}(x)}{W_{t}}\bigg)^{-\theta_{w}}L_{t}
\end{equation}
and the wage index 
\begin{equation} W_{t}=\bigg(\int_{0}^{1}w_{t}(x)^{1-\theta_{w}}\; \mathrm{d}x\bigg)^{1/(1-\theta_{w})}\end{equation}
There is also wage dispersion 
\begin{equation}\Delta^{w}= \int_{x}{\bigg(\frac{p_{x}}{P}\bigg)}^{-\theta_{w}}\; \mathrm{d}\mu_{x}\end{equation}
with evolution analogous to price dispersion (35)
\begin{equation} \Delta^{w}_{t}=\frac{({1-\alpha_{w}(1+\pi^{w}_{t})^{\theta_{w}-1})}^{\theta_{w}/(\theta_{w}-1)}}{(1-\alpha_{w})^{1/(\theta_{w}-1)}} + 
\alpha_{w}(1+\pi^{w}_{t})^{\theta_{w}}\Delta^{w}_{t-1} \end{equation}
There is also a slight alteration to the market clearing condition (39) 
\begin{equation}\Delta^{w}_{t}C_{t}=A_{t}L_{t}
\end{equation}
\subsubsection{Optimization Problem}
The union sets wages to maximize the welfare of its members. It gets to reset wages with probability $\alpha_{w}$ each period. The problem can be written as follows 
\begin{multline} \max_{w^{*}_{t}(x)} \mathcal{L}_{t}= \mathbb{E}_{t} \sum_{T=t}^{\infty}(\alpha_{w} \beta )^{T-t}\Bigg[\psi_{T}u'(C_{T})w^{*}_{t}(x)\Pi^{-1}_{t, \, T}\bigg(\frac{w^{*}_{t}(x)\Pi^{-1}_{t, \, T}}{W_{T}}\bigg)^{-\theta_{w}}L_{T} \\ -\bigg(\frac{w^{*}_{t}(x)\Pi^{-1}_{t, \, T}}{W_{T}}\bigg)^{-\theta_{w}(1+\eta)}\frac{L^{1+\eta}_{T}}{1+\eta}\Bigg]\end{multline}
where I have used the functional form for the disutility of labor.
The first order condition is 
\begin{equation} (w^{*}_{t})^{1+\theta_{w}\eta}=\frac{\theta_{w}}{\theta_{w}-1}\frac{\aleph^{w}_{t}}{\beth^{w}_{t}}\end{equation}
with 
\begin{equation} \aleph^{w}_{t}=\mathbb{E}_{t}\sum_{T=t}^{\infty}(\alpha_{w} \beta )^{T-t} (W_{T})^{\theta_{w}(1+\eta)}\Pi^{\theta_{w}(1+\eta)}_{t, \, T}(\Delta^{w}_{T})^{1+\eta}L^{1+\eta}_{T}
\end{equation}
\begin{equation} \beth^{w}_{t}=\mathbb{E}_{t}\sum_{T=t}^{\infty}(\alpha_{w} \beta )^{T-t} \psi_{T}u'(C_{T})(W_{T})^{\theta_{w}}\Pi^{\theta_{w}-1}_{t, \, T}L_{T}\end{equation}
Optimal reset wage can be viewed as a generalized weighted mean of flexible wages, according to Proposition 27 since
\begin{equation} (w^{f}_{t})^{1+\theta_{w}\eta}=\frac{\theta_{w}}{\theta_{w}-1}
\frac{\nu'(L_{t})}{\psi_{t}u'(C_{t})}
\end{equation}
which are set as a mark up over the marginal rate of substitution between labor and leisure, which is decreasing in the substitutability between labor varieties. Aggregate wages evolve according to 
\begin{equation} W_{t}^{1-\theta_{w}}=(1-\alpha_{w})(w^{*}_{t})^{1-\theta_{w}}+\alpha_{w} \bigg(\frac{W_{t-1}}{1+\pi_{t}}\bigg)^{1-\theta_{w}}\end{equation}
The difference with the price level construction equation is that we are using real wages as opposed to nominal prices. The final and crucial step is to expand out the recursions (369) and (370)
\begin{equation}  \aleph^{w}_{t}=W_{t}^{\theta_{w}(1+\eta)}(\Delta^{w}_{t})^{1+\eta}L_{t}^{1+\eta}+ \alpha_{w}\beta \mathbb{E}_{t}(1+\pi_{t+1})^{\theta_{w}(1+\eta)}\aleph^{w}_{t+1}\end{equation}
\begin{equation}\beth^{w}_{t}= \psi_{t}u'(Y_{t})W_{t}^{\theta_{w}}L_{t} + \alpha_{w}\beta \mathbb{E}_{t}(1+\pi_{t+1})^{\theta_{w}-1}\beth^{w}_{t+1}\end{equation}
Linearizing these equations at ZINSS, it is clear that a common root in the lag polynomials will arise with 
\begin{equation} \mathbb{L}_{\aleph^{w}}=\mathbb{L}_{\beth^{w}}=\alpha_{w}\beta
\end{equation}
This confirms we can apply Theorem 7 and conclude a bifurcation invalidates the linear approximations at ZINSS, without having to go through a detailed solution. Moreover, this analysis extends to larger models typically used in central banks, which feature both price and wage rigidity, like \cite{christiano2005nominal} and \cite{smets2007shocks}.
\subsection{Unconventional Policy Settings}
This final subsection looks at monetary policy at the unconventional settings discussed in Theorem 8, Propositions 23 and 24. The main result housed in the first subsection here is a proof of Proposition 22. Implicitly both rest on Theorem 3 and Proposition 16. There is some further discussion. The special case of no persistence occupies the second part.
\begin{subsubsection}{Proof of Proposition 23}
\begin{proof}
Focus on the case where the central bank seeks to implement Divine Coincidence, with a policy that destroys the present demand shock alongside the present output deviation, in order to stabilize the demand side of the economy. This is without loss of generality, as other arguments would merely generate additional persistence. Appendix E.1.4 implies that it would have to set
$$\hat{i}_{t}=-\hat{\psi}_{t}-\frac{\sigma} {\beta}\hat{y}^{e}_{t}$$ 
This would imply $$\hat{y}^{e}_{t}=-\psi_{t}$$ substituting into (176) would create a Phillips curve, where inflation evolves autonomously facing two non-degenerate shock terms. It is clear from
It is clear from (298), (299) and (302) that
$$b= \beta(1+\alpha) $$
$$\tilde{b}_{3}=\alpha \beta^{2}$$
whilst $\tilde{b}_{0}=1$ is unchanged. 
This implies the characteristic equation takes the form
$$\lambda^{2}-\beta(1+\alpha)\lambda -\alpha =0$$
It is clear this equation factorizes with two roots inside the unit circle ensuring persistence $$\lambda = \alpha \beta, \, \beta $$ In the limit where $\beta \rightarrow 1$, it is clear that there are not enough eigenvalues outside the unit circle for a solution. Elsewhere, for any initial conditions, it will leave the neighborhood of ZINSS at $O(\beta^{-T})\sqrt \varepsilon $ with probability one.
\end{proof}
\begin{remark}
The error structure would have surprising properties at this policy setting. When $\beta \rightarrow 1$, it would take the form
$$-\frac{\alpha \tilde{b}^{\circ}_{4}}{1+\alpha}\hat{\psi}_{t} + \frac{\tilde{b}^{\circ}_{4}}{\alpha(1+\alpha)}\hat{\psi}_{t-1}$$
This would reverse the direction of the initial impulse response. The fact that this case is ruled out ex post is an example of the model preserving equilibriating mechanisms. In this case, it is likely to be inflation rising in response to demand shocks. This asymmetry is possible in opposition to Theorem 6 because monetary policy has created a new source of market failure, by destroying households' inter-temporal substitution possibilities. 
\end{remark}
\begin{subsubsection}{No Persistence Case}    
\end{subsubsection}
\begin{proposition} At standard output smoothing settings, the policy in Proposition 24 that eliminates persistence is feasible.\end{proposition}
\begin{proof}
The persistence eliminating policy is equivalent to setting $a_{\pi}$ so as to make $\tilde{b}_{0}=0$. This creates a forward-looking system with two jump variables. Consider the standard case where $\beta \rightarrow 1$, $\sigma=1$. The policy reaction to inflation can be expressed as 
\begin{equation} \hat{i}_{t}=- \frac{\alpha(1+a_{y})}{(1-\alpha)^{2}(1+\eta)}\pi_{t}\end{equation}
Thus the economy is governed by 
\begin{equation} \mathbb{E}_{t}\pi_{t+1}= \frac{(1 + \alpha + \tilde{\omega})}{\alpha}\pi_{t}- \frac{\tilde{\omega}}{\alpha}\bigg( \alpha(1+a_{y})-1\bigg)\hat{y}^{e}_{t}
\end{equation}
\begin{equation} \mathbb{E}_{t}\hat{y}^{e}_{t+1}=- \frac{\alpha(1+a_{y})}{(1-\alpha)^{2}(1+\eta)}\pi_{t} + (1+a_{y})\hat{y}^{e}_{t}
\end{equation}
Thus the eigenvalue polynomial reads 
\begin{equation} \bigg(\frac{(1 + \alpha + \tilde{\omega})}{\alpha}-\lambda \bigg)\bigg( 1+a_{y}-\lambda\bigg)- \frac{(1 + \alpha + \tilde{\omega})}{(1-\alpha)^{2}(1+\eta)}\bigg( \alpha(1+a_{y})-1\bigg)=0 \end{equation}
Expanding out and removing the substitution (54) yields
\begin{equation}
 \lambda^{2}-\bigg( \frac{1 + \alpha}{\alpha} + \frac{(1-\alpha)^{2}}{\alpha^{2}}\frac{(1+\eta)}{(1+a_{y})} + 1+a_{y}\bigg)\lambda  +  \frac{(2+a_{y})}{\alpha}+ \frac{(1-\alpha)^{2}}{\alpha^{2}}(1+\eta)=0
\end{equation}
Converting to the inverse eigenvalue form 
(49) implies $q_{0}=p/r$, whilst $q_{1}=1/r$, where respectively $$r=\frac{(2+a_{y})}{\alpha}+ \frac{(1-\alpha)^{2}}{\alpha^{2}}(1+\eta)$$
$$p=-\bigg( \frac{1 + \alpha}{\alpha} + \frac{(1-\alpha)^{2}}{\alpha^{2}}\frac{(1+\eta)}{(1+a_{y})} + 1+a_{y}\bigg)$$
Recall (52) and (53) are the conditions for the desired configuration. The latter is met because $r>2$, hence the former condition amounts to \newline $\vert p \vert -r -1$ having to be strictly negative. This follows from the calculation $$\vert p \vert -r -1=-\bigg(\frac{(1-\alpha)}{\alpha}(1+a_{y}) + a_{y}\frac{(1-\alpha)^{2}}{\alpha^{2}}\frac{(1+\eta)}{(1+a_{y})} \bigg)<0$$
which is self-evident for $a_{y} \geq 0$.
\end{proof}
The final result demonstrates robustness to uncertainty that might prevent the government destroying the lag of inflation.
\begin{proposition} Suppose there is uncertainty about the parameter settings $\delta >0$, then a non-persistence policy can be implemented $O(\delta\sqrt{\varepsilon})$, in the case where $1-\beta << \sqrt{\varepsilon} << \delta << \vert \varepsilon \vert $.
\end{proposition}
\begin{proof} In the light of (376), this amounts to computing the limit as $a_{\pi} \rightarrow a^{*}_{\pi}= \alpha(1+a_{y})/(1-\alpha)^{2}(1+\eta)$ and then noting that this implies persistence will be $O(\delta\sqrt{\varepsilon})$, thanks to the epsilon management scheme. All that remains is to check the eigenvalues
$$\bigg( \lambda -\frac{1}{\alpha}\bigg)\bigg( \lambda^{2}- \bigg\{ 2+a_{y} + \frac{(1-\alpha)^{2}}{\alpha}(1+\eta)\bigg\} \lambda + a_{y} \bigg)=0$$
and the same Rouche's theorem argument goes through for $a_{y} \geq 0$.
\end{proof}
In fact, the exact roots could be calculated from the expression in Footnote 102. The next section should make it easy to appreciate the prevalence of parameter uncertainty.
\end{subsubsection}
\section{Empirical Robustness}
This section has two components; the first justifies the source of parameters selected and identifies sources of uncertainty and reasonable grounds for disagreement. The second looks at how robust the Phillips curve is to these different settings and undertakes a brief comparison with the existing benchmark. The findings are broadly supportive of the new solution. 
\subsection{Parameter Selection}
This subsection explains the parametization choices in the paper. 
There are detailed comparisons with econometric evidence and where possible priors commonly used in macroeconomics. It has four divisions, reflecting the three main parameters and a fourth briefly discussing the policy rule.
\subsubsection{$\sigma =1$}
Throughout the calibration we use $\sigma =1$. This is motivated by the balanced growth path refinement proposed in Appendix D.1.2. In the non-stochastic limit, $\sigma =1$ is necessary for the economy to experience a constant growth rate without labour supply permanently increasing or decreasing. In the non-stochastic limit, $\sigma =1$ is necessary for the economy to experience a constant growth rate without labour supply permanently increasing or decreasing. This is a standard assumption in the growth literature and can be seen as the theoretical justification for the ubiquitous practice of filtering or de-trending data before business cycle analysis. Indeed, empirical work in many areas of long-run macroeconomics favours an estimate close to unity, which is the value suggested by the UK government economic service (see \cite{groom2019new}). Crucially, it simplifies calculations significantly by causing inter-temporal substitution and wealth effects on labour supply to cancel out.
\par The choice is plausible but not unproblematic from an econometric standpoint. Macroeconometric studies typically produce estimates that are large but often imprecise and weakly identified (see \cite{hall1988intertemporal}, \cite{yogo2004estimating}, \cite{havranek2015measuring} and \cite{ascari2021empirical}). They often produce more plausible estimates when the focus is on durable goods, as in studies such as \cite{mankiw1982hall} and \cite{ogaki1998measuring}.
\par Nevertheless, it is microeconometric evidence that seems more reliable here. \cite{crump2022subjective} estimate $\sigma=2$, using individual expectations data from a consumer survey. Meta-analysis from \cite{havranek2015measuring} and \cite{havranek2015cross} come to a similar answer. Although, the former favours a higher estimate of between $3$ and $4$, to reflect publication bias, which causes systematic under-reporting of low or negative estimates. His confidence interval rules out $\sigma<1.25$. These estimates only concern those who do not face binding borrowing constraints. It has been recognised since \cite{vissing2002limited} that these consumers will not follow the standard Euler equation. There is widespread evidence of credit constraint.
\par As a refinement to our set-up, imagine there were a substantial fraction of consumers who were shut out of the financial system. Suppose, for simplicity, they behaved as though they had $\sigma \rightarrow 0$. \cite{kaplan2014wealthy} estimate the fraction of households living hand-to-mouth could be as high as one-third in the United States and UK. If I took $\sigma =2$ as the value for the unconstrained, it would imply an average elasticity of substitution of $4/3 \approx 1.3$, which is closer to unity. Indeed, $\sigma=1$ would lie inside the \cite{havranek2015measuring} confidence bounds since the lower bound estimate for the unconstrained of $1.2$ would correspond to an aggregate value of  $\sigma =0.80$ under the  confidence interval.
 \par There are several recent estimates that imply higher values of $\sigma > 4$, such as \cite{best2020estimating} and \cite{landais2021value}, concerning mortgage notches and benefit cut-offs respectively. However, these may be context specific and appear too large to apply to aggregate consumption.\footnote{In the context of risk aversion, \cite{o2018modeling} caution against apply estimates from one market or economic environment to others. The role of $\sigma$ as the coefficient of relative risk aversion has been downplayed here. This is because in the small noise limit it does not come into play, as risk vanishes with the second order terms. This has been widely understood in microeconomics since \cite{arrow1971}. It was via the calibration theorem introduced by \cite{rabin2000expected} and \cite{rabin2001anomalies}, subsequently extended by \cite{safra2008calibration}. This would change with significant noise, although there is no particular evidence of stronger risk aversion than inter-temporal substitution forces. \cite{havranek2015measuring} failed to detect a significant difference between estimates based on risk aversion rather than inter-temporal substitution. Of our two anomalous results, \cite{landais2021value} focuses on risk aversion whilst \cite{best2020estimating} deals with substitution. Consult \cite{barseghyan2018estimating} for a broader base of evidence.} My calibration is within range of the \cite{smets2007shocks} prior centred at 1.5 with standard deviation 0.37.\footnote{This is $\sigma_{c}$ in their notation. Their prior is normal so my choice has a p-value $0.912$, which lies at the lower end of their plausible range, with respect to standard significance levels.} Overall, setting $\sigma =1$ does not seem to be an unreasonable simplification of econometric evidence, although efforts could be made to relax this assumption in future work. 
\subsubsection{$\eta=4$}
This is the problem parameter. In general, microeconometric evidence of low elasticities (and therefore high $\eta$) clashes with macroeconometric studies that favor the opposite conclusion. On the one hand, at the intensive margin, which is technically the only margin in operation in this representative agent framework, labor supply is usually found to be almost unresponsive particularly for primary earners (see \cite{meghir2010labour} and \cite{keane2011labor}).
The standard real business cycle model requires $\eta < 1/2$ to come close to generating plausible labor supply volatility (see \cite{king1999resuscitating} and \cite{whalen2017estimates}). It is currently unclear how Keynesian forces would affect this discrepancy. 
\par When I broaden the evidence base to include the intensive margin, results are frequently more favourable to a synthesis with $\eta$ around $2$ or $2.5$ on average, as argued by \cite{whalen2017estimates} and \cite{elminejad2021publication}. This is because it is often costly to adjust hours in full-time employment,
which makes participation decisions relatively elastic for second earners and those nearing retirement. Unfortunately, publication bias exerts a server downward bias on average estimates for $\eta$.
The issue is once again that it is rational for individual researchers to throw out negative or insignificant elasticity estimates, on theoretical grounds, with adverse consequences for the profession.
\par Natural experiments give conflicting results. \cite{sigurdsson2019labor} exploits the 1987 tax holiday in Iceland to discover a substantive labor supply response consistent with $\eta$ around $2$. His small positive estimates of the extensive margin reaction is within a standard range. The novel finding is a large response at the intensive margin, particularly concentrated amongst men, who took up second jobs. This is unusual because the previous consensus was that labor supply is less responsive for males than females. However, it chimes with findings of elastic labor supply in sectors where hours of work have greater flexibility (see for example, \cite{fehr2007workers}, \cite{farber2015you}, \cite{gine2017labor}, \cite{mas2019labor}, \cite{tazhitdinova2022increasing}  and \cite{angrist2021uber}). 
\par Nevertheless, a couple of similar studies come up with very high values of $\eta$ for tax experiments in Switzerland (\cite{martinez2021intertemporal}) and Argentina (\cite{tortarolo2020takes}). \cite{sigurdsson2021norwegian} argues these estimates can be reconciled by appealing to different labor market structures in these countries, that make labor supply systematically less flexible for example in Switzerland than Iceland. As the authors acknowledge, Argentina is known to have a heavily regulated and unionized labor market.\footnote{This harks back to a literature linking macroeconomic shocks to labor market institutions, associated with \cite{blanchard2000role}, \cite{alesina2005work} and more recently \cite{rogerson2022shocks}, although the emphasis on the short rather than medium run is something of a twist.} The paper focuses on Norway, which has a structure somewhere between Iceland and Switzerland, where the evidence is consistent with $\eta \approx 6$.
\par Finally, turning to the macroeconometric studies, authors like \cite{elminejad2021publication} are surely too critical of macroeconometric methods. These studies are better able to account for general equilibrium effects than microeconometrics based on within life cycle comparisons, as set out by 
\cite{gottlieb2021measurement}. There is a body of recent work explaining labor market dynamics, with some success, using elasticities implying $\eta$ between one and two, such as \cite{krusell2017gross}, \cite{chang20192018}, \cite{park2020consumption} and \cite{kneip2020aggregation}.\footnote{
\cite{smets2007shocks} have a prior for our $\eta$ (their $\sigma_{L}$) with mean $2$ and standard deviation $0.75$ capturing these cases.} \cite{hall2009reconciling} explicitly demonstrate how an economy with a high value of $\eta$ and a shock to labor market frictions can mimic one with a more standard macroeconomic calibration for the parameter.
\par It is important to recognise that objectives may differ between microeconomics and macroeconomics. The paper has demonstrated that pricing frictions make for a complicated dynamic model. In the interests of parsimony one wants to simplify the labor market, as far as possible. For prediction and forecasting purposes the main objective may be to select a value of $\eta$, which closes the model with an accurate Okun's law relationship, even if this is not an "accurate" reflection of microeconomics. After all, this is effectively what frequentist estimation does. The final procedure is to consider a range of estimates between $1$ and $6$. The lower bound is motivated by the recent macroeconomic studies. The upper bound is motivated by the confidence interval of \cite{sigurdsson2021norwegian} and the particular emphasis on $4$ is that is the bottom value considered justifiable by \cite{elminejad2021publication}. Robustness is important; my ad hoc selections are no substitute for rigorous econometric analysis. I predict that the techniques in this paper will help to analyse models with more realistic labor market frictions and heterogeniety.
\subsubsection{$\alpha =2/3$}
This parameter governs the degree of nominal rigidity. Its bounds are the tightest, reflecting the wide availability of large price databases from statistical agencies, as well as the overwhelming evidence of nominal rigidity.
Nevertheless, there are still challenges and some uncertainty in mapping a heterogeneous environment into the single parameter of a benchmark model. 
\par In US data headline prices are highly flexible, changing around once a quarter on average \cite{bils2004some}. Similar results have been reported for Israel (\cite{baharad2004price}) and Britain (\cite{bunn2012individual}). However, they are appreciably less flexible in the Eurozone, with only $15\%$ changing each month (see \cite{alvarez2006sticky} and \cite{dhyne2006price}). The consensus is that headline figures overstate the flexibility of prices.
\par The first issue is with sales prices. There is a general concern that they might be orthogonal to macroeconomic developments and as such they ought to be excluded from calibrations of adjustment frequency here. \cite{nakamura2008five} show that between $60\%$ and $86\%$ of sales prices return to the same level afterwards. There are a couple of theoretical rationales. \cite{kehoe2015prices} use a menu cost model to show how the transitory nature of sales reduces their impact on the overall flexibility of the price level. \cite{guimaraes2011sales} suggests that firms in direct competition have an incentive to stagger sales, smoothing away much of their effect on inflation. \par Using scanner data \cite{eichenbaum2011reference} shows that "reference prices"- defined as the modal price in a given quarter- are considerably more sticky than headline prices. They then describe price-setting by a simple and accurate rule and demonstrate that the degree of nominal rigidity roughly mimics a menu cost model, calibrated to fit the frequency of reference price adjustments, indicating this is probably the right moment to target. Finally, \cite{anderson2017informational} report institutional details that suggest retail sales are predominantly fixed in advance, in addition to econometric evidence that they are unresponsive to economic conditions and are used in some instances to hide permanent price increases. On the other hand, \cite{gorodnichenko2017price}, \cite{anderson2017informational}, \cite{dixon2020impact}, \cite{kryvtsov2021cyclicality} and \cite{carvalho2021price} do find some responsiveness to business cycle conditions.\footnote{There is also some evidence such as \cite{klenow2007sticky}, \cite{sheremirov2020price} and several of the previous papers which uncover some responsiveness of sales prices and the size of reductions to inflationary conditions.} I advocate these effects be ignored in the small noise limit. By way of theoretical justification, consider a model, such as \cite{alvarez2020temporary}, where firms pay a fixed cost to change their price plan. In the small noise limit these small thresholds would never be met. \par Moreover, there are several other arguments in favor of dropping sales. \cite{klenow2010microeconomic} shows that aggregation to quarterly frequency, as is standard in DSGE, depresses any biases. It reduces or removes differences between continental Europe and Anglo-Saxon nations, where discount strategies are less common (see for example \cite{berardi2015more} and \cite{sudo2014micro}).\footnote{\cite{berardi2020everyday} shows that a leading retailer, with a policy of no discounting, has a similar degree of rigidity of regular prices, as competitors who do discount. This seems to further support the idea of removing sales.} Finally, \cite{nakamura2008five} shows that for the US the median frequency matches our calibration of $\alpha =2/3$. 
\par The final issue is with heterogeniety. The frequency distribution across sectors is heavily right skewed so the median is less than the mean. \cite{bils2004some} have favored calibrating to the median for models with no ex ante heterogeniety. In models with greater heterogeniety, overall nominal rigidity is dominated by the slow-adjusting sectors. This favors a longer length of calibration, if one is interested in reflecting these concerns (see \cite{dixon2011contract}). This motivates the robustness checks with $\alpha=4/5$. 
\par The lower bound $\alpha=3/5$ is justified by the recent finding that internet prices are more flexible (see \cite{gorodnichenko2018price}). The overall effect is quite small since differences seem to be concentrated in the subset of online only retailers, according to \cite{cavallo2016billion}. There is also the issue that these innovations do not fit the ergodic structure of our model and are only slowly being incorporated into official statistics. Finally, the lower bound is consistent with previous DSGE priors which favoured price flexibility.\footnote{In Table 1A, \cite{smets2007shocks} use a beta distribution with mean $0.5$ and standard deviation $0.1$ for $\xi_{p}$ their counterpart to $\alpha$. These are associated with location and scale parameters both equal to $12$ and $2/3$ falls just outside its standard one sided confidence set with p-value approximately $0.0448$, whereas $0.6$ is inside at around $0.164$.} 
\subsubsection{Policy Rule}
A few comments are in order regarding the choice of policy rule parameters. I intended to use the standard $(a_{\pi}, \, a_{y})'=(1.5, \, 0.5)'$, associated with the original Taylor rule. I adjusted the inflation reaction down to $0.5$, when I discovered the rule does not yield an equilibrium solution. With previous intuition about the policy rule thoroughly overturned, I decided to aim for maximum robustness. Settings for the output reaction between $0$ and $2.5$ are considered. I suspect a more tightly bounded set of parameters, describing interest rate setting, will become available, as our understanding of optimal monetary policy develops.
\subsection{Phillips Curve}
This subsection is split in two. The first part covers two alternative benchmarks. The second studies robustness at the standard settings and draws a comparison with the existing solution and empirics. Most emphasis is placed on the output slope coefficient in relation to macroeconometric estimates.
\subsubsection{Alternative Cases}
Beyond the benchmark there are two focal cases. The first is where policy is inactive. The second is approaching the point of blowup, where $a_{\pi} \rightarrow 1$. At otherwise standard settings these are 
\begin{equation} \pi_{t} = 0.4\pi_{t-1} -0.111\hat{y}^{e}_{t} - 0.222\hat{\Delta}_{t} +  0.267\mathbb{E}_{t}\pi_{t+1} + 0.333(\hat{\psi}_{t}-\hat{\psi}_{t-1})\end{equation}
\begin{equation} \pi_{t} = 0.7\pi_{t-1} - 0.25\hat{\Delta}_{t} +  0.3\mathbb{E}_{t}\pi_{t+1} + 0.25(\hat{\psi}_{t}-\hat{\psi}_{t-1})\end{equation}
\par Note that in the later case only the lagged inflation coefficient changes substantially from the text. Movements in the other coefficients are much smaller. It is interesting to look at how the cut-off value of the lag varies as the various parameters are varied. Information on the other coefficients in these cases comes in the next part. Note that the cut-off is always less than unity.\footnote{To see why this is true note that I can express the cut off as $$\bar{b}_{0}= 1-\alpha/b $$ } This reflects the fact that inflation has to be stationary. The following triptych of tables alter parameters one at a time, keeping the others at the standard setting.
\newpage
\begin{center} 
 Table (I): Lag Cut-Off at Different Frisch Elasticities 
 \end{center}
 \begin{center}
\begin{tabular}{|p{1.5cm}|p{1.5cm}|} 
\hline 
 $\eta $ &  $\bar{b}_{0}$   \\
  \hline 
  1 & 0.647   \\
  2 & 0.667  \\
  6 & 0.727  \\
  \hline
\end{tabular}
\end{center}
\begin{center} 
 Table (II): Lag Cut-Off at Different Price Rigidities
 \end{center}
 \begin{center}
\begin{tabular}{|p{1.5cm}|p{1.5cm}|} 
\hline 
 $\alpha$ &  $\bar{b}_{0}$   \\
  \hline 
  0.6 & 0.759   \\
  0.8 & 0.593  \\
  \hline
\end{tabular}
\end{center}
\begin{center} 
 Table (III): Lag Cut-Off at Different Output Responses
 \end{center}
 \begin{center}
\begin{tabular}{|p{1.5cm}|p{1.5cm}|} 
\hline 
 $a_{y}$ &  $\bar{b}_{0}$   \\
  \hline 
  0 &  0.733  \\
  1 & 0.68 \\ 
  1.5 & 0.667 \\ 
  2 & 0.657 \\
  2.5 & 0.65 \\ 
  \hline
\end{tabular}
\end{center}
It is clear that, the cut-off is increasing in $\eta$ but decreasing in $\alpha$ and $a_{y}$.\footnote{This result can be proved with elementary calculus. The only difficult case is $\alpha$ but after some rearrangement the convenient form 
$$\frac{\partial b_{0}} {\partial \alpha} \vert_{a_{\pi}=1}=-\big(\tilde{b}_{0} + \frac{\partial b_{0}} {\partial \alpha} \big)\bigg/b^{2}<0$$}
Therefore a robust calibration for the cut-off lies between $0.563$ and $0.942$.
The lower bound is close to its theoretical value $0.5$ (for $a_{\pi} \rightarrow 1$).\footnote{By way of justification as $a_{\pi} \rightarrow 1$
$$b_{0} \rightarrow \frac{1+ \tilde{\omega}}{1+ \alpha + \tilde{\omega}}=1- \frac{\alpha}{1+ \alpha + \tilde{\omega}}$$
This is clearly increasing in $\tilde{\omega}$ and decreasing in $\alpha$. Sending $\alpha \rightarrow 1$ gives the desired lower bound, as it minimizes $\tilde{\omega}$ and maximizes $\alpha$.} The large range reflects the non-linearity of the coefficients and the wide range of parameters considered. A similar exercise could be performed for the other focal case but I leave this for future work.
\subsubsection{Benchmark Parametization}
In this subsection, I study changes in the Phillips curve slope coefficients when structural parameters are varied, over the ranges discussed in the previous section. I then perform a similar exercise for the existing solution. The difference remains stark. This should be a spur for future rigorous econometric research.
\newpage
\begin{center} Table (IV): Slope Coefficients Varying Frisch Elasticity
\end{center}
\begin{center}
\begin{tabular}{|p{1cm}|p{1cm}|p{1cm}|p{1cm}|p{1cm}|p{1cm}|} 
\hline 
   $\eta$    & $b_{0}$  & $b_{1}$& $b_{2}$  & $b_{3}$ & ${b}_{4}$ \\
\hline 
  1 & 0.588 &  0 & -0.074 & 0.353 & 0.118  \\
  2 & 0.583 & 0 & -0.139 & 0.333 & 0.167  \\
  6 & 0.568 & 0 & -0.341 & 0.271 & 0.318  \\
  \hline
\end{tabular}
\end{center}
\par The lagged inflation coefficient seems to be almost invariant to changes in the supply side, which is encouraging. The lead is also relatively unresponsive. This is encouraging from a Keynesian perspective. I anticipate the model will soon be augmented with additional frictions to better model the propagation of monetary shocks to labor and capital markets. I anticipate the model will soon be augmented with additional frictions to better model the propagation of monetary shocks to labor and capital markets. This will rest upon the robust foundation of our new understanding of inflation determination embodied in these coefficients. 
\par The zero in the third column is a figment of the output neutral parameter choice. When I repeat the exercise at different values of $a_{y}$, the variation is somewhat larger but the magnitude never exceeds $0.15$.\footnote{At $a_{y}=0$ the trio of $\eta$ values yield slopes of $-0.056$, $-0.077$ and $-0.137$, whilst when $a_{y}=1$ they are  $(0.022, \, 0.043, \, 0.077 )'$. It appears easier to generate larger slope magnitudes of negative sign.} The error and particularly the price dispersion coefficient seem to increase sharply with $\eta$. Results for the other two structural parameters are displayed below.
\begin{center} Table (V): Slope Coefficients Varying Price Rigidities
\end{center}
\begin{center}
\begin{tabular}{|p{1cm}|p{1cm}|p{1cm}|p{1cm}|p{1cm}|p{1cm}|} 
\hline 
   $\alpha$    & $b_{0}$  & $b_{1}$& $b_{2}$  & $b_{3}$ & ${b}_{4}$ \\
\hline 
  0.6 & 0.580 &  -0.036 & -0.457 & 0.241 & 0.357  \\
  0.8 & 0.551 & 0.017 & -0.046 & 0.406 & 0.085 \\
  \hline
\end{tabular}
\end{center}
\begin{center} Table (VI): Slope Coefficients at Greater Output Responses
\end{center}
\begin{center}
\begin{tabular}{|p{1cm}|p{1cm}|p{1cm}|p{1cm}|p{1cm}|p{1cm}|}
\hline 
   $a_{y}$    & $b_{0}$  & $b_{1}$& $b_{2}$  & $b_{3}$ & ${b}_{4}$ \\
\hline 
  1 & 0.58 & 0.067 & -0.267 & 0.32 & 0.2 \\
  1.5 & 0.583 & 0.111 & -0.278 & 0.333 & 0.167 \\
  2 & 0.586 & 0.143 & -0.286 &  0.343 & 0.143 \\
  2.5 & 0.588 & 0.167 & -0.292 & 0.35 & 0.125 \\ 
  \hline
\end{tabular}
\end{center}
\par The lag coefficient is unresponsive to either change. The output slope is relatively unresponsive to price rigidity but grows somewhat larger than zero when there is a large output response. The price dispersion gradient explodes, by an order of magnitude, when the economy moves from a rigid to a flexible setting. On the other hand, it barely adjusts to changes in the output gap. This disjuncture reflects the composition of $\tilde{b}_{2}$, with additional non-linearity with respect to $\alpha$ and independence from $a_{y}$, coming about from its origin on the supply rather than demand side. By contrast, the forward term nearly doubles in size, as prices become more rigid, but is almost unresponsive to output response. Intuitively, pricing decisions are driven primarily by microeconomic constraints on price adjustment, reflected in $\alpha$, far more than substitution incentives created by central banks through $a_{y}$. Lastly, the error coefficient is also more responsive to resetting than monetary policy, decreasing by a factor of more than four, as prices become more rigid. The last table collates salient results from the present Phillips curve calibration exercise. 
\begin{center} Table (VII): Slope Coefficients Robustness 
\end{center}
\begin{center}
\begin{tabular}{|p{2cm}|p{1cm}|p{1cm}|p{1cm}|p{1cm}|p{1cm}|} 
\hline 
   Statistic   & $b_{0}$  & $b_{1}$& $b_{2}$  & $b_{3}$ & ${b}_{4}$ \\
\hline 
  Max & 0.588 &  0.167 &  -0.046  & 0.406 & 0.085  \\
  Min & 0.551 & -0.137 & -0.074 & 0.241 & 0.357  \\
  Mean & 0.579 & 0.028 & -0.242 & 0.328 & 0.187  \\
  Std Dev. & 0.011 & 0.088 & 0.125 & 0.045 & 0.087  \\
  Benchmark & 0.575 & 0 & -0.25 & 0.3 & 0.25 \\
  Bench Dev. & 0.041 
  & 0.093 & 0.125 & 0.053 & 0.107 \\
  \hline
\end{tabular}
\end{center}
\par Note that I have used the alternative calibrations from the previous footnote, to reflect the dispersion of the slope for $b_{1}$, as they allow for variation with respect to $\eta$. If I used the actual results from Table (IV) its column would read $(0.167, \, -0.137, \, 0.052, \, 0.069, \, 0, \, 0.086 )'$, giving a similar picture. These standard and benchmark deviations reflect considerable stability in my calibrations. All of which differ considerably from the existing benchmark. 
\par They can be used to construct 
quasi-confidence
intervals as well as quasi-hypothesis tests. For example, the standard calibration for the slope coefficient $(0.833)$ lies $8.928$ standard deviations away from the benchmark calibration or 
$9.148$ with the alternative mean and standard deviation measures. With a normal distribution, these would be sufficient for rejection of equality of the two slopes, at a very high level of certainty. It is also possible to approach robustness and testing by varying parameters simultaneously. For example, $b_{3}$ is strictly increasing in $\alpha$ and $a_{y}$ but strictly decreasing in $\eta$. Thus in our robust calibration, it can be confidently bounded below its value when $\alpha$ and $a_{y}$ are maximized at $0.8$ and $2.5$ respectively, whilst $\eta$ is minimized at $1$. This bound equals $0.441$.
This is a small but significant improvement on $0.5$, the theoretical bound derived in Proposition 25. 
In general, for $b_{0}$, $b_{2}$ and $b_{4}$, it is possible to rule out the existing values on the basis of such \emph{a priori} restrictions.
\par This leaves only $b_{1}$; its incarnation under the old solution $\tilde{\omega}$ (54) is decreasing in $\alpha$ but increasing in $\eta$. Changing just one of these parameters, it stays above our confidence range at $0.25$.\footnote{This occurs when I adjust $\alpha$ to $0.8$, when $\eta =1$ it is higher at $0.333$.} Its slope is minimized overall at $0.1$, which lies towards the upper bound of our confidence range of two standard deviations around the benchmark.\footnote{This conclusion is robust to any of the three deviation measures.} Empirical studies are often able to reject a slope this steep. The estimates in Table 3 of \cite{fuhrer2006intrinsic} allow me to reject this hypothesis, in all cases where real marginal costs are used and two of the four cases where a more noisy output gap measure is employed.  
\par Overall, this part has provided a strong body of suggestive evidence that the new approximation provides a better fit than the existing coefficient by coefficient. Moreover, the stability of the critical slope coefficients, in response to plausible structural parameter uncertainty, is a showcase for robustness in macroeconomics. This exercise has surely been
a useful prelude to future rigorous econometric investigation. 
\section{Additional Evidence}
The scope of this section is to provide additional empirical evidence, extending beyond the parameters of the main model. This section is divided into three parts. The first considers other structural parameters not present under Calvo but used with Rotemberg. The second applies these to the claims made in Proposition 11. The third and final subsection looks at the evidence concerning the volatility of trend inflation, supporting Section 9.5 and Point 2.
\subsection{Other Structural Parameters}
There are two sub-divisions, one for the price adjustment parameter and the other for the substitution parameter. Microeconometric studies are emphasised.
\subsubsection{$c_{p}=50$} 
\par This parameter was previously calibrated using the (overturned) equivalence in Proposition 11. Nevertheless, direct evidence is available. Values for $c_{p}$ are derived from estimates of the cost of price changes as a share of output, according to the formula $$c_{p}=\frac{C^{a}}{Y}\frac{2}{\pi^{2}}$$ 
At the upper end are estimates of four percent, derived from calculations by \cite{willis2006magazine}, using magazine price data, originally analysed by \cite{cecchetti1986frequency}. \cite{slade1998optimal} comes to a similar estimate for salted crackers.\footnote{\cite{levy2004real} documents extreme cases of price rigidity including a period of 73 years when the price of a nickel coke remained unchanged. Although, this is likely specific to the price point and its role in marketing.} The most appealing estimate comes from \cite{zbaracki2004managerial} who come to a figure of $1.22\%$. It considers a broader range of costs associated with the price setting process, including information gathering and internal communication costs, as well as customer communication and negotiation, whereas other studies focus mainly on physical menu costs, which are found to constitute a small percentage of the total. It even finds evidence of a (small) portion of  convex adjustment costs, consistent with the Rotemberg framework. 
\par The principle limitation is coverage. There is only one firm, observed for one year 1997-1998; in fact there is just one set of price changes.  This obviously raises concerns with generalization. The emphasis in the paper on negotiations with large customers, which is less common in services, and suggests $1.22\%$ could be an overestimate for the whole economy. 
\par At the lower end, estimates can be as low as $0.5\%$ in some retail contexts (\cite{levy1997magnitude}, \cite{levy1998price}, \cite{dutta1999menu} and \cite{bergen2008little}). These are likely to be biased downwards, as a reflection of the overall cost of price changing, since they include the impact of less costly sales price changes.\footnote{In fact, it is common for prices to revert back to their previous value- consistent with very low adjustment costs (see \cite{anderson2017informational}).} The frequency of sales varies across countries, as documented for example in \cite{berardi2015more}. They are much more common in the United States and Great Britain than France. This is likely the result of regulatory restrictions in France (see \cite{freeman2008supply}). This suggests that firms' adjustment costs might vary between countries.
\par This discussion underscores the importance of incorporating heterogeniety into macroeconomics and using microeconomic data to discipline and test our models. This should remain a research priority. Although, as I explain earlier Calvo should prove a better model to build from than Rotemberg. For quantitative analysis, in keeping with the previous arguments, I will shade the \cite{zbaracki2004managerial} number by setting $$c_{p}=50$$ and $$\pi = 2\%$$ I will contain a cost of nominal rigidity of $1\%$.\footnote{There are issues with determining the "right" inflation rate when your sample contains only one observation. The difficulty is that 1997-1998 was not a period of stable inflation, due probably to the onset of the Asian Financial crisis and an associated collapse in oil prices. In 1998 the rate of growth of US GDP deflator fell back to $1.1\%$ from $1.6\%$ the year before. The movement in CPI was more muted falling from $1.7\%$ to $1.6\%$. This figures are insufficient reason to deviate from the benchmark of 
$2\%$ trend inflation.  Data comes from https://fred.stlouisfed.org/series/GDPDEF and https://fred.stlouisfed.org/series/CPIAUCSL respectively.} 
\subsubsection{$\theta =6$}
There are two main sources of estimates, demand elasticities and markups. For the central estimate, 
I lean closer to the former, whilst using the latter for robustness. 
A meta-analysis by \cite{bajzik2020estimating} propounds a central estimate of $4$, with $6$ the highest plausible level. However, there is an element of extrapolation here, since this model is closed economy and exporters might be systematically different from non-exporters due to selection effects. \cite{de2012markups} suggest this. The authors' preferred estimate $3.8$ derives from annual data.
This is systematically lower than with quarterly observations, the relevant business cycle frequency.\footnote{Nevertheless, they are able to rule out $\theta < 2.5$, which justifies the restrictions in Theorem 2.} 
\par The most direct evidence comes from \cite{rosenthalkay2021seven} who estimate seven million demand elasticities, from a widely used marketing database. They come up with an average estimate a little below $3$. Although their prior mitigates against negative estimates, they still find that around a quarter of firms have elasticities less than unity. Also around $40\%$ are less than $2$. %two.% 
\par These results do not fit the basic theory. Recall that monopolists should always price on the inelastic portion of their demand curve and models based around monopolistic competition blow up when $\theta < 1$. Likely culprits include uniform pricing, where firms set the same price in all markets, as documented by \cite{cavallo2014currency}, \cite{cavallo2017online} and \cite{dellavigna2019uniform}, or fair pricing based on the idea that customers punish firms seen to be unfairly profiting from for example external events (see \cite{rotemberg2011fair}, \cite{cavallo2014prices} and \cite{gagnon2020small}). 
\par Naturally, these extensions are too complicated for a stylized model. I advocate an upward adjustment of the mean, to remove the effect of these low values. In the present informal environment, it is surely adequate to select $4$ as a nearby candidate calibration.
\par For markup estimates the most significant challenge is deciding which costs are fixed. Accounting data distinguishes two main types, Cost of Goods Sold (COGS) and Selling General and Administrative (SG \& A). \cite{de2020rise} assign COGS to variable costs and (SG \& A) to fixed. In the United States this measure increases rapidly from $10\%$ in $1980$ to $60\%$ in $2016$. They find similar patterns 
in other parts of the world in \cite{de2018global}. 
\par Nevertheless, very large increases in profit share do not seem to be consistent with other macroeconomic developments, according to \cite{basu2019price}, \cite{autor2020fall} and \cite{karabarbounis2019accounting}. Indeed, when (SG $\&$ A) are thought of as variable, markups are much smaller between $5$ and $20\%$, typically around $10-15\%$ with much steadier changes (see \cite{hall2018using}, \cite{traina2018aggregate} and \cite{kirov2021measuring}).\footnote{They are also more aligned with evidence, like \cite{de2012markups}, that use price rather than just revenue data, a practice criticised by \cite{bond2021some}.} Moreover, there are concerns that efficiency enhancing processes might be driving markup increases in some sectors or subsets of firms (see \cite{crouzet2019understanding}, \cite{autor2020fall} and \cite{rossi2021diverging}). These forces lie outside the scope of the model. \cite{syverson2019macroeconomics} provides extensive discussion of these and other issues.
Overall, it appears reasonable to consider estimates with $\theta$ equal to 
$4$ and $6$, as well as some with $8$. The lower value seems to better represent the elasticity of substitution. $6$ better represents markups, where it is consistent with a steady state value of $20\%$; $8$ is a robustness check, consistent with the lower values reported when the whole of (SG \& A) is regarded as fixed. 
\subsection{Lucas Critique}
Proposition 11 and the parametizations in Appendix H.1.2 suggest that the observational equivalence part of the Lucas critique applies to the Rotemberg model (at ZINSS) only if the microeconomic evidence supports $ \tilde{\omega} > 1$. At the headline parametization $$\tilde{\omega} =1/2$$
which is too low. At the upper end of the values for $\theta$ this rises to $0.7$. It is close but not quite met if the upper bound for $\eta$ is introduced, when it reaches $0.98$. Given how imprecise the calibration exercise is, this is not a decisive rejection. 
\par Recall this is not a sufficient condition; in fact it is only necessary down the limit where $V_{z}/V_{m} \rightarrow 0$. This parameter is difficult to pin down because the model focuses on shocks that firms do not know, in advance of making price decisions. This contrasts with evidence of firms spending resources accumulating information before taking pricing decisions. Firms may be aware of the realization of many of the shocks in the data. There is a strong body of evidence indicating that firms are relatively inattentive to macroeconomic developments, particularly in stable macroeconomic times. This would favour a relatively low value of $V_{z}/V_{m}$.\footnote{Supportive survey evidence is discussed in \cite{coibion2018formation}, \cite{andrade2022no}, \cite{born2023firm}, \cite{weber2022subjective}, 
and \cite{candia2023perceived}. Consult \cite{sheffrin1996rational} for more extensive discussion of bringing the Lucas model to the data. \cite{ball2012short}, \cite{ball2014long}, \cite{lucas2015stab} and \cite{benati2021international} are recent estimates of money demand systems. \cite{alexopoulos2011read} and \cite{garin2019supply} derive estimates of the volatility of technology. Production function errors, discussed in the previous subsubsection, would constitute another source.}
\par Lastly, the parametization could be effected by the addition of real frictions. These suppress the relationship between real output and marginal costs (see \cite{gopinath2011search}), which would show up as
\begin{equation} \hat{mc}_{t}=\frac{1}{1 +R}(\sigma + \eta) \hat{y}^{e}_{t}\end{equation} where $R>0$ measures the size of the real friction. These have been popular because the slope of the Rotemberg Phillips curve appears too high, from a macroeconometric standpoint. The difficulty is that microeconometric evidence favours smaller values, which are too small to overturn these conclusions (see
\cite{beck2020price}). This is probably as far as one should take this exercise. 
\subsection{Trend Inflation Volatility}
This part presents evidence on the volatility of trend inflation. The focus here is on time series ideas. It matters how you de-trend the data. \par There is significant low frequency volatility in inflation, over typical sample periods. Contrast the Great Depression with the Great Inflation and think of Japan's lost decades. Under inflation targeting it is natural to think of the headline target as the trend inflation rate. This makes trend inflation much less volatile. Nevertheless, there has been discussion about changing the inflation target which should be factored in by rational agents, (see for example \cite{blanchard2010rethinking}, \cite{ball2013case} and \cite{cecchetti2017case}). The European Central Bank's recent move to a symmetric inflation target might work out like this (\cite{angeloni2021new}). 
\par The asymptotic justification is a little delicate though. As shocks shrink, are trend inflation shocks sufficiently large to be considered first order? Typically, trend inflation is not observed but estimated via a statistical decomposition. It is difficult to get away from the influence of tuning parameters set \emph{a priori}. In the popular \cite{hodrick1997postwar} framework, the quadratic loss function weights deviations in trend $\lambda = 1600$ times more than variation in the cycle for quarterly data (\cite{ravn2002adjusting}). This corresponds to a standard deviation ratio of 40, which can be viewed as a Bayesian prior. Take the standard deviation of quarterly output to be $0.5\%$. This is plausible for advanced economies in normal times. It is roughly consistent with a variety of evidence presented by \cite{cogley2002evolving}, \cite{stock2002has}, \cite{stock2005understanding},  \cite{keating2012greater} and \cite{hulseman2017there}. 
\par Throughout the subsequent discussion I use asymptotic notation $x \asymp y$ to mean $x$ and $y$ are of the same order of magnitude. I take this to mean that one variable is on average no more than a tenth of the other.\footnote{Asymptotic notation has a strict mathematical definition. It denotes that the models the two variables are bi-Lipschitz so 
$$x \asymp y \iff \exists C \in \mathbb{R}: \frac{1}{C}\vert y \vert < x < C\vert y \vert $$ I use it figuratively here.} 
\par The requirement for the trend inflation shock $\varepsilon^{\bar{\pi}}$ to be first order is that 
$$\varepsilon^{2}<< \vert \varepsilon^{\bar{\pi}} \vert$$ where $\varepsilon$ here means all the other shocks in the model. With these calibrations in place we have $\hat{\varepsilon}^{2}=2.5 \times 10^{-5}$ and $\vert \hat{\varepsilon}^{\bar{\pi}}\vert =1.25 \times 10^{-4}$ so $\vert \hat{\varepsilon}^{\bar{\pi}}\vert =5\hat{\varepsilon}^{2}$. Thus we can conclude that although
$$\varepsilon^{2}< \vert \varepsilon^{\bar{\pi}} \vert$$
$$\varepsilon^{2} \asymp \vert \varepsilon^{\bar{\pi}} \vert$$
Hence, trend inflation volatility appears second order.
\par The results are very different with the method of \cite{hamilton2018you}. He has no consumer prices in his data set. The nearest variable he has is the GDP deflator, which includes producer prices. In his Table 1 he estimates $\lambda =0.216$, implying a standard deviation ratio of $0.465$, which makes the trend more volatile than the cycle $\vert \hat{\varepsilon}^{\bar{\pi}}\vert \approx 2.15\vert \hat{\varepsilon}\vert $. This implies 
$$ \vert \varepsilon \vert \asymp \vert \varepsilon^{\bar{\pi}} \vert$$
\par It seems an unusual claim that trend inflation is more volatile than the de-trended component. Indeed, this is an extreme conclusion amongst the time series considered. The highest value of $\lambda$ he finds for his 13 series is $9.596$ for the real rate. Round this up to $10$. This implies $\vert \hat{\varepsilon}^{\bar{\pi}}\vert \approx 0.32 \hat{\varepsilon}^{2}$ so it is fair to conclude trend inflation volatility is first order, as this estimate supports
$$ \vert \varepsilon \vert \asymp \vert \varepsilon^{\bar{\pi}} \vert$$
\par Hamilton's technique is controversial. The literature has long been aware of problems with the standard filter. It generates spurious cycles, uses future information to construct present trend components and does not account for events that might be best considered structural breaks; for more on this view consult \cite{harvey1993detrending}, \cite{king1993low}, \cite{cogley1995effects}, \cite{canova1998detrending}, \cite{baxter1999measuring} and \cite{christiano2003band}. Nevertheless, recent work by \cite{hodrick2020exploration} shows that his filter outperforms others, when there are slow-moving changes in persistent components- which might be a plausible consequence of, for example, long-run structural change.\footnote{Real-time estimates from more traditional methods, that attempt to distinguish between supply and demand shocks, favor even smoother trends (see \cite{blanchard1989dynamic} and \cite{coibion2018cyclical}). Although plausible, these will have to be re-examined, in the light of the emergence of demand shocks in the new Phillips curve.}
\par This calibration exercise and surrounding discussion have justified the claim that there is no decisive empirical answer to whet
her shocks to trend inflation are first order or not. This should be an important field for future work. We are in a far better place than before as a discipline, if we are discussing the econometric implementation of our main model, rather than debating its underlying contents. 
\end{document}